%% file: ys-202605-010-no-go-theorem-for-bec-nelson-pf.tex
\documentclass[11pt,a4paper]{article}

\usepackage{fontspec}
\usepackage[titletoc,title,header]{appendix}

\usepackage{amsmath}
\usepackage{amssymb}
\usepackage{amsthm}
\usepackage{mathtools}
\usepackage{xparse}    % for \NewDocumentCommand

\usepackage[margin=1in]{geometry}

\usepackage{graphicx}
\usepackage{xcolor}

\usepackage{hyperref}
\hypersetup{
  colorlinks=true,
  linkcolor=blue,
  citecolor=blue,
  urlcolor=blue,
  pdfauthor={Yoshitsugu Sekine},
  pdftitle={No-Go Theorem for BEC in the Nelson and Pauli--Fierz Models}
}

\usepackage[]{natbib}
\theoremstyle{plain}
\newtheorem{thm}{Theorem}[section]
\newtheorem{lem}[thm]{Lemma}
\newtheorem{prop}[thm]{Proposition}

\theoremstyle{definition}
\newtheorem{defn}[thm]{Definition}

\theoremstyle{remark}
\newtheorem{rem}[thm]{Remark}

\input{mycommands.tex}

\title{No-Go Theorem for BEC in the Nelson and Pauli--Fierz Models}

\author{%
Yoshitsugu Sekine\\{\small\texttt{4429sekine@gmail.com}}%
}

\date{\today}

\begin{document}

\maketitle

\begin{abstract}
We construct KMS states for the Nelson model, the spinless Pauli--Fierz model, and the Pauli--Fierz model with spin by functional integral representations, and study point-source models after removal of the infrared and ultraviolet cutoffs. If the physical test-function space can distinguish the zero mode, the absence of off-diagonal long-range order, the vanishing of the zero-mode form, the vanishing of the condensate density, the order-parameter criterion, and the triviality of the BEC directions and the BEC ideal are equivalent. We also describe the infrared quotient and the BEC ideal in the resolvent algebra uniformly for the three models, and formulate an operator-algebraic sufficient condition for separately given spatially translation-invariant KMS states.

\noindent\textbf{Keywords:} Nelson model, Pauli--Fierz model, functional integral, equilibrium state, KMS state, Bose--Einstein condensation
\end{abstract}

\setcounter{tocdepth}{3}
\tableofcontents

\section{Introduction}\label{introduction}

We extend the no-go theorem for Bose--Einstein condensation (BEC) of quasiparticles obtained for the van Hove model and the spin-boson model \cite{YoshitsuguSekine006,YoshitsuguSekine008} to the Nelson model, the spinless Pauli--Fierz model, and the Pauli--Fierz model with spin. In the Nelson model, a particle and a scalar field are linearly coupled. In the spinless Pauli--Fierz model, we discuss the Hamiltonian with minimal coupling appearing in non-relativistic QED \cite[Section 3.2]{LorincziHiroshimaBetz3}. In the model with spin, the coupling between the magnetic flux density \(\physmfdensity = \varot\opfocksegalradiation\) and the Pauli matrices is further added.

In a free Bose gas, BEC zero-mode covariance appears from the chemical-potential limit of a bounded system. Here we introduce the interaction between particles and fields, remove the infrared and ultraviolet cutoffs, and discuss whether the zero-mode covariance still remains as an independent condensation direction on the physical test-function space. In the process of taking the infinite-volume limit from bounded systems, we divide the covariance of the free field into the non-zero-mode component and the BEC zero-mode component, and decompose the two-point functions into the Gaussian component of the free field and the interaction factor depending on particle paths and spin paths. This decomposition is the basic mechanism for deciding what remains and what disappears in the long-distance limit.

Since point-source interactions have singularities at both infrared and ultraviolet ends, we start with a family of source functions \(\varrho_{\kappa,\Lambda}\) equipped with infrared and ultraviolet cutoffs, and finally take the limit \(\kappa\downarrow0,\Lambda\uparrow\infty\). In the Nelson model, we construct the renormalized kernel of the point-source scalar interaction, and in the Pauli--Fierz model, we handle the cutoff removal for the transverse current source and the external field cross term. In the model with spin, we further add a Pauli term on the spin path, and after confirming the integrability of the time-ordering exponent and the magnetic flux density factor, we construct the local kernel and KMS state after cutoff removal.

The construction of KMS states is based on the functional integral representation in \cite[Chapter 21]{DerezinskiGerard001}. We combine the particle loop measure on the \(\sminvtemperature\)-periodic path space, the spin path measure in the model with spin, and the centered Gaussian random variable of the free field, and define expectations with the Euclidean weight including interactions. These finite-time correlation functions give KMS states by the reconstruction theorem. The assertion proved directly in this paper is the computation of the long-distance limit of two-point functions within the functional integral representation.

The main result is as follows. When the two-point off-diagonal long-range order is computed in the KMS state after cutoff removal, the cross terms involving the non-zero-mode covariance and the interaction factor vanish in the long-distance limit, and only the BEC zero-mode form remains and is detected by off-diagonal long-range order. If the physical test-function space can distinguish the zero mode, the absence of off-diagonal long-range order, the vanishing of the zero-mode form, the vanishing of the condensate density, the order-parameter criterion, and the triviality of the corresponding BEC directions and BEC ideals are equivalent. Finally, when a spatially translation-invariant KMS state is separately given, we organize the order-parameter net of the resolvent algebra and describe how the \(\lp^{1}\) asymptotic abelianness, weak clustering, and primary-state assumptions appearing in an Araki--Haag--Kastler--Takesaki type theorem \cite{ArakiHaagKastlerTakesaki1} serve as sufficient conditions implying the absence of off-diagonal long-range order. Since the state constructed from the particle loop measure with a confining potential is in general not spatially translation invariant, this operator-algebraic sufficient condition is only a supplementary discussion from the viewpoint of stability. Because translation invariance of the time evolution is required, this may provide one perspective when the particle system is extended to a suitable many-body system.

\section{Main Results}\label{expedition0012133}

The models discussed here are the Nelson model, the spinless Pauli--Fierz model, and the Pauli--Fierz model with spin. For each, the settings for both particles and fields are different. Here, the description will be organized mainly around the Nelson model, while the Pauli--Fierz models will be discussed in detail in their respective sections.

\subsection{Space Setting and Basic Operators}\label{space-setting-and-basic-operators}

To discuss BEC in typical situations, we assume the spatial dimension is \(d \geq 3\) in principle. We will always state explicitly when the dimension is restricted to \(d = 3\) for brevity.

The particle Hilbert space is \(\sphilb{H}_{\txtparticle} = \fun{\lp^2}{\fldreal^{d}}\) for the Nelson model and the spinless Pauli--Fierz model, and \(\sphilb{H}_{\txtparticle} = \fun{\lp^{2}}{\fldreal^{d};\ringratint_{2}}\) for the Pauli--Fierz model with spin. The basic form of the particle Hamiltonian is \(\physham[h]_{\txtparticle} = -\onehalf \laplacian + V\). The first term \(\laplacian\) is the Laplacian, and the second term \(V\) is the potential. To define KMS states, we assume the potential is confining, of Kato class, bounded from below, and real-valued. In particular, we assume the validity of the loop measure representation, the heat kernel representation, and the Markov decomposition of the Brownian bridge measure.

Let the basic complex Hilbert space for bosons and its real subspace be \[\sphilb{H}_{\txtbsn} = \fun{\lp^{2}}{\fldreal^{d}}, \quad \sphilb{H}_{\txtbsn,\txtreal} = \fun{\lp_{\txtreal}^{2}}{\fldreal^{d}} = \fun{\lp^{2}}{\fldreal^{d};\fldreal}\] The symplectic form is \(\sigma(f,g) = \opimag \bkt{f}{g}_{\sphilb{H}_{\txtbsn}}\). We choose the realification of the complex Hilbert space \(\sphilb{H}_{\txtbsn}\) as \(X\) for the real symplectic space \((X,\sigma)\), and the inner product of this real Hilbert space is defined by \(\opreal \bkt{f}{g}\).

Let the one-particle Hamiltonian (dispersion relation) defined in momentum space \(\greuctr{k}{d}\) be \(\omega(k) = \abs{k}\), and let the non-negative self-adjoint operator defined for a non-positive chemical potential \(\smchemicalpotential \leq 0\) be \(K_{\sminvtemperature,\smchemicalpotential} = \coth \frac{\sminvtemperature \rbk{\omega - \smchemicalpotential}}{2}\). For the associated non-degenerate non-negative symmetric quasi-bilinear form \(\opform{q}_{\txtbsn,\txtnonzero,\sminvtemperature,\smchemicalpotential}\), let the associated inner product space and its completion be \[\sphilb{D}_{\txtbsn,\sminvtemperature,\smchemicalpotential} = \pairbk{\fun{\opformdomain}{\opform{q}_{\txtbsn,\txtnonzero,\sminvtemperature,\smchemicalpotential}},\opform{q}_{\txtbsn,\txtnonzero,\sminvtemperature,\smchemicalpotential}}, \quad \sphilb{H}_{\txtbsn,\sminvtemperature,\smchemicalpotential} = \gtclos{\sphilb{D}_{\txtbsn,\sminvtemperature,\smchemicalpotential}}^{\opform{q}_{\txtbsn,\txtnonzero,\sminvtemperature,\smchemicalpotential}}\]

In general, the bosonic Fock space over a complex Hilbert space \(\sphilb{H}_{\txtbsn}\) is defined by \(\spfock_{\txtbsn} = \fun{\spfock_{\txtbsn}}{\sphilb{H}_{\txtbsn}} = \bigoplus_{n=0}^{\infty} \bigotimes_{\txtsym}^{n} \sphilb{H}_{\txtbsn}\). This is the bosonic Fock space used particularly in the discussion of the Nelson model; the radiation field used in the Pauli--Fierz model will be defined later. For any \(f \in \sphilb{H}_{\txtbsn}\), let \(\opfockcran_{\txtfock}(f)\) be the creation and annihilation operators on the bosonic Fock space. The Segal field operator is \(\opfocksegal_{\txtfock}(f) = \frac{1}{\sqrt{2}} \rbk{\opfockcr_{\txtfock}(f) + \opfockan_{\txtfock}(f)}\), and the Weyl operator is defined by \(\opfockweyl_{\txtfock}(f) = \napiernum^{\imunit \opfocksegal_{\txtfock}(f)}\). For a non-expansive operator \(V\), the second quantized operator of the second kind \(\fun{\opfocksndqnt_{\txtbsn}}{V}\) is defined by \(\oplus_{n=0}^{\infty} \otimes_{\txtsym}^{n} V\). The second quantized operator of the first kind \(\fun{\opfocksndqntdiff_{\txtbsn}}{\omega}\) is defined as the derivative at the origin of \(\fun{\opfocksndqnt_{\txtbsn}}{\napiernum^{\imunit t \omega}}\) for the unitary operator \(U_t = \napiernum^{\imunit t \omega}\) on the one-particle space. In particular, \(\physham_{\txtbsn,\txtfr} = - \imunit \fnrestr{\opod{t} \fun{\opfocksndqnt_{\txtbsn}}{\napiernum^{\imunit t \omega}}} {t = 0}\) is called the free Hamiltonian of bosons. Furthermore, the free Hamiltonian with chemical potential \(\smchemicalpotential\) is defined by \(\physham_{\txtbsn,\txtfr,\smchemicalpotential} = \fun{\opfocksndqntdiff_{\txtbsn}}{\omega - \smchemicalpotential}\).

The total Hilbert space \(\sphilb{H}_{\txtnelson}\) describing the interacting particle-boson field system is defined by the tensor product \(\sphilb{H}_{\txtnelson} = \sphilb{H}_{\txtparticle} \otimes \spfock_{\txtbsn}\). We use the other isomorphisms, including the direct integral with constant fiber, without explicit mention: \[\sphilb{H}_{\txtnelson} = \sphilb{H}_{\txtparticle} \otimes \spfock_{\txtbsn} = \fun{\lp^{2}}{\fldreal^{d};\spfock_{\txtbsn}} = \int_{\fldreal^{d}} \spfock_{\txtbsn} \opdmsr{x}.\]

\subsection{Point Sources and Cutoff Sources}\label{point-sources-and-cutoff-sources}

Both the Nelson model and the Pauli--Fierz model describe particles and fields coupled via a source function. We introduce the basic formulation concerning this source. Since we will ultimately take the point source limit, we first consider a point source \(\varrho = \diracdelta_{0}\) weighted at the origin, and define a source with infrared and ultraviolet cutoffs by \[\faftr{\varrho}_{\kappa,\Lambda}(k) = \faftr{\varrho}(k) \fndef{\kappa \leq \abs{k} \leq \Lambda}(k), \quad \faftr{\varrho}(k)=1.\] In particular, we sometimes denote the source with only the infrared cutoff removed as \(\varrho_{\Lambda}\), and the source with only the ultraviolet cutoff removed as \(\varrho_{\kappa}\).

For any \(\kappa>0\) and \(\Lambda < \infty\), the cutoff source satisfies \(\int_{\fldreal^{d}} \frac{\abs{\faftr{\varrho}_{\kappa,\Lambda}(k)}^2}{\omega(k)^{3}} \opdmsr{k} < \infty\), and the limit \(\lim_{\kappa \to 0,\Lambda \to \infty} \varrho_{\kappa,\Lambda} \to \varrho\) holds as a tempered distribution. For both the Nelson model and the Pauli--Fierz model, we first define all quantities in the cutoff model and then take the point source limit.

Since the interaction appears in all equations through the source \(\varrho_{\kappa,\Lambda}\) in both the Nelson model and the Pauli--Fierz model, the question of which quantities remain finite in the limit after cutoff removal is reduced to the behavior of the linear functionals associated with the source. In particular, focusing on the fact that the Fourier transform is a function, we define \(\faftr{\mathsf{m}} = \faftr{\varrho} / \omega^{\frac{3}{2}}\) and \(\faftr{\mathsf{m}_{\kappa,\Lambda}} = \faftr{\varrho}_{\kappa,\Lambda} / \omega^{\frac{3}{2}}\), and also use the same symbols for linear functionals on \(\sphilb{H}_{\txtbsn}\). That is, \begin{equation}\label{expedition0012350}
\begin{aligned}
\mathsf{m}(f)
=
\int_{\fldreal^{d}}
\frac{\faftr{f}(k)}{\omega(k)^{3/2}}
\opdmsr{k},
\quad
\mathsf{m}_{\kappa,\Lambda}(f)
=
\int_{\fldreal^{d}}
\frac{\overline{\faftr{\varrho}_{\kappa,\Lambda}(k)}\faftr{f}(k)}
{\omega(k)^{3/2}}
\opdmsr{k}
\end{aligned}
\end{equation} These appear as fundamental quantities that provide correction terms for physical field operators. In particular, for the dispersion relation \(\omega(k) =\abs{k}\), \(\mathsf{m}(f)\) in the point source limit requires the integrability of \(\omega(k)^{-\frac{3}{2}} \faftr{f}(k)\) on the ultraviolet side. Therefore, to discuss up to and including the point source limit, it is natural to keep \(\dom \mathsf{m}\) in mind from the outset, and construct the field operators and one-point functions defined in the cutoff model in a way that sends them to the limit on this domain.

\subsection{Resolvent Algebra}\label{expedition0012083}

Following \cite{DetlevBuchholz001}, we introduce the definition and basic properties of the resolvent algebra. Let \((X,\sigma)\) be a symplectic space. Let \(\oaresolventalgebra_0\) be the universal unital \(\ast\)-algebra generated by the set \(\set{\oaresolvent(\lambda,f)}
{\lambda \in \fldmultiplicativegroup{\fldreal}, f \in \sphilb{H}_{\txtbsn}}\), and assume that it satisfies the following resolvent relations: \begin{align}
\oaresolvent(\lambda,0)
&=
-\frac{\imunit}{\lambda} \idone, \\ % (1)
\faadj{\oaresolvent(\lambda,f)}
&=
\oaresolvent(-\lambda,f), \\ % (2)
\nu \oaresolvent(\nu \lambda, \nu f)
&=
\oaresolvent(\lambda, f), \\ % (3)
\oaresolvent(\lambda,f) - \oaresolvent(\mu,f)
&=
\imunit
(\mu - \lambda)
\oaresolvent(\lambda,f) \cdot \oaresolvent(\mu,f) \\ % (4)
&=
\imunit
(\mu - \lambda)
\oaresolvent(\mu,f) \cdot \oaresolvent(\lambda,f), \\ % (4)
\commutator{\oaresolvent(\lambda,f)}{\oaresolvent(\mu,g)}
&=
\imunit
\sigma(f,g)
\oaresolvent(\lambda,f)
\oaresolvent(\mu,g)^2
\oaresolvent(\lambda,f), \label{expedition0012052} \\ % (5)
\oaresolvent(\lambda,f)
\oaresolvent(\mu,g)
&=
\oaresolvent(\lambda+\mu, f+g)
\cdot
\rbkleft{\oaresolvent(\lambda,f)} \\
&\quad\rbkright{+
\oaresolvent(\mu,g)
+\imunit \sigma(f,g) \oaresolvent(\lambda,f)^2 \oaresolvent(\mu,g)} % (6)
\end{align} In particular, by condition \eqref{expedition0012052}, \(\oaresolvent(\lambda,f)\) and \(\oaresolvent(\mu,f)\) commute when \(f\) is the same.

The \(\ast\)-algebra obtained by introducing an appropriate norm into \(\oaresolventalgebra_0\) and completing it is called the abstract resolvent algebra, or simply the resolvent algebra. For details on the norm, refer to Definition \cite[P.2730, Definition 3.4]{BuchholzGrundling2}. In particular, by Theorem \cite[P.2730, Theorem 3.6 (iii)]{BuchholzGrundling2}, \(\norm{\oaresolvent(\lambda,f)}
= \frac{1}{\abs{\lambda}}\) holds.

As a dense subalgebra, we choose the \(\ast\)-subalgebra generated by finite products of the generators \(\oaresolvent(\lambda,f)\) of \(\oaresolventalgebra(\sphilb{H}_{\txtbsn},\sigma)\), and denote it specifically as: \[\oaresolventalgebra_{\txtfin}
=
\oastaralgebra
\set{\prod^{\txtfin} \oaresolvent(z_j,f_j)}{z_j \in \fldcmp \setminus \fldreal, f_j \in X}\] Furthermore, the \(\ast\)-subalgebra when \(\sphilb{H}_{\txtbsn}\) is restricted to an arbitrary subspace \(\sphilb{D}_{\txtbsn}\) is specifically denoted by \(\oaresolventalgebra_{\txtfin}(\sphilb{D}_{\txtbsn},\sigma)\). In discussions of Bose-Einstein condensation, the first variable can become long, making it difficult to distinguish from the second variable. Therefore, depending on the situation, a semicolon may be used as a variable separator, writing it as \(\oaresolvent(\lambda;f)\).

As is well-known for ordinary resolvents, the resolvent is analytic with respect to the first variable, and a similar property holds for general resolvent algebras. Utilizing this, the relations are obtained by extending \(\lambda
\in \fldreal\) of the resolvent algebra to a complex variable \(z
\in \fldcmp \setminus \imunit \fldreal\): \begin{align}
\oaresolvent(z,0)
&=
-\frac{\imunit}{z} \idone, \\ % (8)
\faadj{\oaresolvent(z,f)}
&=
\oaresolvent(-\cmpconj{z},f), \\ % (9)
\nu \oaresolvent(\nu z, \nu f)
&=
\oaresolvent(z,f),
\quad
\nu \in \fldmultiplicativegroup{\fldreal}, \\ % (10)
\oaresolvent(z,f) - \oaresolvent(w,f)
&=
\imunit
(w - z)
\oaresolvent(z,f) \cdot \oaresolvent(w,f) \\ % (11)
&=
\imunit
(w - z)
\oaresolvent(w,f) \cdot \oaresolvent(z,f), \\ % (11)
\commutator{\oaresolvent(z,f)}{\oaresolvent(w,g)}
&=
\imunit
\sigma(f,g)
\oaresolvent(z,f)
\oaresolvent(w,g)^2
\oaresolvent(z,f), \\ % (12)
\oaresolvent(z,f)
\oaresolvent(w,g)
&=
\oaresolvent(z+w, f+g)
\cdot
\rbkleft{\oaresolvent(z,f)} \\
&\quad\rbkright{+
\oaresolvent(w,g)
+\imunit \sigma(f,g) \oaresolvent(z,f)^2 \oaresolvent(w,g)} % (13)
\end{align} These are also called resolvent relations.

Let \(\oaresolventalgebra(X,\sigma)\) be the resolvent algebra, and let \(S\) be a subset of the symplectic space \(X\). When a representation \(\oarepn
\in \Rep(\oaresolventalgebra(X,\sigma), \sphilb{H}_{\txtbsn,\oarepn})\) satisfies \(\Ker \oarepn(\oaresolvent(1,f))
= \setone{0}\) for any \(f
\in S\), this representation is called a regular representation on \(S\). When the GNS representation of a state \(\oastate\) on the resolvent algebra is a regular representation on \(X\), this state is called a regular state.

\begin{prop}[\cite{BuchholzGrundling2}]\label{expedition0011838}
For a symplectic space $\pairbk{X,\sigma}$ of arbitrary dimension, let $S
\subset X$ be a non-degenerate finite-dimensional space.
\begin{enumerate}
\item
The norms of the full resolvent algebra $\oaresolventalgebra(X,\sigma)$ and the subalgebra $\oaresolventalgebra(S,\sigma)$ agree on the $\ast$-subalgebra $$\oastaralgebra
\set{\oaresolvent(\lambda,f)}
{f \in S, \lambda \in \fldreal \setminus \setone{0}}$$
In particular, $\oaresolventalgebra(S,\sigma)
\subset \oaresolventalgebra(X,\sigma)$ holds.

\item
The full resolvent algebra is the inductive limit of the net $\fml{\oaresolventalgebra(S,\sigma)}
{S \subset X}$ for non-degenerate finite-dimensional spaces $S
\subset X$.

\item
Any regular representation of the full resolvent algebra $\oaresolventalgebra(X,\sigma)$ is faithful.
\end{enumerate}
In particular, the center of the full resolvent algebra is trivial.
\end{prop}

\subsection{Settings for the Bose Field at Finite Temperature}\label{expedition0011278}

Here, we again use the settings adopted in reference \cite{YoshitsuguSekine004}. For inverse temperature \(\sminvtemperature
> 0\), the time interval corresponding to the periodicity of the KMS state is \[S_{\sminvtemperature}
=
\closedinterval{-\frac{\sminvtemperature}{2}}{\frac{\sminvtemperature}{2}}.\]

We consider the setup for BEC in an infinite system, inheriting the notation and settings from reference \cite{AsaoArai28,YoshitsuguSekine004,YoshitsuguSekine006,YoshitsuguSekine007,YoshitsuguSekine008}. Let the density of the condensate at inverse temperature \(\sminvtemperature
> 0\) be \(\smnumberdensity_{\txtbsn,0}(\sminvtemperature)\). We define the non-closed, non-negative symmetric bilinear form corresponding to the condensed component by \begin{equation}\label{expedition0012440}
\begin{aligned}
\opform{q}_{\txtbsn,0,\sminvtemperature}(f)
=
2 (2 \pi)^d \smnumberdensity_{\txtbsn,0}(\sminvtemperature)
\abs{\faftr{f}(0)}^{2},
\quad
\opformdomain(\opform{q}_{\txtbsn,0,\sminvtemperature})
=
\fun{\lp^{1}}{\fldreal^{d}}
\cap
\fun{\lp^{2}}{\fldreal^{d}}
\end{aligned}
\end{equation} and define the subspace \(\sphilb{D}_{\txtbsn,0,\sminvtemperature}
=
\opformdomain(\opform{q}_{\txtbsn,0,\sminvtemperature})
\cap
\sphilb{H}_{\txtbsn,\sminvtemperature}\), and the physical one-particle space as \[\sphilb{D}_{\txtbsn,\txtphys,\sminvtemperature}
=
\dom \mathsf{m}
\cap
\sphilb{D}_{\txtbsn,0,\sminvtemperature}
=
\dom \mathsf{m}
\cap
\opformdomain(\opform{q}_{\txtbsn,0,\sminvtemperature})
\cap
\sphilb{H}_{\txtbsn,\sminvtemperature}.\] Furthermore, for any \(f
\in \sphilb{D}_{\txtbsn,0,\sminvtemperature}\), we define the quasi-bilinear form \[\opform{q}_{\txtbsn,\txtbec,\sminvtemperature}(f)
=
\opform{q}_{\txtbsn,0,\sminvtemperature}(f)
+\opform{q}_{\txtbsn,\txtnonzero,\sminvtemperature}(f).\] In particular, we take \(\sphilb{D}_{\txtbsn,\txtphys,\sminvtemperature}
=
\dom \mathsf{m} \cap \sphilb{D}_{\txtbsn,0,\sminvtemperature}\) as the one-particle subspace for a system where infrared and ultraviolet divergences have been removed. Since the spatial settings need to be changed for the Nelson model and the Pauli--Fierz model considering the radiation field, the radiation field for the Pauli--Fierz model will be defined in detail later.

The KMS state of the free scalar field is described by the singular Gaussian \(\sminvtemperature\)-Markov path space associated with the free Bose field Hamiltonian \(\physham_{\txtbsn,\txtfr}\) \cite{YoshitsuguSekine004}: \[\pairbk{
\prbqspace_{\txtbsn,\sminvtemperature},
\mblfml{S}_{\txtbsn,\sminvtemperature},
\mblfml{S}_{\txtbsn,0,\sminvtemperature},
U_{\txtbsn,t},
R_{\txtbsn},
\msrprb_{\txtbsn,\txtfr,\sminvtemperature}}.\] Furthermore, we set \(\sphilb{K}_{\txtbsn,\sminvtemperature}
=
\fun{\lp^{2}}{S_{\sminvtemperature};\sphilb{H}_{\txtbsn,\txtreal}}\), and realize it as a real Gaussian measure space by utilizing the inclusion of \(\faadjpresharp{{\prbqspace_{\txtbsn,\sminvtemperature}}}
\subset
\sphilb{K}_{\txtbsn,\sminvtemperature}
\subset
\prbqspace_{\txtbsn,\sminvtemperature}\) with the regularization operator \cite{YoshitsuguSekine004}. In particular, let a general element of \(\prbqspace_{\txtbsn,\sminvtemperature}\) be \(\opfocksegal\). For each \(t
\in S_{\sminvtemperature}\), define the isometric operator \[j_t
\colon
\rbk{2\omega\tanh \frac{\sminvtemperature \omega}{2}}^{\onehalf}
\sphilb{H}_{\txtbsn,\txtreal}
\to
\faadjpresharp{{\prbqspace_{\txtbsn,\sminvtemperature}}},
\quad
j_t f
=
\diracdelta_t \otimes f.\] For any \(f
\in
\rbk{2\omega\tanh \frac{\sminvtemperature \omega}{2}}^{\onehalf}
\sphilb{H}_{\txtbsn,\txtreal}\), define the sharp-time field by \[\opfocksegal_{t}(f)
=
\opfocksegal(j_{t} f)
=
\dualbkt{\opfocksegal}{j_t f}.\] This satisfies \(\opfocksegal(j_t f)
\in \bigcap_{1 \leq p < \infty}
\fun{\lp^p}{\prbqspace_{\txtbsn,\sminvtemperature},\msrprb_{\txtbsn,\txtfr,\sminvtemperature}}\). The \(\sigma\)-algebra \(\mblfml{S}_{\txtbsn,0,\sminvtemperature}\) at time \(0\) is generated by \(\set{\opfocksegal(j_{0} f)}{f}\), and the total \(\sigma\)-algebra \(\mblfml{S}_{\txtbsn,\sminvtemperature}\) is generated by its time translations. Furthermore, time reversal and time translation are given by \[\funrbk{R_{\txtbsn} F}{\opfocksegal}
=
\fun{F}{\tilde{r} \opfocksegal},
\quad
\funrbk{U_{\txtbsn,t} F}{\opfocksegal}
=
\fun{F}{\tilde{u}_t \opfocksegal},\] as defined in reference \cite{YoshitsuguSekine004}. The characteristic functional of the free field measure \(\msrprb_{\txtbsn,\txtfr,\sminvtemperature}\) is given by \begin{equation}\label{expedition0012134}
\begin{aligned}
\int_{\prbqspace_{\txtbsn,\sminvtemperature}}
\napiernum^{\imunit \opfocksegal(F)}
\opdmsr{\msrprb_{\txtbsn,\txtfr,\sminvtemperature}}(\opfocksegal)
&=
\fnexp{-\oneoverfour\fun{\opform{q}_{\txtbsn,\txtbec,\sminvtemperature}^{\txteuclid}}{F}},
\quad
F
\in \opformdomain(\opform{q}_{\txtbsn,\txtbec,\sminvtemperature}^{\txteuclid}),
\\ %%%%%%%%%%%%%%%%
\fun{\opform{q}_{\txtbsn,\txtbec,\sminvtemperature}^{\txteuclid}}{j_t f,j_s g}
&=
\fun{\opform{q}_{\txtbsn,0,\sminvtemperature}^{\txteuclid}}{j_t f,j_s g}
+
\fun{\opform{q}_{\txtbsn,\txtnonzero,\sminvtemperature}^{\txteuclid}}{j_t f,j_s g}
\\ %%%%%%%%%%%%%%%%
\fun{\opform{q}_{\txtbsn,0,\sminvtemperature}^{\txteuclid}}{j_t f,j_s g}
&=
\fun{\opform{q}_{\txtbsn,0,\sminvtemperature}}{f,g}
\\ %%%%%%%%%%%%%%%%
\fun{\opform{q}_{\txtbsn,\txtnonzero,\sminvtemperature}^{\txteuclid}}{j_t f,j_s g}
&=
\int_{\fldreal^{d}}
\overline{\faftr{f}(k)}
\frac{\napiernum^{-\abs{t-s}\omega(k)}
+\napiernum^{-(\sminvtemperature-\abs{t-s})\omega(k)}}
{1-\napiernum^{-\sminvtemperature\omega(k)}}
\faftr{g}(k)
\opdmsr{k}
\end{aligned}
\end{equation} Here, the quasi-bilinear form with \(\txteuclid\) is intended to be the Euclideanization of the quasi-bilinear form without \(\txteuclid\).

As can be seen from reference \cite{AsaoArai28,YoshitsuguSekine004,YoshitsuguSekine006,YoshitsuguSekine008}, in models where interaction with the field enters linearly, when taking the limit from a bounded system with an added chemical potential, the zero mode describing BEC separates from the free field component. In discussions assuming BEC, we align the notation such that the equal-time restriction \(\fun{\opform{q}_{\txtbsn,\txtbec,\sminvtemperature}^{\txteuclid}}{j_{t} f,j_{t} g}\) matches \(\opform{q}_{\txtbsn,\txtbec,\sminvtemperature}(f,g)\). In particular, the zero-mode component is also expressed as the spacetime covariance \(\opform{q}_{\txtbsn,0,\sminvtemperature}^{\txteuclid}\), and its value is given by \(\opform{q}_{\txtbsn,0,\sminvtemperature}(f,g)\), independent of the time difference. This is because the zero mode corresponds to the \(k
= 0\) component of the dispersion relation \(\omega(k)
= \abs{k}\) and satisfies \(\napiernum^{-\abs{t-s}\omega(0)}
=
\napiernum^{-(\sminvtemperature-\abs{t-s})\omega(0)}
=
1\). Therefore, when taking the BEC limit from a bounded system and extracting the zero mode as an independent bilinear form, its dependence on the Euclidean time direction becomes trivial, appearing as a component independent of the time difference in the spacetime covariance. Consequently, only the non-zero mode component is carried by the usual finite-temperature kernel.

\subsection{\texorpdfstring{Basic \(\sminvtemperature\)-Markov Path Space for the Particle System}{Basic \textbackslash sminvtemperature-Markov Path Space for the Particle System}}\label{expedition0012140}

The particle system also forms a \(\sminvtemperature\)-Markov path space \cite{DerezinskiGerard001}.

Let \(I_{<,\sminvtemperature}^{n}\) be the set of finite time sequences \(\nfoldvar{t}{n}
= \pairbk{t_1,\cdots,t_n}\) satisfying \(-\frac{\sminvtemperature}{2}
< t_1
< \cdots
< t_{n}
< \frac{\sminvtemperature}{2}\), and let \(I_{\leq,\sminvtemperature}^{n}\) be the set obtained by replacing the inequalities \(<\) with \(\leq\). For any \(\nfoldvar{t}{n}
\in I_{<,\sminvtemperature}^{n}\) and bounded real Borel functions \(f_1,\cdots,f_n\), regarding each \(f_j\) as a multiplication operator, the normalized probability measure \(\msrprb_{\txtparticle,\sminvtemperature}\) is characterized by the finite-dimensional distributions given by the integral representation of the Brownian bridge and the Schrödinger semigroup kernel: \[\begin{aligned}
&\int_{\Omega_{\txtparticle,\sminvtemperature}}
\prod_{j=1}^{n} \fun{f_j}{\prbprocess_{t_j}}
\opdmsr{\msrprb_{\txtparticle,\sminvtemperature}(\prbprocess)}
\\ %%%%%%%%%%%%%%%%
&=
\frac{1}{\smpartitionfunc_{\txtparticle,\sminvtemperature}}
\sqfun{\trace}{\napiernum^{-\rbk{t_{1}+\frac{\sminvtemperature}{2}} \physham[h]_{\txtparticle}} f_1
\napiernum^{-\rbk{t_{2}-t_{1}} \physham[h]_{\txtparticle}} f_2
\cdots
\napiernum^{-\rbk{t_{n}-t_{n-1}} \physham[h]_{\txtparticle}} f_n
\napiernum^{-\rbk{\frac{\sminvtemperature}{2}-t_{n}} \physham[h]_{\txtparticle}}}.
\end{aligned}\]

Define the time translation on the periodic time set \(S_{\sminvtemperature}\) by \(\rbk{u_{\txtparticle,t} \prbprocess}_{s}
=
\prbprocess_{s+t \mod \sminvtemperature}\) and the time reflection by \(\rbk{r_{\txtparticle} \prbprocess}_{s}
= \prbprocess_{-s}\). Based on these maps, define the unitary operators \(U_{\txtparticle,t}\) and \(R_{\txtparticle}\) on the particle-side \(\lp^{2}\) space by \(\funrbk{U_{\txtparticle,t} F}{\prbprocess}
=
F(u_{\txtparticle,t} \prbprocess)\) and \(\funrbk{R_{\txtparticle} F}{\prbprocess}
=
F(r_{\txtparticle} \prbprocess)\). For a time set \(I
\subset S_{\sminvtemperature}\), define \(\mblfml{S}_{\txtparticle,\sminvtemperature,I}
= \mblfmlgenerated{\set{\prbprocess_{t}}{t \in I}}\), and denote the corresponding conditional expectation by \(\prbexp_{\txtparticle,\sminvtemperature,I}\). In particular, let \(\mblfml{S}_{\txtparticle,0,\sminvtemperature}\) be the \(\sigma\)-algebra generated by \(\prbprocess_{0}\).

\begin{prop}\label{expedition0012139}
The sextuple formed by the particle path space,
$\pairbk{\Omega_{\txtparticle,\sminvtemperature},
\mblfml{S}_{\txtparticle,\sminvtemperature},
\mblfml{S}_{\txtparticle,0,\sminvtemperature},
U_{\txtparticle,t},
R_{\txtparticle},
\msrprb_{\txtparticle,\sminvtemperature}}$, is a $\sminvtemperature$-Markov path space \cite{DerezinskiGerard001}.
In particular, the commutativity between time reflection and conditional expectation,
$R_{\txtparticle} \prbexp_{\txtparticle,\sminvtemperature,\setone{0,\frac{\sminvtemperature}{2}}}
=
\prbexp_{\txtparticle,\sminvtemperature,\setone{0,\frac{\sminvtemperature}{2}}} R_{\txtparticle}$, and the $\sminvtemperature$-Markov property
$\prbexp_{\txtparticle,\sminvtemperature,\closedinterval{0}{\frac{\sminvtemperature}{2}}}
\prbexp_{\txtparticle,\sminvtemperature,\closedinterval{-\frac{\sminvtemperature}{2}}{0}}
=
\prbexp_{\txtparticle,\sminvtemperature,\setone{0,\frac{\sminvtemperature}{2}}}$ hold.
\end{prop}

We add a proof for the reader's sake. For finitely many times \(t_1,\ldots,t_n
\in S_{\sminvtemperature}\) and a bounded Borel function \(\Phi
\colon
(\fldreal^d)^n
\to
\fldcmp\), functions expressed as \(F(\prbprocess)
=
\fun{\Phi}
{\prbprocess_{t_1},\ldots,\prbprocess_{t_n}}\) and their finite linear combinations are called bounded cylinder functions on the particle path. If \(\Phi\) is bounded and continuous, then \(F\) is called a bounded continuous cylinder function. In the above representation, a function for which all times \(t_1,\ldots,t_n\) belong to \(I\) is called a bounded cylinder function that is \(\mblfml{S}_{\txtparticle,\sminvtemperature,I}\)-measurable with respect to the time set \(I
\subset S_{\sminvtemperature}\).

\begin{proof}
($\sminvtemperature$-periodicity): By definition, $U_{\txtparticle,0}
= \idone$ is clear.
It remains only to show the strong continuity of $U_{\txtparticle,t}$.

Fix a time sequence $s_1<\cdots<s_n$ and $a
\in S_{\sminvtemperature}$.
Take representatives of each $s_j+a$ on $S_{\sminvtemperature}$ and rearrange them in increasing order.
If no endpoint is crossed, the neighboring time differences are preserved, so invariance follows from the semigroup representation of the finite-dimensional distributions.
If an endpoint is crossed, the rearrangement is a cyclic permutation of the time sequence and the corresponding multiplication operators, and this cyclic permutation does not change the value by the cyclicity of the trace.
Therefore, for any cylinder function $F$,
$\int_{\Omega_{\txtparticle,\sminvtemperature}}
\fun{F}{u_{\txtparticle,a} \prbprocess}
\opdmsr{\msrprb_{\txtparticle,\sminvtemperature}(\prbprocess)}
=
\int F(\prbprocess)
\opdmsr{\msrprb_{\txtparticle,\sminvtemperature}(\prbprocess)}$ holds.

Consider a bounded continuous cylinder function $F(\prbprocess)
= F_0(\prbprocess_{s_1},\cdots,\prbprocess_{s_n})$.
Since each path $\prbprocess
\in
\Omega_{\txtparticle,\sminvtemperature}
= \fun{\conti}{S_{\sminvtemperature};\fldreal^{d}}$ is continuous, $F(u_{\txtparticle,t} \prbprocess)\to F(\prbprocess)$ holds pointwise as $t
\to 0$.
Since this $F$ is bounded, the dominated convergence theorem gives $\twonorm{U_{\txtparticle,t} F - F}
\to 0$.
Bounded continuous cylinder functions are dense in $\fun{\lp^{2}}{\msrprb_{\txtparticle,\sminvtemperature}}$, and therefore the unitarity of the family $\fml{U_{\txtparticle,t}}{}$ extends this strong continuity to all $F
\in \fun{\lp^{2}}{\msrprb_{\txtparticle,\sminvtemperature}}$.
Thus $U_{\txtparticle,t}$ gives a strongly continuous unitary group on $\fun{\lp^{2}}{\msrprb_{\txtparticle,\sminvtemperature}}$.

(Reflection): By the self-adjointness of the Hamiltonian $\physham[h]_{\txtparticle}$, the semigroup kernel $K_t$ is symmetric.
Reading the trace representation of the finite-dimensional distributions in the reverse direction, the time-reflected finite-dimensional distribution agrees with the original distribution.
Therefore $\msrprb_{\txtparticle,\sminvtemperature}$ is invariant under the action of $r_{\txtparticle}$, so the operator $R_{\txtparticle}$ is a unitary reflection operator.

(Commutativity of time reflection and conditional expectation): By periodicity, $r_{\txtparticle}$ preserves the two-point set $\setone{0,\frac{\sminvtemperature}{2}}$, and hence the operator $R_{\txtparticle}$ preserves $\mblfml{S}_{\txtparticle,\sminvtemperature,\setone{0,\frac{\sminvtemperature}{2}}}$.
Next we verify the $r_{\txtparticle}$-invariance of $\msrprb_{\txtparticle,\sminvtemperature}$ on finite-dimensional distributions.
By definition,
$$\int_{\Omega_{\txtparticle,\sminvtemperature}}
\prod_{j=1}^n
\fun{f_j}{(r_{\txtparticle} \prbprocess)_{t_j}}
\opdmsr{\msrprb_{\txtparticle,\sminvtemperature}(\prbprocess)}
=
\int_{\Omega_{\txtparticle,\sminvtemperature}}
\prod_{j=1}^n
f_j(X_{-t_j})
\opdmsr{\msrprb_{\txtparticle,\sminvtemperature}(\prbprocess)}$$
holds.
Let $K_t(x,y)$ be the integral kernel of the heat operator $\napiernum^{-t \physham[h]_{\txtparticle}}$.
Applying the kernel representation of the finite-dimensional distributions to the time sequence $-t_n<\cdots<-t_1$, the right-hand side can be written as
$$\begin{aligned}
\frac{1}{\smpartitionfunc_{\txtparticle,\sminvtemperature}}
\int_{(\fldreal^{d})^n}
&K_{\frac{\sminvtemperature}{2}-t_n}(x_0,x_n)
f_n(x_n)
K_{t_n-t_{n-1}}(x_n,x_{n-1})
f_{n-1}(x_{n-1})
\\ %%%%%%%%%%%%%%%%
&\cdots
K_{t_2-t_1}(x_2,x_1)
f_1(x_1)
K_{t_1+\frac{\sminvtemperature}{2}}(x_1,x_0)
\opdmsr{x_0} \opdmsr{x_1} \cdots \opdmsr{x_n}.
\end{aligned}$$
By the symmetry of the kernel, $K_s(x,y)
= K_s(y,x)$, this can be rewritten as
$$\begin{aligned}
\frac{1}{\smpartitionfunc_{\txtparticle,\sminvtemperature}}
\int_{(\fldreal^{d})^n}
&K_{\frac{\sminvtemperature}{2}+t_1}(x_0,x_1)
f_1(x_1)
K_{t_2-t_1}(x_1,x_2)
f_2(x_2)
\\ %%%%%%%%%%%%%%%%
&\cdots
K_{t_n-t_{n-1}}(x_{n-1},x_n)
f_n(x_n)
K_{\frac{\sminvtemperature}{2}-t_n}(x_n,x_0)
\opdmsr{x_0} \opdmsr{x_1} \cdots \opdmsr{x_n}
\end{aligned}$$
This agrees with the kernel representation of the finite-dimensional distribution
$\int_{\Omega_{\txtparticle,\sminvtemperature}}
\prod_{j=1}^n
\fun{f_j}{\prbprocess_{t_j}}
\opdmsr{\msrprb_{\txtparticle,\sminvtemperature}(\prbprocess)}$.
Therefore
$$\int_{\Omega_{\txtparticle,\sminvtemperature}}
\prod_{j=1}^n
\fun{f_j}{(r_{\txtparticle} \prbprocess)_{t_j}}
\opdmsr{\msrprb_{\txtparticle,\sminvtemperature}(\prbprocess)}
=
\int_{\Omega_{\txtparticle,\sminvtemperature}}
\prod_{j=1}^n
\fun{f_j}{X_{t_j}}
\opdmsr{\msrprb_{\txtparticle,\sminvtemperature}(\prbprocess)}$$
holds.
Since the finite-dimensional cylinder class generated by bounded cylinder functions generates $\mblfml{S}_{\txtparticle,\sminvtemperature}$, the monotone class theorem shows that $\msrprb_{\txtparticle,\sminvtemperature}$ is $r_{\txtparticle}$-invariant.
Since the time reflection $r_{\txtparticle}$ preserves the two-point set $\setone{0,\frac{\sminvtemperature}{2}}$, we have $\fun{\inv{r_{\txtparticle}}}
{S_{\txtparticle,\sminvtemperature,\setone{0,\frac{\sminvtemperature}{2}}}}
= S_{\txtparticle,\sminvtemperature,\setone{0,\frac{\sminvtemperature}{2}}}$.
In general, when a measure-preserving transformation $r$ preserves a sub-$\sigma$-algebra $\mblfml{G}$, the corresponding unitary operator $RF
= F \circ r$ commutes with the conditional expectation.
Hence the desired commutativity
$R_{\txtparticle} \prbexp_{\txtparticle,\sminvtemperature,\setone{0,\frac{\sminvtemperature}{2}}}
=
\prbexp_{\txtparticle,\sminvtemperature,\setone{0,\frac{\sminvtemperature}{2}}} R_{\txtparticle}$ follows.

(The $\sminvtemperature$-Markov property): It suffices to check this on bounded cylinder functions.
In particular, fixing the boundary values $\prbprocess_{0}
= x$ and $\prbprocess_{\frac{\sminvtemperature}{2}}
= y$, it suffices to check whether the conditional expectations on the interval $\closedinterval{0}{\frac{\sminvtemperature}{2}}$ and on the interval $\closedinterval{-\frac{\sminvtemperature}{2}}{0}$ decompose as a product.
Let $F_{\txtneg}$ be a bounded cylinder function that is $\mblfml{S}_{\txtparticle,\sminvtemperature,\closedinterval{-\frac{\sminvtemperature}{2}}{0}}$-measurable, and let $F_{\txtnonneg}$ be a bounded cylinder function that is $\mblfml{S}_{\txtparticle,\sminvtemperature,\closedinterval{0}{\frac{\sminvtemperature}{2}}}$-measurable.
By the representation of the finite-dimensional distributions and the product representation of the Feynman--Kac kernel, under the condition of fixed boundary values the integrations over the two half-intervals can be written separately. Hence
$$\prbexp_{\txtparticle,\sminvtemperature,\setone{0,\frac{\sminvtemperature}{2}}}
(F_{\txtneg} F_{\txtnonneg})
=
\prbexp_{\txtparticle,\sminvtemperature,\setone{0,\frac{\sminvtemperature}{2}}}(F_{\txtneg})
\prbexp_{\txtparticle,\sminvtemperature,\setone{0,\frac{\sminvtemperature}{2}}}(F_{\txtnonneg})$$
holds.
In particular, the path on the half-interval $\closedinterval{-\frac{\sminvtemperature}{2}}{0}$ and the path on the half-interval $\closedinterval{0}{\frac{\sminvtemperature}{2}}$ are conditionally independent.
Therefore, if $F_{\txtneg}$ depends only on the lower half-interval, then
$\begin{aligned}
\prbexp_{\txtparticle,\sminvtemperature,\closedinterval{0}{\frac{\sminvtemperature}{2}}}(F_{\txtneg})
=
\prbexp_{\txtparticle,\sminvtemperature,\setone{0,\frac{\sminvtemperature}{2}}}(F_{\txtneg})
\end{aligned}$
holds.
Similarly, if $F_{\txtnonneg}$ depends only on the upper half-interval, then
$\begin{aligned}
\prbexp_{\txtparticle,\sminvtemperature,\closedinterval{-\frac{\sminvtemperature}{2}}{0}}(F_{\txtnonneg})
=
\prbexp_{\txtparticle,\sminvtemperature,\setone{0,\frac{\sminvtemperature}{2}}}(F_{\txtnonneg})
\end{aligned}$
holds.
Consequently,
$$\begin{aligned}
\prbexp_{\txtparticle,\sminvtemperature,[0,\frac{\sminvtemperature}{2}]}\prbexp_{\txtparticle,\sminvtemperature,[-\frac{\sminvtemperature}{2},0]}(F_{\txtneg}F_{\txtnonneg})
&=
\fun{\prbexp_{\txtparticle,\sminvtemperature,[0,\frac{\sminvtemperature}{2}]}}
{F_{\txtneg} \prbexp_{\txtparticle,\sminvtemperature,\{0,\frac{\sminvtemperature}{2}\}}(F_{\txtnonneg})}
\\ %%%%%%%%%%%%%%%%
&=
\prbexp_{\txtparticle,\sminvtemperature,\setone{0,\frac{\sminvtemperature}{2}}}(F_{\txtneg})
\prbexp_{\txtparticle,\sminvtemperature,\setone{0,\frac{\sminvtemperature}{2}}}(F_{\txtnonneg})
\\ %%%%%%%%%%%%%%%%
&=
\prbexp_{\txtparticle,\sminvtemperature,\{0,\frac{\sminvtemperature}{2}\}}(F_{\txtneg}F_{\txtnonneg})
\end{aligned}$$
holds.
It remains to extend this to the class of cylinder functions spanned by finite sums.
By $L^2$-density and the contractivity of conditional expectation,
$\prbexp_{\txtparticle,\sminvtemperature,[0,\frac{\sminvtemperature}{2}]}\prbexp_{\txtparticle,\sminvtemperature,[-\frac{\sminvtemperature}{2},0]}
=
\prbexp_{\txtparticle,\sminvtemperature,\{0,\frac{\sminvtemperature}{2}\}}$ follows as an operator identity on $\lp^{2}(\msrprb_{\txtparticle,\sminvtemperature})$.
\end{proof}

\subsection{\texorpdfstring{\(\sminvtemperature\)-Markov Path Space for the Total System and the Interaction Model}{\textbackslash sminvtemperature-Markov Path Space for the Total System and the Interaction Model}}\label{expedition0012146}

We consider the total system of the particle and the Bose field. For the particle, we use Subsection \ref{expedition0012140}; for the Bose field, we use the \(\sminvtemperature\)-Markov path space of the free field from Subsection \ref{expedition0011278}. As the \(\sminvtemperature\)-Markov path space associated with the free Hamiltonian of the Nelson model, \(\physham_{\txtnelson,0}
=
\physham[h]_{\txtparticle} \otimes \idone
+\idone \otimes \physham_{\txtbsn,\txtfr}\), we consider \[\begin{aligned}
\pairbk{\prbqspace_{\txtnelson,\sminvtemperature},
\mblfml{S}_{\txtnelson,\sminvtemperature},
\mblfml{S}_{\txtnelson,0,\sminvtemperature},
U_{\txtnelson,t},
R_{\txtnelson},
\msrprb_{\txtnelson,0,\sminvtemperature}}.
\end{aligned}\] The sample space is \(\prbqspace_{\txtnelson,\sminvtemperature}
= \Omega_{\txtparticle,\sminvtemperature} \times \prbqspace_{\txtbsn,\sminvtemperature}\). The \(\sigma\)-algebras are \(\mblfml{S}_{\txtnelson,\sminvtemperature}
= \mblfml{S}_{\txtparticle,\sminvtemperature} \times \mblfml{S}_{\txtbsn,\sminvtemperature}\) and \(\mblfml{S}_{\txtnelson,0,\sminvtemperature}
= \mblfml{S}_{\txtparticle,0,\sminvtemperature} \times \mblfml{S}_{\txtbsn,0,\sminvtemperature}\), the time evolution is \(U_{\txtnelson,t}
= U_{\txtparticle,t} \otimes U_{\txtbsn,t}\), the time reflection is \(R_{\txtnelson}
= R_{\txtparticle} \otimes R_{\txtbsn}\), and the probability measure is defined by \(\msrprb_{\txtnelson,0,\sminvtemperature}
= \msrprb_{\txtparticle,\sminvtemperature} \otimes \msrprb_{\txtbsn,\txtfr,\sminvtemperature}\).

For any measurable \(I
\subset S_{\sminvtemperature}\), define the source on particle paths corresponding to the cutoff source by \(\mathsf{J}_{I,\kappa,\Lambda}(\prbprocess)
=
\int_{I}
\fun{j_t}
{\lambda_{\prbprocess_{t},\kappa,\Lambda}}
\opdmsr{t}\). Following the textbook \cite{DerezinskiGerard001}, define the KMS state of the cutoff Nelson model by \[\begin{aligned}
\oastate[\psi_{\txtnelson,\sminvtemperature,\kappa,\Lambda}](F)
&=
\frac{1}{\smpartitionfunc_{\txtnelson,\sminvtemperature,\kappa,\Lambda}}
\int_{\prbqspace_{\txtnelson,\sminvtemperature}}
F(\prbprocess,\opfocksegal)
\fnexp{-\physcplconst
\fun{\opfocksegal}{\mathsf{J}_{S_{\sminvtemperature},\kappa,\Lambda}(\prbprocess)}}
\opdmsr{\msrprb_{\txtnelson,0,\sminvtemperature}(\prbprocess,\opfocksegal)},
\\ %%%%%%%%%%%%%%%%
\smpartitionfunc_{\txtnelson,\sminvtemperature,\kappa,\Lambda}
&=
\int_{\prbqspace_{\txtnelson,\sminvtemperature}}
\fnexp{-\physcplconst
\fun{\opfocksegal}{\mathsf{J}_{S_{\sminvtemperature},\kappa,\Lambda}(\prbprocess)}}
\opdmsr{\msrprb_{\txtnelson,0,\sminvtemperature}(\prbprocess,\opfocksegal)}
\end{aligned}\] Because of the infrared cutoff, \(\lambda_{x,\kappa,\Lambda}\) satisfies \(\faftr{\lambda}_{x,\kappa,\Lambda}(0)
= 0\). Therefore, for any \(f
\in \sphilb{D}_{\txtbsn,\txtphys,\sminvtemperature}\) and any interval \(I
\subset S_{\sminvtemperature}\), \[\begin{aligned}
\fun{\opform{q}_{\txtbsn,0,\sminvtemperature}^{\txteuclid}}{j_t f,j_s\lambda_{x,\kappa,\Lambda}}
&=0,
\\ %%%%%%%%%%%%%%%%
\fun{\opform{q}_{\txtbsn,0,\sminvtemperature}^{\txteuclid}}{j_t\lambda_{x,\kappa,\Lambda},j_s\lambda_{x,\kappa,\Lambda}}
&=
0,
\\ %%%%%%%%%%%%%%%%
\fun{\opform{q}_{\txtbsn,\txtbec,\sminvtemperature}^{\txteuclid}}{j_t f,\mathsf{J}_{I,\kappa,\Lambda}(\prbprocess)}
&=
\fun{\opform{q}_{\txtbsn,\txtnonzero,\sminvtemperature}^{\txteuclid}}{j_t f,\mathsf{J}_{I,\kappa,\Lambda}(\prbprocess)},
\\ %%%%%%%%%%%%%%%%
\fun{\opform{q}_{\txtbsn,\txtbec,\sminvtemperature}^{\txteuclid}}{\mathsf{J}_{I,\kappa,\Lambda}(\prbprocess)}
&=
\fun{\opform{q}_{\txtbsn,\txtnonzero,\sminvtemperature}^{\txteuclid}}{\mathsf{J}_{I,\kappa,\Lambda}(\prbprocess)}
\end{aligned}\] holds.

\subsection{Bounded System Setting}\label{expedition0012145}

The notation for bounded systems is regarded as a finite-volume regularization of the whole space \(\fldreal^d\). In particular, for \(L>0\), let the hypercube with side length \(L\) and volume \(L^d\) be \[I_L
=
\closedinterval{-\frac{L}{2}}{\frac{L}{2}},
\quad
I_L^d
=
\closedinterval{-\frac{L}{2}}{\frac{L}{2}}^{d}\] We impose periodic boundary conditions on the one-particle Hamiltonian of the bosons. If the one-dimensional lattice is \(\setlattice_{L}
=
\frac{2 \pi}{L} \ringratint\), then the dual momentum space is \(\setlattice_{L}^{d}\). Let the one-particle Hilbert space of the bounded system be \(\sphilb{H}_{\txtbsn,L}
= \fun{\lp^{2}}{I_{L}^{d}}\), and let the Fock space be \(\spfock_{\txtbsn,L}
=
\fun{\spfock_{\txtbsn}}{\sphilb{H}_{\txtbsn,L}}\).

For the above \(L\), let \(P_L\) be the projection from \(\sphilb{H}_{\txtbsn}
\to \sphilb{H}_{\txtbsn,L}\). Below, when constructing objects in the bounded system from a test function \(f\) or a cutoff source \(\lambda\) on \(\fldreal^d\), we first define \(f\) or \(\lambda\) on \(\fldreal^d\) and then project it to \(P_L f\) or \(P_L \lambda\). The free boson Hamiltonian and the number operator of the bounded system are also defined by \(\physham_{\txtbsn,\txtfr,L}
=
\fun{\opfocksndqntdiff_{\txtbsn}}{P_{L}\omega P_{L}}\) and \(N_{\txtbsn,L}
=
\opfocksndqntdiff(P_{L} \idone)\). Under the assumption of periodic boundary conditions, \(\napiernum^{-\sminvtemperature \omega_{L}}\) is trace class for \(\omega_{L}
= P_{L} \omega P_{L}\), and its second quantization \(\fun{\opfocksndqnt_{\txtbsn}}{\napiernum^{-\sminvtemperature \omega_{L}}}\) is also trace class \cite{AsaoArai28}.

To discuss BEC at finite temperature, in the bounded system we take the chemical potential to be \(\smchemicalpotential
< 0\) and control the zero mode. For a fixed total density \(\bar{\smnumberdensity}_{\txtbsn}>0\), define the regularization parameter \(y_L>1\) of the bounded system by \[\frac{1}{L^d}
\sum_{k\in\setlattice_L^d}
\frac{1}{y_L\napiernum^{\sminvtemperature\omega(k)}-1}
=
\bar{\smnumberdensity}_{\txtbsn}\] Then, for the free Hamiltonian with chemical potential, \(\physham_{\txtbsn,\txtfr,\smchemicalpotential,L}
=
\fun{\opfocksndqntdiff_{\txtbsn}}{\omega_L - \smchemicalpotential}\), the operator \(\napiernum^{-\sminvtemperature
\physham_{\txtbsn,\txtfr,\smchemicalpotential,L}}\) is trace class on \(\spfock_{\txtbsn,L}\), and the value of its trace is \(\prod_{k\in\setlattice_L^d}
\frac{1}
{y_{L} - \napiernum^{-\sminvtemperature \rbk{\omega(k) - \smchemicalpotential}}}\).

For any \(t,s
\in S_{\sminvtemperature}\) and \(f,g
\in P_{L} \sphilb{H}_{\txtbsn,\txtreal}\), define the thermal covariance form of the free field in the bounded system by \[\begin{aligned}
\fun{\opform{q}_{\txtbsn,\txtbec,\sminvtemperature,\smchemicalpotential,L}^{\txteuclid}}
{j_t f,j_s g}
&=
\fun{\opform{q}_{\txtbsn,0,\sminvtemperature,\smchemicalpotential,L}^{\txteuclid}}
{j_t f,j_s g}
+\fun{\opform{q}_{\txtbsn,\txtnonzero,\sminvtemperature,\smchemicalpotential,L}^{\txteuclid}}
{j_t f,j_s g},
\\ %%%%%%%%%%%%%%%%
\fun{\opform{q}_{\txtbsn,0,\sminvtemperature,\smchemicalpotential,L}^{\txteuclid}}
{j_t f,j_s g}
&=
\frac{(2\pi)^d}{L^d}
\frac{y_L+1}{y_L-1}
\overline{\faftr{f}(0)}\faftr{g}(0),
\\ %%%%%%%%%%%%%%%%
\fun{\opform{q}_{\txtbsn,\txtnonzero,\sminvtemperature,\smchemicalpotential,L}^{\txteuclid}}
{j_t f,j_s g}
&=
\frac{(2\pi)^d}{L^d}
\sum_{k\in\setlattice_L^d\setminus\setone{0}}
\overline{\faftr{f}(k)}
\frac{y_L \napiernum^{-\abs{t-s} \rbk{\omega(k) - \smchemicalpotential}}
+\napiernum^{-(\sminvtemperature-\abs{t-s}) \rbk{\omega(k) - \smchemicalpotential}}}
{y_L - \napiernum^{-\sminvtemperature \rbk{\omega(k) - \smchemicalpotential}}}
\faftr{g}(k)
\end{aligned}\] and extend it appropriately. Since the dispersion relation satisfies \(\omega(0)
= 0\), the zero-mode component does not depend on the time difference, while the non-zero-mode component has the usual finite-temperature kernel of the free Bose field. The limit \(L
\to \infty\) of this decomposition corresponds to the infinite-volume decomposition into \(\opform{q}_{\txtbsn,0,\sminvtemperature}^{\txteuclid}\) and \(\opform{q}_{\txtbsn,\txtnonzero,\sminvtemperature}^{\txteuclid}\) introduced in the previous subsection \cite{AsaoArai28,YoshitsuguSekine004}.

We represent the KMS state of the bounded system with chemical potential for the free Bose field by using the singular Gaussian \(\sminvtemperature\)-Markov path space \[\pairbk{\prbqspace_{\txtbsn,\sminvtemperature,L},
\mblfml{S}_{\txtbsn,\sminvtemperature,L},
\mblfml{S}_{\txtbsn,0,\sminvtemperature,L},
U_{\txtbsn,L,t},
R_{\txtbsn,L},
\msrprb_{\txtbsn,\txtfr,\sminvtemperature,\smchemicalpotential,L}}.\] For an observable \(A\) of the bounded system, this measure is characterized by the finite-dimensional distributions associated with the state determined from the trace for the free field, \[\fun{\psi_{\txtbsn,\txtfr,\sminvtemperature,\smchemicalpotential,L}^{\txteuclid}}{A}
=
\frac{\sqfun{\trace_{\spfock_{\txtbsn,L}}}
{\napiernum^{-\sminvtemperature \physham_{\txtbsn,\txtfr,\smchemicalpotential,L}} A}}
{\sqfun{\trace_{\spfock_{\txtbsn,L}}}
{\napiernum^{-\sminvtemperature \physham_{\txtbsn,\txtfr,\smchemicalpotential,L}}}}\] That is, for \(t_1,\ldots,t_n
\in S_{\sminvtemperature}\), \(f_1,\ldots,f_n
\in P_{L}\sphilb{H}_{\txtbsn,\txtreal}\), \(s_1,\ldots,s_n
\in\fldreal\), if \(F
= \sum_{j=1}^{n}s_jj_{t_j}f_j\), then \begin{equation}\label{expedition0012340}
\begin{aligned}
\int_{\prbqspace_{\txtbsn,\sminvtemperature,L}}
\napiernum^{\imunit \opfocksegal(F)}
\opdmsr{\msrprb_{\txtbsn,\txtfr,\sminvtemperature,\smchemicalpotential,L}(\opfocksegal)}
=
\fnexp{-\oneoverfour
\fun{\opform{q}_{\txtbsn,\txtbec,\sminvtemperature,\smchemicalpotential,L}^{\txteuclid}}{F}}
\end{aligned}
\end{equation} holds.

We consider the total system of the particle and the Bose field. In the bounded system, it is necessary to take the chemical potential into account. As the \(\sminvtemperature\)-Markov path space associated with the free Hamiltonian \(\physham_{\txtnelson,0,\smchemicalpotential,L}
=
\physham[h]_{\txtparticle} \otimes \idone
+\idone \otimes \physham_{\txtbsn,\txtfr,\smchemicalpotential,L}\), we consider \[\begin{aligned}
\pairbk{\prbqspace_{\txtnelson,\sminvtemperature,L},
\mblfml{S}_{\txtnelson,\sminvtemperature,L},
\mblfml{S}_{\txtnelson,0,\sminvtemperature,L},
U_{\txtnelson,L,t},
R_{\txtnelson,L},
\msrprb_{\txtnelson,0,\sminvtemperature,\smchemicalpotential,L}}.
\end{aligned}\] The sample space is \(\prbqspace_{\txtnelson,\sminvtemperature,L}
= \Omega_{\txtparticle,\sminvtemperature} \times \prbqspace_{\txtbsn,\sminvtemperature,L}\), the \(\sigma\)-algebras are \(\mblfml{S}_{\txtnelson,\sminvtemperature,L}
= \mblfml{S}_{\txtparticle,\sminvtemperature} \times \mblfml{S}_{\txtbsn,\sminvtemperature,L}\) and \(\mblfml{S}_{\txtnelson,0,\sminvtemperature,L}
= \mblfml{S}_{\txtparticle,0,\sminvtemperature} \times \mblfml{S}_{\txtbsn,0,\sminvtemperature,L}\), the time evolution is \(U_{\txtnelson,L,t}
= U_{\txtparticle,t} \otimes U_{\txtbsn,L,t}\), the time reflection is \(R_{\txtnelson,L}
= R_{\txtparticle} \otimes R_{\txtbsn,L}\), and the probability measure is defined by \(\msrprb_{\txtnelson,0,\sminvtemperature,\smchemicalpotential,L}
= \msrprb_{\txtparticle,\sminvtemperature} \otimes \msrprb_{\txtbsn,\txtfr,\sminvtemperature,\smchemicalpotential,L}\).

For any measurable \(I
\subset S_{\sminvtemperature}\), define the source on particle paths corresponding to the cutoff source by \(\mathsf{J}_{I,\kappa,\Lambda,L}(\prbprocess)
=
\int_{I}
j_t P_{L} \lambda_{\prbprocess_{t},\kappa,\Lambda}
\opdmsr{t}\). Following the textbook \cite{DerezinskiGerard001}, define the KMS state of the cutoff Nelson model by \[\begin{aligned}
\oastate[\psi_{\txtnelson,\sminvtemperature,\kappa,\Lambda,\smchemicalpotential,L}](F)
&=
\frac{1}{\smpartitionfunc_{\txtnelson,\sminvtemperature,\kappa,\Lambda,\smchemicalpotential,L}}
\int_{\prbqspace_{\txtnelson,\sminvtemperature,L}}
F(\prbprocess,\opfocksegal)
\fnexp{-\physcplconst
\fun{\opfocksegal}{\mathsf{J}_{S_{\sminvtemperature},\kappa,\Lambda,L}(\prbprocess)}}
\opdmsr{\msrprb_{\txtnelson,0,\sminvtemperature,\smchemicalpotential,L}(\prbprocess,\opfocksegal)},
\\ %%%%%%%%%%%%%%%%
\smpartitionfunc_{\txtnelson,\sminvtemperature,\kappa,\Lambda,\smchemicalpotential,L}
&=
\int_{\prbqspace_{\txtnelson,\sminvtemperature,L}}
\fnexp{-\physcplconst
\fun{\opfocksegal}{\mathsf{J}_{S_{\sminvtemperature},\kappa,\Lambda,L}(\prbprocess)}}
\opdmsr{\msrprb_{\txtnelson,0,\sminvtemperature,\smchemicalpotential,L}(\prbprocess,\opfocksegal)}.
\end{aligned}\] Here each \(\lambda_{x,\kappa,\Lambda,L}
= P_{L} \lambda_{x,\kappa,\Lambda}\) satisfies \(\faftr{\lambda}_{x,\kappa,\Lambda,L}(0)=0\) because of the infrared cutoff. Therefore, for any \(f
\in \sphilb{D}_{\txtbsn,\txtphys,\sminvtemperature}\) and any interval \(I
\subset S_{\sminvtemperature}\), \[\begin{aligned}
\fun{\opform{q}_{\txtbsn,0,\sminvtemperature,\smchemicalpotential,L}^{\txteuclid}}{j_t P_{L}f,j_sP_{L}\lambda_{x,\kappa,\Lambda}}
&=0,
\\ %%%%%%%%%%%%%%%%
\fun{\opform{q}_{\txtbsn,0,\sminvtemperature,\smchemicalpotential,L}^{\txteuclid}}{j_tP_{L}\lambda_{x,\kappa,\Lambda},j_sP_{L}\lambda_{x,\kappa,\Lambda}}
&=
0,
\\ %%%%%%%%%%%%%%%%
\fun{\opform{q}_{\txtbsn,\txtbec,\sminvtemperature,\smchemicalpotential,L}^{\txteuclid}}{j_t P_{L}f,\mathsf{J}_{I,\kappa,\Lambda,L}(\prbprocess)}
&=
\fun{\opform{q}_{\txtbsn,\txtnonzero,\sminvtemperature,\smchemicalpotential,L}^{\txteuclid}}{j_t P_{L}f,\mathsf{J}_{I,\kappa,\Lambda,L}(\prbprocess)},
\\ %%%%%%%%%%%%%%%%
\fun{\opform{q}_{\txtbsn,\txtbec,\sminvtemperature,\smchemicalpotential,L}^{\txteuclid}}{\mathsf{J}_{I,\kappa,\Lambda,L}(\prbprocess)}
&=
\fun{\opform{q}_{\txtbsn,\txtnonzero,\sminvtemperature,\smchemicalpotential,L}^{\txteuclid}}{\mathsf{J}_{I,\kappa,\Lambda,L}(\prbprocess)}
\end{aligned}\] holds.

\subsection{KMS States for Spinless Particles}\label{expedition0012138}

Let the circle representing the periodicity of a KMS state and the time loop be denoted by \(S_{\sminvtemperature}
= \closedinterval{-\frac{\sminvtemperature}{2}}{\frac{\sminvtemperature}{2}}\), and regard all operations in \(S_{\sminvtemperature}\) as operations modulo \(\sminvtemperature\) so that they remain in this range.

For any \(x
\in \fldreal^{d}\), let \(\msrprb_{x,x}^{\sminvtemperature,\txtbrownbridge}\) be the Brownian bridge measure of length \(\sminvtemperature\) whose initial and terminal points are both \(x\), and define the particle loop space by \[\begin{aligned}
\Omega_{\txtparticle,\sminvtemperature}
=
\set{\prbprocess \in \fun{\conti}{S_{\sminvtemperature};\fldreal^{d}}}
{\fun{\prbprocess}{-\frac{\sminvtemperature}{2}}
= \fun{\prbprocess}{\frac{\sminvtemperature}{2}}}.
\end{aligned}\] Define the unnormalized particle loop measure by \begin{equation}\label{expedition0012428}
\begin{aligned}
\opdmsr{\widetilde{\msrprb}_{\txtparticle,\sminvtemperature}}(\prbprocess)
=
\int_{\fldreal^{d}}
\fnexp{-\int_{S_{\sminvtemperature}}
V(\prbprocess_{t})
\opdmsr{t}}
\opdmsr{\msrprb_{x,x}^{\sminvtemperature,\txtbrownbridge}(\prbprocess)}
\opdmsr{x}.
\end{aligned}
\end{equation} Then the Feynman--Kac formula gives \[\widetilde{\msrprb}_{\txtparticle,\sminvtemperature}(\Omega_{\txtparticle,\sminvtemperature})
=
\int_{\fldreal^{d}}
\int_{\Omega_{\txtparticle,\sminvtemperature}}
\napiernum^{-\int_{S_{\sminvtemperature}}V(\prbprocess_{t})\opdmsr{t}}
\opdmsr{\msrprb_{x,x}^{\sminvtemperature,\txtbrownbridge}}(\prbprocess)
\opdmsr{x}
=
\int_{\fldreal^{d}}
\napiernum^{-\sminvtemperature \physham[h]_{\txtparticle}}(x,x)
\opdmsr{x}
=
\sqfun{\trace_{\fldreal^{d}}}{\napiernum^{-\sminvtemperature \physham[h]_{\txtparticle}}}
< \infty.\] The integrand is non-negative and nontrivial, so the total measure is nonzero. Therefore the normalized particle loop measure \begin{equation}\label{expedition0012429}
\msrprb_{\txtparticle,\sminvtemperature}
=
\frac{1}{\smpartitionfunc_{\txtparticle,\sminvtemperature}}
\widetilde{\msrprb}_{\txtparticle,\sminvtemperature}
\end{equation} can be defined. This is the measure corresponding to the KMS state of the particle.

\subsection{Main Theorem}\label{main-theorem}

In the two-point functions after removal of the infrared and ultraviolet cutoffs, the cross terms involving the interaction factor and the non-zero-mode covariance vanish in the long-distance limit. In the Nelson model, \(\onehalf \fun{\opform{q}_{\txtbsn,0,\sminvtemperature}} {f,g}\) appearing on the right-hand side of \eqref{expedition0012413} completely describes the two-point off-diagonal long-range order. In the Pauli--Fierz models, \(\onehalf \fun{\opform{q}_{\txtrad,0,\sminvtemperature}} {f,g}\) appearing on the right-hand sides of \eqref{expedition0012416} and \eqref{expedition0012419} completely describes the two-point off-diagonal long-range order. On the other hand, the order parameters in Definitions \ref{expedition0012159}, \ref{expedition0012168}, and \ref{expedition0012269} detect only the zero-momentum mode in finite volume. Hence, if the physical test-function space can distinguish the zero mode in the sense of Definition \ref{expedition0012433} below, then the absence of off-diagonal long-range order, the vanishing of the zero-mode form, the vanishing of the condensate density, and the order-parameter criterion are equivalent.

\begin{defn}[Zero-mode distinguishability]\label{expedition0012433}
We say that a physical test-function space can distinguish the zero mode when the following conditions are satisfied.
\begin{enumerate}
\item
Nelson model: the zero-mode evaluation $f\mapsto\faftr{f}(0)$ is not identically zero on $\sphilb{D}_{\txtbsn,\txtphys,\sminvtemperature}$. In particular, if $\fun{\opform{q}_{\txtbsn,0,\sminvtemperature}}{f}
= 0$ holds for every $f
\in \sphilb{D}_{\txtbsn,\txtphys,\sminvtemperature}$, then $\smnumberdensity_{\txtbsn,0}(\sminvtemperature)
= 0$.

\item
Pauli--Fierz models: the projected zero-mode evaluation $f\mapsto\Pi_{\txtrad,0}\faftr{f}(0)$ is not identically zero on $\sphilb{D}_{\txtrad,\txtphys,\sminvtemperature}$. In particular, if $\fun{\opform{q}_{\txtrad,0,\sminvtemperature}}{f}
= 0$ holds for every $f
\in \sphilb{D}_{\txtrad,\txtphys,\sminvtemperature}$, then $\smnumberdensity_{\txtrad,0}(\sminvtemperature)
= 0$.
\end{enumerate}
\end{defn}

\begin{thm}[No-Go Theorem for BEC via Off-Diagonal Long-Range Order and Order Parameters]
Consider the Nelson model, the spinless Pauli--Fierz model, and the Pauli--Fierz model with spin in three-dimensional space after removal of the infrared and ultraviolet cutoffs. Assume that the physical test-function space of each model can distinguish the zero mode in the sense of Definition \ref{expedition0012433}. Then, in each of the three models, the following conditions are equivalent.
\begin{enumerate}
\item
The corresponding off-diagonal long-range order in \eqref{expedition0012413}, \eqref{expedition0012416}, or \eqref{expedition0012419} is $0$ for all physical test functions.

\item
The corresponding zero-mode form in \eqref{expedition0012440}, \eqref{eq:beta-pf-covariance}, or \eqref{eq:beta-pf-covariance} is $0$ on the physical test-function space.

\item
The corresponding zero-mode density in \eqref{expedition0012440}, \eqref{eq:beta-pf-covariance}, or \eqref{eq:beta-pf-covariance} is $0$.

\item
The corresponding order parameter in \eqref{expedition0012437}, \eqref{expedition0012438}, or \eqref{expedition0012439} converges to $1$.
\end{enumerate}
Furthermore, when these equivalent conditions hold, the corresponding BEC direction set and BEC ideal defined in Definition \ref{expedition0012163} are trivial.
\end{thm}

The construction and proof for the Nelson model are given in Definition \ref{expedition0012159} and Theorem \ref{expedition0012160}. For the spinless Pauli--Fierz model, Definition \ref{expedition0012168} and Theorem \ref{expedition0012169} are the corresponding results, and for the Pauli--Fierz model with spin, Definition \ref{expedition0012269} and Theorem \ref{expedition0012270} are the corresponding results. The vanishing of the BEC direction set and the BEC ideal in the resolvent algebra is discussed in Proposition \ref{expedition0012164}.

\section{Nelson Model}\label{nelson-model}

We first discuss bounded systems and then take the infinite-volume limit. Unlike the Pauli--Fierz models, which require adjustments for particle spin and the radiation field, the Nelson model can directly apply the setting introduced in Section \ref{expedition0012133}.

\subsection{Definition of the Hamiltonian}\label{definition-of-the-hamiltonian}

We define the cutoff Nelson model in infinite volume and reduce it to the bounded-system setting of Section \ref{expedition0012133}. For the infinite-volume total Hilbert space, isomorphisms such as \[\sphilb{H}_{\txtnelson}
= \sphilb{H}_{\txtparticle} \otimes \fun{\spfock_{\txtbsn}}{\sphilb{H}_{\txtbsn}}
= \fun{\lp^{2}}{\fldreal^{d}; \fun{\spfock_{\txtbsn}}{\sphilb{H}_{\txtbsn}}}
= \int_{\fldreal^{d}} \fun{\spfock_{\txtbsn}}{\sphilb{H}_{\txtbsn}} \opdmsr{x}\] will be used without further mention. The physical bosonic one-particle subspace in the Nelson model, assuming the point-source limit, is \(\sphilb{D}_{\txtnelson,\txtbsn,\txtphys,\sminvtemperature}
=
\sphilb{D}_{\txtbsn,\txtphys,\sminvtemperature}
=
\dom \mathsf{m} \cap \sphilb{D}_{\txtbsn,0,\sminvtemperature}\).

For any \(x
\in\fldreal^d\), set \(\faftr{\lambda}_{x,\kappa,\Lambda}(k)
=
\frac{\faftr{\varrho}_{\kappa,\Lambda}(k)}{\sqrt{\omega(k)}}
\napiernum^{-\imunit kx}\). For the fiberwise interaction Hamiltonian \(\physham_{\txtinteraction,\kappa,\Lambda}(x)
=
\opfocksegal_{\txtfock}(\lambda_{x,\kappa,\Lambda})\), define the interaction on the total Hilbert space by \(\physham_{\txtinteraction,\kappa,\Lambda}
=
\int_{\fldreal^{d}}^{\oplus}
\physham_{\txtinteraction,\kappa,\Lambda}(x)
\opdmsr{x}\), and define the cutoff Nelson Hamiltonian of the infinite system by \[\physham_{\txtnelson,\kappa,\Lambda}
=
\physham[h]_{\txtparticle} \otimes 1
+1 \otimes \physham_{\txtbsn,\txtfr}
+\physcplconst \physham_{\txtinteraction,\kappa,\Lambda}.\] For Hamiltonians with cutoffs removed, we use the notation obtained by removing \(\kappa,\Lambda\), namely \(\physham_{\txtnelson,\Lambda}\), \(\physham_{\txtnelson,\kappa}\), \(\physham_{\txtnelson}\).

Next we define the corresponding objects in the bounded system. Following the bounded-system setting of Section \ref{expedition0012133}, project the source on \(\fldreal^d\) and set \(\lambda_{x,\kappa,\Lambda,L}
=
P_{L}\lambda_{x,\kappa,\Lambda}\). The total Hilbert space of the bounded system is \(\sphilb{H}_{\txtnelson,L}
=
\sphilb{H}_{\txtparticle}\otimes\spfock_{\txtbsn,L}\). For the fiberwise \(\physham_{\txtinteraction,\kappa,\Lambda,L}(x)
=
\opfocksegal_{\txtfock}(\lambda_{x,\kappa,\Lambda,L})\), define the interaction on the total Hilbert space by \(\physham_{\txtinteraction,\kappa,\Lambda,L}
=
\int_{\fldreal^d}^{\oplus}
\physham_{\txtinteraction,\kappa,\Lambda,L}(x)\opdmsr{x}\). The cutoff Nelson Hamiltonian of the bounded system and the Nelson Hamiltonian with chemical potential are defined by \[\begin{aligned}
\physham_{\txtnelson,\kappa,\Lambda,L}
&=
\physham[h]_{\txtparticle} \otimes 1
+1 \otimes \physham_{\txtbsn,\txtfr,L}
+\physcplconst \physham_{\txtinteraction,\kappa,\Lambda,L},
\\ %%%%%%%%%%%%%%%%
\physham_{\txtnelson,\kappa,\Lambda,\smchemicalpotential,L}
&=
\physham[h]_{\txtparticle} \otimes 1
+1 \otimes \physham_{\txtbsn,\txtfr,\smchemicalpotential,L}
+\physcplconst \physham_{\txtinteraction,\kappa,\Lambda,L}.
\end{aligned}\]

\subsection{KMS States with Infrared and Ultraviolet Cutoffs}\label{expedition0012141}

The description in Subsection \ref{expedition0012146} describes the infinite system. As discussed in \cite{AsaoArai28,YoshitsuguSekine004,YoshitsuguSekine006}, the infinite-volume limit for the free Hamiltonian can be treated in the same way as the limit of the free Bose gas: in particular, if one takes the limit \(\smchemicalpotential
\uparrow 0\) after \(L
\to \infty\), the zero-mode component of BEC appears.

We again recall the basic setting for describing the interacting system. The free Hamiltonian \(\physham_{\txtnelson,0}
=
\physham[h]_{\txtparticle} \otimes \idone
+\idone \otimes \physham_{\txtbsn,\txtfr}\) is associated with the \(\sminvtemperature\)-Markov path space \[\begin{aligned}
\pairbk{\prbqspace_{\txtnelson,\sminvtemperature},
\mblfml{S}_{\txtnelson,\sminvtemperature},
\mblfml{S}_{\txtnelson,0,\sminvtemperature},
U_{\txtnelson,t},
R_{\txtnelson},
\msrprb_{\txtnelson,0,\sminvtemperature}}.
\end{aligned}\] The sample space is \(\prbqspace_{\txtnelson,\sminvtemperature}
= \Omega_{\txtparticle,\sminvtemperature} \times \prbqspace_{\txtbsn,\sminvtemperature}\), the \(\sigma\)-algebras are \(\mblfml{S}_{\txtnelson,\sminvtemperature}
= \mblfml{S}_{\txtparticle,\sminvtemperature} \times \mblfml{S}_{\txtbsn,\sminvtemperature}\) and \(\mblfml{S}_{\txtnelson,0,\sminvtemperature}
= \mblfml{S}_{\txtparticle,0,\sminvtemperature} \times \mblfml{S}_{\txtbsn,0,\sminvtemperature}\), the time evolution is \(U_{\txtnelson,t}
= U_{\txtparticle,t} \otimes U_{\txtbsn,t}\), the time reflection is \(R_{\txtnelson}
= R_{\txtparticle} \otimes R_{\txtbsn}\), and the probability measure is defined by \(\msrprb_{\txtnelson,0,\sminvtemperature}
= \msrprb_{\txtparticle,\sminvtemperature} \otimes \msrprb_{\txtbsn,\txtfr,\sminvtemperature}\).

As also defined in Subsection \ref{expedition0012146}, for any measurable \(I
\subset S_{\sminvtemperature}\), define the source on particle paths corresponding to the cutoff source by \begin{equation}\label{expedition0012203}
\begin{aligned}
\mathsf{J}_{I,\kappa,\Lambda}(\prbprocess)
=
\int_{I}
\fun{j_t}
{\lambda_{\prbprocess_{t},\kappa,\Lambda}}
\opdmsr{t}.
\end{aligned}
\end{equation} Each \(\lambda_{x,\kappa,\Lambda}\) satisfies \(\faftr{\lambda}_{x,\kappa,\Lambda}(0)=0\) because of the infrared cutoff. Therefore, for any \(f
\in \sphilb{D}_{\txtbsn,\txtphys,\sminvtemperature}\) and any interval \(I
\subset S_{\sminvtemperature}\), \begin{equation}\label{expedition0012151}
\begin{aligned}
\fun{\opform{q}_{\txtbsn,0,\sminvtemperature}^{\txteuclid}}{j_t f,j_s\lambda_{x,\kappa,\Lambda}}
&=0,
\\ %%%%%%%%%%%%%%%%
\fun{\opform{q}_{\txtbsn,0,\sminvtemperature}^{\txteuclid}}{j_t\lambda_{x,\kappa,\Lambda},j_s\lambda_{x,\kappa,\Lambda}}
&=
0,
\\ %%%%%%%%%%%%%%%%
\fun{\opform{q}_{\txtbsn,\txtbec,\sminvtemperature}^{\txteuclid}}{j_t f,\mathsf{J}_{I,\kappa,\Lambda}(\prbprocess)}
&=
\fun{\opform{q}_{\txtbsn,\txtnonzero,\sminvtemperature}^{\txteuclid}}{j_t f,\mathsf{J}_{I,\kappa,\Lambda}(\prbprocess)},
\\ %%%%%%%%%%%%%%%%
\fun{\opform{q}_{\txtbsn,\txtbec,\sminvtemperature}^{\txteuclid}}{\mathsf{J}_{I,\kappa,\Lambda}(\prbprocess)}
&=
\fun{\opform{q}_{\txtbsn,\txtnonzero,\sminvtemperature}^{\txteuclid}}{\mathsf{J}_{I,\kappa,\Lambda}(\prbprocess)}
\end{aligned}
\end{equation} holds.

The Feynman--Kac--Nelson kernel of the infinite system is defined by \begin{equation}\label{expedition0012338}
\begin{aligned}
F_{\txtnelson,I,\kappa,\Lambda}(\prbprocess,\opfocksegal)
=
\fnexp{-\physcplconst
\fun{\opfocksegal}{\mathsf{J}_{I,\kappa,\Lambda}(\prbprocess)}}.
\end{aligned}
\end{equation} We first check that this is well-defined. Notice that the following statement applies to both the bounded and infinite systems.

\begin{prop}\label{expedition0012137}
Define the constant depending on the cutoff variables by
$$c_{\sminvtemperature,\kappa,\Lambda}
=
\frac{2}{1 - \napiernum^{-\sminvtemperature \kappa}}
\int_{\kappa \leq \abs{k} \leq \Lambda}
\frac{\opdmsr{k}}{\omega(k)}
=
\frac{2\absvol{\mansphere^{d-1}}}{1 - \napiernum^{-\sminvtemperature \kappa}}
\frac{1}{d-1}
\rbk{\Lambda^{d-1} - \kappa^{d-1}}.$$
Then, for any cutoff variables $0
< \kappa
< \Lambda
< \infty$ and any closed interval $I$,
$$\mathsf{J}_{I,\kappa,\Lambda}
\in \opformdomain(\opform{q}_{\txtbsn,\txtbec,\sminvtemperature}^{\txteuclid}),
\quad
\opform{q}_{\txtbsn,\txtbec,\sminvtemperature}^{\txteuclid}
(\mathsf{J}_{I,\kappa,\Lambda}(\prbprocess))
\leq
\absvol{I}^2
c_{\sminvtemperature,\kappa,\Lambda}$$
holds.
In particular, $\opfocksegal(\mathsf{J}_{I,\kappa,\Lambda}(\prbprocess))$ is a centered Gaussian random variable.
\end{prop}

\begin{proof}
First we show that $\mathsf{J}_{I,\kappa,\Lambda}
\in \opformdomain(\opform{q}_{\txtbsn,\txtbec,\sminvtemperature}^{\txteuclid})$.
By \eqref{expedition0012151},
$$\begin{aligned}
\fun{\opform{q}_{\txtbsn,\txtbec,\sminvtemperature}^{\txteuclid}}{\mathsf{J}_{I,\kappa,\Lambda}(\prbprocess)}
=
\fun{\opform{q}_{\txtbsn,\txtnonzero,\sminvtemperature}^{\txteuclid}}{\mathsf{J}_{I,\kappa,\Lambda}(\prbprocess)}
\end{aligned}$$
so it suffices to estimate the non-zero-mode part.
For the integrand,
$$\begin{aligned}
&\abs{\fun{\opform{q}_{\txtbsn,\txtbec,\sminvtemperature}^{\txteuclid}}
{\fun{j_{s}}{\lambda_{\prbprocess_{s},\kappa,\Lambda}},
\fun{j_{u}}{\lambda_{\prbprocess_{u},\kappa,\Lambda}}}}
\leq
\int_{\kappa \leq \abs{k} \leq \Lambda}
\frac{\napiernum^{-\abs{s-u} \omega(k)} + \napiernum^{-\rbk{\sminvtemperature - \abs{s-u}} \omega(k)}}
{1 - \napiernum^{-\sminvtemperature \omega(k)}}
\frac{1}
{\omega(k)}
\opdmsr{k}
\\ %%%%%%%%%%%%%%%%
&\leq
\frac{2}{1 - \napiernum^{-\sminvtemperature \kappa}}
\int_{\kappa \leq \abs{k} \leq \Lambda}
\frac{\opdmsr{k}}{\omega(k)}
=
\frac{2 \absvol{\mansphere^{d-1}}}{1 - \napiernum^{-\sminvtemperature \kappa}}
\int_{\kappa \leq r \leq \Lambda}
r^{d-2} \opdmsr{r}
=
\frac{2\absvol{\mansphere^{d-1}}}{1 - \napiernum^{-\sminvtemperature \kappa}}
\frac{1}{d-1}
\rbk{\Lambda^{d-1} - \kappa^{d-1}}
<
\infty
\end{aligned}$$
holds.
This estimate is independent of $\prbprocess$.
Therefore $\opfocksegal(\mathsf{J}_{I,\kappa,\Lambda}(\prbprocess))$ is defined as a centered Gaussian random variable.
The treatment of the time integral is clear.
\end{proof}

To define the KMS state perturbatively, we further examine the Feynman--Kac--Nelson kernel \eqref{expedition0012338}. Proposition \ref{expedition0012154} can also be applied to the bounded system without change.

\begin{prop}\label{expedition0012154}
For fixed $\kappa,\Lambda$, the family $\fml{F_{\txtnelson,I,\kappa,\Lambda}}
{I \subset S_{\sminvtemperature}}$ defined by \eqref{expedition0012338} is a local Feynman--Kac--Nelson perturbation \cite[Chapter 21]{DerezinskiGerard001}.
In particular, it is adapted to the $\sigma$-algebra of the interval $I$, satisfies
$F_{\txtnelson,I\cup J,\kappa,\Lambda}
=
F_{\txtnelson,I,\kappa,\Lambda}
F_{\txtnelson,J,\kappa,\Lambda}$ for disjoint intervals, is covariant under time translations and reflections, belongs to $\fun{\lp^{p}}
{\prbqspace_{\txtnelson,\sminvtemperature},
\msrprb_{\txtnelson,0,\sminvtemperature}}$ for every $1 \leq p < \infty$, and is $\lp^{p}$-continuous with respect to the interval endpoints.

Moreover, for any $0
< t
< \frac{\sminvtemperature}{2}$,
$$\begin{aligned}
\sphilb{D}_{t,\kappa,\Lambda}^{0}
=
\prbexp_{0,\frac{\sminvtemperature}{2}}
\fun{\splinspan}
{\bigcup_{0 \leq s < \frac{\sminvtemperature}{2} - t}
F_{\txtnelson,\closedinterval{0}{s},\kappa,\Lambda}
\fun{\lp^{\infty}}{\prbqspace_{\txtnelson,\sminvtemperature},
\mblfml{S}_{\closedinterval{0}{\frac{\sminvtemperature}{2}-t}}}}
\end{aligned}$$
is defined, then for any $0
\leq s \leq t \leq \frac{\sminvtemperature}{2}$ and $f
\in \fun{\lp^{2}}{\prbqspace_{\txtnelson,\sminvtemperature},
\msrprb_{\txtnelson,0,\sminvtemperature}}$, there exists a unique
$P_{\kappa,\Lambda,s}^{0}
\colon
\sphilb{D}_{t,\kappa,\Lambda}^{0}
\to \sphilb{D}_{t-s,\kappa,\Lambda}^{0}$ satisfying
$$P_{\kappa,\Lambda,s}^{0}
\prbexp_{\setone{0,\frac{\sminvtemperature}{2}}}
f
=
\prbexp_{\setone{0,\frac{\sminvtemperature}{2}}}
F_{\txtnelson,\closedinterval{0}{s},\kappa,\Lambda}
U_{\txtnelson,s}
f.$$
In particular, $\fml{P_{\kappa,\Lambda,t}^{0}}
{t \in \closedinterval{0}{\frac{\sminvtemperature}{2}}}$ is a local Hermitian semigroup.
\end{prop}

\begin{proof}
(Measurability): The function $F_{\txtnelson,I,\kappa,\Lambda}$ in \eqref{expedition0012338} depends only on the particle path $\prbprocess_s$ with $s\in I$ and the field $\opfocksegal(j_s \cdot)$.
Thus $F_{\txtnelson,I,\kappa,\Lambda}$ is measurable with respect to the interval $\sigma$-algebra of the total system generated by $I$.

(Multiplicativity): By additivity of the integration interval, for disjoint intervals $I$ and $J$,
$\mathsf{J}_{I \cup J,\kappa,\Lambda}(\prbprocess)
=
\mathsf{J}_{I,\kappa,\Lambda}(\prbprocess)
+\mathsf{J}_{J,\kappa,\Lambda}(\prbprocess)$ holds.
Therefore the exponentiated equality
$F_{\txtnelson,I \cup J,\kappa,\Lambda}
=
F_{\txtnelson,I,\kappa,\Lambda}
F_{\txtnelson,J,\kappa,\Lambda}$ holds pointwise.

(Time-translation covariance): The time translation of the total system is $U_{\txtnelson,t}
= U_{\txtparticle,t} \otimes U_{\txtbsn,t}$.
By time translation of the particle path and the sharp-time field,
$$\int_I j_s \lambda_{\prbprocess_{s+t},\kappa,\Lambda}
\opdmsr{s}
=
U_{\txtbsn,-t}
\int_{I+t} j_r \lambda_{\prbprocess_r,\kappa,\Lambda}
\opdmsr{r}$$
holds, and hence $U_{\txtnelson,t}F_{\txtnelson,I,\kappa,\Lambda}
=F_{\txtnelson,I+t,\kappa,\Lambda}$ follows.

(Reflection): The time reflection of the total system is $R_{\txtnelson}
= R_{\txtparticle} \otimes R_{\txtbsn}$.
By reflection of the particle path and of the sharp-time field,
$$\int_I j_s \lambda_{\prbprocess_{-s},\kappa,\Lambda}
\opdmsr{s}
=
R_{\txtbsn}
\int_{-I} j_r \lambda_{\prbprocess_r,\kappa,\Lambda}
\opdmsr{r}$$
holds.
Thus the reflection property of the Feynman--Kac--Nelson kernel follows on the total system as $R_{\txtnelson} F_{\txtnelson,I,\kappa,\Lambda}
= F_{\txtnelson,-I,\kappa,\Lambda}$.

(Exponential integrability): For any $p
\geq 1$, once the particle path is fixed, $\fun{\opfocksegal}
{\mathsf{J}_{I,\kappa,\Lambda}(\prbprocess)}$ is a centered Gaussian random variable.
By the exponential integral formula for centered Gaussian random variables,
$$\sqfun{\prbexp_{\msrprb_{\txtbsn,\txtfr,\sminvtemperature}}}
{\fnexp{-p \physcplconst \opfocksegal(\mathsf{J}_{I,\kappa,\Lambda}(\prbprocess))}}
=
\fnexp{\frac{p^2 \physcplconst^2}{4}
\fun{\opform{q}_{\txtbsn,\txtbec,\sminvtemperature}^{\txteuclid}}
{\mathsf{J}_{I,\kappa,\Lambda}(\prbprocess)}}$$
holds.
By Proposition \ref{expedition0012137}, $\fun{\opform{q}_{\txtbsn,\txtbec,\sminvtemperature}^{\txteuclid}}
{\mathsf{J}_{I,\kappa,\Lambda}(\prbprocess)}
\leq \absvol{I}^2 c_{\sminvtemperature,\kappa,\Lambda}$ holds, so
$\pnorm{F_{\txtnelson,I,\kappa,\Lambda}}^p
\leq
\fnexp{\frac{p^2 \physcplconst^2}{4}
\absvol{I}^2
c_{\sminvtemperature,\kappa,\Lambda}}
< \infty$ follows.

(Continuity with respect to interval endpoints): Here let $I
= \closedinterval{a}{b}$ and $\ep
> 0$.
Then
$\mathsf{J}_{\closedinterval{a}{b+\ep},\kappa,\Lambda}(\prbprocess)
-\mathsf{J}_{\closedinterval{a}{b},\kappa,\Lambda}(\prbprocess)
=
\mathsf{J}_{\closedinterval{b}{b+\ep},\kappa,\Lambda}(\prbprocess)$.
By Proposition \ref{expedition0012137} and Gaussian isometry,
$$\sqfun{\prbexp_{\msrprb_{\txtbsn,\txtfr,\sminvtemperature}}}
{\abs{\opfocksegal(\mathsf{J}_{\closedinterval{b}{b+\ep},\kappa,\Lambda})}^2}
=
\onehalf
\fun{\opform{q}_{\txtbsn,\txtbec,\sminvtemperature}^{\txteuclid}}
{\mathsf{J}_{\closedinterval{b}{b+\ep},\kappa,\Lambda}(\prbprocess)}
\leq
\onehalf \ep^2 c_{\sminvtemperature,\kappa,\Lambda}$$
holds.
Thus the field on the small interval converges to $0$ in $\lp^2$ as $\ep\to0$.
The preceding exponential moment estimate gives a uniform bound on the exponential moments of the small interval, so Hölder's inequality and Vitali's theorem imply $F_{\txtnelson,\closedinterval{a}{b+\ep},\kappa,\Lambda}
\to F_{\txtnelson,\closedinterval{a}{b},\kappa,\Lambda}$ in every $\lp^p$.
The construction of the local Hermitian semigroup follows from \cite[Proposition 21.62]{DerezinskiGerard001}.
\end{proof}

Following the textbook \cite{DerezinskiGerard001}, the KMS state of the cutoff Nelson model is defined by \begin{equation}\label{expedition0012152}
\begin{aligned}
\oastate[\psi_{\txtnelson,\sminvtemperature,\kappa,\Lambda}](F)
&=
\frac{1}{\smpartitionfunc_{\txtnelson,\sminvtemperature,\kappa,\Lambda}}
\int_{\prbqspace_{\txtnelson,\sminvtemperature}}
F(\prbprocess,\opfocksegal)
\fnexp{-\physcplconst
\fun{\opfocksegal}{\mathsf{J}_{S_{\sminvtemperature},\kappa,\Lambda}(\prbprocess)}}
\opdmsr{\msrprb_{\txtnelson,0,\sminvtemperature}(\prbprocess,\opfocksegal)},
\\ %%%%%%%%%%%%%%%%
\smpartitionfunc_{\txtnelson,\sminvtemperature,\kappa,\Lambda}
&=
\int_{\prbqspace_{\txtnelson,\sminvtemperature}}
\fnexp{-\physcplconst
\fun{\opfocksegal}{\mathsf{J}_{S_{\sminvtemperature},\kappa,\Lambda}(\prbprocess)}}
\opdmsr{\msrprb_{\txtnelson,0,\sminvtemperature}(\prbprocess,\opfocksegal)}.
\end{aligned}
\end{equation}

\subsection{KMS States of the Bounded System with Chemical Potential}\label{kms-states-of-the-bounded-system-with-chemical-potential}

We use \(\msrprb_{\txtbsn,\txtfr,\sminvtemperature,\smchemicalpotential,L}\) and \(\opform{q}_{\txtbsn,\txtbec,\sminvtemperature,\smchemicalpotential,L}^{\txteuclid}\) fixed in the bounded-system setting of Subsection \ref{expedition0012145}. Furthermore, the Feynman--Kac--Nelson kernel of the bounded system may be taken as the analogue of \eqref{expedition0012338}, \[\begin{aligned}
F_{\txtnelson,I,\kappa,\Lambda,L}(\prbprocess,\opfocksegal)
=
\fnexp{-\physcplconst
\fun{\opfocksegal}{\mathsf{J}_{I,\kappa,\Lambda,L}(\prbprocess)}}.
\end{aligned}\]

The finite-temperature Feynman--Kac--Nelson formula is expressed, for the probability measure determined by \(\physham_{\txtnelson,\kappa,\Lambda,\smchemicalpotential,L}\), as the perturbation by \(F_{\txtnelson,S_{\sminvtemperature},\kappa,\Lambda,L}\). For a bounded cylinder function \(F\), define the KMS state and the partition function by \begin{equation}\label{expedition0012341}
\begin{aligned}
\psi_{\txtnelson,\sminvtemperature,\kappa,\Lambda,\smchemicalpotential,L}^{\txteuclid}(F)
&=
\frac{1}{\smpartitionfunc_{\txtnelson,\sminvtemperature,\kappa,\Lambda,\smchemicalpotential,L}}
\int_{\prbqspace_{\txtnelson,\sminvtemperature,L}}
F(\prbprocess,\opfocksegal)
F_{\txtnelson,S_{\sminvtemperature},\kappa,\Lambda,L}(\prbprocess,\opfocksegal)
\opdmsr{\msrprb_{\txtnelson,0,\sminvtemperature,\smchemicalpotential,L}}(\prbprocess,\opfocksegal),
\\ %%%%%%%%%%%%%%%%
\smpartitionfunc_{\txtnelson,\sminvtemperature,\kappa,\Lambda,\smchemicalpotential,L}
&=
\int_{\prbqspace_{\txtnelson,\sminvtemperature,L}}
F_{\txtnelson,S_{\sminvtemperature},\kappa,\Lambda,L}(\prbprocess,\opfocksegal)
\opdmsr{\msrprb_{\txtnelson,0,\sminvtemperature,\smchemicalpotential,L}(\prbprocess,\opfocksegal)}.
\end{aligned}
\end{equation} The definition of cylinder functions is given in the discussion after the statement of Proposition \ref{expedition0012139}. This is the functional integral representation of the bounded-system KMS state associated with \(\physham_{\txtnelson,\kappa,\Lambda,\smchemicalpotential,L}\).

\begin{prop}[Characteristic function of the interacting KMS state in the bounded system]\label{expedition0012342}
Fix cutoff variables $0
< \kappa
< \Lambda
< \infty$ and $L
> 0$, and define the weighted particle loop measure of the bounded system by
\begin{equation}\label{expedition0012343}
\begin{aligned}
\opdmsr{\widetilde{\msrprb}_{\txtnelson,\txtparticle,\sminvtemperature,\kappa,\Lambda,\smchemicalpotential,L}}(\prbprocess)
&=
\frac{1}
{\smpartitionfunc_{\txtnelson,\sminvtemperature,\kappa,\Lambda,\smchemicalpotential,L}}
\fnexp{\frac{\physcplconst^{2}}{4}
\fun{\opform{q}_{\txtbsn,\txtbec,\sminvtemperature,\smchemicalpotential,L}^{\txteuclid}}
{\mathsf{J}_{S_{\sminvtemperature},\kappa,\Lambda,L}(\prbprocess)}}
\opdmsr{\msrprb_{\txtparticle,\sminvtemperature}(\prbprocess)}.
\end{aligned}
\end{equation}
For any $f
\in \sphilb{D}_{\txtbsn,\txtphys,\sminvtemperature}$,
$t\in S_{\sminvtemperature}$,
$s\in\fldreal$, define
\begin{equation}\label{expedition0012344}
\begin{aligned}
\mathsf{S}_{\txtnelson,\sminvtemperature,\kappa,\Lambda,\smchemicalpotential,L;t}^{(1)}(s f)
&=
\int_{\Omega_{\txtparticle,\sminvtemperature}}
\fnexp{\frac{\imunit s\physcplconst}{2}
\fun{\opform{q}_{\txtbsn,\txtbec,\sminvtemperature,\smchemicalpotential,L}^{\txteuclid}}
{j_t P_{L} f,
\mathsf{J}_{S_{\sminvtemperature},\kappa,\Lambda,L}(\prbprocess)}}
\opdmsr{\widetilde{\msrprb}_{\txtnelson,\txtparticle,\sminvtemperature,\kappa,\Lambda,\smchemicalpotential,L}}(\prbprocess)
\\ %%%%%%%%%%%%%%%%
&=
\int_{\Omega_{\txtparticle,\sminvtemperature}}
\fnexp{\frac{\imunit s\physcplconst}{2}
\fun{\opform{q}_{\txtbsn,\txtnonzero,\sminvtemperature,\smchemicalpotential,L}^{\txteuclid}}
{j_t P_{L} f,
\mathsf{J}_{S_{\sminvtemperature},\kappa,\Lambda,L}(\prbprocess)}}
\opdmsr{\widetilde{\msrprb}_{\txtnelson,\txtparticle,\sminvtemperature,\kappa,\Lambda,\smchemicalpotential,L}
(\prbprocess)}.
\end{aligned}
\end{equation}
Then the characteristic function of the interacting KMS state for the Segal field is
\begin{equation}\label{expedition0012345}
\begin{aligned}
\fun{\psi_{\txtnelson,\sminvtemperature,\kappa,\Lambda,\smchemicalpotential,L}^{\txteuclid}}
{\fnexp{\imunit s\opfocksegal(j_t P_{L} f)}}
=
\fnexp{-\frac{s^{2}}{4}
\fun{\opform{q}_{\txtbsn,\txtbec,\sminvtemperature,\smchemicalpotential,L}^{\txteuclid}}
{j_t P_{L} f}}
\mathsf{S}_{\txtnelson,\sminvtemperature,\kappa,\Lambda,\smchemicalpotential,L;t}^{(1)}(s f).
\end{aligned}
\end{equation}
By definition, $\abs{\mathsf{S}_{\txtnelson,\sminvtemperature,\kappa,\Lambda,\smchemicalpotential,L;t}^{(1)}(s f)}
\leq 1$.
Thus the correction due to the interaction appears only through the non-zero-mode form $\opform{q}_{\txtbsn,\txtnonzero,\sminvtemperature,\smchemicalpotential,L}^{\txteuclid}$ and the weighted particle loop measure \eqref{expedition0012343}.
In particular, the estimate of the one-point characteristic function in the bounded system reduces to an estimate on weighted particle paths, as in the van Hove model and the spin-boson model.
\end{prop}

\begin{proof}
For brevity of notation, use $h_{t,L}
= j_t P_{L} f$ and $\mathsf{J}_{L}(\prbprocess)
= \mathsf{J}_{S_{\sminvtemperature},\kappa,\Lambda,L}(\prbprocess)$.
By the defining formula \eqref{expedition0012341}, we obtain
\begin{equation}\label{expedition0012346}
\begin{aligned}
&\fun{\psi_{\txtnelson,\sminvtemperature,\kappa,\Lambda,\smchemicalpotential,L}^{\txteuclid}}
{\fnexp{\imunit s\opfocksegal(h_{t,L})}}
\\ %%%%%%%%%%%%%%%%
&=
\frac{1}
{\smpartitionfunc_{\txtnelson,\sminvtemperature,\kappa,\Lambda,\smchemicalpotential,L}}
\int_{\Omega_{\txtparticle,\sminvtemperature}}
\int_{\prbqspace_{\txtbsn,\sminvtemperature,L}}
\fnexp{\imunit s\opfocksegal(h_{t,L})
-\physcplconst\fun{\opfocksegal}{\mathsf{J}_{L}(\prbprocess)}}
\opdmsr{\msrprb_{\txtbsn,\txtfr,\sminvtemperature,\smchemicalpotential,L}}(\opfocksegal)
\opdmsr{\msrprb_{\txtparticle,\sminvtemperature}}(\prbprocess).
\end{aligned}
\end{equation}
Fix the particle path $\prbprocess$.
Regard the bounded-system Gaussian characteristic functional \eqref{expedition0012340} as the exponential-moment formula for finitely many centered real Gaussian variables.
Applying this formula to the linear functional $\imunit s h_{t,L}-\physcplconst\mathsf{J}_{L}(\prbprocess)$ gives
\begin{equation}\label{expedition0012347}
\begin{aligned}
&\int_{\prbqspace_{\txtbsn,\sminvtemperature,L}}
\fnexp{\imunit s\opfocksegal(h_{t,L})
-\physcplconst\fun{\opfocksegal}{\mathsf{J}_{L}(\prbprocess)}}
\opdmsr{\msrprb_{\txtbsn,\txtfr,\sminvtemperature,\smchemicalpotential,L}}(\opfocksegal)
\\ %%%%%%%%%%%%%%%%
&=
\fnexp{-\frac{s^{2}}{4}
\fun{\opform{q}_{\txtbsn,\txtbec,\sminvtemperature,\smchemicalpotential,L}^{\txteuclid}}
{h_{t,L}}
+\frac{\imunit s\physcplconst}{2}
\fun{\opform{q}_{\txtbsn,\txtbec,\sminvtemperature,\smchemicalpotential,L}^{\txteuclid}}
{h_{t,L},\mathsf{J}_{L}(\prbprocess)}
+\frac{\physcplconst^{2}}{4}
\fun{\opform{q}_{\txtbsn,\txtbec,\sminvtemperature,\smchemicalpotential,L}^{\txteuclid}}
{\mathsf{J}_{L}(\prbprocess)}}.
\end{aligned}
\end{equation}
Setting $s=0$ in the same formula, the partition function is represented as an integral over particle paths:
\begin{equation}\label{expedition0012348}
\begin{aligned}
\smpartitionfunc_{\txtnelson,\sminvtemperature,\kappa,\Lambda,\smchemicalpotential,L}
=
\int_{\Omega_{\txtparticle,\sminvtemperature}}
\fnexp{\frac{\physcplconst^{2}}{4}
\fun{\opform{q}_{\txtbsn,\txtbec,\sminvtemperature,\smchemicalpotential,L}^{\txteuclid}}
{\mathsf{J}_{L}(\prbprocess)}}
\opdmsr{\msrprb_{\txtparticle,\sminvtemperature}(\prbprocess)}.
\end{aligned}
\end{equation}
Because $\mathsf{J}_{L}(\prbprocess)$ has no zero-mode component by the infrared cutoff,
\begin{equation}\label{expedition0012349}
\begin{aligned}
\fun{\opform{q}_{\txtbsn,\txtbec,\sminvtemperature,\smchemicalpotential,L}^{\txteuclid}}
{h_{t,L},\mathsf{J}_{L}(\prbprocess)}
&=
\fun{\opform{q}_{\txtbsn,\txtnonzero,\sminvtemperature,\smchemicalpotential,L}^{\txteuclid}}
{h_{t,L},\mathsf{J}_{L}(\prbprocess)},
\\ %%%%%%%%%%%%%%%%
\fun{\opform{q}_{\txtbsn,\txtbec,\sminvtemperature,\smchemicalpotential,L}^{\txteuclid}}
{\mathsf{J}_{L}(\prbprocess)}
&=
\fun{\opform{q}_{\txtbsn,\txtnonzero,\sminvtemperature,\smchemicalpotential,L}^{\txteuclid}}
{\mathsf{J}_{L}(\prbprocess)}.
\end{aligned}
\end{equation}
Substituting \eqref{expedition0012347} into \eqref{expedition0012346} and normalizing by the partition function representation \eqref{expedition0012348}, we obtain \eqref{expedition0012345} through \eqref{expedition0012343} and \eqref{expedition0012344}.
The integrand in \eqref{expedition0012344} has absolute value $1$, and $\widetilde{\msrprb}_{\txtnelson,\txtparticle,\sminvtemperature,\kappa,\Lambda,\smchemicalpotential,L}$ is a probability measure, so
$\abs{\mathsf{S}_{\txtnelson,\sminvtemperature,\kappa,\Lambda,\smchemicalpotential,L;t}^{(1)}(s f)}
\leq1$.
Moreover, by \eqref{expedition0012349}, the zero-mode form does not appear in the interaction correction.
This proves the claim.
\end{proof}

The infinite-volume cutoff state is obtained for these finite-time correlation functions by taking \(L \to \infty\) and then \(\smchemicalpotential
\uparrow 0\).

In the following discussion, we consider only the infinite system.

\subsection{Removal of Infrared and Ultraviolet Cutoffs}\label{removal-of-infrared-and-ultraviolet-cutoffs}

We remove the infrared and ultraviolet cutoffs for finite field correlation functions obtained from the KMS state of the cutoff Nelson model. In particular, we must discuss the limit of the dynamics and the existence of KMS states for that limit. When discussing cutoff objects, the cutoff variables are fixed. To simplify the discussion, in the point-source limit and in the construction of the renormalized local semigroup we take the spatial dimension to be \(d
= 3\). The only place where three-dimensionality is used essentially in this subsection is the cutoff removal for the point-source singular part, and from Proposition \ref{expedition0012157} onward we may use the cutoff-removal limit as an assumption.

The singular Gaussian \(\sminvtemperature\)-Markov path space for the free Hamiltonians of the particle and the Bose field was defined in Subsection \ref{expedition0012141}.

\subsubsection{Renormalized Kernel and Generator Representation}\label{renormalized-kernel-and-generator-representation}

For the Feynman--Kac--Nelson kernel \(F_{\txtnelson,I,\kappa,\Lambda}(\prbprocess,\opfocksegal)\) defined by \eqref{expedition0012338}, following the textbook \cite[p.513]{LorincziHiroshimaBetz3}, use the constant \(\physenergy_{\txtnelson,\kappa,\Lambda}^{\txtrenormalization}\) correcting the ultraviolet divergence and define \begin{equation}\label{expedition0012177}
\begin{aligned}
F_{\txtnelson,\closedinterval{a}{b},\kappa,\Lambda}^{\txtrenormalization}
=
\napiernum^{(b-a) \physenergy_{\txtnelson,\kappa,\Lambda}^{\txtrenormalization}}
F_{\txtnelson,\closedinterval{a}{b},\kappa,\Lambda},
\quad
\physenergy_{\txtnelson,\kappa,\Lambda}^{\txtrenormalization}
=
-\frac{\physcplconst^2}{2}
\int_{\kappa \leq \abs{k} \leq \Lambda}
\frac{1}{\omega(k)}
\frac{1}{\omega(k) + \onehalf \omega(k)^2}
\opdmsr{k}
\end{aligned}
\end{equation}

\begin{prop}\label{expedition0012153}
For any $0
< \kappa
< \Lambda
< \infty$, the family $\fml{F_{\txtnelson,\closedinterval{a}{b},\kappa,\Lambda}^{\txtrenormalization}}
{\closedinterval{a}{b} \subset S_{\sminvtemperature}}$ is a local Feynman--Kac--Nelson kernel \cite{DerezinskiGerard001}.
In particular, it satisfies interval measurability, multiplicativity, reflection, $\lp^{p}$-continuity with respect to interval endpoints, and exponential integrability.

Moreover, for any $0
< t
< \frac{\sminvtemperature}{2}$, define
\begin{equation}\label{expedition0012178}
\begin{aligned}
\sphilb{D}_{t,\kappa,\Lambda}^{\txtrenormalization}
=
\prbexp_{0,\frac{\sminvtemperature}{2}}
\fun{\splinspan}
{\bigcup_{0 \leq s < \frac{\sminvtemperature}{2} - t}
F_{\txtnelson,\closedinterval{0}{s},\kappa,\Lambda}^{\txtrenormalization}
\fun{\lp^{\infty}}{\prbqspace_{\txtnelson,\sminvtemperature},
\mblfml{S}_{\closedinterval{0}{\frac{\sminvtemperature}{2}-t}}}}.
\end{aligned}
\end{equation}
Then, for any $0
\leq s \leq t \leq \frac{\sminvtemperature}{2}$ and $f
\in \fun{\lp^{2}}{\prbqspace_{\txtnelson,\sminvtemperature},
\msrprb_{\txtnelson,0,\sminvtemperature}}$,
\begin{equation}\label{expedition0012179}
\begin{aligned}
P_{\kappa,\Lambda,s}^{\txtrenormalization}
\prbexp_{\setone{0,\frac{\sminvtemperature}{2}}}
f
=
\prbexp_{\setone{0,\frac{\sminvtemperature}{2}}}
F_{\txtnelson,\closedinterval{0}{s},\kappa,\Lambda}^{\txtrenormalization}
U_{\txtnelson,s}
f,
\end{aligned}
\end{equation}
there exists a unique $P_{\kappa,\Lambda,s}^{\txtrenormalization}
\colon
\sphilb{D}_{t,\kappa,\Lambda}^{\txtrenormalization}
\to \sphilb{D}_{t-s,\kappa,\Lambda}^{\txtrenormalization}$ satisfying \eqref{expedition0012179}.
In particular, $\fml{P_{\kappa,\Lambda,t}^{\txtrenormalization}}
{t \in \closedinterval{0}{\frac{\sminvtemperature}{2}}}$ is a local Hermitian semigroup.
\end{prop}

\begin{proof}
Proposition \ref{expedition0012154} applies to the unrenormalized kernel $F_{\txtnelson,\closedinterval{a}{b},\kappa,\Lambda}$.
The renormalized kernel in \eqref{expedition0012177} only multiplies it by the deterministic scalar factor $\napiernum^{(b-a)\physenergy_{\txtnelson,\kappa,\Lambda}^{\txtrenormalization}}$, so interval measurability and exponential integrability are preserved immediately.
For adjacent intervals $I
= \closedinterval{a}{b}$ and $J
= \closedinterval{b}{c}$,
$$\napiernum^{(c-a) \physenergy_{\txtnelson,\kappa,\Lambda}^{\txtrenormalization}}
=
\napiernum^{(b-a)\physenergy_{\txtnelson,\kappa,\Lambda}^{\txtrenormalization}}
\napiernum^{(c-b)\physenergy_{\txtnelson,\kappa,\Lambda}^{\txtrenormalization}}$$
holds, so multiplicativity is also preserved.
Reflection follows from reflection invariance of the interval length.
$\lp^p$-continuity with respect to interval endpoints follows by combining the $\lp^p$-continuity of the unrenormalized kernel with the ordinary continuity of the scalar factor.
Therefore $\fml{F_{\txtnelson,\closedinterval{a}{b},\kappa,\Lambda}^{\txtrenormalization}}{}$ is a local Feynman--Kac--Nelson kernel.

The construction of the local Hermitian semigroup from the domain in \eqref{expedition0012178} and the action relation in \eqref{expedition0012179} follows from \cite[Proposition 21.62]{DerezinskiGerard001}.
\end{proof}

In the following discussion, it is enough to restrict to tensor-product-type functions corresponding to generators, rather than general \(F,G
\in
\fun{\lp^{2}}
{\prbqspace_{\txtnelson,\sminvtemperature},
\mblfml{S}_{\txtnelson,\sminvtemperature},
\msrprb_{\txtnelson,0,\sminvtemperature}}\).

For bounded time-zero particle functions \(F_{\txtparticle}\) and \(G_{\txtparticle}\) and one-particle-space elements \(f,g
\in \sphilb{D}_{\txtbsn,\txtphys,\sminvtemperature}\) of the Bose field, consider, as time-zero generators, \(F(\prbprocess,\opfocksegal)
=
F_{\txtparticle}(\prbprocess_{0})
\napiernum^{\imunit \opfocksegal(j_0 f)}\) and \(G(\prbprocess,\opfocksegal)
= G_{\txtparticle}(\prbprocess_0)
\napiernum^{\imunit \opfocksegal(j_0 g)}\).

\begin{prop}[Kernel representation on generators]
For the above $F,G$,
\begin{equation}\label{expedition0012180}
\begin{aligned}
&\bkt{F}
{P_{\kappa,\Lambda,t}^{\txtrenormalization} G}
=
\int_{\Omega_{\txtparticle,\sminvtemperature}}
\cmpconj{F_{\txtparticle}(\prbprocess_{0})}
G_{\txtparticle}(\prbprocess_{t})
\\ %%%%%%%%%%%%%%%%
&\quad\times
\fnexp{
t \physenergy_{\txtnelson,\kappa,\Lambda}^{\txtrenormalization}
-\oneoverfour
\opform{q}_{\txtbsn,\txtbec,\sminvtemperature}^{\txteuclid}(j_t g - j_0 f,j_t g - j_0 f)
-\frac{\imunit \physcplconst}{2}
\opform{q}_{\txtbsn,\txtbec,\sminvtemperature}^{\txteuclid}
(j_t g - j_0 f,\mathsf{J}_{\closedinterval{0}{t},\kappa,\Lambda}(\prbprocess))}
\\ %%%%%%%%%%%%%%%%
&\quad\times
\fnexp{\frac{\physcplconst^2}{4}
\opform{q}_{\txtbsn,\txtbec,\sminvtemperature}^{\txteuclid}
(\mathsf{J}_{\closedinterval{0}{t},\kappa,\Lambda}(\prbprocess),\mathsf{J}_{\closedinterval{0}{t},\kappa,\Lambda}(\prbprocess))}
\opdmsr{\msrprb_{\txtparticle,\sminvtemperature}(\prbprocess)}
\end{aligned}
\end{equation}
holds.
In particular, when $f = g = 0$, this reduces to a representation only by the effective weight on particle paths.
Furthermore, using \eqref{expedition0012151},
$$\begin{aligned}
\fun{\opform{q}_{\txtbsn,\txtbec,\sminvtemperature}^{\txteuclid}}
{j_t g - j_0 f,\mathsf{J}_{\closedinterval{0}{t},\kappa,\Lambda}(\prbprocess)}
&=
\fun{\opform{q}_{\txtbsn,\txtnonzero,\sminvtemperature}^{\txteuclid}}
{j_t g - j_0 f,\mathsf{J}_{\closedinterval{0}{t},\kappa,\Lambda}(\prbprocess)},
\\ %%%%%%%%%%%%%%%%
\fun{\opform{q}_{\txtbsn,\txtbec,\sminvtemperature}^{\txteuclid}}
{\mathsf{J}_{\closedinterval{0}{t},\kappa,\Lambda}(\prbprocess)}
&=
\fun{\opform{q}_{\txtbsn,\txtnonzero,\sminvtemperature}^{\txteuclid}}
{\mathsf{J}_{\closedinterval{0}{t},\kappa,\Lambda}(\prbprocess)}
\end{aligned}$$
shows that the zero-mode component does not appear in the parts depending on the cutoff source.
\end{prop}

\begin{proof}
For the time-zero generator $G$,
$P_{\kappa,\Lambda,t}^{\txtrenormalization} G
=
\prbexp_{\setone{0,\frac{\sminvtemperature}{2}}}
F_{\txtnelson,\closedinterval{0}{t},\kappa,\Lambda}^{\txtrenormalization}
U_{\txtnelson,t}G$.
Since $U_{\txtnelson,t}
=U_{\txtparticle,t} \otimes U_{\txtbsn,t}$,
$U_{\txtnelson,t} G
=
G_{\txtparticle}(\prbprocess_t)
\napiernum^{\imunit \opfocksegal(j_t g)}$ holds.
Moreover, recalling that $F_{\txtnelson,\closedinterval{0}{t},\kappa,\Lambda}^{\txtrenormalization}
=
\napiernum^{t\physenergy_{\txtnelson,\kappa,\Lambda}^{\txtrenormalization}}
\napiernum^{-\physcplconst
\opfocksegal(\mathsf{J}_{\closedinterval{0}{t},\kappa,\Lambda}(\prbprocess))}$, we obtain
$$\begin{aligned}
&\bkt{F}{P_{\kappa,\Lambda,t}^{\txtrenormalization}G}
\\ %%%%%%%%%%%%%%%%
&=
\int_{\Omega_{\txtparticle,\sminvtemperature}
\times \prbqspace_{\txtbsn,\sminvtemperature}}
\cmpconj{F_{\txtparticle}(\prbprocess_0)}
\napiernum^{-\imunit\opfocksegal(j_0 f)}
\napiernum^{t\physenergy_{\txtnelson,\kappa,\Lambda}^{\txtrenormalization}}
\napiernum^{-\physcplconst
\opfocksegal(\mathsf{J}_{\closedinterval{0}{t},\kappa,\Lambda}(\prbprocess))}
G_{\txtparticle}(\prbprocess_t)
\napiernum^{\imunit\opfocksegal(j_t g)}
\opdmsr{\msrprb_{\txtparticle,\sminvtemperature}(\prbprocess)}
\opdmsr{\msrprb_{\txtbsn,\txtfr,\sminvtemperature}(\opfocksegal)}
\\ %%%%%%%%%%%%%%%%
&=
\int_{\Omega_{\txtparticle,\sminvtemperature}}
\cmpconj{F_{\txtparticle}(\prbprocess_0)}
G_{\txtparticle}(\prbprocess_t)
\napiernum^{t\physenergy_{\txtnelson,\kappa,\Lambda}^{\txtrenormalization}}
\sqfun{\prbexp_{\msrprb_{\txtbsn,\txtfr,\sminvtemperature}}}
{\napiernum^{-\imunit \opfocksegal(j_0 f)}
\napiernum^{\imunit \opfocksegal(j_t g)}
\napiernum^{-\physcplconst
\opfocksegal(\mathsf{J}_{\closedinterval{0}{t},\kappa,\Lambda}(\prbprocess))}}
\opdmsr{\msrprb_{\txtparticle,\sminvtemperature}(\prbprocess)}.
\end{aligned}$$
It remains only to fix the particle path $\prbprocess$ and integrate over the Bose field.

Under the abbreviation $B_{\closedinterval{0}{t}}(f,g)
=
j_t g-j_0 f$, consider the real Gaussian random variables $X
=
\opfocksegal(B_{\closedinterval{0}{t}}(f,g))$ and $Y
=
\opfocksegal(\mathsf{J}_{\closedinterval{0}{t},\kappa,\Lambda}(\prbprocess))$.
Then the Bose-field part containing the interaction can be rewritten as
$$\begin{aligned}
\sqfun{\prbexp_{\msrprb_{\txtbsn,\txtfr,\sminvtemperature}}}
{\napiernum^{-\imunit \opfocksegal(j_0 f)}
\napiernum^{\imunit \opfocksegal(j_t g)}
\napiernum^{-\physcplconst
\opfocksegal(\mathsf{J}_{\closedinterval{0}{t},\kappa,\Lambda}(\prbprocess))}}
=
\prbexp_{\msrprb_{\txtbsn,\txtfr,\sminvtemperature}}
\sqbk{\napiernum^{\imunit X-\physcplconst Y}}.
\end{aligned}$$
Differentiating \eqref{expedition0012134} as a characteristic functional with real coefficients gives
$$\begin{aligned}
&\sqfun{\prbexp_{\msrprb_{\txtbsn,\txtfr,\sminvtemperature}}}{X^2}
=
\onehalf
\fun{\opform{q}_{\txtbsn,\txtbec,\sminvtemperature}^{\txteuclid}}
{B_{\closedinterval{0}{t}}(f,g)},
\\ %%%%%%%%%%%%%%%%
&\sqfun{\prbexp_{\msrprb_{\txtbsn,\txtfr,\sminvtemperature}}}{XY}
=
\onehalf
\fun{\opform{q}_{\txtbsn,\txtbec,\sminvtemperature}^{\txteuclid}}
{B_{\closedinterval{0}{t}}(f,g),
\mathsf{J}_{\closedinterval{0}{t},\kappa,\Lambda}(\prbprocess)},
\\ %%%%%%%%%%%%%%%%
&\sqfun{\prbexp_{\msrprb_{\txtbsn,\txtfr,\sminvtemperature}}}{Y^2}
=
\onehalf
\fun{\opform{q}_{\txtbsn,\txtbec,\sminvtemperature}^{\txteuclid}}
{\mathsf{J}_{\closedinterval{0}{t},\kappa,\Lambda}(\prbprocess)}.
\end{aligned}$$
Applying the general formula for a centered real Gaussian vector $(X,Y)$,
$$\sqfun{\prbexp_{\msrprb_{\txtbsn,\txtfr,\sminvtemperature}}}{\napiernum^{aX+bY}}
=
\fnexp{\onehalf
\sqfun{\prbexp_{\msrprb_{\txtbsn,\txtfr,\sminvtemperature}}}{(aX+bY)^2}}$$
with $a
= \imunit$ and $b
= -\physcplconst$, we obtain
$$\begin{aligned}
\prbexp_{\msrprb_{\txtbsn,\txtfr,\sminvtemperature}}
\sqbk{\napiernum^{\imunit X-\physcplconst Y}}
&=
\fnexp{
-\oneoverfour
\fun{\opform{q}_{\txtbsn,\txtbec,\sminvtemperature}^{\txteuclid}}
{B_{\closedinterval{0}{t}}(f,g)}
-\frac{\imunit\physcplconst}{2}
\fun{\opform{q}_{\txtbsn,\txtbec,\sminvtemperature}^{\txteuclid}}
{B_{\closedinterval{0}{t}}(f,g),
\mathsf{J}_{\closedinterval{0}{t},\kappa,\Lambda}(\prbprocess)}}
\\ %%%%%%%%%%%%%%%%
&\quad\times
\fnexp{\frac{\physcplconst^2}{4}
\fun{\opform{q}_{\txtbsn,\txtbec,\sminvtemperature}^{\txteuclid}}
{\mathsf{J}_{\closedinterval{0}{t},\kappa,\Lambda}(\prbprocess)}}.
\end{aligned}$$
Returning to the original formula and expanding the abbreviation gives \eqref{expedition0012180}.
\end{proof}

\subsubsection{Effective Action and Regular Part}\label{effective-action-and-regular-part}

Define the effective action \(\mathsf{U}_{\kappa,\Lambda,t}(\prbprocess)\) appearing in the particle-only semigroup kernel and the field covariance kernel \(W_{\kappa,\Lambda}(\tau,x)\) by \begin{equation}\label{expedition0012182}
\begin{aligned}
\mathsf{U}_{\kappa,\Lambda,t}(\prbprocess)
&=
t \physenergy_{\txtnelson,\kappa,\Lambda}^{\txtrenormalization}
+\frac{\physcplconst^2}{4}
\fun{\opform{q}_{\txtbsn,\txtbec,\sminvtemperature}^{\txteuclid}}
{\mathsf{J}_{\closedinterval{0}{t},\kappa,\Lambda}(\prbprocess),
\mathsf{J}_{\closedinterval{0}{t},\kappa,\Lambda}(\prbprocess)},
\\ %%%%%%%%%%%%%%%%
W_{\kappa,\Lambda}(\tau,x)
&=
\int_{\kappa \leq \abs{k} \leq \Lambda}
\frac{\napiernum^{-\abs{\tau} \omega(k)} + \napiernum^{-(\sminvtemperature-\abs{\tau}) \omega(k)}}
{1 - \napiernum^{-\sminvtemperature \omega(k)}}
\frac{\napiernum^{-ikx}}{\omega(k)}
\opdmsr{k}.
\end{aligned}
\end{equation} Then \begin{equation}\label{expedition0012183}
\begin{aligned}
\mathsf{U}_{\kappa,\Lambda,t}(\prbprocess)
=
t \physenergy_{\txtnelson,\kappa,\Lambda}^{\txtrenormalization}
+\frac{\physcplconst^2}{4}
\int_0^t
\int_0^t
W_{\kappa,\Lambda}(r-s,\prbprocess_r - \prbprocess_s)
\opdmsr{r} \opdmsr{s}
\end{aligned}
\end{equation} holds. Furthermore, decomposing the field covariance kernel into the temperature-free singular part \(W_{0,\kappa,\Lambda}\) and the finite-temperature regular part \(W_{\sminvtemperature,\kappa,\Lambda}\) gives \begin{equation}\label{expedition0012184}
\begin{aligned}
W_{\kappa,\Lambda}
&=
W_{0,\kappa,\Lambda}
+W_{\sminvtemperature,\kappa,\Lambda},
\\ %%%%%%%%%%%%%%%%
W_{0,\kappa,\Lambda}(\tau,x)
&=
\int_{\kappa\leq \abs{k} \leq\Lambda}
\napiernum^{-\abs{\tau} \omega(k)}
\frac{\napiernum^{-ikx}}{\omega(k)}
\opdmsr{k},
\\ %%%%%%%%%%%%%%%%
W_{\sminvtemperature,\kappa,\Lambda}(\tau,x)
&=
\int_{\kappa \leq \abs{k} \leq \Lambda}
\frac{\napiernum^{-(\sminvtemperature-\abs{\tau}) \omega(k)}
+\napiernum^{-\rbk{\sminvtemperature + \abs{\tau}}\omega(k)}}
{1-\napiernum^{-\sminvtemperature\omega(k)}}
\frac{\napiernum^{-ikx}}{\omega(k)}
\opdmsr{k}.
\end{aligned}
\end{equation} The infrared integrability condition in the next Proposition \ref{expedition0012143} is indeed satisfied for the three-dimensional Nelson model with dispersion relation \(\omega(k)
= \abs{k}\).

\begin{prop}[Domination and regularity of the regular part]\label{expedition0012143}
Assume the infrared integrability condition $\int_{\omega(k)\leq 1}
\omega(k)^{-2}\opdmsr{k}<\infty$, and let the time satisfy $0
\leq \abs{\tau}
\leq \frac{\sminvtemperature}{2}$.
The regular part $W_{\sminvtemperature,\kappa,\Lambda}$ in \eqref{expedition0012184} converges locally uniformly as $\kappa
\downarrow 0$ and $\Lambda
\uparrow \infty$; define its limit $W_{\sminvtemperature}$ by
\begin{equation}\label{expedition0012201}
\begin{aligned}
W_{\sminvtemperature}(\tau,x)
=
\int_{\fldreal^{d}}
\frac{\napiernum^{-(\sminvtemperature-\abs{\tau}) \omega(k)}
+\napiernum^{-\rbk{\sminvtemperature + \abs{\tau}}\omega(k)}}
{1-\napiernum^{-\sminvtemperature\omega(k)}}
\frac{\napiernum^{-ikx}}{\omega(k)}
\opdmsr{k}
\end{aligned}
\end{equation}
Moreover, for any multi-index $\alpha$ and integer $m
\geq 0$, when $\tau
\ne 0$, the derivative $\partial_{\tau}^{m}
\partial_{x}^{\alpha}
W_{\sminvtemperature,\kappa,\Lambda}(\tau,x)$ also has an integrable bound independent of the cutoffs.
In particular, cutoff removal for the regular part can be handled by the ordinary dominated convergence theorem.
\end{prop}

\begin{proof}
(Exponential decay on the ultraviolet side): In particular, if $\omega(k)
\geq 1$, then for some constant $c_1
> 0$ we have the estimate
$$0
\leq
\frac{\napiernum^{-(\sminvtemperature-\abs{\tau}) \omega(k)}
+\napiernum^{-\sminvtemperature\omega(k)}
\napiernum^{-\abs{\tau}\omega(k)}}
{1-\napiernum^{-\sminvtemperature\omega(k)}}
\leq
\frac{\napiernum^{-\frac{\sminvtemperature}{2}\omega(k)}
+\napiernum^{-\sminvtemperature\omega(k)}}
{1-\napiernum^{-\sminvtemperature\omega(k)}}
\leq
c_1
\napiernum^{-\frac{\sminvtemperature}{2} \omega(k)}.$$
Therefore, on the ultraviolet side,
$\int_{\omega(k) \geq 1}
\napiernum^{-\frac{\sminvtemperature}{2} \omega(k)}
\frac{\opdmsr{k}}{\omega(k)}
< \infty$ holds.

(Infrared side): Here as well, for some constant $c_2
> 0$,
$$1 - \napiernum^{-\sminvtemperature \omega(k)}
\geq c_2 \omega(k),
\quad
\napiernum^{-(\sminvtemperature-\abs{\tau}) \omega(k)}
+\napiernum^{-\sminvtemperature \omega(k)}
\napiernum^{-\abs{\tau} \omega(k)}
\leq 2$$
holds, and the infrared integrability condition gives
$c_2
\int_{\omega(k)\leq 1}
\frac{\opdmsr{k}}{\omega(k)^{2}}
<\infty$.
Since this estimate is independent of the cutoffs, cutoff removal for the regular part $W_{\sminvtemperature,\kappa,\Lambda}$ can be handled by the dominated convergence theorem.

(Regularity): For a multi-index $\alpha$ and integer $m
\geq 0$, formal differentiation at $\tau \ne 0$ gives
$$\begin{aligned}
\partial_{\tau}^{m}
\partial_{x}^{\alpha}
W_{\sminvtemperature,\kappa,\Lambda}(\tau,x)
=
\int_{\kappa \leq \abs{k} \leq \Lambda}
\sum_{\varepsilon=\pm 1}
P_{m,\alpha,\varepsilon}(\sgn \tau,k)
\frac{\napiernum^{-(\sminvtemperature+\varepsilon\abs{\tau}) \omega(k)}}
{1 - \napiernum^{-\sminvtemperature\omega(k)}}
\frac{\napiernum^{-\imunit kx}}{\omega(k)}
\opdmsr{k}.
\end{aligned}$$
Here $P_{m,\alpha,\varepsilon}(\sgn\tau,k)$ collects the factors produced when $\partial_{\tau}^{m}$ acts on $\napiernum^{-(\sminvtemperature+\varepsilon\abs{\tau})\omega(k)}$ and when $\partial_x^{\alpha}$ acts on $\napiernum^{-\imunit kx}$.
Concretely, apart from signs depending on $\sgn\tau$ and $\varepsilon$ and powers of $\imunit$, it is $\omega(k)^m k^{\alpha}$.
For fixed $m,\alpha$, for some constant $c_3
> 0$, $\abs{P_{m,\alpha,\varepsilon}(\sgn\tau,k)}
\leq c_3\omega(k)^{m+\abs{\alpha}}$ holds.
Therefore, for $0
\leq \abs{\tau}
\leq \frac{\sminvtemperature}{2}$, for some constant $c_4
> 0$,
$$\abs{\partial_{\tau}^{m}
\partial_{x}^{\alpha}
W_{\sminvtemperature,\kappa,\Lambda}(\tau,x)}
\leq
c_4
\int_{\omega(k)\leq 1}
\omega(k)^{m+\abs{\alpha}-2}\opdmsr{k}
+c_4
\int_{\omega(k) \geq 1}
\omega(k)^{m+\abs{\alpha}-1}
\napiernum^{-\frac{\sminvtemperature}{2}\omega(k)}
\opdmsr{k}$$
is obtained.
The first term is finite because $m+\abs{\alpha}\geq 0$ and by the infrared integrability condition.
The second term is also finite by exponential decay.
In particular, for any cutoff sequence, $W_{\sminvtemperature,\kappa,\Lambda}$ and its finite-order time and spatial derivatives have locally uniform integrable bounds, and cutoff removal can be handled by the ordinary dominated convergence theorem.
\end{proof}

\subsubsection{Singular Part and Centering Term}\label{singular-part-and-centering-term}

We discuss the singular part \(W_{0,\kappa,\Lambda}\) in \eqref{expedition0012184}. Computing \(\frac{\physcplconst^2}{4}
\int\int W_{0,\kappa,\Lambda}\) in the effective action \(\mathsf{U}_{\kappa,\Lambda,t}\) in \eqref{expedition0012183}, we obtain \[\begin{aligned}
\int_0^t
\int_0^r
\napiernum^{-(r-s) \rbk{\omega(k) + \onehalf \omega(k)^2}}
\opdmsr{s} \opdmsr{r}
=
\frac{t}{\omega(k) + \onehalf \omega(k)^2}
-\frac{1 - \napiernum^{-t \rbk{\omega(k) + \onehalf \omega(k)^2}}}{\rbk{\omega(k) + \onehalf \omega(k)^2}^2},
\end{aligned}\] and the first term on the right-hand side is cancelled by \(t \physenergy_{\txtnelson,\kappa,\Lambda}^{\txtrenormalization}\). Based on this, define the corrected action of the singular part by \begin{equation}\label{expedition0012185}
\begin{aligned}
\mathsf{U}_{0,\kappa,\Lambda,t}(\prbprocess)
&=
\frac{\physcplconst^2}{2}
\int_0^t
\opdmsr{r}
\int_0^r
\opdmsr{s}
\int_{\kappa \leq \abs{k} \leq \Lambda}
\frac{\napiernum^{-(r-s) \omega(k)}}{\omega(k)}
\rbk{\napiernum^{-ik(\prbprocess_r - \prbprocess_s)} - \napiernum^{-(r-s) \onehalf \omega(k)^2}}
\opdmsr{k}
\\ %%%%%%%%%%%%%%%%
&\quad
-\frac{\physcplconst^2}{2}
\int_{\kappa\leq \abs{k} \leq\Lambda}
\frac{1 - \napiernum^{-t \rbk{\omega(k) + \onehalf \omega(k)^2}}}
{\omega(k)\rbk{\omega(k) + \onehalf \omega(k)^2}^2}
\opdmsr{k}.
\end{aligned}
\end{equation} Note that the coefficient is \(\onehalf\) because the double integral has been reduced to a triangular region by symmetry.

In the following lemma, three-dimensionality is used for integrability of the comparison kernel \(\rbk{\omega(k)
+\onehalf \omega(k)^2}^{-2}\) on both the low-momentum and high-momentum sides, and for the fact that the contributions from the respective cutoff edges vanish.

\begin{lem}[$\lp^{2}$ estimate for the centering term]\label{expedition0012142}
Let the spatial dimension be $d=3$.
For a measurable set $A
\subset \fldreal^{d}$, set
\begin{equation}\label{expedition0012186}
\begin{aligned}
\mathsf{C}_{A,t}(\prbprocess)
=
\int_0^t
\opdmsr{r}
\int_0^r
\opdmsr{s}
\int_{A}
\frac{\napiernum^{-(r-s) \omega(k)}}{\omega(k)}
\rbk{\napiernum^{-\imunit k(\prbprocess_r - \prbprocess_s)} - \napiernum^{-\onehalf (r-s) \omega(k)^2}}
\opdmsr{k}.
\end{aligned}
\end{equation}
Then, for some constant $c_1>0$,
$$\begin{aligned}
\sqfun{\prbexp_{\msrprb_{\txtparticle,\sminvtemperature}}}
{\abs{\mathsf{C}_{A,t}}^2}
\leq
c_1
\int_A \frac{dk}{\rbk{\omega(k) + \onehalf \omega(k)^2}^2}
\end{aligned}$$
holds.
In particular, the contribution of the high-momentum shell $A
= \setone{\abs{k}\geq\Lambda}$ converges to $0$ as $\Lambda\uparrow\infty$, and the contribution of the low-momentum ball $A
= \{\abs{k}\leq\kappa\}$ also converges to $0$ as $\kappa\downarrow0$.
\end{lem}

\begin{proof}
First we discuss the estimate for the unperturbed Brownian bridge measure, and then discuss the estimate for the particle loop measure with confining potential obtained by perturbation with the potential.

(Brownian bridge measure): For the Brownian bridge measure $\msrprb_{x,x}^{\sminvtemperature,\txtbrownbridge}$ of length $\sminvtemperature$ with initial and terminal point fixed at $x$, and for time variables $0
\leq s
\leq r
\leq t
\leq \frac{\sminvtemperature}{2}$, let $\tau
=r-s$, and set the square root of the integrand appearing in the final estimate to be $q(k)
=
\rbk{\omega(k)+\onehalf\omega(k)^2}^{-1}$.

The Brownian bridge increment $\prbprocess_r-\prbprocess_s$ is a Gaussian variable with mean $0$ and variance $\tau(1-\tau/\sminvtemperature)$, and satisfies
$$\sqfun{\prbexp_{\msrprb_{x,x}^{\sminvtemperature,\txtbrownbridge}}}
{\napiernum^{-\imunit k \rbk{\prbprocess_r - \prbprocess_s}}}
=
\fnexp{-\onehalf \tau \rbk{1 - \frac{\tau}{\sminvtemperature}} \omega(k)^2},$$
so the expression in parentheses in the integrand of $\mathsf{C}_{A,t}$ cannot be regarded as the centered quantity as it stands.
Therefore we decompose it as
$$\begin{aligned}
\napiernum^{-\imunit k(\prbprocess_r-\prbprocess_s)}
-\napiernum^{-\onehalf\tau\omega(k)^2}
&=
\mathsf{M}_{r,s}(k)
+\mathsf{D}_{r,s}(k),
\\ %%%%%%%%%%%%%%%%
\mathsf{M}_{r,s}(k)
&=
\fnexp{-\imunit k(\prbprocess_r-\prbprocess_s)}
-\fnexp{-\onehalf \tau \rbk{1 - \frac{\tau}{\sminvtemperature}} \omega(k)^2},
\\ %%%%%%%%%%%%%%%%
\mathsf{D}_{r,s}(k)
&=
\fnexp{-\onehalf \tau \rbk{1 - \frac{\tau}{\sminvtemperature}} \omega(k)^2}
-\fnexp{-\onehalf \tau \omega(k)^2}.
\end{aligned}$$

(Deterministic bridge correction $\mathsf{D}_{r,s}$): To simplify notation, set $a
= \onehalf\tau\omega(k)^2$ and $b
= \onehalf\tau \rbk{1 - \frac{\tau}{\sminvtemperature}}\omega(k)^2$.
Since the condition $\tau
\leq \frac{\sminvtemperature}{2}$ implies $\tau \rbk{1 - \frac{\tau}{\sminvtemperature}}
\geq \frac{\tau}{2}$, these satisfy $0
\leq b
\leq a$ and $a-b
=
\frac{\tau^2}{2\sminvtemperature}\omega(k)^2$.
Applying the mean value theorem to the function $x
\mapsto \napiernum^{-x}$, there exists $\theta
\in \closedinterval{b}{a}$ such that
$$\abs{\mathsf{D}_{r,s}(k)}
=
\abs{\napiernum^{-b}-\napiernum^{-a}}
=
\rbk{a-b}\napiernum^{-\theta}
\leq
\frac{\tau^2}{2\sminvtemperature}\omega(k)^2
\napiernum^{-b}.$$
Moreover, by $\tau
\leq \frac{\sminvtemperature}{2}$,
we have
$b
\geq \frac14\tau\omega(k)^2$.
Therefore, for suitable constants $c_2,c_3>0$,
$$\begin{aligned}
\abs{\mathsf{D}_{r,s}(k)}
\leq
c_2
\tau^2 \omega(k)^2
\fnexp{-c_3 \tau \omega(k)^2}
\end{aligned}$$
is obtained, and we substitute this into the time integral.
First, for fixed $r$, if $\tau
=r-s$, then $s=0$ corresponds to $\tau=r$ and $s=r$ corresponds to $\tau=0$, so
$$\begin{aligned}
\int_0^r
\frac{\napiernum^{-(r-s)\omega(k)}}{\omega(k)}
\abs{\mathsf{D}_{r,s}(k)}
\opdmsr{s}
\leq
c_2
\int_0^r
\tau^2\omega(k)
\napiernum^{-\tau\rbk{\omega(k)+c_3\omega(k)^2}}
\opdmsr{\tau}.
\end{aligned}$$
Furthermore, integrating with respect to $r$ gives
$$\begin{aligned}
&\int_0^t
\int_0^r
\tau^2\omega(k)
\napiernum^{-\tau\rbk{\omega(k)+c_3\omega(k)^2}}
\opdmsr{\tau}\opdmsr{r}
=
\int_0^t
\rbk{\int_{\tau}^{t}\opdmsr{r}}
\tau^2\omega(k)
\napiernum^{-\tau\rbk{\omega(k)+c_3\omega(k)^2}}
\opdmsr{\tau}
\\ %%%%%%%%%%%%%%%%
&=
\int_0^t
(t-\tau)
\tau^2\omega(k)
\napiernum^{-\tau\rbk{\omega(k)+c_3\omega(k)^2}}
\opdmsr{\tau}
\leq
t
\int_0^t
\tau^2\omega(k)
\napiernum^{-\tau\rbk{\omega(k)+c_3\omega(k)^2}}
\opdmsr{\tau}.
\end{aligned}$$
Absorbing this $t$ and $c_2$ into the constant, for some constant $c_4>0$,
$$\begin{aligned}
\int_0^t
\int_0^r
\frac{\napiernum^{-(r-s)\omega(k)}}{\omega(k)}
\abs{\mathsf{D}_{r,s}(k)}
\opdmsr{s}\opdmsr{r}
\leq
c_4
\int_0^t
\tau^2\omega(k)
\napiernum^{-\tau\rbk{\omega(k)+c_3\omega(k)^2}}
\opdmsr{\tau}
\end{aligned}$$
holds.
Moreover, setting $\mu(k)
= \omega(k) + c_3\omega(k)^2$, we get
$$\begin{aligned}
\int_0^t
\tau^2\omega(k)
\napiernum^{-\tau\mu(k)}
\opdmsr{\tau}
\leq
\omega(k)
\int_0^\infty
\tau^2
\napiernum^{-\tau\mu(k)}
\opdmsr{\tau}
=
\frac{2\omega(k)}{\mu(k)^3}.
\end{aligned}$$
This estimate is used on the high-momentum side.
Indeed, if $\omega(k)
\geq 1$, then $\mu(k)
\geq c_3\omega(k)^2$, so for some constant $c_5>0$,
$$\begin{aligned}
\frac{2\omega(k)}{\mu(k)^3}
\leq
\frac{c_5}{\omega(k)^5}
\leq
c_5q(k)
\end{aligned}$$
holds.

On the low-momentum side $\omega(k)
\leq 1$, we use $\tau^2
\leq t^2$ rather than exponential decay.
Then $\napiernum^{-\tau \rbk{\omega(k)+c_3\omega(k)^2}}
\leq 1$ gives
$$\begin{aligned}
\int_0^t
\tau^2
\omega(k)
\napiernum^{-\tau \rbk{\omega(k) + c_3 \omega(k)^2}}
\opdmsr{\tau}
\leq
\omega(k)\int_0^t \tau^2\opdmsr{\tau}
=
\frac{t^3}{3}\omega(k)
\leq
t^3\omega(k).
\end{aligned}$$
Furthermore, for $0<\omega(k)\leq1$,
$$\begin{aligned}
\omega(k)\rbk{\omega(k)+\onehalf\omega(k)^2}
=
\omega(k)^2+\onehalf\omega(k)^3
\leq
\frac{3}{2},
\end{aligned}$$
so $t^3\omega(k)
\leq \frac{3t^3}{2}
q(k)$.
Thus, for some constant $c_6>0$,
$$\begin{aligned}
\int_0^t
\tau^2
\omega(k)
\napiernum^{-\tau \rbk{\omega(k) + c_3 \omega(k)^2}}
\opdmsr{\tau}
\leq
t^3\omega(k)
\leq
c_6q(k)
\end{aligned}$$
is obtained.
Combining the low-momentum and high-momentum sides, for some constant $c_7>0$,
$$\begin{aligned}
\int_0^t
\tau^2\omega(k)
\napiernum^{-\tau\rbk{\omega(k)+c_3\omega(k)^2}}
\opdmsr{\tau}
\leq
c_7q(k)
\end{aligned}$$
is obtained.

(Centered bridge fluctuation $\mathsf{M}_{r,s}$): The Brownian bridge is a Gaussian Markov process with diffusion coefficient $1$.
Let $M$ denote the natural martingale part of the bridge, and write the Euclidean inner product as a dot, $\vainnprod$.
By the martingale representation for the bridge semigroup, it can be represented as
$\begin{aligned}
\mathsf{M}_{r,s}(k)
=
\int_s^r
\vagrad_y \Phi_{r,s,k}(u,\prbprocess_u;\prbprocess_s)
\vainnprod \opdmsr{M_u}
\end{aligned}$,
where $\Phi_{r,s,k}$ is defined, for $0
\leq s
\leq u
\leq r
\leq t$, by
$$\begin{aligned}
\Phi_{r,s,k}(u,y;z)
&=
\sqfuncond{\prbexp_{\msrprb_{x,x}^{\sminvtemperature,\txtbrownbridge}}}
{\napiernum^{-\imunit k(\prbprocess_r-z)}}
{\prbprocess_u=y}
=
\int_{\fldreal^{d}}
\napiernum^{-\imunit k(w-z)}
\frac{p_{r-u}(y,w)p_{\sminvtemperature-r}(w,x)}
{p_{\sminvtemperature-u}(y,x)}
\opdmsr{w},
\\ %%%%%%%%%%%%%%%%
p_a(y,w)
&=
\frac{1}{(2\pi a)^{\frac{d}{2}}}
\fnexp{-\frac{\abs{y-w}^{2}}{2a}}.
\end{aligned}$$

We estimate the Gaussian integral above.
When $\prbprocess_u
= y$ is fixed at time $u$, the conditional distribution of $\prbprocess_r$ is Gaussian with mean and variance
$$\begin{aligned}
m_{u,r}(y)
&=
y+\frac{r-u}{\sminvtemperature-u}(x-y)
=
\frac{\sminvtemperature-r}{\sminvtemperature-u}y
+\frac{r-u}{\sminvtemperature-u}x,
\\ %%%%%%%%%%%%%%%%
\sigma_{u,r}^{2}
&=
\frac{(r-u)(\sminvtemperature-r)}
{\sminvtemperature-u},
\end{aligned}$$
so $\Phi_{r,s,k}(u,y;z)
=
\napiernum^{-\imunit k(m_{u,r}(y)-z)}
\napiernum^{-\onehalf\sigma_{u,r}^{2}\omega(k)^2}$.
Since the variance $\sigma_{u,r}^{2}$ does not depend on $y$, differentiating with respect to $y$ gives
$\vagrad_y\Phi_{r,s,k}(u,y;z)
=
-\imunit
\frac{\sminvtemperature-r}{\sminvtemperature-u}
k
\Phi_{r,s,k}(u,y;z)$.
Under the condition $0
\leq u
\leq r
\leq t
\leq \frac{\sminvtemperature}{2}$,
$$\begin{aligned}
0
\leq
\frac{\sminvtemperature-r}{\sminvtemperature-u}
\leq 1,
\quad
\frac{\sminvtemperature-r}{\sminvtemperature-u}
\geq \onehalf,
\quad
\sigma_{u,r}^{2}
\geq
\frac{r-u}{2}
\end{aligned}$$
holds, and therefore $\abs{\vagrad_y\Phi_{r,s,k}(u,y;z)}
\leq
\omega(k)
\napiernum^{-\frac{1}{4}(r-u)\omega(k)^2}$.
Thus, for suitable constants $c_8,c_9>0$,
$$\abs{\vagrad_y \Phi_{r,s,k}(u,y;z)}
\leq
c_8
\omega(k)
\napiernum^{-c_9(r-u)\omega(k)^2}$$
holds.

Let the whole centered part be
$$\begin{aligned}
\mathcal{M}_{A,t}
=
\int_0^t
\opdmsr{r}
\int_0^r
\opdmsr{s}
\int_A
\frac{\napiernum^{-(r-s)\omega(k)}}{\omega(k)}
\mathsf{M}_{r,s}(k)
\opdmsr{k}.
\end{aligned}$$
Substituting the martingale representation above and interchanging the order of integration over the region $0
\leq s
\leq u
\leq r
\leq t$ by Fubini's theorem for stochastic integrals gives
$$\begin{aligned}
\mathcal{M}_{A,t}
&=
\int_0^t
\mathcal{V}_{A,t}(u,\prbprocess)
\vainnprod\opdmsr{M_u},
\\ %%%%%%%%%%%%%%%%
\mathcal{V}_{A,t}(u,\prbprocess)
&=
\int_u^t
\opdmsr{r}
\int_0^u
\opdmsr{s}
\int_A
\frac{\napiernum^{-(r-s)\omega(k)}}{\omega(k)}
\vagrad_y\Phi_{r,s,k}(u,\prbprocess_u;\prbprocess_s)
\opdmsr{k},
\end{aligned}$$
and Itô isometry gives
$$\begin{aligned}
\sqfun{\prbexp_{\msrprb_{x,x}^{\sminvtemperature,\txtbrownbridge}}}
{\abs{\mathcal{M}_{A,t}}^2}
=
\sqfun{\prbexp_{\msrprb_{x,x}^{\sminvtemperature,\txtbrownbridge}}}
{\int_0^t
\abs{\mathcal{V}_{A,t}(u,\prbprocess)}^2
\opdmsr{u}}.
\end{aligned}$$

Next, to estimate the time integral for fixed $u,k$, set $\tau
= r-s$ and $\rho=r-u$.
Then $0 \leq s \leq u \leq r \leq t$ implies $0
\leq \rho
\leq \tau$.
The time part obtained from the gradient estimate can be bounded, with a suitable constant $c$, as
$$\begin{aligned}
&\int_u^t
\int_0^u
\napiernum^{-(r-s)\omega(k)}
\napiernum^{-c_9(r-u)\omega(k)^2}
\opdmsr{s}\opdmsr{r}
\leq
\int_0^t
\napiernum^{-\tau\omega(k)}
\rbk{\int_0^\tau
\napiernum^{-c_9\rho\omega(k)^2}
\opdmsr{\rho}}
\opdmsr{\tau}
\\ %%%%%%%%%%%%%%%%
&\leq
c
\int_0^t
\napiernum^{-\tau\omega(k)}
\rbk{\tau \wedge \omega(k)^{-2}}
\opdmsr{\tau}.
\end{aligned}$$
By isometry of stochastic integrals, Fubini's theorem, and Minkowski's inequality, for some constant $c_{10}>0$,
$$\begin{aligned}
\sqfun{\prbexp_{\msrprb_{x,x}^{\sminvtemperature,\txtbrownbridge}}}
{\abs{\mathcal{M}_{A,t}}^2}
\leq
c_{10}
\int_A
\rbk{
\int_0^t
\napiernum^{-\tau\omega(k)}
\rbk{\tau \wedge \omega(k)^{-2}}
\opdmsr{\tau}}^2
\opdmsr{k}
\end{aligned}$$
is obtained.
The expression in parentheses can be estimated as follows.
Under the condition $\omega(k)
\leq 1$,
$$\begin{aligned}
\int_0^t
\napiernum^{-\tau\omega(k)}
\rbk{\tau \wedge \omega(k)^{-2}}
\opdmsr{\tau}
\leq
\int_0^t \tau\opdmsr{\tau}
\leq
\frac{t^2}{2}
\leq
c_{11}q(k),
\end{aligned}$$
while under the condition $\omega(k)\geq1$,
$$\begin{aligned}
\int_0^t
\napiernum^{-\tau\omega(k)}
\rbk{\tau \wedge \omega(k)^{-2}}
\opdmsr{\tau}
\leq
\omega(k)^{-2}
\int_0^\infty
\napiernum^{-\tau\omega(k)}
\opdmsr{\tau}
=
\omega(k)^{-3}
\leq
c_{12}q(k).
\end{aligned}$$
Increasing the constant if necessary, we obtain the contribution of the centered part:
$$\begin{aligned}
\sqfun{\prbexp_{\msrprb_{x,x}^{\sminvtemperature,\txtbrownbridge}}}
{\abs{\mathcal{M}_{A,t}}^2}
\leq
c_{10}
\int_A q(k)^2\opdmsr{k}.
\end{aligned}$$

(Summary): Let the deterministic bridge correction part be
$$\begin{aligned}
\mathcal{D}_{A,t}
=
\int_0^t
\opdmsr{r}
\int_0^r
\opdmsr{s}
\int_A
\frac{\napiernum^{-(r-s)\omega(k)}}{\omega(k)}
\mathsf{D}_{r,s}(k)
\opdmsr{k}.
\end{aligned}$$
Writing the estimate for the deterministic bridge correction obtained above separately once more on the low-momentum and high-momentum sides, under the condition $0<\omega(k)\leq1$ we have
$$\begin{aligned}
&c_4
\int_0^t
\tau^2\omega(k)
\napiernum^{-\tau\rbk{\omega(k)+c_3\omega(k)^2}}
\opdmsr{\tau}
\leq
c_4t^3\omega(k)
\\ %%%%%%%%%%%%%%%%
&=
c_4t^3\omega(k)
\rbk{\omega(k)+\onehalf\omega(k)^2}^2 q(k)^2
\leq
\frac{9c_4t^3}{4} q(k)^2,
\end{aligned}$$
and under the condition $\omega(k)
\geq 1$,
$$\begin{aligned}
c_4
\int_0^t
\tau^2\omega(k)
\napiernum^{-\tau\rbk{\omega(k)+c_3\omega(k)^2}}
\opdmsr{\tau}
\leq
\frac{c_4c_5}{\omega(k)^5}
=
c_4c_5
\frac{\rbk{\omega(k)+\onehalf\omega(k)^2}^2}{\omega(k)^5}
q(k)^2
\leq
\frac{9c_4c_5}{4} q(k)^2
\end{aligned}$$
is obtained.
Therefore, for some constant $c_{13}>0$,
$$\begin{aligned}
\abs{\mathcal{D}_{A,t}}
\leq
\int_A
\int_0^t
\int_0^r
\frac{\napiernum^{-(r-s)\omega(k)}}{\omega(k)}
\abs{\mathsf{D}_{r,s}(k)}
\opdmsr{s}\opdmsr{r}\opdmsr{k}
\leq
c_{13}
\int_A q(k)^2\opdmsr{k}
\end{aligned}$$
is obtained.
Here, by $Q
= \int_{\fldreal^{d}}q(k)^2\opdmsr{k}
< \infty$,
$$\begin{aligned}
\abs{\mathcal{D}_{A,t}}^2
\leq
c_{13}^2
\rbk{\int_A q(k)^2\opdmsr{k}}^2
\leq
c_{13}^2 Q
\int_A q(k)^2\opdmsr{k}
\end{aligned}$$
is obtained.

For the centered part, we have already shown that $\sqfun{\prbexp_{\msrprb_{x,x}^{\sminvtemperature,\txtbrownbridge}}}
{\abs{\mathcal{M}_{A,t}}^2}
\leq
c_{10}
\int_A q(k)^2\opdmsr{k}$.
By the decomposition $\mathsf{C}_{A,t}(\prbprocess)
=
\mathcal{M}_{A,t}(\prbprocess)+\mathcal{D}_{A,t}$
and the inequality $\abs{u+v}^2
\leq
2 \abs{u}^2 + 2 \abs{v}^2$,
$$\begin{aligned}
&\sqfun{\prbexp_{\msrprb_{x,x}^{\sminvtemperature,\txtbrownbridge}}}
{\abs{\mathsf{C}_{A,t}(\prbprocess)}^2}
\leq
2
\sqfun{\prbexp_{\msrprb_{x,x}^{\sminvtemperature,\txtbrownbridge}}}
{\abs{\mathcal{M}_{A,t}}^2}
+2\abs{\mathcal{D}_{A,t}}^2
\\ %%%%%%%%%%%%%%%%
&\leq
\rbk{2c_{10}+2c_{13}^2Q}
\int_A q(k)^2\opdmsr{k}
=
c_{14}
\int_A q(k)^2\opdmsr{k},
\end{aligned}$$
and the required estimate for Brownian bridges with fixed endpoints follows.

(Particle loop measure with confining potential): First set
$$\begin{aligned}
F_A(\prbprocess)
=
\abs{\mathsf{C}_{A,t}(\prbprocess)}^2,
\quad
W(\prbprocess)
=
\fnexp{-\int_0^{\sminvtemperature}
V(\prbprocess_u)\opdmsr{u}}.
\end{aligned}$$
By the definition of the particle loop measure,
$$\begin{aligned}
\sqfun{\prbexp_{\msrprb_{\txtparticle,\sminvtemperature}}}{F_A}
&=
\frac{1}{\smpartitionfunc_{\txtparticle,\sminvtemperature}}
\int_{\fldreal^{d}}
\sqfun{\prbexp_{\msrprb_{x,x}^{\sminvtemperature,\txtbrownbridge}}}
{W F_A}
\opdmsr{x}
\end{aligned}$$
is obtained.
For a fixed exponent $p
> 1$, let the Hölder conjugate exponent be $p^{\prime}
= \frac{p}{p-1}$.
Hölder's inequality gives, for each $x$,
$$\begin{aligned}
\sqfun{\prbexp_{\msrprb_{x,x}^{\sminvtemperature,\txtbrownbridge}}}
{W F_A}
&\leq
\sqfun{\prbexp_{\msrprb_{x,x}^{\sminvtemperature,\txtbrownbridge}}}
{W^p}^{\frac{1}{p}}
\sqfun{\prbexp_{\msrprb_{x,x}^{\sminvtemperature,\txtbrownbridge}}}
{F_A^{p^{\prime}}}^{\frac{1}{p^{\prime}}}.
\end{aligned}$$
Applying the Brownian bridge estimate to the $2p^{\prime}$-th moment by the heat-kernel gradient estimate in the same way as for the BDG inequality, there is a constant $c_{14,p}>0$ such that, for all $x$ and $A$,
$\sqfun{\prbexp_{\msrprb_{x,x}^{\sminvtemperature,\txtbrownbridge}}}
{F_A^{p^{\prime}}}^{\frac{1}{p^{\prime}}}
\leq
c_{14,p}
\int_A q(k)^2\opdmsr{k}$.

By the heat-kernel estimate for the confining potential,
$$\begin{aligned}
K_{p,V,\sminvtemperature}
=
\int_{\fldreal^{d}}
\sqfun{\prbexp_{\msrprb_{x,x}^{\sminvtemperature,\txtbrownbridge}}}
{W^p}^{\frac{1}{p}}
\opdmsr{x}
<\infty
\end{aligned}$$
holds, and therefore
$$\begin{aligned}
&\sqfun{\prbexp_{\msrprb_{\txtparticle,\sminvtemperature}}}{F_A}
\leq
\frac{c_{14,p}}{\smpartitionfunc_{\txtparticle,\sminvtemperature}}
\int_{\fldreal^{d}}
\sqfun{\prbexp_{\msrprb_{x,x}^{\sminvtemperature,\txtbrownbridge}}}
{W^p}^{\frac{1}{p}}
\opdmsr{x}
\int_A q(k)^2\opdmsr{k}
\\ %%%%%%%%%%%%%%%%
&=
\frac{c_{14,p}K_{p,V,\sminvtemperature}}
{\smpartitionfunc_{\txtparticle,\sminvtemperature}}
\int_A q(k)^2\opdmsr{k}
=
c_{15}
\int_A q(k)^2\opdmsr{k}
\end{aligned}$$
is obtained.
Taking this $c_{15}$, after a suitable adjustment, as the $c_1$ in the statement gives the desired estimate.

Finally, $\int_{\abs{k} \leq \kappa}
q(k)^2\opdmsr{k}
\to 0$ and $\int_{\abs{k}\geq\Lambda}
q(k)^2\opdmsr{k}
\to 0$ give the claims for the low-momentum ball and high-momentum shell.
\end{proof}

\subsubsection{Renormalized Action and the Limit of Local Semigroups}\label{renormalized-action-and-the-limit-of-local-semigroups}

In Proposition \ref{expedition0012383}, three-dimensionality is used for cutoff removal in the singular part and as the infrared integrability condition of Proposition \ref{expedition0012143} for the regular part. Concretely, we use the deterministic correction term and the low- and high-momentum integrability of the comparison kernel in Lemma \ref{expedition0012142}.

\begin{prop}[Convergence of the renormalized action]\label{expedition0012383}
Let the spatial dimension be $d=3$.
For any $0
< t
\leq \frac{\sminvtemperature}{2}$, $\mathsf{U}_{\kappa,\Lambda,t}(\prbprocess)$ in \eqref{expedition0012182} converges in the topology of $\fun{\lp^{2}}
{\Omega_{\txtparticle,\sminvtemperature},
\msrprb_{\txtparticle,\sminvtemperature}}$ as $\kappa\downarrow0$ and $\Lambda\uparrow\infty$.
Using $\mathsf{C}_{A,t}$ in \eqref{expedition0012186}, define its limit $\mathsf{U}_{t}$ by
\begin{equation}\label{expedition0012202}
\begin{aligned}
\mathsf{U}_{t}(\prbprocess)
&=
\mathsf{U}_{\sminvtemperature,t}(\prbprocess)
+\frac{\physcplconst^{2}}{2}\mathsf{C}_{t}(\prbprocess)
-\frac{\physcplconst^{2}}{2}B_t,
\quad
\mathsf{U}_{\sminvtemperature,t}(\prbprocess)
=
\frac{\physcplconst^{2}}{4}
\int_{0}^{t}
\int_{0}^{t}
W_{\sminvtemperature}
(r-s,\prbprocess_r-\prbprocess_s)
\opdmsr{r}\opdmsr{s},
\\ %%%%%%%%%%%%%%%%
\mathsf{C}_{t}
&=
\fun{\lp^{2}}{\msrprb_{\txtparticle,\sminvtemperature}}\text{-}
\lim_{\kappa\downarrow0,\Lambda\uparrow\infty}
\int_0^t
\opdmsr{r}
\int_0^r
\opdmsr{s}
\int_{\kappa \leq \abs{k} \leq \Lambda}
\frac{\napiernum^{-(r-s) \omega(k)}}{\omega(k)}
\rbk{\napiernum^{-\imunit k(\prbprocess_r - \prbprocess_s)} - \napiernum^{-\onehalf (r-s) \omega(k)^2}}
\opdmsr{k},
\\ %%%%%%%%%%%%%%%%
B_t
&=
\int_{\fldreal^{d}}
\frac{1 - \napiernum^{-t \rbk{\omega(k) + \onehalf \omega(k)^2}}}
{\omega(k)\rbk{\omega(k) + \onehalf \omega(k)^2}^2}
\opdmsr{k}.
\end{aligned}
\end{equation}
For any $p < \infty$,
$$\begin{aligned}
\sup_{0<\kappa<\Lambda<\infty}
\int_{\Omega_{\txtparticle,\sminvtemperature}}
\napiernum^{p \mathsf{U}_{\kappa,\Lambda,t}(\prbprocess)}
\opdmsr{\msrprb_{\txtparticle,\sminvtemperature}(\prbprocess)}
<
\infty
\end{aligned}$$
holds.
\end{prop}

\begin{proof}
We split the action into the regular part and the singular part.
In particular, for $\mathsf{U}_{0,\kappa,\Lambda,t}$ in \eqref{expedition0012185}, set
\begin{equation}\label{expedition0012187}
\begin{aligned}
\mathsf{U}_{\kappa,\Lambda,t}
=
\mathsf{U}_{0,\kappa,\Lambda,t}
+\mathsf{U}_{\sminvtemperature,\kappa,\Lambda,t},
\quad
\mathsf{U}_{\sminvtemperature,\kappa,\Lambda,t}(\prbprocess)
=
\frac{\physcplconst^{2}}{4}
\int_{0}^{t}
\int_{0}^{t}
W_{\sminvtemperature,\kappa,\Lambda}
(r-s,\prbprocess_r-\prbprocess_s)
\opdmsr{r}\opdmsr{s}.
\end{aligned}
\end{equation}
(Regular part): By Proposition \ref{expedition0012143}, $W_{\sminvtemperature,\kappa,\Lambda}$ has a bounded estimate independent of the cutoffs for $0
\leq r,s
\leq t
\leq \frac{\sminvtemperature}{2}$, and converges locally uniformly to the cutoff-removal limit $W_{\sminvtemperature}$ defined in \eqref{expedition0012201}.
Therefore, for any path $\prbprocess$,
\begin{equation}\label{expedition0012188}
\mathsf{U}_{\sminvtemperature,\kappa,\Lambda,t}(\prbprocess)
\to
\mathsf{U}_{\sminvtemperature,t}(\prbprocess)
=
\frac{\physcplconst^{2}}{4}
\int_{0}^{t}
\int_{0}^{t}
W_{\sminvtemperature}
(r-s,\prbprocess_r-\prbprocess_s)
\opdmsr{r}\opdmsr{s}
\end{equation}
holds.
Moreover, since $\sup_{\kappa,\Lambda,\prbprocess}
\abs{\mathsf{U}_{\sminvtemperature,\kappa,\Lambda,t}(\prbprocess)}
< \infty$, \eqref{expedition0012188} gives $\lp^{2}$ convergence of the regular part and uniform boundedness of its exponential moments.

(Singular part): For the measurable set $A_{\kappa,\Lambda}
= \setone{\kappa\leq\abs{k}\leq\Lambda}$, use the notation $\mathsf{C}_{A_{\kappa,\Lambda},t}(\prbprocess)$ from \eqref{expedition0012186}.
The corrected representation \eqref{expedition0012185} gives
\begin{equation}\label{expedition0012189}
\begin{aligned}
\mathsf{U}_{0,\kappa,\Lambda,t}(\prbprocess)
=
\frac{\physcplconst^{2}}{2}
\mathsf{C}_{A_{\kappa,\Lambda},t}(\prbprocess)
-\frac{\physcplconst^{2}}{2}
B_{\kappa,\Lambda,t},
\quad
B_{\kappa,\Lambda,t}
=
\int_{A_{\kappa,\Lambda}}
\frac{1 - \napiernum^{-t \rbk{\omega(k) + \onehalf \omega(k)^2}}}
{\omega(k)\rbk{\omega(k) + \onehalf \omega(k)^2}^2}
\opdmsr{k}.
\end{aligned}
\end{equation}
(Deterministic term): This converges absolutely.
In particular, the following argument holds for suitable constants $c_1,c_2>0$.
Using the estimate $1-\napiernum^{-t \rbk{\omega+\onehalf\omega^2}}
\leq t(\omega+\onehalf\omega^2)$ under the condition $\omega(k)
\leq 1$, the integrand is bounded by $c_1\omega(k)^{-2}$, and in three dimensions
$\int_{\omega(k) \leq 1}
\omega(k)^{-2}
\opdmsr{k}
< \infty$ holds.
On the other hand, when $\omega(k)\geq1$, the integrand is bounded by $c_2\omega(k)^{-5}$, and even after multiplying by the three-dimensional measure, in the ultraviolet direction
$\int_1^\infty
r^{-3}
\opdmsr{r}
< \infty$ holds.
Therefore $B_{\kappa,\Lambda,t}$ converges to $B_t$ in \eqref{expedition0012202}.

(Centering term): Taking two cutoffs $A_{\kappa,\Lambda}$ and $A_{\kappa',\Lambda'}$, their difference is the integral over the symmetric difference $A_{\kappa,\Lambda}
\triangle A_{\kappa',\Lambda'}$.
By Lemma \ref{expedition0012142}, for some constant $c_3>0$,
\begin{equation}\label{expedition0012190}
\sqfun{\prbexp_{\msrprb_{\txtparticle,\sminvtemperature}}}
{\abs{\mathsf{C}_{A_{\kappa,\Lambda},t}
-\mathsf{C}_{A_{\kappa',\Lambda'},t}}^{2}}
\leq
c_3
\int_{A_{\kappa,\Lambda}\triangle A_{\kappa',\Lambda'}}
\frac{\opdmsr{k}}
{\rbk{\omega(k)+\onehalf\omega(k)^2}^{2}}
\end{equation}
is obtained.
The dominating function on the right-hand side is integrable on $\fldreal^{3}$, and the right-hand side converges to $0$ as the cutoffs satisfy $\kappa,\kappa'\downarrow0$ and $\Lambda,\Lambda'\uparrow\infty$.
Therefore, by \eqref{expedition0012190}, $\mathsf{C}_{A_{\kappa,\Lambda},t}$ is a Cauchy sequence in $\fun{\lp^{2}}{\msrprb_{\txtparticle,\sminvtemperature}}$ and converges in $\lp^{2}$ to $\mathsf{C}_t$ defined in \eqref{expedition0012202}.
The above argument and \eqref{expedition0012187}, \eqref{expedition0012189} imply that $\mathsf{U}_{\kappa,\Lambda,t}
\to
\mathsf{U}_{t}$ in $\fun{\lp^{2}}{\msrprb_{\txtparticle,\sminvtemperature}}$.

(Exponential moments): By the estimate for the regular part in Proposition \ref{expedition0012143}, for any $p<\infty$, with a suitable constant $c_4$, for all cutoffs and paths,
$\napiernum^{p\mathsf{U}_{\sminvtemperature,\kappa,\Lambda,t}(\prbprocess)}
\leq
c_4$ holds.
By absolute convergence of the deterministic term $B_{\kappa,\Lambda,t}$, for some constant $c_5$,
$\napiernum^{-p\frac{\physcplconst^2}{2}B_{\kappa,\Lambda,t}}
\leq
c_5$ holds uniformly.

The remaining issue is the exponential moment of $\mathsf{C}_{A_{\kappa,\Lambda},t}$.
We reuse the decomposition
$\mathsf{C}_{A_{\kappa,\Lambda},t}=M_{\kappa,\Lambda,t}+D_{\kappa,\Lambda,t}$ used in the proof of Lemma \ref{expedition0012142}.
For $D_{\kappa,\Lambda,t}$ coming from the deterministic bridge correction, the deterministic bridge correction estimate in Lemma \ref{expedition0012142} gives, with a suitable constant $c_6$,
$\napiernum^{p\frac{\physcplconst^2}{2}D_{\kappa,\Lambda,t}(\prbprocess)}
\leq
c_6$ uniformly.

We discuss the centered bridge fluctuation $M_{\kappa,\Lambda,t}$.
Applying the Brownian bridge martingale representation in the proof of Lemma \ref{expedition0012142} to $0 \leq s \leq u \leq r \leq t$, for each $(r,s,k)$ the fluctuation can be represented as
$$\int_s^r
\vagrad_y\Phi_{r,s,k}(u,\prbprocess_u;\prbprocess_s)
\vainnprod\opdmsr{M_u}.$$
Substituting this into the time-momentum integral for $\mathsf{C}_{A_{\kappa,\Lambda},t}$, Fubini's theorem for stochastic integrals is applicable by the square integrability discussed below.
Thus, moving the $u$-integral outside, we can write
$$\begin{aligned}
M_{\kappa,\Lambda,t}
&=
\int_0^t
\mathcal{V}_{\kappa,\Lambda,t}(u,\prbprocess)
\vainnprod\opdmsr{M_u},
\\ %%%%%%%%%%%%%%%%
\mathcal{V}_{\kappa,\Lambda,t}(u,\prbprocess)
&=
\int_u^t
\int_0^u
\int_{A_{\kappa,\Lambda}}
\frac{\napiernum^{-(r-s)\omega(k)}}{\omega(k)}
\vagrad_y\Phi_{r,s,k}(u,\prbprocess_u;\prbprocess_s)
\opdmsr{k}\opdmsr{s}\opdmsr{r}.
\end{aligned}$$
The square integrability needed for this application follows from the heat-kernel gradient estimate used in the estimate of the centering term in Lemma \ref{expedition0012142}.
That is, for suitable constants $c_7,c_8>0$,
\begin{equation}\label{expedition0012192}
\abs{\vagrad_y\Phi_{r,s,k}(u,y;z)}
\leq
c_7
\omega(k)
\napiernum^{-c_8(r-u)\omega(k)^2}
\end{equation}
holds.
By \eqref{expedition0012192} and the time-integral estimate in Lemma \ref{expedition0012142}, for some constant $c_9>0$,
$$\int_0^t
\abs{\mathcal{V}_{\kappa,\Lambda,t}(u,\prbprocess)}^2
\opdmsr{u}
\leq
c_9
\int_{A_{\kappa,\Lambda}}
\frac{\opdmsr{k}}
{\rbk{\omega(k)+\onehalf\omega(k)^2}^{2}}$$
holds.
The passage to the particle loop measure uses the Hölder estimate on finite time intervals and the heat-kernel estimate, as in the final part of the proof of Lemma \ref{expedition0012142}.
Therefore, for some constant $c_{10}>0$, under the particle loop measure,
$\prbdbkquadvar{M_{\kappa,\Lambda,t}}_{t}
\leq
c_{10}
\int_{\fldreal^{3}}
\frac{\opdmsr{k}}
{\rbk{\omega(k)+\onehalf\omega(k)^2}^{2}}$ holds.
Hereafter, take $c_{11}
=
c_{10}
\int_{\fldreal^{3}}
\frac{\opdmsr{k}}
{\rbk{\omega(k)+\onehalf\omega(k)^2}^{2}}
<\infty$.
The exponential martingale $\napiernum^{aM_{\kappa,\Lambda,t}
-\onehalf a^{2}\prbdbkquadvar{M_{\kappa,\Lambda,t}}_{t}}$ of a continuous local martingale is a positive local martingale with expectation at most $1$, so for any $a
\in \fldreal$,
$$\sqfun{\prbexp_{\msrprb_{\txtparticle,\sminvtemperature}}}
{\napiernum^{aM_{\kappa,\Lambda,t}}}
\leq
\napiernum^{\onehalf a^2 c_{11}}$$
holds uniformly in the cutoffs.

Apply the above argument with $a
= \onehalf p\physcplconst^2$.
With the constant $c_{12}
=
c_4c_5c_6
\napiernum^{\frac{p^2\physcplconst^4}{8}c_{11}}$, we have
$\sqfun{\prbexp_{\msrprb_{\txtparticle,\sminvtemperature}}}
{\napiernum^{p\mathsf{U}_{\kappa,\Lambda,t}}}
\leq
c_{12}$ for all cutoffs.
In particular, for any $p<\infty$,
$$\sup_{0<\kappa<\Lambda<\infty}
\sqfun{\prbexp_{\msrprb_{\txtparticle,\sminvtemperature}}}
{\napiernum^{p \mathsf{U}_{\kappa,\Lambda,t}}}
< \infty$$
holds.
\end{proof}

Restricting to the physical space \(\sphilb{D}_{\txtbsn,\txtphys,\sminvtemperature}\), we discuss the limit of the semigroup kernels. Proposition \ref{expedition0012144} uses integrability in the sense of belonging to the domain of \(\mathsf{m}\), and also assumes the cutoff removal of Proposition \ref{expedition0012383}. However, in the estimate of this proposition itself, what is used is the weighted integrability obtained from \(f
\in \sphilb{D}_{\txtbsn,\txtphys,\sminvtemperature}
\subset \dom \mathsf{m}\).

\begin{prop}\label{expedition0012144}
Assume that the function $h_t$ is a finite linear combination of terms of the form $j_{s} f$ with $f
\in \sphilb{D}_{\txtbsn,\txtphys,\sminvtemperature}$.
Then $\opform{q}_{\txtbsn,\txtbec,\sminvtemperature}^{\txteuclid}(h_t,\mathsf{J}_{\closedinterval{0}{t},\kappa,\Lambda}(\prbprocess))$ converges for each $\prbprocess$ as $\kappa\downarrow0$ and $\Lambda\uparrow\infty$, and has the uniform estimate needed for integration.
\end{prop}

\begin{proof}
It suffices to prove the case of a single $h_t
= j_{a} f$ for any $0 \leq a \leq t$.
By \eqref{expedition0012151}, the zero-mode component vanishes in the cross term with the cutoff source.
Writing it explicitly as an integral, we obtain
\begin{equation}\label{expedition0012195}
\begin{aligned}
\fun{\opform{q}_{\txtbsn,\txtbec,\sminvtemperature}^{\txteuclid}}
{j_{a} f, \mathsf{J}_{\closedinterval{0}{t},\kappa,\Lambda}(\prbprocess)}
=
\int_0^t
\int_{\fldreal^{d}}
\faftr{f}(k)
\frac{\overline{\faftr{\varrho}_{\kappa,\Lambda}(k)}}{\omega(k)^{\onehalf}}
\frac{\napiernum^{-\abs{a-s}\omega(k)}
+\napiernum^{-(\sminvtemperature-\abs{a-s})\omega(k)}}
{1-\napiernum^{-\sminvtemperature\omega(k)}}
\napiernum^{-\imunit k \prbprocess_s}
\opdmsr{k} \opdmsr{s}.
\end{aligned}
\end{equation}
Taking the absolute value removes the exponential $\napiernum^{\imunit k \prbprocess_s}$, and after interchanging the order of integration in that form, for some constant $c_{1}$,
$$\begin{aligned}
\int_0^t
\frac{\napiernum^{-\abs{a-s}\omega(k)}
+\napiernum^{-(\sminvtemperature-\abs{a-s})\omega(k)}}
{1-\napiernum^{-\sminvtemperature\omega(k)}}
\opdmsr{s}
\leq
c_{1}
\int_{S_{\sminvtemperature}}
\frac{\napiernum^{-\abs{r}\omega(k)}
+\napiernum^{-(\sminvtemperature-\abs{r})\omega(k)}}
{1-\napiernum^{-\sminvtemperature\omega(k)}}
\opdmsr{r}
=
\frac{2c_{1}}{\omega(k)}
\end{aligned}$$
holds.
Therefore, for some constant $c_{2}$,
$$\begin{aligned}
\abs{\fun{\opform{q}_{\txtbsn,\txtbec,\sminvtemperature}^{\txteuclid}}
{j_{a}f, \mathsf{J}_{\closedinterval{0}{t},\kappa,\Lambda}(\prbprocess)}}
\leq
c_{2}
\int_{\fldreal^{d}}
\frac{\abs{\faftr{f}(k)}}
{\omega(k)^{\frac{3}{2}}}
\opdmsr{k}
<
\infty
\end{aligned}$$
is obtained.
The right-hand side is finite because $f
\in \sphilb{D}_{\txtbsn,\txtphys,\sminvtemperature}
\subset \dom \mathsf{m}$.
By assumption, $\overline{\faftr{\varrho}_{\kappa,\Lambda}(k)}
\to 1$, and as an integrable dominating function independent of the cutoffs on $[0,t] \times \fldreal^{d}$ in \eqref{expedition0012195}, we can take $\frac{\abs{\faftr{f}(k)}}{\omega(k)^{\onehalf}}$ multiplied by the finite-temperature kernel.
Thus the cutoff-removal limit follows from the dominated convergence theorem.
For finite linear combinations, apply the same argument to each term and add the results.
\end{proof}

We assume the cutoff removal of Proposition \ref{expedition0012383}. No new dimension-dependent estimate appears in the limiting procedure for the following theorem; we use the convergence and uniform integrability from Proposition \ref{expedition0012144} and Proposition \ref{expedition0012383}.

\begin{thm}[Renormalized local semigroup of the point-source Nelson model]\label{expedition0012156}
For any $f,g
\in \sphilb{D}_{\txtbsn,\txtphys,\sminvtemperature}$, consider the tensor-product-type generators $F
=
F_{\txtparticle}(\prbprocess_{0})
\napiernum^{\imunit \opfocksegal(j_0 f)}$ and $G
= G_{\txtparticle}(\prbprocess_{0})
\napiernum^{\imunit \opfocksegal(j_0 g)}$.
Then there exists a local Hermitian semigroup $P^{\txtrenormalization}
= \fml{P_t^{\txtrenormalization}}{t \in \closedinterval{0}{\frac{\sminvtemperature}{2}}}$ satisfying
$\bkt{F}{P_{t}^{\txtrenormalization} G}
=
\lim_{\kappa \downarrow 0, \Lambda \uparrow \infty}
\bkt{F}{P_{\kappa,\Lambda,t}^{\txtrenormalization} G}$.
The limit inherits the local semigroup property, Hermiticity, and reflection property of fixed cutoffs in the sense of bilinear forms.
\end{thm}

\begin{proof}
Use the kernel representation on generators in \eqref{expedition0012180}.
The particle-only effective action in \eqref{expedition0012182} converges in $\lp^{2}$, and its exponential weights are uniformly integrable.
The cross term between the field and elements of the form $j_{s} f$ can be handled by the dominated convergence theorem using \eqref{expedition0012195} and Proposition \ref{expedition0012144}.
Therefore the limit of the sesquilinear forms can be taken by Hölder's inequality and Vitali's theorem.
Writing the local semigroup property, Hermiticity, and reflection property that hold for fixed cutoffs as bilinear-form identities and taking the limit, the corresponding properties are inherited by the limiting kernel.
\end{proof}

\subsubsection{Limit of KMS States and Physical Fields}\label{limit-of-kms-states-and-physical-fields}

We assume the local semigroup limit of Theorem \ref{expedition0012156}. This proposition itself uses no new dimension-dependent estimate.

\begin{prop}[KMS state after cutoff removal]\label{expedition0012157}
For times $-\frac{\sminvtemperature}{2}
\leq t_1
\leq \cdots
\leq t_n
\leq \frac{\sminvtemperature}{2}$, bounded measurable functions $F_j$, and $f_j
\in \sphilb{D}_{\txtbsn,\txtphys,\sminvtemperature}$, set $A_j
=
F_j(\prbprocess_{t_j})
\napiernum^{\imunit \opfocksegal(j_{t_j}f_j)}$.
Then the finite-time correlation functions after cutoff removal are defined by
$$\begin{aligned}
\oastate[\psi_{\txtnelson,\sminvtemperature}](A_1 \cdots A_n)
=
\lim_{\kappa\downarrow0,\Lambda\uparrow\infty}
\oastate[\psi_{\txtnelson,\sminvtemperature,\kappa,\Lambda}](A_1 \cdots A_n).
\end{aligned}$$
This limit defines the KMS state of the point-source Nelson model as a functional-integral representation.
\end{prop}

\begin{proof}
First express the cutoff correlation functions as products of semigroup kernels.
For each $j
= 1,\ldots,n-1$, set the time difference $\tau_j
= t_{j+1}-t_j$, and set the remaining time on the circle to $\tau_n
= \sminvtemperature-t_n+t_1$.
Also denote the time-zero multiplication operator by $\mathsf{M}(F,f)
= F(\prbprocess_0)\napiernum^{\imunit\opfocksegal(j_0 f)}$.
If necessary, by splitting intervals whose length exceeds $\frac{\sminvtemperature}{2}$ in the middle and inserting the identity operator, each local semigroup kernel can be applied only at times in its domain.
By reconstruction from local semigroups in the textbook \cite{DerezinskiGerard001}, take the partition $0
= r_0
< r_1
< \cdots
< r_m
= \sminvtemperature$ used in the denominator so that each difference $r_{\ell}
-r_{\ell-1}$ belongs to $\closedinterval{0}{\frac{\sminvtemperature}{2}}$.
For the cutoff state,
\begin{equation}\label{expedition0012197}
\begin{aligned}
\oastate[\psi_{\txtnelson,\sminvtemperature,\kappa,\Lambda}]
(A_1\cdots A_n)
=
\frac{\bkt{1}
{P_{\kappa,\Lambda,\tau_1}^{\txtrenormalization}\mathsf{M}(F_2,f_2)
P_{\kappa,\Lambda,\tau_2}^{\txtrenormalization}
\cdots
\mathsf{M}(F_n,f_n)
P_{\kappa,\Lambda,\tau_n}^{\txtrenormalization}
\mathsf{M}(F_1,f_1) 1}}
{\bkt{1}{
P_{\kappa,\Lambda,r_1-r_0}^{\txtrenormalization}
P_{\kappa,\Lambda,r_2-r_1}^{\txtrenormalization}
\cdots
P_{\kappa,\Lambda,r_m-r_{m-1}}^{\txtrenormalization}1}}
\end{aligned}
\end{equation}
can be written.

By Theorem \ref{expedition0012156}, each factor $\bkt{F}{P_{\kappa,\Lambda,\tau}^{\txtrenormalization}G}$ has a cutoff-removal limit on generators whose field direction lies in $\sphilb{D}_{\txtbsn,\txtphys,\sminvtemperature}$.
Since only finitely many factors appear, sending the semigroup kernels in \eqref{expedition0012197} to the limit from the left in order gives, for the numerator,
$$\begin{aligned}
&\bkt{1}
{P_{\kappa,\Lambda,\tau_1}^{\txtrenormalization}\mathsf{M}(F_2,f_2)
P_{\kappa,\Lambda,\tau_2}^{\txtrenormalization}\cdots
\mathsf{M}(F_n,f_n)
P_{\kappa,\Lambda,\tau_n}^{\txtrenormalization}\mathsf{M}(F_1,f_1)1}
\\ %%%%%%%%%%%%%%%%
&\to
\bkt{1}
{P_{\tau_1}^{\txtrenormalization}\mathsf{M}(F_2,f_2)
P_{\tau_2}^{\txtrenormalization}\cdots
\mathsf{M}(F_n,f_n)
P_{\tau_n}^{\txtrenormalization}\mathsf{M}(F_1,f_1)1}.
\end{aligned}$$
Applying the same argument in the case $F_j=1$ and $f_j=0$, the denominator also has a limit.
The denominator corresponds to the limit of the cutoff partition function and is used as a positive normalization factor.

Next, verify circular cyclicity, which corresponds to the KMS property.
In the semigroup representation of the cutoff correlation function, moving the observable $A_1$ to the end cyclically changes the sequence of time differences from $\tau_1,\ldots,\tau_n$ to $\tau_2,\ldots,\tau_n,\tau_1$.
By the product representation of local semigroups and associativity of the inner product, this is only the representation of the correlation function on the same circle $S_{\sminvtemperature}$ viewed from a different starting point.
Therefore the circular cyclicity of the form
$$\begin{aligned}
\oastate[\psi_{\txtnelson,\sminvtemperature,\kappa,\Lambda}]
(A_1A_2\cdots A_n)
=
\oastate[\psi_{\txtnelson,\sminvtemperature,\kappa,\Lambda}]
(A_2\cdots A_n A_1)
\end{aligned}$$
that holds for the cutoff state is preserved in the cutoff-removal limit.
Similarly, the reflection property and time-translation covariance that hold for fixed cutoffs are also preserved in the sense of bilinear forms by the limit in Theorem \ref{expedition0012156}.
Hence the limiting correlation functions give a system of Euclidean correlation functions with $\sminvtemperature$-periodicity, reflection, and Markov property.
This is the KMS state by functional integration in the sense of this section.

For normalized correlation functions on the time circle $S_{\sminvtemperature}$, looking explicitly at the contribution of the scalar factor inserted into the renormalized kernel, since the sum of the lengths of all intervals is $\sum_{j=1}^{n}\tau_j
= \sminvtemperature$, the numerator contains
$$\begin{aligned}
\prod_{j=1}^{n}
\napiernum^{\tau_j
\physenergy_{\txtnelson,\kappa,\Lambda}^{\txtrenormalization}}
=
\napiernum^{\sminvtemperature
\physenergy_{\txtnelson,\kappa,\Lambda}^{\txtrenormalization}}.
\end{aligned}$$
The same factor appears in the denominator, so it cancels in the normalized correlation functions.
Therefore the renormalization constant is not an arbitrary additive constant left in the KMS state, but the normalization required for the Feynman--Kac--Nelson kernel and the local semigroup kernel to converge in the point-source limit.
The preceding argument shows that the Euclidean KMS state of the point-source Nelson model is not the formal limit of the unrenormalized kernel, but is obtained as the limit of the local Hermitian semigroups constructed from the renormalized Feynman--Kac--Nelson kernel according to the definition in the textbook \cite{DerezinskiGerard001}.
\end{proof}

We use the state after cutoff removal obtained in Proposition \ref{expedition0012157}. This proposition itself uses no new dimension-dependent estimate, and the required cutoff removal is contained in the cutoff removal of the source in Proposition \ref{expedition0012360} and the convergence of the state in Proposition \ref{expedition0012157}.

\begin{prop}[Source on particle paths after cutoff removal]\label{expedition0012360}
Fix a closed interval $I
\subset S_{\sminvtemperature}$.
For the cutoff source $\mathsf{J}
_{I,\kappa,\Lambda}(\prbprocess)$ defined in \eqref{expedition0012203}, define the source $\mathsf{J}_{I}(\prbprocess)$ with infrared and ultraviolet cutoffs removed by the cross form for any finite linear combination $h
=
\sum_{\ell=1}^{n}
c_\ell
j_{a_\ell} f_\ell$,
$a_\ell
\in S_{\sminvtemperature}$,
$f_\ell
\in \sphilb{D}_{\txtbsn,\txtphys,\sminvtemperature}$:
\begin{equation}\label{expedition0012361}
\begin{aligned}
\fun{\opform{q}_{\txtbsn,\txtbec,\sminvtemperature}^{\txteuclid}}
{h,\mathsf{J}_{I}(\prbprocess)}
=
\sum_{\ell=1}^{n}
c_\ell
\int_I
\opdmsr{r}
\int_{\fldreal^d}
\faftr{f_\ell}(k)
\frac{\napiernum^{-\abs{a_\ell-r}\omega(k)}
+\napiernum^{-(\sminvtemperature-\abs{a_\ell-r})\omega(k)}}
{1-\napiernum^{-\sminvtemperature\omega(k)}}
\frac{\napiernum^{-\imunit k\prbprocess_r}}
{\omega(k)^{\onehalf}}
\opdmsr{k}.
\end{aligned}
\end{equation}
Then, for each particle path $\prbprocess$,
\begin{equation}\label{expedition0012362}
\begin{aligned}
\lim_{\kappa\downarrow0,\Lambda\uparrow\infty}
\fun{\opform{q}_{\txtbsn,\txtbec,\sminvtemperature}^{\txteuclid}}
{h,\mathsf{J}_{I,\kappa,\Lambda}(\prbprocess)}
=
\fun{\opform{q}_{\txtbsn,\txtbec,\sminvtemperature}^{\txteuclid}}
{h,\mathsf{J}_{I}(\prbprocess)}
\end{aligned}
\end{equation}
holds.
Furthermore, the right-hand side and left-hand side are bounded by an integrable dominating function independent of the cutoffs.
\end{prop}

\begin{proof}
By linearity, it suffices to prove the case $h
= j_a f$.
By \eqref{expedition0012151}, the zero-mode component vanishes in the cross term with the cutoff source, so
\begin{equation}\label{expedition0012363}
\begin{aligned}
\fun{\opform{q}_{\txtbsn,\txtbec,\sminvtemperature}^{\txteuclid}}
{j_a f,\mathsf{J}_{I,\kappa,\Lambda}(\prbprocess)}
=
\int_I
\opdmsr{r}
\int_{\fldreal^{d}}
\faftr{f}(k)
\frac{\overline{\faftr{\varrho}_{\kappa,\Lambda}(k)}}{\omega(k)^{\onehalf}}
\frac{\napiernum^{-\abs{a-r}\omega(k)}
+\napiernum^{-(\sminvtemperature-\abs{a-r})\omega(k)}}
{1-\napiernum^{-\sminvtemperature\omega(k)}}
\napiernum^{-\imunit k \prbprocess_r}
\opdmsr{k}
\end{aligned}
\end{equation}
holds.
Taking the absolute value removes the phase depending on the particle path.
If $\abs{I}$ denotes the length of the closed interval $I$, then integrating the finite-temperature kernel over $S_{\sminvtemperature}$ gives, for some constant $C_I$,
\begin{equation}\label{expedition0012364}
\begin{aligned}
\abs{\fun{\opform{q}_{\txtbsn,\txtbec,\sminvtemperature}^{\txteuclid}}
{j_a f,\mathsf{J}_{I,\kappa,\Lambda}(\prbprocess)}}
\leq
C_I
\int_{\fldreal^d}
\frac{\abs{\faftr{f}(k)}}{\omega(k)^{\frac32}}
\opdmsr{k}
<\infty.
\end{aligned}
\end{equation}
The right-hand side is finite because $f
\in \sphilb{D}_{\txtbsn,\txtphys,\sminvtemperature}
\subset \dom\mathsf{m}$.
By assumption, $\overline{\faftr{\varrho}_{\kappa,\Lambda}(k)}
\to 1$, and the dominating function in \eqref{expedition0012364} can be used, so \eqref{expedition0012362} follows from the dominated convergence theorem.
\end{proof}

Define the weighted particle loop measure after cutoff removal by \begin{equation}\label{expedition0012365}
\begin{aligned}
\opdmsr{\widetilde{\msrprb}_{\txtparticle,\sminvtemperature}(\prbprocess)}
=
\frac{\fnexp{\frac{\physcplconst^2}{4}
\fun{\opform{q}_{\txtbsn,\txtbec,\sminvtemperature}^{\txteuclid}}
{\mathsf{J}_{S_{\sminvtemperature}}(\prbprocess)}}}
{\int_{\Omega_{\txtparticle,\sminvtemperature}}
\fnexp{\frac{\physcplconst^2}{4}
\fun{\opform{q}_{\txtbsn,\txtbec,\sminvtemperature}^{\txteuclid}}
{\mathsf{J}_{S_{\sminvtemperature}}(\mathit Y)}}
\opdmsr{\msrprb_{\txtparticle,\sminvtemperature}}(\mathit Y)}
\opdmsr{\msrprb_{\txtparticle,\sminvtemperature}}(\prbprocess).
\end{aligned}
\end{equation}

\begin{prop}[Physical field after cutoff removal]\label{expedition0012158}
For any $f
\in \sphilb{D}_{\txtbsn,\txtphys,\sminvtemperature}$, define
\begin{equation}\label{expedition0012351}
\begin{aligned}
\mathsf{Y}_{\txtnelson,\sminvtemperature;f}(\prbprocess)
&=
-\frac{\physcplconst}{2}
\fun{\opform{q}_{\txtbsn,\txtbec,\sminvtemperature}^{\txteuclid}}
{j_0 f,\mathsf{J}_{S_{\sminvtemperature}}(\prbprocess)},
\\ %%%%%%%%%%%%%%%%
\ell_{\txtnelson,\sminvtemperature}(f)
&=
\int_{\Omega_{\txtparticle,\sminvtemperature}}
\mathsf{Y}_{\txtnelson,\sminvtemperature;f}(\prbprocess)
\opdmsr{\widetilde{\msrprb}_{\txtparticle,\sminvtemperature}}(\prbprocess).
\end{aligned}
\end{equation}
If the physical field after cutoff removal is defined by
\begin{equation}\label{expedition0012200}
\opfocksegal_{\txtphys,\sminvtemperature}(j_0 f)
=
\opfocksegal(j_0 f)
+\ell_{\txtnelson,\sminvtemperature}(f)
\end{equation}
then $\fun{\oastate[\psi_{\txtnelson,\sminvtemperature}]}
{\opfocksegal_{\txtphys,\sminvtemperature}(j_0 f)}
= 0$ holds.
\end{prop}

\begin{proof}
Let the cutoff one-point characteristic function be $\chi_{\kappa,\Lambda,f}(s)
=
\fun{\oastate[\psi_{\txtnelson,\sminvtemperature,\kappa,\Lambda}]}
{\fnexp{\imunit s\opfocksegal(j_0 f)}}$.
If we fix the particle path and integrate the field in the defining formula \eqref{expedition0012152} of the cutoff KMS state, the additional factor coming from the interaction and depending on $s$ is $$\fnexp{\frac{\imunit s\physcplconst}{2}
\fun{\opform{q}_{\txtbsn,\txtbec,\sminvtemperature}^{\txteuclid}}
{j_0 f,\mathsf{J}_{S_{\sminvtemperature},\kappa,\Lambda}(\prbprocess)}}.$$
Applying Proposition \ref{expedition0012360} with $I=S_{\sminvtemperature}$ and $h
= j_0 f$, $\fun{\opform{q}_{\txtbsn,\txtbec,\sminvtemperature}^{\txteuclid}}
{j_0 f,\mathsf{J}_{S_{\sminvtemperature},\kappa,\Lambda}(\prbprocess)}$ has a cutoff-removal limit for each particle path, and its absolute value is bounded by an integrable dominating function independent of the cutoffs.
Combining this with the same exponential-moment estimate used for convergence of the renormalized action, when $s$ is restricted to a neighborhood of the origin, $\chi_{\kappa,\Lambda,f}(s)$ and its $s$-derivative are dominated by the same integrable dominating function.
By the dominated convergence theorem and Vitali's theorem, $\chi_{\kappa,\Lambda,f}$ converges locally uniformly to a limit in a neighborhood of the origin, and the derivative at $s = 0$ can also be interchanged with the limit.

Using the state after cutoff removal defined in Proposition \ref{expedition0012157}, the one-point characteristic function after cutoff removal
$\chi_f(s)
=
\fun{\oastate[\psi_{\txtnelson,\sminvtemperature}]}
{\fnexp{\imunit s\opfocksegal(j_0 f)}}$
satisfies, in a neighborhood of the origin,
$$\chi_f(s)
=
\fnexp{-\frac{s^2}{4}
\fun{\opform{q}_{\txtbsn,\txtbec,\sminvtemperature}^{\txteuclid}}{j_0 f}}
\int_{\Omega_{\txtparticle,\sminvtemperature}}
\fnexp{-\imunit s\mathsf{Y}_{\txtnelson,\sminvtemperature;f}(\prbprocess)}
\opdmsr{\widetilde{\msrprb}_{\txtparticle,\sminvtemperature}}(\prbprocess).$$
Differentiating this formula at $s=0$, the first derivative of the Gaussian factor vanishes, and
$\fun{\oastate[\psi_{\txtnelson,\sminvtemperature}]}
{\opfocksegal(j_0 f)}
=
-\ell_{\txtnelson,\sminvtemperature}(f)$ is obtained.
In particular, the expectation of the physical field defined in \eqref{expedition0012200} is zero.
\end{proof}

\begin{rem}
The $\opfocksegal_{\txtphys,\sminvtemperature}(j_0 f)$ introduced here is not an operation replacing the covariance of the boson field with another covariance.
The covariance part is still described by $\opform{q}_{\txtbsn,\txtbec,\sminvtemperature}
=\opform{q}_{\txtbsn,0,\sminvtemperature}
+\opform{q}_{\txtbsn,\txtnonzero,\sminvtemperature}$.
The correction term $\ell_{\txtnelson,\sminvtemperature}(f)$ for the physical field is a linear renormalization for removing the first moment generated by the point-source Nelson interaction.
Therefore, the physical field after cutoff removal is a procedure on the test-function space $\sphilb{D}_{\txtbsn,\txtphys,\sminvtemperature}$, which still has meaning after removing the infrared and ultraviolet cutoffs, for normalizing the field mean to zero and separating the fluctuation part of correlations from the covariance of the BEC zero mode.
This is the point-source Nelson version of the selection criterion for the field operator discussed by Yukalov \cite{VIYukalov001}: the physically relevant field is selected after subtracting the deterministic one-point part so that its expectation vanishes on the physical test-function space.
\end{rem}

\subsubsection{Off-Diagonal Long-Range Order, Order Parameters, and the No-Go Theorem for BEC}\label{off-diagonal-long-range-order-order-parameters-and-the-no-go-theorem-for-bec}

\begin{defn}[Order Parameter of the Nelson Model]\label{expedition0012159}
Let $I_L^3$ be the cube centered at the origin with side length $L$, and set $V_L=L^3$. Define the functions approximating the zero-momentum mode in finite volume by $$\mathsf{b}_L^{(0)}
=
\frac{1}{V_L^{\onehalf}}
\fndef{I_L^3},
\quad
\mathsf{b}_L^{(1)}
=
\frac{1}{V_L}\fndef{I_L^3}.$$
For the bounded-system Nelson model state $\psi_{\txtnelson,\sminvtemperature,\smchemicalpotential,L}$, define
\begin{equation}\label{expedition0012437}
\begin{aligned}
\mathsf{o}_{\txtnelson,\sminvtemperature,L}^{(\#)}
=
\imunit
\fun{\psi_{\txtnelson,\sminvtemperature,\smchemicalpotential,L}}
{\fun{\oaresolvent}{1,\mathsf{b}_L^{(\#)}}}
\end{aligned}
\end{equation}
and call it the order parameter of the Nelson model.
\end{defn}

\begin{prop}[Off-Diagonal Long-Range Order of the Nelson Model]\label{expedition0012412}
For the Nelson-model KMS state after cutoff removal constructed in Proposition \ref{expedition0012157} and $f,g \in \sphilb{D}_{\txtbsn,\txtphys,\sminvtemperature}$, the two-point off-diagonal long-range order is given by
\begin{equation}\label{expedition0012413}
\lim_{\abs{x}\to\infty}
\fun{\oastate[\psi_{\txtnelson,\sminvtemperature}]}
{\opfocksegal(j_0 f)\opfocksegal(j_0\tau_x g)}
=
\onehalf\fun{\opform{q}_{\txtbsn,0,\sminvtemperature}}{f,g}.
\end{equation}
\end{prop}

\begin{proof}
Apply \eqref{expedition0012197} of Proposition \ref{expedition0012157} to a two-point exponential observable. Under spatial translation, $\faftr{\tau_xg}(k)=\napiernum^{-\imunit kx}\faftr{g}(k)$. The zero-mode form is independent of the spatial variable because $\faftr{\tau_xg}(0)=\faftr{g}(0)$, while the cross term of the non-zero-mode covariance converges to $0$ by the Riemann--Lebesgue lemma. The same oscillatory factor appears in the cross term with the source in Proposition \ref{expedition0012360}, and the dominating function in \eqref{expedition0012364} is available; hence the cross term between the translated external field and the particle source vanishes by the dominated convergence theorem. Differentiating the two-point characteristic functional in two coefficients shows that the only cross term remaining in the long-distance limit is the zero-mode form coming from the free Bose gas.
\end{proof}

\begin{prop}[Order-Parameter Criterion for the Nelson Model]\label{expedition0012414}
For the order parameter of Definition \ref{expedition0012159}, the following assertions hold. The condition $\smnumberdensity_{\txtbsn,0}(\sminvtemperature)>0$ is equivalent to
$$\lim_{L\to\infty}
\mathsf{o}_{\txtnelson,\sminvtemperature,L}^{(1)}
=
\int_0^\infty
\fnexp{-r-\frac12\smnumberdensity_{\txtbsn,0}(\sminvtemperature)r^2}
\opdmsr{r}
<1.$$
Moreover, $\smnumberdensity_{\txtbsn,0}(\sminvtemperature)=0$ is equivalent to $$\lim_{L\to\infty}\mathsf{o}_{\txtnelson,\sminvtemperature,L}^{(1)}=1.$$
\end{prop}

\begin{proof}
The finite-volume function $\mathsf{b}_L^{(1)}$ has the normalization that detects only the zero-momentum mode. In the finite-volume approximation of Proposition \ref{expedition0012157}, the zero-momentum component of the interaction source does not cross with the external zero mode because of the compensation in \eqref{expedition0012151}. Therefore, in the Laplace-transform representation of the resolvent, only the zero-mode evaluation of the free Bose gas remains. This evaluation is the same as the computation for the free Bose gas in \cite{YoshitsuguSekine004}, and the displayed limit formula follows.
\end{proof}

\begin{thm}[No-Go Theorem for BEC in the Nelson Model via Off-Diagonal Long-Range Order]\label{expedition0012160}
Assume that the physical test-function space $\sphilb{D}_{\txtbsn,\txtphys,\sminvtemperature}$ can distinguish the zero mode in the sense of Definition \ref{expedition0012433}. Then the following assertions are equivalent for the cutoff-removed KMS state constructed in Proposition \ref{expedition0012157}.
\begin{enumerate}
\item
For all $f,g \in \sphilb{D}_{\txtbsn,\txtphys,\sminvtemperature}$, the off-diagonal long-range order in \eqref{expedition0012413} is $0$.

\item
The equality $\opform{q}_{\txtbsn,0,\sminvtemperature}=0$ holds.

\item
The condition $\smnumberdensity_{\txtbsn,0}(\sminvtemperature)=0$ holds.

\item
The order parameter of Definition \ref{expedition0012159} satisfies $\lim_{L\to\infty}\mathsf{o}_{\txtnelson,\sminvtemperature,L}^{(1)}=1$.
\end{enumerate}
In particular, the absence of off-diagonal long-range order and the order-parameter criterion are equivalent to the disappearance of the BEC zero mode on the physical test-function space after cutoff removal.
\end{thm}

\begin{proof}
Proposition \ref{expedition0012412} gives the equivalence between assertions (1) and (2). Zero-mode distinguishability and the defining formula $\fun{\opform{q}_{\txtbsn,0,\sminvtemperature}}{f}
=
2(2\pi)^3\smnumberdensity_{\txtbsn,0}(\sminvtemperature)\abs{\faftr{f}(0)}^2$ give the equivalence between assertions (2) and (3). Proposition \ref{expedition0012414} gives the equivalence between assertions (3) and (4).
\end{proof}

\section{Spinless Pauli--Fierz Model}\label{expedition0012278}

\subsection{Definition of the Hamiltonian}\label{definition-of-the-hamiltonian-1}

To avoid confusion with the chemical potential, in the Pauli--Fierz model the indices for spatial components are \(1
\leq \mathrm{i},\mathrm{j}
\leq d\), and the polarization index is \(\mathrm{j}_{\txtrad}\in\ringratint_2=\setone{1,2}\). Following the textbook \cite[Section 3.2.1]{LorincziHiroshimaBetz3}, the total Hilbert space of the interacting system consisting of one spinless electron and the radiation field is \[\begin{aligned}
\mathcal{H}_{\txtpaulifierz}
=
\sphilb{H}_{\txtparticle}
\otimes
\spfock_{\txtrad},
\quad
\sphilb{H}_{\txtparticle}
=
\fun{\lp^2}{\fldreal^{d}},
\quad
\spfock_{\txtrad}
=
\spfock_{\txtbsn}\rbk{\fun{\lp^2}{\fldreal^{d} \times \ringratint_{2}}}.
\end{aligned}\] The factor \(\ringratint_{2}\) in the radiation field represents the two polarization degrees of freedom of photons. The dispersion relation of the radiation field is \(\omega(k)
= \abs{k}\), and the free Hamiltonian of the radiation field is given by the second-quantization operator \(\opfocksndqntdiff_{\txtrad}\) on the radiation field as \(\physham_{\txtrad,\txtfr}
= \opfocksndqntdiff_{\txtrad}(\omega \oplus \omega)\).

First consider the standard Pauli--Fierz model with a regular charge distribution \(\varphi\). As in the textbook \cite{LorincziHiroshimaBetz3}, assume that its Fourier transform \(\faftr{\varphi}\) satisfies \(\faftr{\varphi}(-k)=\overline{\faftr{\varphi}(k)}\), and that \(\sqrt{\omega}\faftr{\varphi}\), \(\faftr{\varphi}/\sqrt{\omega}\), \(\faftr{\varphi}/\omega\) belong to \(\fun{\lp^2}{\fldreal^{d}}\). For \(k \neq 0\), define the transverse projection \(\delta^{\perp}(k)\) by \begin{equation}\label{expedition0012315}
\begin{aligned}
\delta^{\perp}(k)
\colon \fldreal^{d}
\to \fldreal^{d};
\quad
\delta^\perp(k)v
= v-\frac{k\vainnprod v}{\abs{k}^{2}}k.
\end{aligned}
\end{equation} This is the orthogonal projection onto the hyperplane \(k^\perp
=\set{v \in \fldreal^d}{v \cdot k=0}\). For any \(v
\in \fldreal^d\), it satisfies \(\delta^{\perp}(k)v \vainnprod k
= 0\), and for any \(u
\in k{^{\perp}}\), \(\rbk{v - \delta^{\perp}(k) v} \vainnprod u
= 0\) holds. Furthermore, for each \(k
\neq 0\), choose \(e^{1}(k)\) and \(e^{2}(k)\) as a real orthonormal basis of \(k^\perp\), and write the components as \(e^{\mathrm{j}_{\txtrad}}(k)
= \vecbk{e_{\mathrm{i}}^{\mathrm{j}_{\txtrad}}(k)}_{\mathrm{i}=1}^{d}\). Here the lower index \(\mathrm{i}\) of \(e_{\mathrm{i}}^{\mathrm{j}_{\txtrad}}(k)\) denotes the spatial component, and the upper index \(\mathrm{j}_{\txtrad}\) denotes the polarization number. The Coulomb gauge condition is \(k\vainnprod e^{\mathrm{j}_{\txtrad}}(k)=0\), and we choose the convention so that \(e_{\mathrm{i}}^{\mathrm{j}_{\txtrad}}(-k)=e_{\mathrm{i}}^{\mathrm{j}_{\txtrad}}(k)\) under inversion. Under this convention, assume that \[\sum_{\mathrm{j}_{\txtrad}\in\ringratint_2}
e^{\mathrm{j}_{\txtrad}}(k)\otimes e^{\mathrm{j}_{\txtrad}}(k)
=
\delta^\perp(k)\] holds.

As with \(\lambda_{x,\kappa,\Lambda}\) entering the source in \eqref{expedition0012203} for the Nelson model, denote by \(\lambda_{\txtrad,\mathrm{i},x}\) the one-particle coupling function corresponding to the particle position \(x
\in \fldreal^d\) and the spatial component \(1\leq\mathrm{i}\leq d\). In particular, since the Pauli--Fierz model has radiation-field components and polarizations, distinguish them as \begin{equation}\label{expedition0012377}
\begin{aligned}
\lambda_{\txtrad,\mathrm{i},x}^{\mathrm{j}_{\txtrad}}(k)
=
\frac{\faftr{\varphi}(k)}{\sqrt{\omega(k)}}e_{\mathrm{i}}^{\mathrm{j}_{\txtrad}}(k)\napiernum^{-\imunit kx},
\quad
1\leq\mathrm{i}\leq d,
\quad
\mathrm{j}_{\txtrad}\in\ringratint_2.
\end{aligned}
\end{equation} By the reality condition and the convention for polarization vectors, \[\overline{\lambda_{\txtrad,\mathrm{i},x}^{\mathrm{j}_{\txtrad}}(k)}
=
\frac{\faftr{\varphi}(-k)e_{\mathrm{i}}^{\mathrm{j}_{\txtrad}}(k)\napiernum^{\imunit kx}}
{\sqrt{\omega(k)}},\] and the field operator of the vector potential is defined by the creation and annihilation operators for the radiation field as \[A_{\mathrm{i}}(x)
=
\frac{1}{\sqrt{2}}
\sum_{\mathrm{j}_{\txtrad}\in \ringratint_{2}}
\rbk{\fun{\opfockcr_{\txtrad}}{\lambda_{\txtrad,\mathrm{i},x}^{\mathrm{j}_{\txtrad}},\mathrm{j}_{\txtrad}}
+\fun{\opfockan_{\txtrad}}{\overline{\lambda_{\txtrad,\mathrm{i},x}^{\mathrm{j}_{\txtrad}}},\mathrm{j}_{\txtrad}}}.\] Under the Coulomb gauge, \(\sum_{\mathrm{i}=1}^{d}[\vagrad_{\mathrm{i}},A_{\mathrm{i}}]=0\) holds.

To ensure existence of KMS states, the Pauli--Fierz Hamiltonian with an electron potential \(V\) satisfying properties comparable to those in the Nelson model is defined by \[\physham_{\txtpaulifierz}
=
\onehalf\rbk{-\imunit \vagrad \otimes 1 - \physcharge A}^{2}
+ V \otimes 1
+ 1 \otimes \physham_{\txtrad,\txtfr}.\] Under the above assumptions on the charge distribution and the conditions on \(V\) in Section \ref{expedition0012133}, \(\physham_{\txtpaulifierz}\) is self-adjoint and bounded below \cite{LorincziHiroshimaBetz3}.

As in the Nelson model, the final goal is the discussion for the point charge \(\varphi
= \diracdelta_{0}\). Since the point charge does not satisfy the above assumptions on regular charge distributions, the assumptions for the regular model cannot be applied directly to the point-source limit. To include the point-source limit as in the Nelson model, we first start from the charge distribution with infrared and ultraviolet cutoffs \[\faftr{\varphi}_{\kappa,\Lambda}(k)
=
\fndef{\kappa\leq \abs{k}\leq \Lambda}(k).\] For any \(0
< \kappa
< \Lambda
< \infty\), \(\faftr{\varphi}_{\kappa,\Lambda}\) satisfies the assumptions on regular charge distributions, so set \begin{equation}\label{eq:PF-cutoff-polarization-coupling}
\lambda_{\txtrad,\mathrm{i},x,\kappa,\Lambda}^{\mathrm{j}_{\txtrad}}(k)
=
\frac{\faftr{\varphi}_{\kappa,\Lambda}(k)}{\sqrt{\omega(k)}}
e_{\mathrm{i}}^{\mathrm{j}_{\txtrad}}(k)\napiernum^{-\imunit kx}
\end{equation} For the \(\mathrm{i}\)-th standard basis vector \(e_{\mathrm{i}}\) of \(\fldreal^d\), define the corresponding \(\fldcmp^d\)-valued coupling function in the transverse-projection representation by \begin{equation}\label{eq:PF-cutoff-transverse-coupling}
\lambda_{\txtrad,\mathrm{i},x,\kappa,\Lambda}(k)
=
\frac{\faftr{\varphi}_{\kappa,\Lambda}(k)}{\sqrt{\omega(k)}}
e_{\mathrm{i}}\napiernum^{-\imunit kx}
\end{equation} and let the transverse projection \(\delta^\perp(k)\) act in the covariance, extracting the same transverse-wave part as the polarization sum in the polarization representation above. Then the cutoff vector potential is \[A_{\mathrm{i},\kappa,\Lambda}(x)
=
\frac{1}{\sqrt{2}}
\sum_{\mathrm{j}_{\txtrad}\in \ringratint_{2}}
\rbk{\fun{\opfockcr_{\txtrad}}{\lambda_{\txtrad,\mathrm{i},x,\kappa,\Lambda}^{\mathrm{j}_{\txtrad}},\mathrm{j}_{\txtrad}}
+\fun{\opfockan_{\txtrad}}{\overline{\lambda_{\txtrad,\mathrm{i},x,\kappa,\Lambda}^{\mathrm{j}_{\txtrad}}},\mathrm{j}_{\txtrad}}}.\] Thus the spinless Pauli--Fierz Hamiltonian with cutoffs is defined as an infinite-volume object by \[\physham_{\txtpaulifierz,\kappa,\Lambda}
=
\onehalf\rbk{-\imunit \vagrad\otimes 1-\physcharge A_{\kappa,\Lambda}}^{2}
+V\otimes 1+1\otimes \physham_{\txtrad,\txtfr}.\]

\subsection{Euclidean Representation of the Free Radiation Field and Physical Test-Function Space}\label{euclidean-representation-of-the-free-radiation-field-and-physical-test-function-space}

For common use in the discussion of the cutoff bounded system and the cutoff infinite-volume KMS state, we define the basic setting of the finite-temperature Euclidean representation of the free radiation field, the sesquilinear forms, and the physical test-function space. The textbook \cite[Section 3.2.2]{LorincziHiroshimaBetz3} constructs the \(Q\)-space representation of the radiation field. Since the field of the Pauli--Fierz model is a vector field with transverse components, below we use the covariance needed for finite-temperature sharp-time fields, and the transverse projection \(\delta^{\perp}(k)\) defined in \eqref{expedition0012315} is used for non-zero momentum components.

For non-zero momentum, use the representation by polarization vectors. For a test function \(f
= \vecbk{f_1,\ldots,f_d}\) written in vector-potential components, define its polarization components by \[\begin{aligned}
\faftr{f}^{\mathrm{j}_{\txtrad}}(k)
=
\sum_{\mathrm{i}=1}^{d}
e_{\mathrm{i}}^{\mathrm{j}_{\txtrad}}(k)\faftr{f}_{\mathrm{i}}(k),
\quad
\mathrm{j}_{\txtrad}\in\ringratint_2,
\end{aligned}\] and define the polarization sum by \[\begin{aligned}
\sum_{\mathrm{j}_{\txtrad}\in\ringratint_2}
\overline{\faftr{f}^{\mathrm{j}_{\txtrad}}(k)}
\faftr{g}^{\mathrm{j}_{\txtrad}}(k)
&=
\bkt{\faftr{f}(k)}
{\rbk{\sum_{\mathrm{j}_{\txtrad}\in\ringratint_2}
e^{\mathrm{j}_{\txtrad}}(k)\otimes e^{\mathrm{j}_{\txtrad}}(k)}
\faftr{g}(k)}_{\fldcmp^d}
\\ %%%%%%%%%%%%%%%%
&=
\bkt{\faftr{f}(k)}
{\delta^\perp(k)\faftr{g}(k)}_{\fldcmp^d}.
\end{aligned}\] The transverse-projection representation is notation obtained by summing over polarization components.

On the other hand, at \(k
= 0\), \(k/\abs{k}\) is not defined, so substituting \(k
= 0\) into \(\delta^{\perp}(k)\) does not define the zero mode. Thus the zero mode is handled separately by first extracting the zero-momentum Fourier coefficient in the bounded system. For a vector-valued test function \(f
= \vecbk{f_1,\ldots,f_d}\), the zero-momentum coefficient is \(\faftr{f}(0)
\in \fldcmp^d\), and we specify the real directions among these coefficients that remain as condensate components; let the space of those real directions be \(\mathcal{E}_{\txtrad,0}
\subset \fldreal^d\). For example, if all components are retained, set \(\mathcal{E}_{\txtrad,0}
= \fldreal^d\), while if only the direction of a unit vector \(e_0\) is retained, set \(\mathcal{E}_{\txtrad,0}
= \fldreal e_0\). Let \(\Pi_{\txtrad,0}
\colon \fldreal^d
\to \mathcal{E}_{\txtrad,0}\) be the orthogonal projection onto this real subspace, and when it acts on \(\faftr{f}(0),\faftr{g}(0)\in\fldcmp^d\) in the covariance, extend it complex linearly. Below, \(\delta^{\perp}(k)\) is used for non-zero momentum components, and \(\Pi_{\txtrad,0}\) is used for zero-mode components. In particular, \(\Pi_{\txtrad,0}\) is not the value of \(\delta^{\perp}\) at the origin, but a projection expressing which zero-momentum polarization directions are retained as condensate directions in the bounded system.

Following the textbook \cite[Chapter 21]{DerezinskiGerard001}, we define the basic setting for the functional-integral representation. As in the Nelson model, the Bose field is described by the singular Gaussian \(\sminvtemperature\)-Markov path space \cite{YoshitsuguSekine004} associated with the free Hamiltonian \(\physham_{\txtrad,\txtfr}
= \opfocksndqntdiff(\omega \oplus \omega)\) of the radiation field: \[\pairbk{
\prbqspace_{\txtrad,\sminvtemperature},
\mblfml{S}_{\txtrad,\sminvtemperature},
\mblfml{S}_{\txtrad,0,\sminvtemperature},
U_{\txtrad,t},
R_{\txtrad},
\msrprb_{\txtrad,\sminvtemperature}}.\] Here, for \(\sphilb{K}_{\txtrad,\sminvtemperature}
=
\fun{\lp^2}{S_{\sminvtemperature};
\bigoplus^{d}\fun{\lp^2}{\fldreal^d;\fldreal}}\), using a regularizing operator \cite{YoshitsuguSekine004}, realize \(\faadjpresharp{{\prbqspace_{\txtrad,\sminvtemperature}}}
\subset
\sphilb{K}_{\txtrad,\sminvtemperature}
\subset
\prbqspace_{\txtrad,\sminvtemperature}\) as a real Gaussian measure space, and let a generic element of \(\prbqspace_{\txtrad,\sminvtemperature}\) be \(\opfocksegalradiation\). The time-zero \(\sigma\)-algebra \(\mblfml{S}_{\txtrad,0,\sminvtemperature}\) is generated by the time-zero sharp-time fields defined below, and the full \(\sigma\)-algebra \(\mblfml{S}_{\txtrad,\sminvtemperature}\) is generated by their time translations. The operator \(U_{\txtrad,t}\) is time translation along the time circle \(S_{\sminvtemperature}\), and \(R_{\txtrad}\) is time reflection. This probability space is used for the Euclidean representation of the quasifree KMS state \(\oastate[\psi_{\txtrad,\txtfr,\sminvtemperature}]\).

Define the test-function space for sharp-time fields directly so that both the finite-temperature non-zero-mode weight and the zero-mode value have meaning. First set \[\begin{aligned}
\sphilb{D}_{\txtrad,\txtnonzero,\sminvtemperature}
&=
\set{f=(f_{\mathrm{i}})_{\mathrm{i}=1}^{d}\in
\bigoplus^d\fun{\lp^2}{\fldreal^d}}
{
\int_{\fldreal^d}
\frac{1+\napiernum^{-\sminvtemperature\omega(k)}}
{1-\napiernum^{-\sminvtemperature\omega(k)}}
\sum_{\mathrm{j}_{\txtrad}\in\ringratint_2}
\abs{\faftr{f}^{\mathrm{j}_{\txtrad}}(k)}^2
\opdmsr{k}
<\infty},
\\ %%%%%%%%%%%%%%%%
\sphilb{D}_{\txtrad,0,\sminvtemperature}
&=
\bigoplus^{d}
\rbk{\fun{\lp^{1}}{\fldreal^d}\cap\fun{\lp^2}{\fldreal^d}},
\\ %%%%%%%%%%%%%%%%
\sphilb{D}_{\txtrad,\sminvtemperature}
&=
\sphilb{D}_{\txtrad,0,\sminvtemperature}
\cap
\sphilb{D}_{\txtrad,\txtnonzero,\sminvtemperature},
\end{aligned}\] and define the finite-temperature nonnegative symmetric sesquilinear form \(\opform{q}_{\txtrad,\txtbec,\sminvtemperature}^{\txteuclid}\) on this space. As in the free Bose gas, let \(\smnumberdensity_{\txtrad,0}(\sminvtemperature)\) be the nonnegative constant \cite{AsaoArai28,YoshitsuguSekine004} representing the condensate density of the radiation field obtained as the chemical-potential limit of the bounded system. Then, by the infinite-volume limit discussion below, for any \(t,s
\in S_{\sminvtemperature}\) it is given as the sum of the BEC zero-mode component and the non-zero-mode thermal fluctuations: \begin{equation}\label{eq:beta-pf-covariance}
\begin{aligned}
\fun{\opform{q}_{\txtrad,\txtbec,\sminvtemperature}^{\txteuclid}}
{j_t f,j_s g}
&=
\fun{\opform{q}_{\txtrad,0,\sminvtemperature}^{\txteuclid}}
{j_t f,j_s g}
+\fun{\opform{q}_{\txtrad,\txtnonzero,\sminvtemperature}^{\txteuclid}}
{j_t f,j_s g},
\\ %%%%%%%%%%%%%%%%
\fun{\opform{q}_{\txtrad,0,\sminvtemperature}^{\txteuclid}}
{j_t f,j_s g}
&=
\fun{\opform{q}_{\txtrad,0,\sminvtemperature}}{f,g}
=
2(2\pi)^d
\smnumberdensity_{\txtrad,0}(\sminvtemperature)
\bkt{\faftr{f}(0)}
{\Pi_{\txtrad,0}
\faftr{g}(0)}_{\fldcmp^{d}},
\\ %%%%%%%%%%%%%%%%
\fun{\opform{q}_{\txtrad,\txtnonzero,\sminvtemperature}^{\txteuclid}}
{j_t f,j_s g}
&=
\int_{\fldreal^{d}}
\frac{\napiernum^{-\abs{t-s}\omega(k)}
+\napiernum^{-(\sminvtemperature-\abs{t-s})\omega(k)}}
{1-\napiernum^{-\sminvtemperature \omega(k)}}
\sum_{\mathrm{j}_{\txtrad}\in\ringratint_2}
\overline{\faftr{f}^{\mathrm{j}_{\txtrad}}(k)}
\faftr{g}^{\mathrm{j}_{\txtrad}}(k)
\opdmsr{k}
\end{aligned}
\end{equation} is obtained.

For any \(t
\in S_{\sminvtemperature}\) and \(f
\in \sphilb{D}_{\txtrad,\sminvtemperature}\), define the sharp-time field by \[\opfocksegalradiation_t(f)
=
\opfocksegalradiation(j_t f)
=
\dualbkt{\opfocksegalradiation}{j_t f}.\] This is a centered Gaussian random variable with the above finite-temperature covariance. We formulate a sufficient condition for the non-zero-mode bilinear form with a field test function and a current source as arguments to survive in the point-source limit as a radiation-field-side \(\sphilb{D}\)-type space corresponding to the domain of \(\mathsf{m}\) defined by \eqref{expedition0012350} for the Nelson model. First, define the controlling space for the non-zero-mode bilinear form used in this paper by \begin{equation}\label{eq:PF-physical-test-space}
\begin{aligned}
\sphilb{D}_{\txtrad,\txtinteraction,\sminvtemperature}
=
\set{f\in\bigoplus^d\fun{\lp^2}{\fldreal^d}}
{\int_{\fldreal^d}
\frac{1+\napiernum^{-\sminvtemperature\omega(k)}}
{1-\napiernum^{-\sminvtemperature\omega(k)}}
\frac{\abs{\delta^\perp(k)\faftr{f}(k)}_{\fldcmp^d}}
{\sqrt{\omega(k)}}
\rbk{1+\frac{1}{\omega(k)+\onehalf\omega(k)^2}}
\opdmsr{k}<\infty}
\end{aligned}
\end{equation} This space expresses the additional condition required for the non-zero-mode cross term with the point-source current.

\begin{rem}
The second term $\rbk{\omega(k)+\onehalf\omega(k)^2}^{-1}$ in the last parentheses is not a term corresponding to the additive renormalization constants appearing in \eqref{expedition0012182} and \eqref{expedition0012185} for the Nelson model.
It is a momentum weight that appears when estimating the conditional mean and centered fluctuation of a Brownian bridge, and is introduced to dominate the cross term between a field test function and a stochastic-integral current by an integrable function independent of the cutoffs.
The first term in the parentheses corresponds to the part that bounds the finite-temperature kernel directly in absolute value, and the second term corresponds to the part through the conditional mean and centered fluctuation of the Brownian bridge.
Thus $\sphilb{D}_{\txtrad,\txtinteraction,\sminvtemperature}$ is not the maximal form domain, but is a sufficient test-function space that guarantees cutoff removal for the cross terms needed in the point-source limit.
For the transverse component for any $t
\in S_{\sminvtemperature}$,
\begin{equation}\label{eq:PF-physical-cross-kernel}
\begin{aligned}
\frac{\delta^\perp(k)\faftr{f}(k)}
{\sqrt{\omega(k)}}
\frac{\napiernum^{-\abs{t}\omega(k)}
+\napiernum^{-(\sminvtemperature-\abs{t})\omega(k)}}
{1-\napiernum^{-\sminvtemperature\omega(k)}},
\end{aligned}
\end{equation}
this condition requires the stochastic-integral pairing with the full-circle current source to have an $\lp^{1}$ limit, and imposes cutoff-independent integrability on the same type of mixed term appearing in the generator representation of the local semigroup kernel.
\end{rem}

The space \(\sphilb{D}_{\txtrad,\sminvtemperature}
=
\sphilb{D}_{\txtrad,0,\sminvtemperature}
\cap
\sphilb{D}_{\txtrad,\txtnonzero,\sminvtemperature}\) is the space on which both the zero-mode covariance \(\opform{q}_{\txtrad,0,\sminvtemperature}\) and the non-zero-mode covariance \(\opform{q}_{\txtrad,\txtnonzero,\sminvtemperature}\) are defined. This is the test-function space for the BEC sharp-time field of the free radiation field used after cutoff removal, and corresponds to the physical test-function space \(\sphilb{D}_{\txtbsn,\txtphys,\sminvtemperature}=\dom\mathsf{m}\cap\sphilb{D}_{\txtbsn,0,\sminvtemperature}\) for the Nelson model, namely to the domain of \(\mathsf{m}\) defined in \eqref{expedition0012350} and of the zero-mode covariance. Furthermore, define the test-function space used to describe the physical field operators and the cutoff removal limit as the intersection, as in the Nelson model, obtained by adding the cross condition with the point-source interaction to the BEC free-field space: \begin{equation}\label{eq:PF-physical-test-space-majorant}
\begin{aligned}
\sphilb{D}_{\txtrad,\txtphys,\sminvtemperature}
=
\sphilb{D}_{\txtrad,\sminvtemperature}
\cap
\sphilb{D}_{\txtrad,\txtinteraction,\sminvtemperature}
=
\sphilb{D}_{\txtrad,0,\sminvtemperature}
\cap
\sphilb{D}_{\txtrad,\txtnonzero,\sminvtemperature}
\cap
\sphilb{D}_{\txtrad,\txtinteraction,\sminvtemperature}
\end{aligned}
\end{equation} This \(\sphilb{D}_{\txtrad,\txtphys,\sminvtemperature}\) is the physical test-function space used after removal of the infrared and ultraviolet cutoffs. In this definition, \(\sphilb{D}_{\txtrad,\sminvtemperature}\) first preserves both the zero-mode covariance and the non-zero-mode covariance, and \(\sphilb{D}_{\txtrad,\txtinteraction,\sminvtemperature}\) further restricts the directions to those for which the non-zero-mode cross term with the point-source current is still defined after cutoff removal. The condition is imposed on the non-zero-momentum component through the transverse projection \(\delta^\perp(k)\), and no cross condition with the point-source current is imposed on the zero-mode projection \(\Pi_{\txtrad,0}\). Thus \(\sphilb{D}_{\txtrad,\txtphys,\sminvtemperature}\) is the space that preserves the test-function space of the BEC free radiation field appearing after cutoff removal and selects directions in which the cross term between the point-source Pauli--Fierz interaction and a field test function remains finite.

Consider the total system of the particle and the Bose field. Define the (singular Gaussian) \(\sminvtemperature\)-Markov path space associated with the free Hamiltonian \(\physham_{\txtpaulifierz,0}
=\physham[h]_{\txtparticle}\otimes 1+1\otimes \physham_{\txtrad,\txtfr}\) by \[\begin{aligned}
\pairbk{\prbqspace_{\txtpaulifierz,\sminvtemperature},
\mblfml{S}_{\txtpaulifierz,\sminvtemperature},
\mblfml{S}_{\txtpaulifierz,0,\sminvtemperature},
U_{\txtpaulifierz,t},
R_{\txtpaulifierz},
\msrprb_{\txtpaulifierz,0,\sminvtemperature}}.
\end{aligned}\] The sample space is \(\prbqspace_{\txtpaulifierz,\sminvtemperature}
= \Omega_{\txtparticle,\sminvtemperature}\times\prbqspace_{\txtrad,\sminvtemperature}\), the \(\sigma\)-algebras are \(\mblfml{S}_{\txtpaulifierz,\sminvtemperature}
= \mblfml{S}_{\txtparticle,\sminvtemperature}\times\mblfml{S}_{\txtrad,\sminvtemperature}\) and \(\mblfml{S}_{\txtpaulifierz,0,\sminvtemperature}
= \mblfml{S}_{\txtparticle,0,\sminvtemperature}\times\mblfml{S}_{\txtrad,0,\sminvtemperature}\), the time evolution is \(U_{\txtpaulifierz,t}
= U_{\txtparticle,t}\otimes U_{\txtrad,t}\), the time reversal is \(R_{\txtpaulifierz}
= R_{\txtparticle}\otimes R_{\txtrad}\), and the probability measure is \(\msrprb_{\txtpaulifierz,0,\sminvtemperature}
= \msrprb_{\txtparticle,\sminvtemperature}\otimes\msrprb_{\txtrad,\sminvtemperature}\).

\subsection{KMS States of the Bounded System with Chemical Potential}\label{expedition0012299}

We use the finite-temperature Euclidean representation, sesquilinear forms, and physical test-function spaces for the free radiation field, defined according to the textbook \cite[Chapter 21]{DerezinskiGerard001} and \cite{YoshitsuguSekine004}. As in the Nelson model, for the Pauli--Fierz model it is also necessary to verify from the bounded system how the BEC component appears. In this subsection, following the bounded-system arguments for the free Bose gas, the van Hove model, and the spin-boson model used in \cite{AsaoArai28,YoshitsuguSekine004,YoshitsuguSekine006}, we introduce the bounded-system version of the cutoff KMS states and compute exponential expectations of fields and the zero-mode decomposition in the bounded system.

We construct a functional integral representation of the trace state for the bounded-system Hamiltonian including the chemical potential under periodic boundary conditions. Below, fix cutoffs \(0
< \kappa
< \Lambda
< \infty\) and the side length \(L
> 0\) of the bounded system; the bounded-system setting is summarized in Subsection \ref{expedition0012145}. The Hilbert space of the bounded system is \[\begin{aligned}
\mathcal{H}_{\txtpaulifierz,L}
=
\sphilb{H}_{\txtparticle}\otimes\spfock_{\txtrad,L},
\quad
\spfock_{\txtrad,L}
=
\spfock_{\txtbsn}
\rbk{P_{L} \fun{\lp^2}{\fldreal^d} \otimes \fun{\lp^{2}}{\ringratint_2}}.
\end{aligned}\] As for the free Bose gas, for the total radiation-particle density \(\bar{\smnumberdensity}_{\txtrad}
> 0\), fix the bounded-system regularization parameter \(y_L
> 1\) by \begin{equation}\label{expedition0012387}
\begin{aligned}
\frac{1}{L^d}
\sum_{k\in\setlattice_L^d}
\frac{1}{y_L\napiernum^{\sminvtemperature\omega(k)}-1}
=
\bar{\smnumberdensity}_{\txtrad}.
\end{aligned}
\end{equation} This \(y_L\) can be viewed as defining the chemical potential \(\smchemicalpotential
< 0\) by \(y_L
= \napiernum^{-\sminvtemperature\smchemicalpotential}\). Let \(\physham_{\txtrad,L}\) be the free Hamiltonian of the bounded radiation field. Using the cutoff coupling functions in \eqref{eq:PF-cutoff-polarization-coupling} and \eqref{eq:PF-cutoff-transverse-coupling}, define the bounded-system cutoff coupling functions by \begin{equation}\label{expedition0012378}
\begin{aligned}
\lambda_{\txtrad,\mathrm{i},x,\kappa,\Lambda,L}^{\mathrm{j}_{\txtrad}}
=
P_{L}
\lambda_{\txtrad,\mathrm{i},x,\kappa,\Lambda}^{\mathrm{j}_{\txtrad}},
\quad
\lambda_{\txtrad,\mathrm{i},x,\kappa,\Lambda,L}
=
P_{L}\lambda_{\txtrad,\mathrm{i},x,\kappa,\Lambda},
\end{aligned}
\end{equation} where \(P_{L}\) is defined by a suitable extension in the expression on the right. For the bounded-system cutoff vector potential \[A_{\mathrm{i},\kappa,\Lambda,L}(x)
=
\frac{1}{\sqrt{2}}
\sum_{\mathrm{j}_{\txtrad}\in \ringratint_{2}}
\rbk{
\fun{\opfockcr_{\txtrad}}{\lambda_{\txtrad,\mathrm{i},x,\kappa,\Lambda,L}^{\mathrm{j}_{\txtrad}},\mathrm{j}_{\txtrad}}
+\fun{\opfockan_{\txtrad}}{\overline{\lambda_{\txtrad,\mathrm{i},x,\kappa,\Lambda,L}^{\mathrm{j}_{\txtrad}}},\mathrm{j}_{\txtrad}}
},\] set \(A_{\kappa,\Lambda,L}
= \vecbk{A_{\mathrm{i},\kappa,\Lambda,L}}_{\mathrm{i}=1}^{d}\), and define the corresponding bounded-system Hamiltonian and Hamiltonian with chemical potential by \[\begin{aligned}
\physham_{\txtpaulifierz,\kappa,\Lambda,L}
&=
\onehalf
\rbk{-\imunit\vagrad\otimes\idone-\physcharge A_{\kappa,\Lambda,L}}^2
+V \otimes \idone
+\idone \otimes \physham_{\txtrad,L},
\\ %%%%%%%%%%%%%%%%
\physham_{\txtpaulifierz,\kappa,\Lambda,\smchemicalpotential,L}
&=
\onehalf
\rbk{-\imunit\vagrad\otimes\idone-\physcharge A_{\kappa,\Lambda,L}}^2
+V \otimes \idone
+\idone \otimes \physham_{\txtrad,\smchemicalpotential,L},
\\ %%%%%%%%%%%%%%%%
\physham_{\txtrad,\smchemicalpotential,L}
&=
\fun{\opfocksndqntdiff_{\txtrad}}{\omega \oplus \omega - \smchemicalpotential}.
\end{aligned}\] Consider the free Hamiltonian with coupling constant \(\physcharge
= 0\): \[\physham_{\txtpaulifierz,0,\smchemicalpotential,L}
= \physham_{\txtparticle} \otimes \idone
+\idone \otimes \physham_{\txtrad,\smchemicalpotential,L}
= (-\laplacian + V) \otimes \idone
+\idone \otimes \physham_{\txtrad,\smchemicalpotential,L}.\] By assumption, \(\napiernum^{-\sminvtemperature
\physham_{\txtparticle}}\) is of trace class, and by the definition of the dispersion relation and by boundedness and periodicity, as in the Nelson model, \(\napiernum^{-\sminvtemperature
\physham_{\txtrad,\smchemicalpotential,L}}\) is of trace class. In particular, the heat operator \(\napiernum^{-\sminvtemperature \physham_{\txtpaulifierz,0,\smchemicalpotential,L}}\) generated by the bounded-system free Hamiltonian is of trace class. It remains to discuss whether the interaction weight gives a finite positive partition function and hence defines a normalized bounded-system interacting KMS functional, and the probabilistic argument below guarantees this well-definedness. In particular, we first represent the free radiation field, which has already been defined as a bounded-system trace state, by a probability space, and then define the interacting KMS state by using the Feynman--Kac--Nelson kernel argument.

The bounded-system free radiation field can be described by the singular Gaussian \(\sminvtemperature\)-Markov path space associated with the bounded-system trace state with chemical potential: \[\begin{aligned}
\pairbk{\prbqspace_{\txtrad,\sminvtemperature,L},
\mblfml{S}_{\txtrad,\sminvtemperature,L},
\mblfml{S}_{\txtrad,0,\sminvtemperature,L},
U_{\txtrad,L,t},
R_{\txtrad,L},
\msrprb_{\txtrad,\sminvtemperature,\smchemicalpotential,L}}.
\end{aligned}\] For the space of physical test functions in the bounded system, use \(\sphilb{D}
_{\txtrad,\txtphys,\sminvtemperature,L}\) obtained by suitably applying \(P_{L}\) to \eqref{eq:PF-physical-test-space-majorant}. For any \(h
\in P_{L} \sphilb{D}_{\txtrad,\txtphys,\sminvtemperature}\), let the sharp-time field be \(\opfocksegalradiation(j_t h)\), and let \(\opform{q}_{\txtrad,\txtbec,\sminvtemperature,\smchemicalpotential,L}^{\txteuclid}\) denote the sesquilinear form associated with the covariance of this measure. For \(t,s
\in S_{\sminvtemperature}\) and \(f,g
\in \sphilb{D}_{\txtrad,\txtphys,\sminvtemperature}\), it is defined by \[\begin{aligned}
\fun{\opform{q}_{\txtrad,\txtbec,\sminvtemperature,\smchemicalpotential,L}^{\txteuclid}}
{j_t P_{L} f,j_s P_{L} g}
&=
\fun{\opform{q}_{\txtrad,0,\sminvtemperature,\smchemicalpotential,L}^{\txteuclid}}
{j_t P_{L} f,j_s P_{L} g}
+\fun{\opform{q}_{\txtrad,\txtnonzero,\sminvtemperature,\smchemicalpotential,L}^{\txteuclid}}
{j_t P_{L} f,j_s P_{L} g},
\\ %%%%%%%%%%%%%%%%
\fun{\opform{q}_{\txtrad,0,\sminvtemperature,\smchemicalpotential,L}^{\txteuclid}}
{j_t P_{L} f,j_s P_{L} g}
&=
\fun{\opform{q}_{\txtrad,0,\sminvtemperature,\smchemicalpotential,L}}{P_{L} f,P_{L} g},
\\ %%%%%%%%%%%%%%%%
&=
\frac{(2\pi)^{d}}{L^d}
\frac{y_L + 1}{y_L - 1}
\bkt{\faftr{f}(0)}
{\Pi_{\txtrad,0}\faftr{g}(0)}_{\fldcmp^d},
\\ %%%%%%%%%%%%%%%%
\fun{\opform{q}_{\txtrad,\txtnonzero,\sminvtemperature,\smchemicalpotential,L}^{\txteuclid}}
{j_t P_{L} f,j_s P_{L} g}
&=
\frac{(2\pi)^{d}}{L^d}
\sum_{k \in \setlattice_L^d\setminus\setone{0}}
\frac{y_L \napiernum^{-\abs{t-s}\omega(k)}
+\napiernum^{-(\sminvtemperature-\abs{t-s})\omega(k)}}
{y_L - \napiernum^{-\sminvtemperature\omega(k)}}
\bkt{\faftr{f}(k)}
{\delta^\perp(k)\faftr{g}(k)}_{\fldcmp^d}.
\end{aligned}\] The limit of this bounded-system covariance is handled in the same way as in the Nelson model: namely, for finite-time correlation functions, first take \(L\to\infty\) and then take \(\smchemicalpotential\uparrow0\). In the zero-mode part, only the bounded-system \(k=0\) Fourier coefficient is extracted by \(\Pi_{\txtrad,0}\), and this projection appears in the infinite-volume zero-mode form \(\opform{q}_{\txtrad,0,\sminvtemperature}\). The non-zero-mode part is the Riemann-sum limit in which the sum over \(\setlattice_L^d\setminus\setone{0}\) converges to the momentum integral over \(\fldreal^d
\setminus
\setone{0}\). For finitely many \(s_j
\in \fldreal\), \(t_j
\in S_{\sminvtemperature}\), \(f_j
\in \sphilb{D}_{\txtrad,\txtphys,\sminvtemperature}\), if \(G
=
\sum_{j=1}^{n}
s_j
\fun{j_{t_j}}
{P_{L} f_j}\), then the bounded-system Gaussian measure is characterized by \[\begin{aligned}
\int_{\prbqspace_{\txtrad,\sminvtemperature,L}}
\napiernum^{\imunit\opfocksegalradiation(G)}
\opdmsr{\msrprb_{\txtrad,\sminvtemperature,\smchemicalpotential,L}}
(\opfocksegalradiation)
=
\fnexp{-\oneoverfour
\fun{\opform{q}_{\txtrad,\txtbec,\sminvtemperature,\smchemicalpotential,L}^{\txteuclid}}{G}}.
\end{aligned}\]

The free system combining the particle and the bounded-system free radiation field is represented by \[\pairbk{\prbqspace_{\txtpaulifierz,\sminvtemperature,L},
\mblfml{S}_{\txtpaulifierz,\sminvtemperature,L},
\mblfml{S}_{\txtpaulifierz,0,\sminvtemperature,L},
U_{\txtpaulifierz,L,t},
R_{\txtpaulifierz,L},
\msrprb_{\txtpaulifierz,0,\sminvtemperature,\smchemicalpotential,L}},\] where \(\prbqspace_{\txtpaulifierz,\sminvtemperature,L}
=\Omega_{\txtparticle,\sminvtemperature}
\times\prbqspace_{\txtrad,\sminvtemperature,L}\), \(\mblfml{S}_{\txtpaulifierz,\sminvtemperature,L}
=\mblfml{S}_{\txtparticle,\sminvtemperature}
\times\mblfml{S}_{\txtrad,\sminvtemperature,L}\), \(\mblfml{S}_{\txtpaulifierz,0,\sminvtemperature,L}
=\mblfml{S}_{\txtparticle,0,\sminvtemperature}
\times\mblfml{S}_{\txtrad,0,\sminvtemperature,L}\), \(U_{\txtpaulifierz,L,t}=U_{\txtparticle,t}\otimes U_{\txtrad,L,t}\), \(R_{\txtpaulifierz,L}=R_{\txtparticle}\otimes R_{\txtrad,L}\), \(\msrprb_{\txtpaulifierz,0,\sminvtemperature,\smchemicalpotential,L}
=\msrprb_{\txtparticle,\sminvtemperature}
\otimes\msrprb_{\txtrad,\sminvtemperature,\smchemicalpotential,L}\).

The current along the particle path is defined by the same idea as in the infinite system. For \(\lambda
_{\txtrad,\mathrm{i},\prbprocess_t,\kappa,\Lambda,L}\) in \eqref{expedition0012378} and any \(I
\subset S_{\sminvtemperature}\), set \begin{equation}\label{expedition0012300}
\begin{aligned}
\mathsf{I}_{\txtrad,I,\kappa,\Lambda,L}(\prbprocess)
=
\bigoplus_{\mathrm{i}=1}^{d}
\int_{I}
\fun{j_t}
{\lambda_{\txtrad,\mathrm{i},\prbprocess_t,\kappa,\Lambda,L}}
\opdmsr{\prbprocess_{\mathrm{i},t}}.
\end{aligned}
\end{equation}

\begin{prop}[Test-Function Property of the Bounded-System Current Source]\label{expedition0012325}
Fix the bounded-system length $L$ and cutoffs $0<\kappa<\Lambda<\infty$, and let $I
\subset S_{\sminvtemperature}$ be any time interval.
Then the current source $\mathsf{I}
_{\txtrad,I,\kappa,\Lambda,L}(\prbprocess)$ along the particle path $\prbprocess$ is almost surely with respect to the particle measure a test function for the Segal field operator $\opfocksegalradiation$ of the bounded-system radiation field, and satisfies
\begin{equation}\label{expedition0012326}
\begin{aligned}
\sqfun{\prbexp_{\msrprb_{\txtparticle,\sminvtemperature}}}
{\fun{\opform{q}_{\txtrad,\txtbec,\sminvtemperature,\smchemicalpotential,L}^{\txteuclid}}
{\mathsf{I}_{\txtrad,I,\kappa,\Lambda,L}(\prbprocess)}}
<\infty.
\end{aligned}
\end{equation}
Because of the infrared cutoff, $\mathsf{I}_{\txtrad,I,\kappa,\Lambda,L}(\prbprocess)$ has no $k=0$ component, and hence
\begin{equation}\label{expedition0012327}
\begin{aligned}
\fun{\opform{q}_{\txtrad,\txtbec,\sminvtemperature,\smchemicalpotential,L}^{\txteuclid}}
{\mathsf{I}_{\txtrad,I,\kappa,\Lambda,L}(\prbprocess)}
=
\fun{\opform{q}_{\txtrad,\txtnonzero,\sminvtemperature,\smchemicalpotential,L}^{\txteuclid}}
{\mathsf{I}_{\txtrad,I,\kappa,\Lambda,L}(\prbprocess)}
\end{aligned}
\end{equation}
holds.
In particular, $\opfocksegalradiation(\mathsf{I}
_{\txtrad,I,\kappa,\Lambda,L}(\prbprocess))$ is defined as a centered Gaussian random variable of the bounded-system radiation field.
\end{prop}

\begin{proof}
Under the assumptions, $\lambda_{\txtrad,\mathrm{i},x,\kappa,\Lambda,L}$ is supported on the finitely many lattice points $\setlattice_L^d
\cap
\set{k}{\kappa\leq\abs{k}\leq\Lambda}$, and for any $t,x,\mathrm{i}$, $\fun{j_t}
{\lambda_{\txtrad,\mathrm{i},x,\kappa,\Lambda,L}}$ belongs to the domain of the bounded-system covariance form $\opform{q}
_{\txtrad,\txtbec,\sminvtemperature,\smchemicalpotential,L}^{\txteuclid}$.
By boundedness of the support, there exists a constant $C_{\kappa,\Lambda,L}
< \infty$ such that
$$\fun{\opform{q}_{\txtrad,\txtbec,\sminvtemperature,\smchemicalpotential,L}^{\txteuclid}}
{\fun{j_t}
{\lambda_{\txtrad,\mathrm{i},x,\kappa,\Lambda,L}}}
\leq
C_{\kappa,\Lambda,L}$$
holds uniformly in $t,x,\mathrm{i}$.

Represent the Brownian bridge by the semimartingale decomposition
$$\opdmsr{\prbprocess_{\mathrm{i}}(t)}
=
\opdmsr{M_{\mathrm{i}}(t)}
 +b_{\mathrm{i}}(t,\prbprocess)\opdmsr{t}.$$
The stochastic integral in \eqref{expedition0012300} is constructed as the sum of Itô integrals and time integrals over the finitely many lattice-point components.
In the bounded system, $\setlattice_L^d
\cap
\set{k}{\kappa\leq\abs{k}\leq\Lambda}$ is a finite set, so the covariance form reduces to a finite-dimensional nonnegative Hermitian form.
As the Itô isometry for this finite-dimensional form, we have
\begin{equation}\label{expedition0012367}
\begin{aligned}
&\sqfun{\prbexp}
{\fun{\opform{q}_{\txtrad,\txtbec,\sminvtemperature,\smchemicalpotential,L}^{\txteuclid}}
{\int_I
\fun{j_t}{\lambda_{\txtrad,\mathrm{i},\prbprocess_t,\kappa,\Lambda,L}}
\opdmsr{M_{\mathrm{i}}(t)}}}
\\ %%%%%%%%%%%%%%%%
&=
\sqfun{\prbexp}
{\int_I
\fun{\opform{q}_{\txtrad,\txtbec,\sminvtemperature,\smchemicalpotential,L}^{\txteuclid}}
{\fun{j_t}{\lambda_{\txtrad,\mathrm{i},\prbprocess_t,\kappa,\Lambda,L}}}
\opdmsr{\prbdbkquadvar{M_{\mathrm{i}}}_t}}.
\end{aligned}
\end{equation}
Since the martingale part of the Brownian bridge satisfies $\opdmsr{\prbdbkquadvar{M_{\mathrm{i}}}_t}
=\opdmsr{t}$, substituting the preceding uniform estimate
$$\fun{\opform{q}_{\txtrad,\txtbec,\sminvtemperature,\smchemicalpotential,L}^{\txteuclid}}
{\fun{j_t}{\lambda_{\txtrad,\mathrm{i},x,\kappa,\Lambda,L}}}
\leq C_{\kappa,\Lambda,L}$$
into \eqref{expedition0012367} gives
\begin{equation}\label{expedition0012368}
\begin{aligned}
\sqfun{\prbexp_{\msrprb_{\txtparticle,\sminvtemperature}}}
{\fun{\opform{q}_{\txtrad,\txtbec,\sminvtemperature,\smchemicalpotential,L}^{\txteuclid}}
{\int_I
\fun{j_t}
{\lambda_{\txtrad,\mathrm{i},\prbprocess_t,\kappa,\Lambda,L}}
\opdmsr{M_{\mathrm{i}}(t)}}}
\leq
C_{\kappa,\Lambda,L}\abs{I}.
\end{aligned}
\end{equation}
For the drift part, the triangle inequality in the finite-dimensional Hilbert space gives
\begin{equation}\label{expedition0012369}
\begin{aligned}
&\fun{\opform{q}_{\txtrad,\txtbec,\sminvtemperature,\smchemicalpotential,L}^{\txteuclid}}
{\int_I
\fun{j_t}{\lambda_{\txtrad,\mathrm{i},\prbprocess_t,\kappa,\Lambda,L}}
b_{\mathrm{i}}(t,\prbprocess)\opdmsr{t}}
\\ %%%%%%%%%%%%%%%%
&\leq
\rbk{
\int_I
\rbk{\fun{\opform{q}_{\txtrad,\txtbec,\sminvtemperature,\smchemicalpotential,L}^{\txteuclid}}
{\fun{j_t}{\lambda_{\txtrad,\mathrm{i},\prbprocess_t,\kappa,\Lambda,L}}}}^{\onehalf}
\abs{b_{\mathrm{i}}(t,\prbprocess)}
\opdmsr{t}}^{2}
\leq
C_{\kappa,\Lambda,L}
\rbk{\int_I\abs{b_{\mathrm{i}}(t,\prbprocess)}\opdmsr{t}}^{2}.
\end{aligned}
\end{equation}
Taking expectation,
\begin{equation}\label{expedition0012370}
\begin{aligned}
\sqfun{\prbexp_{\msrprb_{\txtparticle,\sminvtemperature}}}
{\fun{\opform{q}_{\txtrad,\txtbec,\sminvtemperature,\smchemicalpotential,L}^{\txteuclid}}
{\int_I
\fun{j_t}
{\lambda_{\txtrad,\mathrm{i},\prbprocess_t,\kappa,\Lambda,L}}
b_{\mathrm{i}}(t,\prbprocess)\opdmsr{t}}}
\leq
C_{\kappa,\Lambda,L}
\sqfun{\prbexp_{\msrprb_{\txtparticle,\sminvtemperature}}}
{\rbk{\int_I\abs{b_{\mathrm{i}}(t,\prbprocess)}\opdmsr{t}}^2}
\end{aligned}
\end{equation}
and therefore integrability of the drift part reduces to
$\sqfun{\prbexp_{\msrprb_{\txtparticle,\sminvtemperature}}}
{\rbk{\int_I\abs{b_{\mathrm{i}}(t,\prbprocess)}\opdmsr{t}}^2}
< \infty$.

Cut the time circle at one point and regard it as a Brownian bridge from $x$ to $x$ on $\closedinterval{0}{\sminvtemperature}$.
Taking the endpoint to be $\sminvtemperature$, for any $0<t<\sminvtemperature$ the drift of the Brownian bridge is
$b(t,\prbprocess)
=
-\frac{\prbprocess_t-x}{\sminvtemperature-t}$.
To write the heat-kernel representation explicitly, let the Brownian heat kernel be
$p_t(\xi)
=
(2\pi t)^{-\frac{d}{2}}
\fnexp{-\frac{\abs{\xi}^{2}}{2t}}$.
Under the normalization of standard Brownian motion in this paper, each coordinate of $p_a$ has variance $a$.
The distribution of the Brownian bridge at time $t$ is
\begin{equation}\label{expedition0012371}
\begin{aligned}
\sqfun{\prbexp_{\msrprb_{x,x}^{\sminvtemperature,\txtbrownbridge}}}
{F(\prbprocess_t)}
&=
\frac{1}{p_{\sminvtemperature}(0)}
\int_{\fldreal^d}
F(y)
p_t(y-x)
p_{\sminvtemperature-t}(x-y)
\opdmsr{y}
\\ %%%%%%%%%%%%%%%%
&=
\int_{\fldreal^d}
F(y)
p_{\frac{t(\sminvtemperature-t)}{\sminvtemperature}}(y-x)
\opdmsr{y}.
\end{aligned}
\end{equation}
In particular, taking $F(y)
= \abs{y-x}^{2}$ gives
\begin{equation}\label{expedition0012372}
\begin{aligned}
\sqfun{\prbexp_{\msrprb_{x,x}^{\sminvtemperature,\txtbrownbridge}}}
{\abs{\prbprocess_t-x}^{2}}
=
d\frac{t(\sminvtemperature-t)}{\sminvtemperature}.
\end{aligned}
\end{equation}
In the componentwise calculation, $b_{\mathrm{i}}(t,\prbprocess)
=
-(\prbprocess_t^{\mathrm{i}}-x^{\mathrm{i}})/(\sminvtemperature-t)$, and hence
\begin{equation}\label{expedition0012373}
\begin{aligned}
\sqfun{\prbexp_{\msrprb_{x,x}^{\sminvtemperature,\txtbrownbridge}}}
{\abs{b_{\mathrm{i}}(t,\prbprocess)}^{2}}
=
\frac{1}{(\sminvtemperature-t)^2}
\sqfun{\prbexp_{\msrprb_{x,x}^{\sminvtemperature,\txtbrownbridge}}}
{\abs{\prbprocess_t^{\mathrm{i}}-x^{\mathrm{i}}}^{2}}
=
\frac{t}{\sminvtemperature(\sminvtemperature-t)}
\leq
\frac{1}{\sminvtemperature-t}
\end{aligned}
\end{equation}
holds.
Using Minkowski's inequality and the interval representation with the cutting point at $0$, square integrability is obtained as
$$\begin{aligned}
\rbk{\sqfun{\prbexp_{\msrprb_{x,x}^{\sminvtemperature,\txtbrownbridge}}}
{\rbk{\int_I\abs{b_{\mathrm{i}}(t,\prbprocess)}\opdmsr{t}}^2}}^{\onehalf}
\leq
\int_I
\rbk{\sqfun{\prbexp_{\msrprb_{x,x}^{\sminvtemperature,\txtbrownbridge}}}
{\abs{b_{\mathrm{i}}(t,\prbprocess)}^{2}}}^{\onehalf}
\opdmsr{t}
\leq
\int_I
\frac{\opdmsr{t}}{\sqrt{\sminvtemperature-t}}
<\infty.
\end{aligned}$$
When the original interval $I
\subset S_{\sminvtemperature}$ on the circle is opened by this cut, $I$ is either one interval in $\closedinterval{0}{\sminvtemperature}$ or, if it crosses the cutting point, splits into two intervals $I_{\txtneg}
\cup I_{\txtnonneg}$.
For one interval, the estimate above applies directly.
If the interval crosses the cutting point, it suffices to estimate the two intervals separately by using
$$\rbk{\int_I\abs{b_{\mathrm{i}}(t,\prbprocess)}\opdmsr{t}}^2
\leq
2\rbk{\int_{I_{\txtneg}}\abs{b_{\mathrm{i}}(t,\prbprocess)}\opdmsr{t}}^2
+2\rbk{\int_{I_{\txtnonneg}}\abs{b_{\mathrm{i}}(t,\prbprocess)}\opdmsr{t}}^2.$$
The part touching the right endpoint $\sminvtemperature$ is handled by the integrability of $(\sminvtemperature-t)^{-\onehalf}$ above.
For the part touching the left endpoint $0$, use the time-reversal map $(R\prbprocess)_t
= \prbprocess_{\sminvtemperature-t}$; the Brownian bridge measure $\msrprb_{x,x}^{\sminvtemperature,\txtbrownbridge}$ is invariant under $R$.
When the right-endpoint drift representation after time reversal is returned to the original time, the singular coefficient to be estimated on the left endpoint side reduces to $\frac{\prbprocess_t-x}{t}$.
Since the same variance estimate \eqref{expedition0012372} follows from \eqref{expedition0012371}, for each component
\begin{equation}\label{expedition0012374}
\begin{aligned}
\sqfun{\prbexp_{\msrprb_{x,x}^{\sminvtemperature,\txtbrownbridge}}}
{\abs{\frac{\prbprocess_t^{\mathrm{i}}-x^{\mathrm{i}}}{t}}^{2}}
=
\frac{1}{t^2}
\sqfun{\prbexp_{\msrprb_{x,x}^{\sminvtemperature,\txtbrownbridge}}}
{\abs{\prbprocess_t^{\mathrm{i}}-x^{\mathrm{i}}}^{2}}
=
\frac{\sminvtemperature-t}{\sminvtemperature t}
\leq
\frac{1}{t}
\end{aligned}
\end{equation}
satisfies the required estimate.
Then $\int_0^\delta
t^{-\onehalf}
\opdmsr{t}
< \infty$ gives the same square integrability at the left endpoint.
Taking $F(y)=\abs{y-x}^{q}$ in the same heat-kernel representation gives
\begin{equation}\label{expedition0012375}
\begin{aligned}
\sqfun{\prbexp_{\msrprb_{x,x}^{\sminvtemperature,\txtbrownbridge}}}
{\abs{\prbprocess_t-x}^{q}}
\leq
\rbk{\int_{\fldreal^d}\abs{y}^{q}
(2\pi)^{-d/2}\fnexp{-\frac{\abs{y}^{2}}{2}}
\opdmsr{y}}
\rbk{\frac{t(\sminvtemperature-t)}{\sminvtemperature}}^{\frac{q}{2}}.
\end{aligned}
\end{equation}
On the right endpoint side,
$$\sqfun{\prbexp_{\msrprb_{x,x}^{\sminvtemperature,\txtbrownbridge}}}
{\abs{b_{\mathrm{i}}(t,\prbprocess)}^q}^{\frac{1}{q}}
\leq
\rbk{\int_{\fldreal}\abs{u}^{q}
(2\pi)^{-\onehalf}
\fnexp{-\frac{u^{2}}{2}}\opdmsr{u}}^{\frac{1}{q}}
(\sminvtemperature-t)^{-\onehalf},$$
and on the left endpoint side,
$$\sqfun{\prbexp_{\msrprb_{x,x}^{\sminvtemperature,\txtbrownbridge}}}
{\abs{\frac{\prbprocess_t^{\mathrm{i}}-x^{\mathrm{i}}}{t}}^q}^{\frac{1}{q}}
\leq
\rbk{\int_{\fldreal}
\abs{u}^{q}
(2\pi)^{-\onehalf}\fnexp{-\frac{u^{2}}{2}}\opdmsr{u}}^{\frac{1}{q}}
t^{-\onehalf}.$$
Substituting these estimates into Minkowski's inequality, for any finite $q>1$ we obtain
\begin{equation}\label{expedition0012329}
\sup_{x\in\fldreal^d}
\sqfun{\prbexp_{\msrprb_{x,x}^{\sminvtemperature,\txtbrownbridge}}}
{\rbk{\int_I\abs{b_{\mathrm{i}}(t,\prbprocess)}\opdmsr{t}}^q}
< \infty.
\end{equation}
Next pass from the Brownian bridge measure to the particle loop measure.
The particle loop measure is given by
$$\opdmsr{\msrprb_{\txtparticle,\sminvtemperature}}(\prbprocess)
=
\frac{1}{\smpartitionfunc_{\txtparticle,\sminvtemperature}}
\int_{\fldreal^d}
\napiernum^{-\int_0^{\sminvtemperature}V(\prbprocess_u)\opdmsr{u}}
\opdmsr{\msrprb_{x,x}^{\sminvtemperature,\txtbrownbridge}}(\prbprocess)
\opdmsr{x}.$$
By the assumption on the particle potential, for some $p
> 1$,
$$\int_{\fldreal^d}
\rbk{\sqfun{\prbexp_{\msrprb_{x,x}^{\sminvtemperature,\txtbrownbridge}}}
{\napiernum^{-p\int_0^{\sminvtemperature}V(\prbprocess_u)\opdmsr{u}}}}^{\frac{1}{p}}
\opdmsr{x}
<\infty$$
holds; let $r$ be the Hölder conjugate exponent of this $p$.
Applying \eqref{expedition0012329} with $q=2r$ gives
$$\sup_{x\in\fldreal^d}
\sqfun{\prbexp_{\msrprb_{x,x}^{\sminvtemperature,\txtbrownbridge}}}
{\rbk{\int_I\abs{b_{\mathrm{i}}(t,\prbprocess)}\opdmsr{t}}^{2r}}
<\infty,$$
and therefore Hölder's inequality for each $x$ gives
$$\begin{aligned}
&\sqfun{\prbexp_{\msrprb_{\txtparticle,\sminvtemperature}}}
{\rbk{\int_I\abs{b_{\mathrm{i}}(t,\prbprocess)}\opdmsr{t}}^{2}}
\\ %%%%%%%%%%%%%%%%
&\leq
\frac{1}{\smpartitionfunc_{\txtparticle,\sminvtemperature}}
\int_{\fldreal^d}
\rbk{\sqfun{\prbexp_{\msrprb_{x,x}^{\sminvtemperature,\txtbrownbridge}}}
{\napiernum^{-p\int_0^{\sminvtemperature}V(\prbprocess_u)\opdmsr{u}}}}^{\frac{1}{p}}
\rbk{\sqfun{\prbexp_{\msrprb_{x,x}^{\sminvtemperature,\txtbrownbridge}}}
{\rbk{\int_I\abs{b_{\mathrm{i}}(t,\prbprocess)}\opdmsr{t}}^{2r}}}^{\frac{1}{r}}
\opdmsr{x}
<\infty.
\end{aligned}$$
Thus, after changing the constant, the endpoint estimates obtained for the Brownian bridge also hold under $\msrprb_{\txtparticle,\sminvtemperature}$.
Summing over the components $\mathrm{i}$ gives \eqref{expedition0012326}.

(Behavior of the zero mode): Because of the infrared cutoff, each $\lambda_{\txtrad,\mathrm{i},x,\kappa,\Lambda,L}$ has no $k=0$ component.
Hence its stochastic integral $\mathsf{I}_{\txtrad,I,\kappa,\Lambda,L}(\prbprocess)$ also has no $k=0$ component and vanishes under the zero-mode covariance $\opform{q}_{\txtrad,0,\sminvtemperature,\smchemicalpotential,L}^{\txteuclid}$.
Thus \eqref{expedition0012327} holds, and the sharp-time field of the bounded-system radiation field is well-defined.
\end{proof}

The bounded-system interaction is defined by perturbing the free measure \(\msrprb_{\txtpaulifierz,0,\sminvtemperature,\smchemicalpotential,L}\) with the local Feynman--Kac--Nelson kernel \begin{equation}\label{expedition0012376}
\begin{aligned}
F_{\txtpaulifierz,I,\kappa,\Lambda,L}(\prbprocess,\opfocksegalradiation)
=
\fnexp{-\imunit\physcharge
\opfocksegalradiation(\mathsf{I}_{\txtrad,I,\kappa,\Lambda,L}(\prbprocess))}.
\end{aligned}
\end{equation}

\begin{prop}[Bounded-System Perturbation Conditions for the Pauli--Fierz Kernel]\label{expedition0012328}
For each fixed $L$, the family $\fml{F_{\txtpaulifierz,I,\kappa,\Lambda,L}}
{I \subset S_{\sminvtemperature}}$ defined by \eqref{expedition0012376} is a local Feynman--Kac--Nelson perturbation \cite[Chapter 21]{DerezinskiGerard001}.
In particular, $F_{\txtpaulifierz,I,\kappa,\Lambda,L}$ is measurable with respect to the $\sigma$-algebra corresponding to the interval $I$, and for disjoint intervals $I,J
\subset S_{\sminvtemperature}$ it satisfies
$$F_{\txtpaulifierz,I\cup J,\kappa,\Lambda,L}(\prbprocess,\opfocksegalradiation)
=
F_{\txtpaulifierz,I,\kappa,\Lambda,L}(\prbprocess,\opfocksegalradiation)
\cdot
F_{\txtpaulifierz,J,\kappa,\Lambda,L}(\prbprocess,\opfocksegalradiation).$$
Moreover, it is covariant under time translations and reflection and belongs to $\lp^{p}$ for any $1
\leq p
< \infty$.
\end{prop}

\begin{proof}
By Proposition \ref{expedition0012325}, $\mathsf{I}
_{\txtrad,I,\kappa,\Lambda,L}(\prbprocess)$ in \eqref{expedition0012300} is a test function for the sharp-time Gaussian field of the bounded-system radiation field, and $\opfocksegalradiation(\mathsf{I}_{\txtrad,I,\kappa,\Lambda,L}(\prbprocess))$ is defined as a centered Gaussian random variable.
The stochastic integral $\mathsf{I}_{\txtrad,I,\kappa,\Lambda,L}(\prbprocess)$ is adapted to the interval $I$ and additive with respect to disjoint intervals, so the exponential kernel is interval-measurable and multiplicative.

Before integrating the Gaussian field, $\abs{F_{\txtpaulifierz,I,\kappa,\Lambda,L}(\prbprocess,\opfocksegalradiation)}
= 1$ holds, giving the $\lp^{p}$ estimate for all $1
\leq p < \infty$.

Covariance under time translations and reflection follows from the corresponding covariance of the particle loop measure, the radiation Segal field, and the stochastic integral.
Finiteness of the sesquilinear form obtained after first integrating the Gaussian field follows from \eqref{expedition0012326}.
\end{proof}

\begin{prop}[Local Semigroup in the Bounded System]\label{expedition0012320}
Fix the bounded-system length variable $L$ and cutoffs $0<\kappa<\Lambda<\infty$.
For any $0
< t
< \frac{\sminvtemperature}{2}$, define
\begin{equation}\label{expedition0012321}
\begin{aligned}
\sphilb{D}_{\txtpaulifierz,t,\kappa,\Lambda,\smchemicalpotential,L}
=
\prbexp_{0,\frac{\sminvtemperature}{2}}
\fun{\splinspan}
{\bigcup_{0\leq s<\frac{\sminvtemperature}{2}-t}
F_{\txtpaulifierz,\closedinterval{0}{s},\kappa,\Lambda,L}
\fun{\lp^{\infty}}
{\prbqspace_{\txtpaulifierz,\sminvtemperature,L},
\mblfml{S}_{\txtpaulifierz,\closedinterval{0}{\frac{\sminvtemperature}{2}-t},L}}}.
\end{aligned}
\end{equation}
Then there exists a local Hermitian semigroup $\fml{P_{\txtpaulifierz,\kappa,\Lambda,\smchemicalpotential,L,t}}
{t\in\closedinterval{0}{\frac{\sminvtemperature}{2}}}$ such that, for any $0\leq s\leq t\leq\frac{\sminvtemperature}{2}$,
\begin{equation}\label{expedition0012322}
\begin{aligned}
P_{\txtpaulifierz,\kappa,\Lambda,\smchemicalpotential,L,s}
\prbexp_{\setone{0,\frac{\sminvtemperature}{2}}}f
=
\prbexp_{\setone{0,\frac{\sminvtemperature}{2}}}
F_{\txtpaulifierz,\closedinterval{0}{s},\kappa,\Lambda,L}
U_{\txtpaulifierz,L,s}f
\end{aligned}
\end{equation}
holds.
\end{prop}

\begin{proof}
By Proposition \ref{expedition0012328}, the family $\fml{F_{\txtpaulifierz,I,\kappa,\Lambda,L}}
{I \subset S_{\sminvtemperature}}$ is a local Feynman--Kac--Nelson perturbation.
The local semigroup reconstruction in the textbook \cite[Chapter 21]{DerezinskiGerard001} applies, and the local Hermitian semigroup is obtained with the domain in \eqref{expedition0012321} and the action in \eqref{expedition0012322}.
\end{proof}

\begin{prop}[Kernel Representation on the Bounded-System Generator]\label{expedition0012323}
For bounded time-zero particle functions $F_{\txtparticle}$ and $G_{\txtparticle}$ and radiation-field test functions $f,g
\in\sphilb{D}_{\txtrad,\txtphys,\sminvtemperature}$, set
$$F
=
F_{\txtparticle}(\prbprocess_0)
\napiernum^{\imunit\opfocksegalradiation(j_0P_{L}f)},
\quad
G
=
G_{\txtparticle}(\prbprocess_0)
\napiernum^{\imunit\opfocksegalradiation(j_0P_{L}g)}.$$
Then
\begin{equation}\label{expedition0012324}
\begin{aligned}
&\bkt{F}
{P_{\txtpaulifierz,\kappa,\Lambda,\smchemicalpotential,L,t}G}
=
\int_{\Omega_{\txtparticle,\sminvtemperature}}
\cmpconj{F_{\txtparticle}(\prbprocess_0)}
G_{\txtparticle}(\prbprocess_t)
\\ %%%%%%%%%%%%%%%%
&\quad\times
\fnexp{-\oneoverfour
\fun{\opform{q}_{\txtrad,\txtbec,\sminvtemperature,\smchemicalpotential,L}^{\txteuclid}}
{j_tP_{L}g-j_0P_{L}f}
+\frac{\physcharge}{2}
\fun{\opform{q}_{\txtrad,\txtbec,\sminvtemperature,\smchemicalpotential,L}^{\txteuclid}}
{j_tP_{L}g-j_0P_{L}f,
\mathsf{I}_{\txtrad,\closedinterval{0}{t},\kappa,\Lambda,L}(\prbprocess)}}
\\ %%%%%%%%%%%%%%%%
&\quad\times
\fnexp{
-\frac{\physcharge^2}{4}
\fun{\opform{q}_{\txtrad,\txtbec,\sminvtemperature,\smchemicalpotential,L}^{\txteuclid}}
{\mathsf{I}_{\txtrad,\closedinterval{0}{t},\kappa,\Lambda,L}(\prbprocess)}}
\opdmsr{\msrprb_{\txtparticle,\sminvtemperature}}(\prbprocess)
\end{aligned}
\end{equation}
holds.
Since the cutoff current source has no $k=0$ component, the zero-mode part does not contribute to the two covariance forms depending on the current source in \eqref{expedition0012324}.
\end{prop}

\begin{proof}
By \eqref{expedition0012322}, $P_{\txtpaulifierz,\kappa,\Lambda,\smchemicalpotential,L,t}G$ is given by
$F_{\txtpaulifierz,\closedinterval{0}{t},\kappa,\Lambda,L}
U_{\txtpaulifierz,L,t}G$, and the field exponent is $\imunit\opfocksegalradiation(j_tP_{L}g-j_0P_{L}f)
-\imunit\physcharge
\opfocksegalradiation(\mathsf{I}_{\txtrad,\closedinterval{0}{t},\kappa,\Lambda,L}(\prbprocess))$.
Fixing the particle path and applying the centered Gaussian characteristic functional of the bounded-system free radiation field gives \eqref{expedition0012324}.
The assertion about the zero mode follows from the fact that $\lambda
_{\txtrad,\mathrm{i},x,\kappa,\Lambda,L}$ contains no $k=0$ component because of the infrared cutoff.
\end{proof}

The corresponding partition function of the interacting system and the representation obtained by integrating the field first are \[\begin{aligned}
&\smpartitionfunc_{\txtpaulifierz,\sminvtemperature,\kappa,\Lambda,\smchemicalpotential,L}
=
\int_{\prbqspace_{\txtpaulifierz,\sminvtemperature,L}}
F_{\txtpaulifierz,S_{\sminvtemperature},\kappa,\Lambda,L}(\prbprocess,\opfocksegalradiation)
\opdmsr{\msrprb_{\txtpaulifierz,0,\sminvtemperature,\smchemicalpotential,L}}
\\ %%%%%%%%%%%%%%%%
&=
\int_{\Omega_{\txtparticle,\sminvtemperature}}
\fnexp{
-\frac{\physcharge^2}{4}
\fun{\opform{q}_{\txtrad,\txtbec,\sminvtemperature,\smchemicalpotential,L}^{\txteuclid}}
{\mathsf{I}_{\txtrad,S_{\sminvtemperature},\kappa,\Lambda,L}(\prbprocess)}}
\opdmsr{\msrprb_{\txtparticle,\sminvtemperature}}(\prbprocess).
\end{aligned}\]

\begin{defn}[Weighted Particle Path Probability Measure in the Bounded System]
For fixed $L$ and cutoffs $0<\kappa<\Lambda<\infty$, define the interacting probability measure $\msrprb_{\txtpaulifierz,\sminvtemperature,\kappa,\Lambda,\smchemicalpotential,L}$ in the bounded system and the weighted particle path probability measure $\msrprb_{\txtpaulifierz,\txtparticle,\sminvtemperature,\kappa,\Lambda,\smchemicalpotential,L}$ in the bounded system by
$$\begin{aligned}
\opdmsr{\msrprb_{\txtpaulifierz,\sminvtemperature,\kappa,\Lambda,\smchemicalpotential,L}}
&=
\frac{1}
{\smpartitionfunc_{\txtpaulifierz,\sminvtemperature,\kappa,\Lambda,\smchemicalpotential,L}}
F_{\txtpaulifierz,S_{\sminvtemperature},\kappa,\Lambda,L}(\prbprocess,\opfocksegalradiation)
\opdmsr{\msrprb_{\txtpaulifierz,0,\sminvtemperature,\smchemicalpotential,L}},
\\ %%%%%%%%%%%%%%%%
\opdmsr{\msrprb_{\txtpaulifierz,\txtparticle,\sminvtemperature,\kappa,\Lambda,\smchemicalpotential,L}}(\prbprocess)
&=
\frac{1}{\smpartitionfunc_{\txtpaulifierz,\sminvtemperature,\kappa,\Lambda,\smchemicalpotential,L}}
\fnexp{-\frac{\physcharge^2}{4}
\fun{\opform{q}_{\txtrad,\txtbec,\sminvtemperature,\smchemicalpotential,L}^{\txteuclid}}
{\mathsf{I}_{\txtrad,S_{\sminvtemperature},\kappa,\Lambda,L}(\prbprocess)}}
\opdmsr{\msrprb_{\txtparticle,\sminvtemperature}}(\prbprocess).
\end{aligned}$$
\end{defn}

For the definition of cylinder functions, see the discussion after the statement of Proposition \ref{expedition0012139}.

\begin{defn}[Functional Integral Representation of the Bounded-System KMS State]\label{expedition0012330}
For a bounded cylinder function $F$, define the bounded-system KMS state by
\begin{equation}\label{eq:PF-equilibrium-functional-finite-volume}
\psi_{\txtpaulifierz,\sminvtemperature,\kappa,\Lambda,\smchemicalpotential,L}^{\txteuclid}(F)
=
\int_{\prbqspace_{\txtpaulifierz,\sminvtemperature,L}}
F(\prbprocess,\opfocksegalradiation)
\opdmsr{\msrprb_{\txtpaulifierz,\sminvtemperature,\kappa,\Lambda,\smchemicalpotential,L}
(\prbprocess,\opfocksegalradiation)}.
\end{equation}
\end{defn}

\begin{prop}[Expectation of the Field Exponential Operator in the Bounded System]\label{expedition0012107}
For any $f
\in \sphilb{D}_{\txtrad,\txtphys,\sminvtemperature}$ and $t
\in S_{\sminvtemperature}$, define $\mathsf{S}_{\txtpaulifierz,\sminvtemperature,\kappa,\Lambda,\smchemicalpotential,L;t}
(P_{L} f)$ by
$$\begin{aligned}
\mathsf{S}_{\txtpaulifierz,\sminvtemperature,\kappa,\Lambda,\smchemicalpotential,L;t}(P_{L} f)
=
\int_{\Omega_{\txtparticle,\sminvtemperature}}
\fnexp{\frac{\physcharge}{2}
\fun{\opform{q}_{\txtrad,\txtbec,\sminvtemperature,\smchemicalpotential,L}^{\txteuclid}}
{j_t P_{L} f,\mathsf{I}_{\txtrad,S_{\sminvtemperature},\kappa,\Lambda,L}(\prbprocess)}}
\opdmsr{\msrprb_{\txtpaulifierz,\txtparticle,\sminvtemperature,\kappa,\Lambda,\smchemicalpotential,L}}(\prbprocess).
\end{aligned}$$
Then the characteristic function of the Segal field in the interacting system is
$$\begin{aligned}
\fun{\psi_{\txtpaulifierz,\sminvtemperature,\kappa,\Lambda,\smchemicalpotential,L}^{\txteuclid}}
{\fnexp{\imunit \opfocksegalradiation(j_t P_{L} f)}}
=
\fnexp{-\oneoverfour
\fun{\opform{q}_{\txtrad,\txtbec,\sminvtemperature,\smchemicalpotential,L}^{\txteuclid}}{j_t P_{L} f}}
\mathsf{S}_{\txtpaulifierz,\sminvtemperature,\kappa,\Lambda,\smchemicalpotential,L;t}(P_{L} f).
\end{aligned}$$
\end{prop}

\begin{proof}
By the definition \eqref{eq:PF-equilibrium-functional-finite-volume},
$$\begin{aligned}
&\fun{\psi_{\txtpaulifierz,\sminvtemperature,\kappa,\Lambda,\smchemicalpotential,L}^{\txteuclid}}
{\fnexp{\imunit \opfocksegalradiation(j_t P_{L} f)}}
\\ %%%%%%%%%%%%%%%%
&=
\frac{\int_{\Omega_{\txtparticle,\sminvtemperature}}
\int_{\prbqspace_{\txtrad,\sminvtemperature,L}}
\fnexp{\imunit \opfocksegalradiation(j_t P_{L} f)
-\imunit\physcharge \opfocksegalradiation(\mathsf{I}_{\txtrad,S_{\sminvtemperature},\kappa,\Lambda,L}(\prbprocess))}
\opdmsr{\msrprb_{\txtrad,\sminvtemperature,\smchemicalpotential,L}}(\opfocksegalradiation)
\opdmsr{\msrprb_{\txtparticle,\sminvtemperature}}(\prbprocess)}
{\smpartitionfunc_{\txtpaulifierz,\sminvtemperature,\kappa,\Lambda,\smchemicalpotential,L}}.
\end{aligned}$$
For the integral with respect to the field, apply the characteristic-function estimate
$$\begin{aligned}
\int_{\prbqspace_{\txtrad,\sminvtemperature,L}}
\fnexp{\imunit \opfocksegalradiation(F)}
\opdmsr{\msrprb_{\txtrad,\sminvtemperature,\smchemicalpotential,L}}
=
\fnexp{-\oneoverfour
\fun{\opform{q}_{\txtrad,\txtbec,\sminvtemperature,\smchemicalpotential,L}^{\txteuclid}}{F}}
\end{aligned}$$
with $F
= j_t P_{L} f - \physcharge \mathsf{I}_{\txtrad,S_{\sminvtemperature},\kappa,\Lambda,L}(\prbprocess)$ to obtain
$$\begin{aligned}
&\int_{\prbqspace_{\txtrad,\sminvtemperature,L}}
\fnexp{\imunit \opfocksegalradiation(j_t P_{L} f)
-\imunit\physcharge \opfocksegalradiation(\mathsf{I}_{\txtrad,S_{\sminvtemperature},\kappa,\Lambda,L}(\prbprocess))}
\opdmsr{\msrprb_{\txtrad,\sminvtemperature,\smchemicalpotential,L}}
\\ %%%%%%%%%%%%%%%%
&=
\fnexp{-\oneoverfour
\fun{\opform{q}_{\txtrad,\txtbec,\sminvtemperature,\smchemicalpotential,L}^{\txteuclid}}{j_t P_{L} f}
+\frac{\physcharge}{2}
\fun{\opform{q}_{\txtrad,\txtbec,\sminvtemperature,\smchemicalpotential,L}^{\txteuclid}}
{j_t P_{L} f,\mathsf{I}_{\txtrad,S_{\sminvtemperature},\kappa,\Lambda,L}(\prbprocess)}}
\\ %%%%%%%%%%%%%%%%
&\quad\times
\fnexp{-\frac{\physcharge^2}{4}
\fun{\opform{q}_{\txtrad,\txtbec,\sminvtemperature,\smchemicalpotential,L}^{\txteuclid}}
{\mathsf{I}_{\txtrad,S_{\sminvtemperature},\kappa,\Lambda,L}(\prbprocess)}}.
\end{aligned}$$
The partition function is represented as
$$\begin{aligned}
\smpartitionfunc_{\txtpaulifierz,\sminvtemperature,\kappa,\Lambda,\smchemicalpotential,L}
=
\int_{\Omega_{\txtparticle,\sminvtemperature}}
\fnexp{-\frac{\physcharge^2}{4}
\fun{\opform{q}_{\txtrad,\txtbec,\sminvtemperature,\smchemicalpotential,L}^{\txteuclid}}
{\mathsf{I}_{\txtrad,S_{\sminvtemperature},\kappa,\Lambda,L}(\prbprocess)}}
\opdmsr{\msrprb_{\txtparticle,\sminvtemperature}(\prbprocess)}.
\end{aligned}$$
It remains only to rearrange these expressions.
\end{proof}

In this proposition, the only quadratic Gaussian factor depending on \(f\) is \(\fun{\opform{q}_{\txtrad,\txtbec,\sminvtemperature,\smchemicalpotential,L}^{\txteuclid}}{j_t P_{L} f}\), and the zero-mode information is contained in this sesquilinear form. In particular, for \(t
= 0\), \[\begin{aligned}
\fun{\opform{q}_{\txtrad,\txtbec,\sminvtemperature,\smchemicalpotential,L}^{\txteuclid}}{j_0 P_{L} f}
&=
\fun{\opform{q}_{\txtrad,0,\sminvtemperature,\smchemicalpotential,L}}{P_{L} f}
+\fun{\opform{q}_{\txtrad,\txtnonzero,\sminvtemperature,\smchemicalpotential,L}}{P_{L} f},
\\ %%%%%%%%%%%%%%%%
\fun{\opform{q}_{\txtrad,0,\sminvtemperature,\smchemicalpotential,L}}{P_{L} f}
&=
\frac{(2\pi)^{d}}{L^d}
\frac{y_L + 1}{y_L - 1}
\bkt{\faftr{f}(0)}
{\Pi_{\txtrad,0}\faftr{f}(0)}_{\fldcmp^d},
\\ %%%%%%%%%%%%%%%%
\fun{\opform{q}_{\txtrad,\txtnonzero,\sminvtemperature,\smchemicalpotential,L}}{P_{L} f}
&=
\frac{(2\pi)^{d}}{L^d}
\sum_{k \in \setlattice_L^d\setminus\setone{0}}
\frac{y_L + \napiernum^{-\sminvtemperature\omega(k)}}
{y_L - \napiernum^{-\sminvtemperature\omega(k)}}
\bkt{\faftr{f}(k)}
{\delta^\perp(k)\faftr{f}(k)}_{\fldcmp^d}.
\end{aligned}\] Under the condition for BEC to occur in the free Bose gas \cite{AsaoArai28,YoshitsuguSekine004}, if the condensate density is denoted by \(\smnumberdensity_{\txtrad,0}(\sminvtemperature)\), then \(\lim_{y_L \downarrow 1}
\frac{1}{L^d(y_L-1)}
=
\smnumberdensity_{\txtrad,0}(\sminvtemperature)\) gives \[\begin{aligned}
\fun{\opform{q}_{\txtrad,0,\sminvtemperature,\smchemicalpotential,L}}{P_{L} f}
\to
2(2\pi)^{d} \smnumberdensity_{\txtrad,0}(\sminvtemperature)
\bkt{\faftr{f}(0)}
{\Pi_{\txtrad,0}\faftr{f}(0)}_{\fldcmp^d}
=
\fun{\opform{q}_{\txtrad,0,\sminvtemperature}}{f}.
\end{aligned}\] By the Riemann-sum limit, \[\begin{aligned}
\fun{\opform{q}_{\txtrad,\txtnonzero,\sminvtemperature,\smchemicalpotential,L}}{P_{L} f}
\to
\int_{\fldreal^{d}}
\frac{1+\napiernum^{-\sminvtemperature\omega(k)}}
{1-\napiernum^{-\sminvtemperature\omega(k)}}
\bkt{\faftr{f}(k)}
{\delta^\perp(k)\faftr{f}(k)}_{\fldcmp^d}
\opdmsr{k}
=
\fun{\opform{q}_{\txtrad,\txtnonzero,\sminvtemperature}}{f}
\end{aligned}\] holds, and therefore \begin{equation}\label{eq:PF-zero-nonzero-decomposition}
\fun{\opform{q}_{\txtrad,\txtbec,\sminvtemperature}^{\txteuclid}}{j_0 f}
=
\fun{\opform{q}_{\txtrad,0,\sminvtemperature}}{f}
+\fun{\opform{q}_{\txtrad,\txtnonzero,\sminvtemperature}}{f}
\end{equation} is obtained. In particular, \(\opform{q}_{\txtrad,0,\sminvtemperature}\) represents the zero-mode component corresponding to BEC, and \(\opform{q}_{\txtrad,\txtnonzero,\sminvtemperature}\) represents the usual thermal fluctuation.

\subsection{KMS States with Infrared and Ultraviolet Cutoffs}\label{kms-states-with-infrared-and-ultraviolet-cutoffs}

We define the infinite-volume cutoff KMS state with fixed infrared and ultraviolet cutoffs \(\kappa,\Lambda\). The bounded-system covariance and its zero-mode decomposition were computed explicitly according to \cite{AsaoArai28,YoshitsuguSekine004,YoshitsuguSekine006}. The following discussion concerns the object obtained after taking \(L\to\infty\) with the cutoffs \(\kappa,\Lambda\) fixed.

We use the finite-temperature Euclidean representation of the free radiation field, the zero-mode projection, the sesquilinear forms, the physical test-function space \(\sphilb{D}_{\txtrad,\txtphys,\sminvtemperature}\), and the free path space combining the particle and the radiation field, as defined according to the textbook \cite[Chapter 21]{DerezinskiGerard001} and \cite{YoshitsuguSekine004}, to define the cutoff current source and the interaction.

In the finite-temperature Euclidean representation, using the cutoff transverse coupling functions defined in \eqref{eq:PF-cutoff-transverse-coupling}, the minimal-coupling interaction appears through the stochastic integral along the particle loop \(\prbprocess\): \begin{equation}\label{expedition0012223}
\mathsf{I}_{\txtrad,I,\kappa,\Lambda}(\prbprocess)
=
\bigoplus_{\mathrm{i}=1}^{d}
\int_{I}
\fun{j_t}
{\lambda_{\txtrad,\mathrm{i},\prbprocess_t,\kappa,\Lambda}}
\opdmsr{\prbprocess_{\mathrm{i},t}}.
\end{equation}

\begin{prop}[Test-Function Property of the Cutoff Current Source]\label{expedition0012221}
Fix any $0<\kappa<\Lambda<\infty$ and any time interval $I
\subset S_{\sminvtemperature}$.
Then $\mathsf{I}_{\txtrad,I,\kappa,\Lambda}(\prbprocess)$, defined for the particle path $\prbprocess$, is almost surely with respect to the particle measure a test function belonging to the completion for the covariance form $\opform{q}_{\txtrad,\txtbec,\sminvtemperature}^{\txteuclid}$ of the radiation field, and satisfies
$$\begin{aligned}
\sqfun{\prbexp_{\msrprb_{\txtparticle,\sminvtemperature}}}
{\fun{\opform{q}_{\txtrad,\txtbec,\sminvtemperature}^{\txteuclid}}
{\mathsf{I}_{\txtrad,I,\kappa,\Lambda}(\prbprocess)}}
< \infty.
\end{aligned}$$
Because of the infrared cutoff, the zero-mode part does not contribute, and
$$\begin{aligned}
\fun{\opform{q}_{\txtrad,\txtbec,\sminvtemperature}^{\txteuclid}}
{\mathsf{I}_{\txtrad,I,\kappa,\Lambda}(\prbprocess)}
=
\fun{\opform{q}_{\txtrad,\txtnonzero,\sminvtemperature}^{\txteuclid}}
{\mathsf{I}_{\txtrad,I,\kappa,\Lambda}(\prbprocess)}
\end{aligned}$$
holds.
In particular, $\fun{\opfocksegalradiation}
{\mathsf{I}_{\txtrad,I,\kappa,\Lambda}(\prbprocess)}$ is meaningful as a Gaussian random variable of the radiation field.
\end{prop}

\begin{proof}
The test-function property follows from Proposition \ref{expedition0012325} by first taking the limit $L
\to \infty$ and then taking $\smchemicalpotential
\uparrow 0$.

Under fixed cutoffs $0<\kappa<\Lambda<\infty$, $\fun{j_t}
{\lambda_{\txtrad,\mathrm{i},x,\kappa,\Lambda}}$ is supported in the momentum region $\kappa\leq\abs{k}\leq\Lambda$, so the finite-temperature covariance kernel and $\omega(k)^{-1}$ are uniformly bounded on this support.
Therefore the covariance norm estimate for the bounded-system current source is preserved in the infinite-volume limit, and the Itô isometry and drift estimate for the semimartingale decomposition of the Brownian bridge give
$$\sqfun{\prbexp_{\msrprb_{\txtparticle,\sminvtemperature}}}
{\fun{\opform{q}_{\txtrad,\txtnonzero,\sminvtemperature}^{\txteuclid}}
{\mathsf{I}_{\txtrad,I,\kappa,\Lambda}(\prbprocess)}}
<\infty.$$
Because of the infrared cutoff, the cutoff current source has no $k=0$ component, so it has no cross term with the zero-mode covariance.
Hence the estimate in $\opform{q}_{\txtrad,\txtbec,\sminvtemperature}^{\txteuclid}$ is given only by the non-zero-mode part, and $\opfocksegalradiation(\mathsf{I}_{\txtrad,I,\kappa,\Lambda}(\prbprocess))$ is defined as a centered Gaussian random variable.
\end{proof}

For a time subset \(I
\subset S_{\sminvtemperature}\), define the Feynman--Kac--Nelson kernel by \begin{equation}\label{expedition0012379}
\begin{aligned}
F_{\txtpaulifierz,I,\kappa,\Lambda}(\prbprocess,\opfocksegalradiation)
=
\fnexp{-\imunit
\physcharge
\opfocksegalradiation(\mathsf{I}_{\txtrad,I,\kappa,\Lambda}(\prbprocess))}
\end{aligned}
\end{equation} This kernel acts as the Euclidean weight used to perturb the product measure \(\msrprb_{\txtpaulifierz,0,\sminvtemperature}\) and define the interacting system.

\begin{prop}[Cutoff Pauli--Fierz Local Kernel]\label{expedition0012206}
For any $0<\kappa<\Lambda<\infty$, the family $\fml{F_{\txtpaulifierz,I,\kappa,\Lambda}}
{I\subset S_{\sminvtemperature}}$ is a local Feynman--Kac--Nelson kernel.
In particular, it satisfies interval measurability, multiplicativity, time-translation covariance, reflection symmetry, $\lp^p$-continuity with respect to interval endpoints, and exponential integrability.

Furthermore, for any $0<t<\frac{\sminvtemperature}{2}$, if
\begin{equation}\label{expedition0012207}
\begin{aligned}
\sphilb{D}_{\txtpaulifierz,t,\kappa,\Lambda}
=
\prbexp_{0,\frac{\sminvtemperature}{2}}
\fun{\splinspan}
{\bigcup_{0\leq s<\frac{\sminvtemperature}{2}-t}
F_{\txtpaulifierz,\closedinterval{0}{s},\kappa,\Lambda}
\fun{\lp^{\infty}}
{\prbqspace_{\txtpaulifierz,\sminvtemperature},
\mblfml{S}_{\txtpaulifierz,\closedinterval{0}{\frac{\sminvtemperature}{2}-t}}}}
\end{aligned}
\end{equation}
is defined, then for any $0
\leq s
\leq t
\leq \frac{\sminvtemperature}{2}$ and $f\in
\fun{\lp^2}{\prbqspace_{\txtpaulifierz,\sminvtemperature},
\msrprb_{\txtpaulifierz,0,\sminvtemperature}}$
there exists a unique $P_{\txtpaulifierz,\kappa,\Lambda,s}
\colon
\sphilb{D}_{\txtpaulifierz,t,\kappa,\Lambda}
\to
\sphilb{D}_{\txtpaulifierz,t-s,\kappa,\Lambda}$ satisfying
\begin{equation}\label{expedition0012208}
\begin{aligned}
P_{\txtpaulifierz,\kappa,\Lambda,s}
\prbexp_{\setone{0,\frac{\sminvtemperature}{2}}}f
=
\prbexp_{\setone{0,\frac{\sminvtemperature}{2}}}
F_{\txtpaulifierz,\closedinterval{0}{s},\kappa,\Lambda}
U_{\txtpaulifierz,s}f.
\end{aligned}
\end{equation}
In particular, $\fml{P_{\txtpaulifierz,\kappa,\Lambda,t}}
{t\in\closedinterval{0}{\frac{\sminvtemperature}{2}}}$ is a local Hermitian semigroup.
\end{prop}

\begin{proof}
It suffices to take the limits $L
\to \infty$ and $\smchemicalpotential
\uparrow 0$ in the result of the bounded-system Proposition \ref{expedition0012320} under fixed cutoffs.
By Proposition \ref{expedition0012221}, the infinite-system current source is defined as a Gaussian test function for the radiation field, and interval measurability, multiplicativity, time-translation covariance, and reflection symmetry remain as limits of the corresponding properties in the bounded system.
Also, $\abs{F_{\txtpaulifierz,I,\kappa,\Lambda}}
= 1$ makes exponential integrability immediate, and $\lp^p$-continuity with respect to interval endpoints follows from the $\lp^2$-continuity of the current source at fixed cutoffs.
The assertion about the local Hermitian semigroup is the general reconstruction of a local Feynman--Kac--Nelson kernel from the domain in \eqref{expedition0012207} and the action in \eqref{expedition0012208}.
\end{proof}

Below, denote the local semigroup defined by \eqref{expedition0012208} by \(P_{\txtpaulifierz,\kappa,\Lambda,t}\). This is not the kernel that directly gives the KMS state on the full time circle, but the semigroup obtained from the local kernel corresponding to the subinterval \(\closedinterval{0}{t}\). The interaction factor on the full circle is denoted by \(F_{\txtpaulifierz,S_{\sminvtemperature},\kappa,\Lambda}\), while the bilinear form of the local semigroup uses \(F
_{\txtpaulifierz,\closedinterval{0}{t},\kappa,\Lambda}\) and \(\mathsf{I}
_{\txtrad,\closedinterval{0}{t},\kappa,\Lambda}\). This distinction is maintained after cutoff removal.

For bounded time-zero particle functions \(F_{\txtparticle}\) and \(G_{\txtparticle}\) and radiation-field test functions \(f,g
\in\sphilb{D}_{\txtrad,\sminvtemperature}\), consider the time-zero generators \(F(\prbprocess,\opfocksegalradiation)
=
F_{\txtparticle}(\prbprocess_0)
\napiernum^{\imunit\opfocksegalradiation(j_0f)}\) and \(G(\prbprocess,\opfocksegalradiation)
=
G_{\txtparticle}(\prbprocess_0)
\napiernum^{\imunit\opfocksegalradiation(j_0g)}\).

\begin{prop}[Pauli--Fierz Kernel Representation on Generators]\label{expedition0012209}
For the above $F,G$,
\begin{equation}\label{expedition0012210}
\begin{aligned}
&\bkt{F}{P_{\txtpaulifierz,\kappa,\Lambda,t}G}
=
\int_{\Omega_{\txtparticle,\sminvtemperature}}
\cmpconj{F_{\txtparticle}(\prbprocess_0)}
G_{\txtparticle}(\prbprocess_t)
\\ %%%%%%%%%%%%%%%%
&\quad\times
\fnexp{-\oneoverfour
\fun{\opform{q}_{\txtrad,\txtbec,\sminvtemperature}^{\txteuclid}}
{j_tg-j_0f}
+\frac{\physcharge}{2}
\fun{\opform{q}_{\txtrad,\txtbec,\sminvtemperature}^{\txteuclid}}
{j_tg-j_0f,
\mathsf{I}_{\txtrad,\closedinterval{0}{t},\kappa,\Lambda}(\prbprocess)}}
\\ %%%%%%%%%%%%%%%%
&\quad\times
\fnexp{-\frac{\physcharge^2}{4}
\fun{\opform{q}_{\txtrad,\txtbec,\sminvtemperature}^{\txteuclid}}
{\mathsf{I}_{\txtrad,\closedinterval{0}{t},\kappa,\Lambda}(\prbprocess)}}
\opdmsr{\msrprb_{\txtparticle,\sminvtemperature}}(\prbprocess)
\end{aligned}
\end{equation}
holds.
Furthermore, since the cutoff current source has no zero-momentum component because of the infrared cutoff, at all places depending on the cutoff current source,
\begin{equation}\label{expedition0012211}
\begin{aligned}
\fun{\opform{q}_{\txtrad,\txtbec,\sminvtemperature}^{\txteuclid}}
{h,\mathsf{I}_{\txtrad,\closedinterval{0}{t},\kappa,\Lambda}(\prbprocess)}
&=
\fun{\opform{q}_{\txtrad,\txtnonzero,\sminvtemperature}^{\txteuclid}}
{h,\mathsf{I}_{\txtrad,\closedinterval{0}{t},\kappa,\Lambda}(\prbprocess)},
\\ %%%%%%%%%%%%%%%%
\fun{\opform{q}_{\txtrad,\txtbec,\sminvtemperature}^{\txteuclid}}
{\mathsf{I}_{\txtrad,\closedinterval{0}{t},\kappa,\Lambda}(\prbprocess)}
&=
\fun{\opform{q}_{\txtrad,\txtnonzero,\sminvtemperature}^{\txteuclid}}
{\mathsf{I}_{\txtrad,\closedinterval{0}{t},\kappa,\Lambda}(\prbprocess)}
\end{aligned}
\end{equation}
holds.
\end{prop}

\begin{proof}
Take $L
\to \infty$ and then $\smchemicalpotential
\uparrow 0$ in the bounded-system generator representation \eqref{expedition0012324} under fixed cutoffs.
The free radiation-field covariance converges to $\opform{q}
_{\txtrad,\txtbec,\sminvtemperature}^{\txteuclid}$ by the Riemann-sum limit of the bounded system and the limit of the zero-mode projection, and the current source $\mathsf{I}
_{\txtrad,\closedinterval{0}{t},\kappa,\Lambda,L}$ converges to $\mathsf{I}
_{\txtrad,\closedinterval{0}{t},\kappa,\Lambda}$.
Substituting this limit into \eqref{expedition0012324} gives \eqref{expedition0012210}.
Equation \eqref{expedition0012211} follows because the momentum representation of the current source has no $k=0$ component due to the infrared cutoff and thus annihilates the zero-momentum form.
\end{proof}

Using the Feynman--Kac--Nelson kernel \(F_{\txtpaulifierz,S_{\sminvtemperature},\kappa,\Lambda}(\prbprocess,\opfocksegalradiation)\) in \eqref{expedition0012379} corresponding to the full time circle \(S_{\sminvtemperature}\), define the cutoff partition function by \[\begin{aligned}
\smpartitionfunc_{\txtpaulifierz,\sminvtemperature,\kappa,\Lambda}
&=
\int_{\prbqspace_{\txtpaulifierz,\sminvtemperature}}
F_{\txtpaulifierz,S_{\sminvtemperature},\kappa,\Lambda}(\prbprocess,\opfocksegalradiation)
\opdmsr{\msrprb_{\txtpaulifierz,0,\sminvtemperature}}(\prbprocess,\opfocksegalradiation)
\\ %%%%%%%%%%%%%%%%
&=
\int_{\Omega_{\txtparticle,\sminvtemperature}}
\fnexp{-\frac{\physcharge^2}{4}
\opform{q}_{\txtrad,\txtbec,\sminvtemperature}^{\txteuclid}
\rbk{\mathsf{I}_{\txtrad,S_{\sminvtemperature},\kappa,\Lambda}(\prbprocess)}}
\opdmsr{\msrprb_{\txtparticle,\sminvtemperature}}(\prbprocess).
\end{aligned}\]

For the definition of cylinder functions, see the discussion after the statement of Proposition \ref{expedition0012139}.

\begin{defn}[KMS State of the Cutoff Pauli--Fierz Model]\label{expedition0012106}
For any bounded cylinder function $F$, define the KMS state of the cutoff Pauli--Fierz model at inverse temperature $\sminvtemperature
> 0$ by
\begin{equation}\label{eq:PF-equilibrium-functional}
\fun{\oastate[\psi_{\txtpaulifierz,\sminvtemperature,\kappa,\Lambda}]}{F}
=
\frac{1}{\smpartitionfunc_{\txtpaulifierz,\sminvtemperature,\kappa,\Lambda}}
\int_{\prbqspace_{\txtpaulifierz,\sminvtemperature}}
F(\prbprocess,\opfocksegalradiation)
F_{\txtpaulifierz,S_{\sminvtemperature},\kappa,\Lambda}(\prbprocess,\opfocksegalradiation)
\opdmsr{\msrprb_{\txtpaulifierz,0,\sminvtemperature}(\prbprocess,\opfocksegalradiation)}.
\end{equation}
\end{defn}

\begin{prop}[Limit to the Cutoff KMS State]\label{expedition0012301}
Fix infrared and ultraviolet cutoffs $0
< \kappa
< \Lambda
< \infty$.
For any finitely many times $t_j
\in S_{\sminvtemperature}$, bounded particle functions $F_j$, and $f_j
\in \sphilb{D}_{\txtrad,\txtphys,\sminvtemperature}$, define the bounded-system observables and the limiting observables by
$$A_{j,L}
=
F_j(\prbprocess_{t_j})
\fnexp{\imunit\opfocksegalradiation(j_{t_j}P_Lf_j)},
\quad
A_j
=
F_j(\prbprocess_{t_j})
\fnexp{\imunit\opfocksegalradiation(j_{t_j}f_j)}.$$
Then, for the bounded-system KMS state defined in Definition \ref{expedition0012330},
\begin{equation}\label{expedition0012302}
\begin{aligned}
\lim_{\smchemicalpotential\uparrow0}
\lim_{L\to\infty}
\fun{\psi_{\txtpaulifierz,\sminvtemperature,\kappa,\Lambda,\smchemicalpotential,L}^{\txteuclid}}
{A_{1,L}\cdots A_{n,L}}
=
\fun{\oastate[\psi_{\txtpaulifierz,\sminvtemperature,\kappa,\Lambda}]}
{A_1\cdots A_n}
\end{aligned}
\end{equation}
holds.
\end{prop}

\begin{proof}
To simplify notation, set
$$B(\prbprocess)=\prod_{j=1}^{n}F_j(\prbprocess_{t_j}),
\quad
G_L=\sum_{j=1}^{n}j_{t_j}P_Lf_j,
\quad
G=\sum_{j=1}^{n}j_{t_j}f_j.$$
Since $B(\prbprocess)$ is a product of bounded functions, there exists a constant $C_B
< \infty$ such that $\abs{B(\prbprocess)}
\leq C_B$.

Substitute $A_{1,L}
\cdots A_{n,L}$ into the definition \eqref{eq:PF-equilibrium-functional-finite-volume} of the bounded-system KMS state, fix the particle path $\prbprocess$, and first integrate the bounded-system radiation field.
Writing the bounded-system KMS state as
$\fun{\psi_{\txtpaulifierz,\sminvtemperature,\kappa,\Lambda,\smchemicalpotential,L}^{\txteuclid}}
{A_{1,L}\cdots A_{n,L}}
=
\frac{N_{L,\smchemicalpotential}}{Z_{L,\smchemicalpotential}}$, the centered Gaussian characteristic functional of the bounded-system free radiation field gives the numerator and denominator respectively as
$$\begin{aligned}
N_{L,\smchemicalpotential}
&=
\int_{\Omega_{\txtparticle,\sminvtemperature}}
B(\prbprocess)
\fnexp{-\oneoverfour
\fun{\opform{q}_{\txtrad,\txtbec,\sminvtemperature,\smchemicalpotential,L}^{\txteuclid}}
{G_L-\physcharge
\mathsf{I}_{\txtrad,S_{\sminvtemperature},\kappa,\Lambda,L}(\prbprocess)}}
\opdmsr{\msrprb_{\txtparticle,\sminvtemperature}(\prbprocess)},
\\ %%%%%%%%%%%%%%%%
Z_{L,\smchemicalpotential}
&=
\int_{\Omega_{\txtparticle,\sminvtemperature}}
\fnexp{-\frac{\physcharge^2}{4}
\fun{\opform{q}_{\txtrad,\txtbec,\sminvtemperature,\smchemicalpotential,L}^{\txteuclid}}
{\mathsf{I}_{\txtrad,S_{\sminvtemperature},\kappa,\Lambda,L}(\prbprocess)}}
\opdmsr{\msrprb_{\txtparticle,\sminvtemperature}(\prbprocess)}.
\end{aligned}$$

For fixed cutoffs $0<\kappa<\Lambda<\infty$, $P_L
\lambda_{\txtrad,\mathrm{i},x,\kappa,\Lambda}$ converges to $\lambda
_{\txtrad,\mathrm{i},x,\kappa,\Lambda}$ in \eqref{eq:PF-cutoff-transverse-coupling}.
Next we verify the limit of the sesquilinear forms.
First consider the term containing only external fields.
For any $j,\ell$, the zero-mode part in the bounded system is
$$\begin{aligned}
\fun{\opform{q}_{\txtrad,0,\sminvtemperature,\smchemicalpotential,L}^{\txteuclid}}
{j_{t_j}P_Lf_j,j_{t_\ell}P_Lf_\ell}
=
\frac{(2\pi)^d}{L^d}
\frac{y_L+1}{y_L-1}
\bkt{\faftr{f_j}(0)}
{\Pi_{\txtrad,0}\faftr{f_\ell}(0)}_{\fldcmp^d}.
\end{aligned}$$
Since $L^{-d}(y_L-1)^{-1}
\to \smnumberdensity_{\txtrad,0}(\sminvtemperature)$ holds in the bounded-system limit of the free Bose gas, this converges to
$$2(2\pi)^d\smnumberdensity_{\txtrad,0}(\sminvtemperature)
\bkt{\faftr{f_j}(0)}
{\Pi_{\txtrad,0}\faftr{f_\ell}(0)}_{\fldcmp^d}
=
\fun{\opform{q}_{\txtrad,0,\sminvtemperature}^{\txteuclid}}
{j_{t_j}f_j,j_{t_\ell}f_\ell}.$$

For the non-zero-mode part,
$$\begin{aligned}
&\fun{\opform{q}_{\txtrad,\txtnonzero,\sminvtemperature,\smchemicalpotential,L}^{\txteuclid}}
{j_{t_j}P_Lf_j,j_{t_\ell}P_Lf_\ell}
\\ %%%%%%%%%%%%%%%%
&=
\frac{(2\pi)^d}{L^d}
\sum_{k\in\setlattice_L^d\setminus\setone{0}}
\frac{y_L\napiernum^{-\abs{t_j-t_\ell}\omega(k)}
+\napiernum^{-(\sminvtemperature-\abs{t_j-t_\ell})\omega(k)}}
{y_L-\napiernum^{-\sminvtemperature\omega(k)}}
\bkt{\faftr{f_j}(k)}
{\delta^\perp(k)\faftr{f_\ell}(k)}_{\fldcmp^d},
\end{aligned}$$
as $y_L
\downarrow 1$, the fractional factor $\frac{y_L\napiernum^{-\abs{t_j-t_\ell}\omega(k)}
+\napiernum^{-(\sminvtemperature-\abs{t_j-t_\ell})\omega(k)}}
{y_L-\napiernum^{-\sminvtemperature\omega(k)}}$ converges pointwise to the corresponding fractional factor appearing in the non-zero-mode part of \eqref{eq:beta-pf-covariance}.
Furthermore, since $f_j,f_\ell
\in \sphilb{D}_{\txtrad,\txtphys,\sminvtemperature}
\subset\sphilb{D}_{\txtrad,\txtnonzero,\sminvtemperature}$, the Riemann sum on the right converges to $\fun{\opform{q}_{\txtrad,\txtnonzero,\sminvtemperature}^{\txteuclid}}
{j_{t_j}f_j,j_{t_\ell}f_\ell}$.
Applying bilinearity to the finite sum $G_L
=
\sum_{j=1}^{n}j_{t_j}P_Lf_j$ gives $\fun{\opform{q}_{\txtrad,\txtbec,\sminvtemperature,\smchemicalpotential,L}^{\txteuclid}}
{G_L}
\to
\fun{\opform{q}_{\txtrad,\txtbec,\sminvtemperature}^{\txteuclid}}
{G}$.

Next verify the terms containing the current source.
Because of the infrared cutoff, $\mathsf{I}_{\txtrad,S_{\sminvtemperature},\kappa,\Lambda,L}(\prbprocess)$ and $\mathsf{I}_{\txtrad,S_{\sminvtemperature},\kappa,\Lambda}(\prbprocess)$ have no zero-momentum component, so the zero-mode part does not contribute in forms containing the current source.
For each $j$ and each component $\mathrm{i}$, the cross term is represented as
$$\begin{aligned}
&\fun{\opform{q}_{\txtrad,\txtnonzero,\sminvtemperature,\smchemicalpotential,L}^{\txteuclid}}
{j_{t_j}P_Lf_j,
\int_{S_{\sminvtemperature}}
j_sP_L\lambda_{\txtrad,\mathrm{i},\prbprocess_s,\kappa,\Lambda}
\opdmsr{\prbprocess_{\mathrm{i},s}}}
\\ %%%%%%%%%%%%%%%%
&=
\int_{S_{\sminvtemperature}}
\frac{(2\pi)^d}{L^d}
\sum_{k\in\setlattice_L^d\cap\set{k}{\kappa\leq\abs{k}\leq\Lambda}}
\frac{y_L\napiernum^{-\abs{t_j-s}\omega(k)}
+\napiernum^{-(\sminvtemperature-\abs{t_j-s})\omega(k)}}
{y_L-\napiernum^{-\sminvtemperature\omega(k)}}
\frac{\bkt{\faftr{f_j}(k)}
{\delta^\perp(k)e_{\mathrm{i}}}_{\fldcmp^d}}
{\sqrt{\omega(k)}}
\napiernum^{-\imunit k\prbprocess_s}
\opdmsr{\prbprocess_{\mathrm{i},s}}.
\end{aligned}$$
Since the wave numbers are restricted to the fixed compact set $\set{k}{\kappa\leq\abs{k}\leq\Lambda}$, each term of the sum
$$\frac{y_L\napiernum^{-\abs{t_j-s}\omega(k)}
+\napiernum^{-(\sminvtemperature-\abs{t_j-s})\omega(k)}}
{y_L-\napiernum^{-\sminvtemperature\omega(k)}}
\frac{\bkt{\faftr{f_j}(k)}
{\delta^\perp(k)e_{\mathrm{i}}}_{\fldcmp^d}}
{\sqrt{\omega(k)}}
\napiernum^{-\imunit k\prbprocess_s}$$
is uniformly bounded in $L,\smchemicalpotential,s$, and the Riemann sum converges uniformly to the corresponding momentum integral.
Applying the Itô-part estimate and drift-part estimate from Proposition \ref{expedition0012325} to the difference in the semimartingale decomposition of the Brownian bridge, this cross term converges in $\fun{\lp^{1}}
{\msrprb_{\txtparticle,\sminvtemperature}}$ to $\fun{\opform{q}_{\txtrad,\txtnonzero,\sminvtemperature}^{\txteuclid}}
{j_{t_j}f_j,
\int_{S_{\sminvtemperature}}
j_s\lambda_{\txtrad,\mathrm{i},\prbprocess_s,\kappa,\Lambda}
\opdmsr{\prbprocess_{\mathrm{i},s}}}$.
Summing over finitely many $j,\mathrm{i}$ gives convergence in $\fun{\lp^{1}}
{\msrprb_{\txtparticle,\sminvtemperature}}$:
$$\fun{\opform{q}_{\txtrad,\txtbec,\sminvtemperature,\smchemicalpotential,L}^{\txteuclid}}
{G_L,
\mathsf{I}_{\txtrad,S_{\sminvtemperature},\kappa,\Lambda,L}(\prbprocess)}
\to
\fun{\opform{q}_{\txtrad,\txtbec,\sminvtemperature}^{\txteuclid}}
{G,
\mathsf{I}_{\txtrad,S_{\sminvtemperature},\kappa,\Lambda}(\prbprocess)}.$$

The self term is treated similarly: substituting the two current sources into the non-zero-mode form and splitting into finitely many components gives a sum of bounded Riemann-sum kernels on the fixed cutoff support and double stochastic integrals.
Applying the estimate of Proposition \ref{expedition0012325} to the difference of the two current sources and summing over finitely many components yields convergence in $\lp^1(\msrprb_{\txtparticle,\sminvtemperature})$:
$$\fun{\opform{q}_{\txtrad,\txtbec,\sminvtemperature,\smchemicalpotential,L}^{\txteuclid}}
{\mathsf{I}_{\txtrad,S_{\sminvtemperature},\kappa,\Lambda,L}(\prbprocess)}
\to
\fun{\opform{q}_{\txtrad,\txtbec,\sminvtemperature}^{\txteuclid}}
{\mathsf{I}_{\txtrad,S_{\sminvtemperature},\kappa,\Lambda}(\prbprocess)}.$$
For the subsequent argument on limits of integrals, this $\lp^{1}$ convergence and the uniform boundedness discussed below are sufficient.

We now pass the convergence obtained so far to the integrals.
For absolute-value estimates, only the real part of the exponent contributes, so it suffices to estimate the real part.
By nonnegativity of the sesquilinear form,
$0
\leq
\opreal
\fun{\opform{q}_{\txtrad,\txtbec,\sminvtemperature,\smchemicalpotential,L}^{\txteuclid}}
{G_L-\physcharge
\mathsf{I}_{\txtrad,S_{\sminvtemperature},\kappa,\Lambda,L}(\prbprocess)}$
holds, and expanding it gives
$$\begin{aligned}
&\opreal
\rbk{\frac{\physcharge}{2}
\fun{\opform{q}_{\txtrad,\txtbec,\sminvtemperature,\smchemicalpotential,L}^{\txteuclid}}
{G_L,
\mathsf{I}_{\txtrad,S_{\sminvtemperature},\kappa,\Lambda,L}(\prbprocess)}}
-\frac{\physcharge^2}{4}
\fun{\opform{q}_{\txtrad,\txtbec,\sminvtemperature,\smchemicalpotential,L}^{\txteuclid}}
{\mathsf{I}_{\txtrad,S_{\sminvtemperature},\kappa,\Lambda,L}(\prbprocess)}
\\ %%%%%%%%%%%%%%%%
&\leq
\oneoverfour
\fun{\opform{q}_{\txtrad,\txtbec,\sminvtemperature,\smchemicalpotential,L}^{\txteuclid}}
{G_L}.
\end{aligned}$$
The right-hand side has a finite limit as $L\to\infty,\ \smchemicalpotential\uparrow0$, and is therefore uniformly bounded for sufficiently large $L$ and $\smchemicalpotential$ sufficiently close to $0$.
The remaining finitely many indices can be absorbed by changing the constant.
Thus there exists a constant $C_G
< \infty$ such that, for all $L,\smchemicalpotential$,
$$\abs{B(\prbprocess)
\fnexp{-\oneoverfour
\fun{\opform{q}_{\txtrad,\txtbec,\sminvtemperature,\smchemicalpotential,L}^{\txteuclid}}
{G_L-\physcharge
\mathsf{I}_{\txtrad,S_{\sminvtemperature},\kappa,\Lambda,L}(\prbprocess)}}}
\leq
C_B C_G.$$
Moreover, the integrand of the denominator is between $0$ and $1$.
Therefore the dominated convergence theorem gives
$$\begin{aligned}
N_{L,\smchemicalpotential}
&\to
N
=
\int_{\Omega_{\txtparticle,\sminvtemperature}}
B(\prbprocess)
\fnexp{-\oneoverfour
\fun{\opform{q}_{\txtrad,\txtbec,\sminvtemperature}^{\txteuclid}}
{G-\physcharge
\mathsf{I}_{\txtrad,S_{\sminvtemperature},\kappa,\Lambda}(\prbprocess)}}
\opdmsr{\msrprb_{\txtparticle,\sminvtemperature}}(\prbprocess),
\\ %%%%%%%%%%%%%%%%
Z_{L,\smchemicalpotential}
&\to
Z
=
\int_{\Omega_{\txtparticle,\sminvtemperature}}
\fnexp{-\frac{\physcharge^2}{4}
\fun{\opform{q}_{\txtrad,\txtbec,\sminvtemperature}^{\txteuclid}}
{\mathsf{I}_{\txtrad,S_{\sminvtemperature},\kappa,\Lambda}(\prbprocess)}}
\opdmsr{\msrprb_{\txtparticle,\sminvtemperature}}(\prbprocess).
\end{aligned}$$
By Proposition \ref{expedition0012221}, $\fun{\opform{q}
_{\txtrad,\txtbec,\sminvtemperature}^{\txteuclid}}{\mathsf{I}_{\txtrad,S_{\sminvtemperature},\kappa,\Lambda}(\prbprocess)}<\infty$ holds almost surely, so the integrand is positive and $Z>0$ follows.

Finally, by the same treatment of the Gaussian integral in the definition \eqref{eq:PF-equilibrium-functional}, the cutoff infinite-volume state on the right-hand side is exactly given by $N / \smpartitionfunc$.
Therefore
$$\lim_{\smchemicalpotential\uparrow0}
\lim_{L\to\infty}
\fun{\psi_{\txtpaulifierz,\sminvtemperature,\kappa,\Lambda,\smchemicalpotential,L}^{\txteuclid}}
{A_{1,L}\cdots A_{n,L}}
=
\frac{N}{Z}
=
\fun{\oastate[\psi_{\txtpaulifierz,\sminvtemperature,\kappa,\Lambda}]}
{A_1\cdots A_n}$$
is obtained.
\end{proof}

\subsection{Removal of Infrared and Ultraviolet Cutoffs}\label{removal-of-infrared-and-ultraviolet-cutoffs-1}

As in \eqref{expedition0012180} and \eqref{expedition0012182} of Theorem \ref{expedition0012156} for the Nelson model, and as in \eqref{expedition0012361}--\eqref{expedition0012364} of Proposition \ref{expedition0012360}, for the spinless Pauli--Fierz model we also first fix cutoff KMS states and define the point-source model as the limit of finite-time correlation functions. Since the interaction couples to the radiation field through the particle current \(\opdmsr{\prbprocess_{t}}\), rather than through the particle position itself, the object to be controlled in cutoff removal is not \(\mathsf{J}_{I,\kappa,\Lambda}\) in \eqref{expedition0012203} for the Nelson model, but the stochastic integral \(\mathsf{I}_{\txtrad,I,\kappa,\Lambda}(\prbprocess)\) with transverse projection. For fixed cutoffs, the local Feynman--Kac--Nelson kernel, the local Hermitian semigroup, and the kernel representation on generators have already been constructed in Propositions \ref{expedition0012206} and \ref{expedition0012209}. The local semigroup appearing here is \(P_{\txtpaulifierz,\kappa,\Lambda,t}\) in \eqref{expedition0012208}, and the local semigroup after cutoff removal is also regarded as this limit. Thus the new argument required in this subsection is the removal of the infrared and ultraviolet cutoffs in the self term of the current source and the external-field cross term appearing in the sesquilinear forms defining these objects.

As with the point-source singular part on the Nelson model side, in this subsection the spatial dimension is \(d
=3\). The places where three-dimensionality is used are stated immediately before each proposition or lemma. Below we distinguish the terms for the KMS state on the full time circle from the terms for subintervals appearing in the local semigroup. For the former argument we use the loop condition \(\int_{S_{\sminvtemperature}}\opdmsr{\prbprocess_t}
= 0\). For the latter, \(\int_{\closedinterval{a}{b}}\opdmsr{\prbprocess_t}
=\prbprocess_b-\prbprocess_a\) remains, but the fixed-cutoff local current source has no zero-momentum component by \eqref{expedition0012211}. Thus, even after cutoff removal, we do not add a new zero-mode local interaction, and instead treat it as the limit of the non-zero-mode sesquilinear form.

\subsubsection{Current Sources and Zero Modes}\label{current-sources-and-zero-modes}

\begin{prop}[Zero Mode of the Current Source and Non-Zero-Mode Reduction on Local Intervals]\label{expedition0012331}
For any $0
< \kappa
< \Lambda
< \infty$ and closed interval $I
\subset S_{\sminvtemperature}$, define the cutoff current source by $\mathsf{I}_{\txtrad,I,\kappa,\Lambda}(\prbprocess)
=
\bigoplus_{\mathrm{i}=1}^{d}
\int_I
\fun{j_t}
{\lambda_{\txtrad,\mathrm{i},\prbprocess_t,\kappa,\Lambda}}
\opdmsr{\prbprocess_{\mathrm{i},t}}$.

\begin{enumerate}
\item
This current source has no zero-momentum component.
For any $I$, $\faftr{\mathsf{I}}_{\txtrad,I,\kappa,\Lambda}(0)
= 0$ holds.
In particular, for any $h
\in \sphilb{D}_{\txtrad,\txtphys,\sminvtemperature}$,
$$\begin{aligned}
\fun{\opform{q}_{\txtrad,\txtbec,\sminvtemperature}^{\txteuclid}}
{h,\mathsf{I}_{\txtrad,I,\kappa,\Lambda}(\prbprocess)}
=
\fun{\opform{q}_{\txtrad,\txtnonzero,\sminvtemperature}^{\txteuclid}}
{h,\mathsf{I}_{\txtrad,I,\kappa,\Lambda}(\prbprocess)}
\end{aligned}$$
holds.

\item
For a closed particle loop on the full time circle, the formal zero-momentum component obtained by removing the cutoff functions vanishes: for $\mathsf{I}_{\txtrad,S_{\sminvtemperature},\kappa,\Lambda}$ in \eqref{expedition0012223},
\begin{equation}\label{expedition0012224}
\begin{aligned}
\faftr{\mathsf{I}}_{\txtrad,S_{\sminvtemperature},\kappa,\Lambda}(0)
=0,
\quad
\int_{-\frac{\sminvtemperature}{2}}^{\frac{\sminvtemperature}{2}}
\opdmsr{\prbprocess_{t}}
=
\prbprocess_{\frac{\sminvtemperature}{2}}
-\prbprocess_{-\frac{\sminvtemperature}{2}}
=0
\end{aligned}
\end{equation}
holds.
In particular, in the partition function of the KMS state and in the full-circle interaction, the BEC zero-mode component $\opform{q}
_{\txtrad,0,\sminvtemperature}$ does not cross with the current source.

\item
For the local kernel on a subinterval $I
= \closedinterval{a}{b}$,
\begin{equation}\label{expedition0012333}
\int_a^b
\opdmsr{\prbprocess_{t}}
= \prbprocess_b-\prbprocess_a
\end{equation}
may remain.
On the other hand, the fixed-cutoff current source $\mathsf{I}_{\txtrad,I,\kappa,\Lambda}(\prbprocess)$ has no zero-momentum component by (1).
When evaluating a covariance form containing the current source on a fixed-cutoff local interval, the zero-mode contribution is $0$ because the zero-momentum component of the current source vanishes, and only the non-zero-mode form
$$\begin{aligned}
\fun{\opform{q}_{\txtrad,\txtbec,\sminvtemperature}^{\txteuclid}}
{h,\mathsf{I}_{\txtrad,I,\kappa,\Lambda}(\prbprocess)}
&=
\fun{\opform{q}_{\txtrad,\txtnonzero,\sminvtemperature}^{\txteuclid}}
{h,\mathsf{I}_{\txtrad,I,\kappa,\Lambda}(\prbprocess)},
\\
\fun{\opform{q}_{\txtrad,\txtbec,\sminvtemperature}^{\txteuclid}}
{\mathsf{I}_{\txtrad,I,\kappa,\Lambda}(\prbprocess)}
&=
\fun{\opform{q}_{\txtrad,\txtnonzero,\sminvtemperature}^{\txteuclid}}
{\mathsf{I}_{\txtrad,I,\kappa,\Lambda}(\prbprocess)}
\end{aligned}$$
remains.
At the fixed-cutoff level, no new zero-mode local interaction appears.
\end{enumerate}
\end{prop}

\begin{proof}
(1): In the momentum representation in \eqref{expedition0012223}, the support of the coupling function $\lambda_{\txtrad,\mathrm{i},x,\kappa,\Lambda}$ is contained in $\kappa\leq\abs{k}\leq\Lambda$, so the desired identity $\faftr{\mathsf{I}}_{\txtrad,I,\kappa,\Lambda}(0)
= 0$ holds.
Since the zero-mode form acts only on the $k
= 0$ component, the equality for the sesquilinear forms also follows.

(2): On the full time circle, the particle path is closed, so $\int_{-\frac{\sminvtemperature}{2}}^{\frac{\sminvtemperature}{2}}
\opdmsr{\prbprocess_t}
=
\prbprocess_{\frac{\sminvtemperature}{2}}
-\prbprocess_{-\frac{\sminvtemperature}{2}}
= 0$ holds, and the two desired identities are obtained.
In particular, the current source appearing in the full-circle partition function and two-point function does not cross with the zero-mode form $\opform{q}
_{\txtrad,0,\sminvtemperature}$.

(3): On a local interval, the endpoint difference in \eqref{expedition0012333} may remain.
However, by \eqref{expedition0012211}, all covariance forms depending on the current source reduce to non-zero-mode forms at fixed cutoffs.
Since the zero-momentum component of the current source entering the zero-mode form is $0$, there is no need to add a separate zero-mode term on a fixed-cutoff local interval.
\end{proof}

\subsubsection{Cutoff Removal for the Current Kernel}\label{cutoff-removal-for-the-current-kernel}

The self term obtained by integrating out the radiation field in the current coupling of the Pauli--Fierz model appears at the fixed-cutoff level as the nonpositive Gaussian factor \begin{equation}\label{expedition0012334}
-\frac{\physcharge^2}{4}
\opform{q}_{\txtrad,\txtnonzero,\sminvtemperature}^{\txteuclid}
\rbk{\mathsf{I}_{\txtrad,I,\kappa,\Lambda}(\prbprocess)}.
\end{equation} The normalized correlation functions in this paper do not include additive renormalization constants of the \(E_{\txtnelson,\kappa,\Lambda}^{\txtrenormalization}\) type introduced in \eqref{expedition0012182} and \eqref{expedition0012185} for the Nelson model. The \(A^2\) term of the minimal coupling is included in the current-current self term in \eqref{expedition0012334} when matching the fixed-cutoff Hamiltonian with the Feynman--Kac--Itô representation. Below we handle the infrared and ultraviolet cutoffs in the non-zero-mode sesquilinear form in \eqref{expedition0012334} and in the external-field-current-source cross term.

In Proposition \ref{expedition0012335}, three-dimensionality is used to fix the notation of the current source after cutoff removal by taking the momentum space to be \(\fldreal^3\) and the current components to be \(\mathrm{i}=1,2,3\). Integrability of the cutoff limit is handled in the subsequent lemmas.

\begin{prop}[Current-Source Representation and Non-Zero-Mode Form after Cutoff Removal]\label{expedition0012335}
Let the spatial dimension be $d=3$.
For any closed interval $I
\subset S_{\sminvtemperature}$, represent the current source after cutoff removal by
\begin{equation}\label{expedition0012336}
\begin{aligned}
\mathsf{I}_{\txtrad,I}(\prbprocess)
=
\bigoplus_{\mathrm{i}=1}^{3}
\int_{I}
\fun{j_t}
{\lambda_{\txtrad,\mathrm{i},\prbprocess_t}}
\opdmsr{\prbprocess_{\mathrm{i},t}}
=
\bigoplus_{\mathrm{i}=1}^{3}
\int_{I}
\fun{j_{t}}
{\frac{1}{\sqrt{\omega(k)}}e_{\mathrm{i}}^{\mathrm{j}_{\txtrad}}(k)
\napiernum^{-\imunit k \prbprocess_t}}
\opdmsr{\prbprocess_{\mathrm{i},t}}
\end{aligned}
\end{equation}
This is not a definition of a test function containing an isolated zero mode; rather, \eqref{expedition0012336} is notation representing the cutoff limits of the non-zero-mode bilinear forms
\begin{equation}\label{expedition0012337}
\begin{aligned}
\fun{\opform{q}_{\txtrad,\txtnonzero,\sminvtemperature}^{\txteuclid}}
{\mathsf{I}_{\txtrad,I}(\prbprocess)}
&=
\lim_{\kappa\downarrow0,\ \Lambda\uparrow\infty}
\fun{\opform{q}_{\txtrad,\txtnonzero,\sminvtemperature}^{\txteuclid}}
{\mathsf{I}_{\txtrad,I,\kappa,\Lambda}(\prbprocess)},
\\
\fun{\opform{q}_{\txtrad,\txtnonzero,\sminvtemperature}^{\txteuclid}}
{j_t f,\mathsf{I}_{\txtrad,I}(\prbprocess)}
&=
\lim_{\kappa\downarrow0,\ \Lambda\uparrow\infty}
\fun{\opform{q}_{\txtrad,\txtnonzero,\sminvtemperature}^{\txteuclid}}
{j_t f,\mathsf{I}_{\txtrad,I,\kappa,\Lambda}(\prbprocess)}.
\end{aligned}
\end{equation}
\end{prop}

\begin{proof}
For the Nelson model, \eqref{expedition0012361} and \eqref{expedition0012362} of Proposition \ref{expedition0012360} define the source on the particle path after cutoff removal as a cutoff limit with the covariance form; in the present proposition, that role is replaced by the current source and the non-zero-mode bilinear form.
Removing the cutoff functions from the momentum representation
$\mathsf{I}_{\txtrad,I,\kappa,\Lambda}(\prbprocess)
=
\bigoplus_{\mathrm{i}=1}^{3}
\int_I
\fun{j_t}
{\lambda_{\txtrad,\mathrm{i},\prbprocess_t,\kappa,\Lambda}}
\opdmsr{\prbprocess_{\mathrm{i},t}}$
in \eqref{expedition0012223} gives the representation in \eqref{expedition0012336}.
However, as verified in Proposition \ref{expedition0012331}, the fixed-cutoff local current source has no zero-momentum component because of the infrared cutoff, and on the full time circle even the formal zero-momentum component vanishes by the loop condition.
Indeed, for the full time circle $I
= S_{\sminvtemperature}$, $\int_{-\frac{\sminvtemperature}{2}}^{\frac{\sminvtemperature}{2}}
\opdmsr{\prbprocess_t}
= 0$ holds.
In the full-circle partition function and correlation functions, no cross term between the current source and $\opform{q}
_{\txtrad,0,\sminvtemperature}$ appears.
For a subinterval $I= \closedinterval{a}{b}$, the endpoint difference $\prbprocess_b-\prbprocess_a$ may remain, but the fixed-cutoff current source has no zero-momentum component.
Thus the terms appearing in the generator representation are interpreted as non-zero-mode forms,
$$\begin{aligned}
\fun{\opform{q}_{\txtrad,\txtbec,\sminvtemperature}^{\txteuclid}}
{h,\mathsf{I}_{\txtrad,I,\kappa,\Lambda}(\prbprocess)}
&=
\fun{\opform{q}_{\txtrad,\txtnonzero,\sminvtemperature}^{\txteuclid}}
{h,\mathsf{I}_{\txtrad,I,\kappa,\Lambda}(\prbprocess)},
\\
\fun{\opform{q}_{\txtrad,\txtbec,\sminvtemperature}^{\txteuclid}}
{\mathsf{I}_{\txtrad,I,\kappa,\Lambda}(\prbprocess)}
&=
\fun{\opform{q}_{\txtrad,\txtnonzero,\sminvtemperature}^{\txteuclid}}
{\mathsf{I}_{\txtrad,I,\kappa,\Lambda}(\prbprocess)},
\end{aligned}$$
and as the cutoff limits in \eqref{expedition0012337}.
\end{proof}

The estimates needed in the main cutoff-removal proposition below are of three types. First, for the full-circle current source, the loop condition \(\int_{S_{\sminvtemperature}}\opdmsr{\prbprocess_t}=0\) is used to replace a constant phase by a difference factor, creating a gain of \(\abs{k}^{2}\) in the infrared current-current kernel. Second, the conditional characteristic function of the Brownian bridge is split into a mean part and a centered fluctuation part, and the centered fluctuation is controlled by Fubini's theorem for stochastic integrals and quadratic-variation estimates. Third, the cross term with a field test function is handled, by the definition of the physical test-function space, with the same dominating function as the transverse component in \eqref{eq:PF-physical-cross-kernel}. For the subinterval kernels appearing in the local semigroup, after decomposing the generator representation of Proposition \ref{expedition0012209} into the same three types of terms, the terms containing endpoint differences of subintervals are passed to the limit as cutoff bilinear forms of the non-zero-mode kernel.

In the next lemma, three-dimensionality is used when treating the particle path and momentum as \(\fldreal^3\)-valued and obtaining \(\abs{k}\) and \(\abs{k}^{2}\) gains at low momentum from the loop condition for the full-circle current. This gain is used later in the infrared integral of the current self term together with the three-dimensional volume element.

\begin{lem}[Infrared Compensation for the Full-Circle Current]\label{expedition0012304}
For the particle loop $\prbprocess$, let the reference point be $x_0
= \prbprocess_{-\frac{\sminvtemperature}{2}}
= \prbprocess_{\frac{\sminvtemperature}{2}}$.
Then, for any $k
\in \fldreal^3$,
$$\begin{aligned}
\int_{S_{\sminvtemperature}}
\napiernum^{-\imunit k\prbprocess_t}
\opdmsr{\prbprocess_t}
=
\napiernum^{-\imunit kx_0}
\int_{S_{\sminvtemperature}}
\rbk{\napiernum^{-\imunit k(\prbprocess_t-x_0)}-1}
\opdmsr{\prbprocess_t}
\end{aligned}$$
holds.
Moreover, for each part of the Brownian bridge decomposition $\opdmsr{\prbprocess_t}
=\opdmsr{M_t}+b(t,\prbprocess)\opdmsr{t}$, there exists a positive constant $C>0$ such that, at low momentum,
$$\begin{aligned}
\sqfun{\prbexp_{\msrprb_{x_0,x_0}^{\sminvtemperature,\txtbrownbridge}}}
{\abs{\int_{S_{\sminvtemperature}}
\rbk{\napiernum^{-\imunit k(\prbprocess_t-x_0)}-1}
\opdmsr{M_t}}^2}
&\leq C\abs{k}^2,
\\ %%%%%%%%%%%%%%%%
\sqfun{\prbexp_{\msrprb_{x_0,x_0}^{\sminvtemperature,\txtbrownbridge}}}
{\abs{\int_{S_{\sminvtemperature}}
\rbk{\napiernum^{-\imunit k(\prbprocess_t-x_0)}-1}
b(t,\prbprocess)\opdmsr{t}}}
&\leq C\abs{k}
\end{aligned}$$
holds.
\end{lem}

\begin{proof}
The first equality follows from the full-circle loop condition $\int_{S_{\sminvtemperature}}\opdmsr{\prbprocess_t}=0$.
The corresponding probabilistic computation for the Nelson model is the proof of \eqref{expedition0012186} and \eqref{expedition0012192} in Lemma \ref{expedition0012142}.
In Lemma \ref{expedition0012142}, the centered term in \eqref{expedition0012186} is split into the conditional mean part of the Brownian bridge and the centered fluctuation, and the centered part is controlled by the heat-kernel gradient estimate and Itô isometry in \eqref{expedition0012192}.
In the present lemma, we apply the same Brownian bridge heat-kernel calculation to the difference factor arising from the loop condition for the full-circle current.

The difference factor is bounded by $\abs{\napiernum^{-\imunit k(\prbprocess_t-x_0)}-1}
\leq \abs{k}\abs{\prbprocess_t-x_0}$.
The martingale part reduces to the estimate of the centered bridge fluctuation in \eqref{expedition0012186} of Lemma \ref{expedition0012142} and to the Itô isometry leading to \eqref{expedition0012192}.
Indeed, by the variance estimate obtained from the Brownian bridge heat-kernel representation in the proof of Lemma \ref{expedition0012142},
$$\begin{aligned}
&\sqfun{\prbexp_{\msrprb_{x_0,x_0}^{\sminvtemperature,\txtbrownbridge}}}
{\abs{\int_{S_{\sminvtemperature}}
\rbk{\napiernum^{-\imunit k(\prbprocess_t-x_0)}-1}
\opdmsr{M_t}}^2}
=
\sqfun{\prbexp_{\msrprb_{x_0,x_0}^{\sminvtemperature,\txtbrownbridge}}}
{\int_{S_{\sminvtemperature}}
\abs{\napiernum^{-\imunit k(\prbprocess_t-x_0)}-1}^2\opdmsr{t}}
\\
&\leq
\abs{k}^{2}
\int_{S_{\sminvtemperature}}
\sqfun{\prbexp_{\msrprb_{x_0,x_0}^{\sminvtemperature,\txtbrownbridge}}}
{\abs{\prbprocess_t-x_0}^{2}}
\opdmsr{t}
\leq
C\abs{k}^{2}
\end{aligned}$$
is obtained.

The drift part also reduces to the Brownian bridge heat-kernel representation of the conditional-mean part separated in \eqref{expedition0012186} of Lemma \ref{expedition0012142}.
Cutting the circle at $x_0$ and treating it as a Brownian bridge from $x_0$ to $x_0$ on $\closedinterval{0}{\sminvtemperature}$, the drift on the right endpoint side is $-(\prbprocess_t-x_0)/(\sminvtemperature-t)$, and the left endpoint side reduces to the same estimate by time reversal.
Combining the variance estimate obtained from the same heat-kernel representation with the Cauchy--Schwarz inequality gives
$$\begin{aligned}
&\sqfun{\prbexp_{\msrprb_{x_0,x_0}^{\sminvtemperature,\txtbrownbridge}}}
{\abs{\int_{S_{\sminvtemperature}}
\rbk{\napiernum^{-\imunit k(\prbprocess_t-x_0)}-1}
b(t,\prbprocess)\opdmsr{t}}}
\\
&\leq
\abs{k}
\int_{S_{\sminvtemperature}}
\sqfun{\prbexp_{\msrprb_{x_0,x_0}^{\sminvtemperature,\txtbrownbridge}}}
{\abs{\prbprocess_t-x_0}^{2}}^{\onehalf}
\sqfun{\prbexp_{\msrprb_{x_0,x_0}^{\sminvtemperature,\txtbrownbridge}}}
{\abs{b(t,\prbprocess)}^{2}}^{\onehalf}
\opdmsr{t}
\leq
C\abs{k}.
\end{aligned}$$
The finiteness of the last time integral follows by applying the Brownian bridge heat-kernel estimate used in the proof of Lemma \ref{expedition0012142} near the endpoints.
\end{proof}

In the next lemma, three-dimensionality is used to check the correspondence between the \(\omega(k)^{-2}\) infrared singularity appearing at low momentum in the transverse finite-temperature kernel and the volume element \(\abs{k}^{2}\opdmsr{\abs{k}}\) in \(d=3\). For the regular part this correspondence gives infrared domination, while for the singular part the same infrared singularity is canceled by the loop compensation of Lemma \ref{expedition0012304}.

\begin{lem}[Regular-Singular Decomposition of the Transverse Finite-Temperature Kernel]\label{expedition0012307}
Let the spatial dimension be $d=3$, and fix $0<\kappa<\Lambda<\infty$.
Define the transverse current-current kernel by
\begin{equation}\label{expedition0012308}
\begin{aligned}
W^{\perp}_{\sminvtemperature,\kappa,\Lambda,\mathrm{i}\mathrm{j}}(r,x)
=
\int_{\kappa\leq\abs{k}\leq\Lambda}
\frac{\napiernum^{-\abs{r}\omega(k)}
+\napiernum^{-(\sminvtemperature-\abs{r})\omega(k)}}
{1-\napiernum^{-\sminvtemperature\omega(k)}}
\frac{\delta^\perp_{\mathrm{i}\mathrm{j}}(k)}
{\omega(k)}
\napiernum^{-\imunit kx}
\opdmsr{k}
\end{aligned}
\end{equation}
and define its singular and regular parts by
\begin{equation}\label{expedition0012309}
\begin{aligned}
W^{\perp}_{0,\kappa,\Lambda,\mathrm{i}\mathrm{j}}(r,x)
&=
\int_{\kappa\leq\abs{k}\leq\Lambda}
\napiernum^{-\abs{r}\omega(k)}
\frac{\delta^\perp_{\mathrm{i}\mathrm{j}}(k)}
{\omega(k)}
\napiernum^{-\imunit kx}
\opdmsr{k},
\\ %%%%%%%%%%%%%%%%
W^{\perp}_{\sminvtemperature,\txtregular,\kappa,\Lambda,\mathrm{i}\mathrm{j}}(r,x)
&=
\int_{\kappa\leq\abs{k}\leq\Lambda}
\frac{
\napiernum^{-(\sminvtemperature-\abs{r})\omega(k)}
+\napiernum^{-\rbk{\sminvtemperature + \abs{r}}\omega(k)}}
{1-\napiernum^{-\sminvtemperature\omega(k)}}
\frac{\delta^\perp_{\mathrm{i}\mathrm{j}}(k)}
{\omega(k)}
\napiernum^{-\imunit kx}
\opdmsr{k}
\end{aligned}
\end{equation}
so that $W^{\perp}_{\sminvtemperature,\kappa,\Lambda,\mathrm{i}\mathrm{j}}
=
W^{\perp}_{0,\kappa,\Lambda,\mathrm{i}\mathrm{j}}
+W^{\perp}_{\sminvtemperature,\txtregular,\kappa,\Lambda,\mathrm{i}\mathrm{j}}$.
This decomposition holds, and for the regular part the cutoff removal limit and $r$-differentiation can be interchanged locally uniformly on compact sets of $r$.
\end{lem}

\begin{proof}
The finite-temperature factor has the same decomposition as \eqref{expedition0012184} for the Nelson model.
Indeed, for any $\omega>0$ and $0\leq\abs{r}\leq\sminvtemperature$,
$$\begin{aligned}
\frac{\napiernum^{-\abs{r}\omega}
+\napiernum^{-(\sminvtemperature-\abs{r})\omega}}
{1-\napiernum^{-\sminvtemperature\omega}}
=
\napiernum^{-\abs{r}\omega}
+
\frac{\napiernum^{-\sminvtemperature\omega}
\napiernum^{-\abs{r}\omega}
+\napiernum^{-(\sminvtemperature-\abs{r})\omega}}
{1-\napiernum^{-\sminvtemperature\omega}}.
\end{aligned}$$
The first term is the singular part in \eqref{expedition0012309}, and the second term is the regular part.

The interchange of cutoff removal and $r$-differentiation for the regular part follows by applying Proposition \ref{expedition0012143} componentwise.
Since the transverse projection is an orthogonal projection, $\abs{\delta^\perp_{\mathrm{i}\mathrm{j}}(k)}
\leq 1$ for each component.
Apart from this inequality, the momentum factor of the regular part is the same as the regular part $W_{\sminvtemperature,\kappa,\Lambda}$ in \eqref{expedition0012184} for the Nelson model and the limiting kernel in \eqref{expedition0012201}.
In spatial dimension $d
= 3$, the infrared integrability condition $\int_{\omega(k)\leq1}
\omega(k)^{-2}
\opdmsr{k}
< \infty$ in Proposition \ref{expedition0012143} holds.
By the dominated-convergence estimate of Proposition \ref{expedition0012143}, $W^{\perp}
_{\sminvtemperature,\txtregular,\kappa,\Lambda,\mathrm{i}\mathrm{j}}$ converges locally uniformly to the cutoff removal limit, and the interchange with $r$-differentiation also holds locally uniformly.

The singular part is the component obtained by multiplying the zero-temperature-type singular kernel $W_{0,\kappa,\Lambda}$ in \eqref{expedition0012184} for the Nelson model by the transverse projection.
The Nelson model estimates corresponding to the short-time singularity on the high-momentum side are \eqref{expedition0012186}, \eqref{expedition0012190}, and \eqref{expedition0012192} in Lemma \ref{expedition0012142}.
In Lemma \ref{expedition0012142}, the centered term in \eqref{expedition0012186} is split into the conditional mean part and the centered bridge fluctuation.
The conditional mean part is dominated by the comparison kernel $\rbk{\omega(k)+\onehalf\omega(k)^2}^{-2}$ in \eqref{expedition0012190}, and the centered bridge fluctuation is dominated by the square root of the same comparison kernel by the heat-kernel gradient estimate and Itô isometry in \eqref{expedition0012192}.
Since the transverse projection satisfies $\abs{\delta^\perp_{\mathrm{i}\mathrm{j}}(k)}\leq1$, the calculations in \eqref{expedition0012186}, \eqref{expedition0012190}, and \eqref{expedition0012192} of Lemma \ref{expedition0012142} can be applied componentwise.

On the low-momentum side when the singular part is inserted into the closed full-circle current, we use not only the kernel decomposition itself but also the difference estimate of Lemma \ref{expedition0012304}.
The loop compensation on the low-momentum side is precisely the calculation in Lemma \ref{expedition0012304}.
When Lemma \ref{expedition0012304} converts the constant phase of the full-circle current into a difference, each current factor is accompanied by a difference factor of the form $\napiernum^{-\imunit k(\prbprocess_t-x_0)}-1$.
Applying the difference-factor estimate of Lemma \ref{expedition0012304} to the two current factors gives a gain of $\abs{k}^{2}$ in the current-current form.
The $\omega(k)^{-2}$ infrared singularity appearing on the low-momentum side of the full finite-temperature kernel is canceled by this gain, and since $\omega(k)=\abs{k}$ and $d=3$, the remaining low-momentum integral is finite of the form $\int_0^1 r^2\opdmsr{r}$.
Thus the treatment of the short-time singularity inherited from \eqref{expedition0012186}, \eqref{expedition0012190}, and \eqref{expedition0012192} in Lemma \ref{expedition0012142} for the Nelson model, and the low-momentum compensation specific to the Pauli--Fierz current, are verified separately.
\end{proof}

\begin{rem}[Type of Current Integral]
The stochastic integral with respect to $\opdmsr{\prbprocess_{\mathrm{i},t}}$ used in this paper is the sum of the Itô integral part and the drift time-integral part after the Brownian bridge is decomposed as a semimartingale.
In accordance with this type of stochastic integral, the double stochastic integrals used in the cutoff removal estimates are first constructed as Itô integrals at fixed cutoffs and then treated as the Feynman--Kac--Itô type kernel representation obtained after integrating out the field as a Gaussian.
No separate Stratonovich-type correction term is introduced.
\end{rem}

The conversion of the constant phase into a difference and the cancellation of the infrared singularity at low momentum in Lemma \ref{expedition0012304} are handled not by the decomposition of the transverse finite-temperature kernel alone, but as the low-momentum tail estimate for the current self term in the next estimate. In the next lemma, three-dimensionality is used on both the infrared and ultraviolet sides of the current self term. On the infrared side, the \(\abs{k}^{2}\) gain of Lemma \ref{expedition0012304} cancels the \(\omega(k)^{-2}\) singularity of the finite-temperature kernel and gives integrability through the low-momentum volume element in \(d=3\). On the ultraviolet side, \(\rbk{\omega(k) + \onehalf \omega(k)^2}^{-2}\), which comes from the heat-kernel estimate and Itô isometry, is integrable in \(d
= 3\). This is used to show that the tails in the low-momentum ball and high-momentum shell vanish. For a measurable set \(A
\subset \fldreal^3\), define the \(A\)-part of the current self-interaction by \begin{equation}\label{expedition0012238}
\begin{aligned}
\mathsf{U}_{A}(\prbprocess)
&=
-\frac{\physcharge^2}{4}
\int_{S_{\sminvtemperature}}
\opdmsr{t}
\int_{S_{\sminvtemperature}}
\opdmsr{s}
\int_{A}
K_{\sminvtemperature}(t-s,k)
\frac{\napiernum^{-\imunit k(\prbprocess_t-\prbprocess_s)}}{\omega(k)}
\bkt{\opdmsr{\prbprocess_t}}
{\delta^\perp(k)\opdmsr{\prbprocess_s}}_{\fldreal^3}
\opdmsr{k},
\\ %%%%%%%%%%%%%%%%
K_{\sminvtemperature}(r,k)
&=
\frac{\napiernum^{-\abs{r}\omega(k)}
+\napiernum^{-(\sminvtemperature-\abs{r})\omega(k)}}
{1-\napiernum^{-\sminvtemperature\omega(k)}}.
\end{aligned}
\end{equation} We call the contribution obtained by changing the measurable set \(A\) in \eqref{expedition0012238} from a cutoff region to a tail region a cutoff difference. In particular, for two cutoffs \(0
< \kappa'
< \kappa
< \Lambda
< \Lambda'
< \infty\), this refers to objects such as \[\begin{aligned}
\mathsf{U}_{\set{k}{\kappa'\leq\abs{k}\leq\Lambda'}}
-
\mathsf{U}_{\set{k}{\kappa\leq\abs{k}\leq\Lambda}}
=
\mathsf{U}_{\set{k}{\kappa'\leq\abs{k}<\kappa}}
+
\mathsf{U}_{\set{k}{\Lambda<\abs{k}\leq\Lambda'}}.
\end{aligned}\] When comparing with the cutoff removal limit, it is enough to estimate the infrared tail \(A=\set{k}{\abs{k}<\kappa}\) and the ultraviolet tail \(A=\set{k}{\abs{k}>\Lambda}\). The next lemma dominates this tail contribution \(\mathsf{U}_A\) by a cutoff-independent integrable function and shows that the current self term is a Cauchy sequence in \(\lp^1\) as the cutoff region is enlarged.

\begin{lem}[Cutoff-Difference Domination for the Transverse Current Kernel]\label{expedition0012305}
For a measurable set $A
\subset\fldreal^3$, consider $\mathsf{U}_{A}$ in \eqref{expedition0012238}.
Then there exists a constant $C>0$ such that
$$\begin{aligned}
\sqfun{\prbexp_{\msrprb_{\txtparticle,\sminvtemperature}}}
{\abs{\mathsf{U}_A}}
\leq
C\rbk{
\int_A
\frac{\opdmsr{k}}
{\rbk{\omega(k)+\onehalf\omega(k)^2}^{2}}
+
\rbk{\int_A
\frac{\opdmsr{k}}
{\rbk{\omega(k)+\onehalf\omega(k)^2}^{2}}}^{\onehalf}}
\end{aligned}$$
holds.
In particular, if $A
= \set{k}{\abs{k}<\kappa}$ or $A
= \set{k}{\abs{k}>\Lambda}$, then the right-hand side converges to $0$ as $\kappa\downarrow0$ or $\Lambda\uparrow\infty$, respectively.
\end{lem}

\begin{proof}
The corresponding cutoff-difference estimates for the Nelson model are \eqref{expedition0012186}, \eqref{expedition0012190}, and \eqref{expedition0012192} of Lemma \ref{expedition0012142} and \eqref{expedition0012201} of Proposition \ref{expedition0012143}.
For the regular part, use \eqref{expedition0012201} of Proposition \ref{expedition0012143} and the dominated-convergence estimate.
For the singular part, use the centered term in \eqref{expedition0012186} of Lemma \ref{expedition0012142}, the cutoff-difference domination in \eqref{expedition0012190}, and the heat-kernel gradient estimate in \eqref{expedition0012192}.
The additional low-momentum compensation on the Pauli--Fierz side is the difference conversion from the loop condition given in Lemma \ref{expedition0012304}.

Separate the low-momentum side and the high-momentum side.
The low-momentum side is the loop compensation specific to the Pauli--Fierz current.
By Lemma \ref{expedition0012304}, each factor of the full-circle current can be written with the difference factor $\napiernum^{-\imunit k(\prbprocess_t-x_0)}-1$ relative to the reference point $x_0$.
Applying the martingale-part and drift-part estimates of Lemma \ref{expedition0012304} to the two current factors, together with the Cauchy--Schwarz inequality, the two current factors on the low-momentum side have a gain of $\abs{k}^{2}$.
On the other hand, when $\omega(k)\leq1$, there is a constant $c_1>0$ such that $\frac{K_{\sminvtemperature}(t-s,k)}{\omega(k)}
\leq
\frac{c_1}{\omega(k)^2}$.
Since the dispersion relation is $\omega(k)=\abs{k}$, this $\omega(k)^{-2}$ infrared singularity is canceled by the $\abs{k}^{2}$ coming from the two current factors.
Thus the remaining momentum integral over the low-momentum ball is finite by the three-dimensional volume element.

The high-momentum side is handled according to the singular-regular decomposition of Lemma \ref{expedition0012307}.
For the regular part, apply the domination and regularity of the regular part in Proposition \ref{expedition0012143} componentwise.
Apart from boundedness of the transverse projection, the momentum factor of the regular part is the same as the regular part in \eqref{expedition0012184} for the Nelson model and the limiting kernel in \eqref{expedition0012201}.
Therefore the cutoff difference is dominated by the dominated-convergence estimate of Proposition \ref{expedition0012143}.

The singular part reduces to the centered-term estimates of Lemma \ref{expedition0012142} for the Nelson model, namely the estimates in \eqref{expedition0012186}, \eqref{expedition0012190}, and \eqref{expedition0012192}.
The reduction of the time circle to finitely many intervals of length at most $\frac{\sminvtemperature}{2}$ and then to integrals of the form $0\leq s\leq r\leq t\leq\frac{\sminvtemperature}{2}$ is the same as in the proof of $\mathsf{C}_{A,t}$ defined in \eqref{expedition0012186} of Lemma \ref{expedition0012142}.
Since the transverse projection is an orthogonal projection, $\abs{\bkt{u}{\delta^\perp(k)v}_{\fldreal^3}}\leq \abs{u}\abs{v}$.
Apart from this inequality, the estimates for the Brownian bridge increments in the singular part are the same as in the proofs of \eqref{expedition0012186} and \eqref{expedition0012192} in Lemma \ref{expedition0012142}.

Concretely, split the conditional characteristic function of the Brownian bridge into the conditional mean part and the centered bridge fluctuation.
For the conditional mean part, apply the argument in the proof of Lemma \ref{expedition0012142} that splits the integrand in \eqref{expedition0012186} into $\mathsf{M}_{r,s}(k)$ and $\mathsf{D}_{r,s}(k)$.
By the heat-kernel representation and time-integral estimate in Lemma \ref{expedition0012142} for the deterministic bridge correction $\mathsf{D}_{r,s}(k)$, the time integral is estimated by $\frac{c_2}
{\omega(k)+\onehalf\omega(k)^2}$.
For the centered bridge fluctuation, apply the martingale representation in the proof of Lemma \ref{expedition0012142} and the heat-kernel gradient estimate in \eqref{expedition0012192}.
Changing the order of integration by Fubini's theorem for stochastic integrals and estimating the quadratic variation by the Itô isometry, the $\lp^{1}$ norm of the centered part is bounded, as in the estimate of the centered part in Lemma \ref{expedition0012142}, by
$$\begin{aligned}
c_3
\rbk{\int_A
\frac{\opdmsr{k}}
{\rbk{\omega(k)+\onehalf\omega(k)^2}^{2}}}^{\onehalf}.
\end{aligned}$$
The conditional mean part is bounded by
$$\begin{aligned}
c_4
\int_A
\frac{\opdmsr{k}}
{\rbk{\omega(k)+\onehalf\omega(k)^2}^{2}}.
\end{aligned}$$
These two estimates give the same comparison kernel $\rbk{\omega(k)
+\onehalf\omega(k)^2}^{-2}$ as the cutoff-difference domination in \eqref{expedition0012190}.
Here $c_2,c_3,c_4$ depend only on $\sminvtemperature$ and the potential estimate for the particle loop measure, not on the measurable set $A$.

The passage to the particle loop measure is the same as the final part of the proof that transfers the Brownian bridge estimates \eqref{expedition0012186}--\eqref{expedition0012192} of Lemma \ref{expedition0012142} to the particle loop measure with confining potential.
Namely, combine the estimates under the Brownian bridge measure with Hölder's inequality and heat-kernel estimates for the potential weight
$\napiernum^{-\int_0^{\sminvtemperature}V(\prbprocess_u)\opdmsr{u}}$
and only change the constant.
Combining the loop compensation estimate on the low-momentum side with the centered-term estimates \eqref{expedition0012186}, \eqref{expedition0012190}, and \eqref{expedition0012192} of Lemma \ref{expedition0012142} on the high-momentum side gives the stated estimate.
Finally, $\rbk{\omega(k)+\onehalf\omega(k)^2}^{-2}$ is integrable in $d=3$ on both the low-momentum and high-momentum sides, so the right-hand side for $A=\set{k}{\abs{k}<\kappa}$ and $A=\set{k}{\abs{k}>\Lambda}$ converges to $0$, respectively.
\end{proof}

In the next lemma, three-dimensionality is used as the domination condition for the momentum integral of the cross term between the external field and the current source. Concretely, the physical test-function space in \eqref{eq:PF-physical-test-space} requires a dominating function for the transverse component that is integrable in the \(d=3\) momentum integral.

\begin{lem}[Cutoff-Difference Domination for the Field Test-Function Cross Term]\label{expedition0012306}
For any $f
\in \sphilb{D}_{\txtrad,\txtphys,\sminvtemperature}$, the cross term
$$\fun{\opform{q}_{\txtrad,\txtnonzero,\sminvtemperature}^{\txteuclid}}
{j_0 f,\mathsf{I}_{\txtrad,S_{\sminvtemperature},\kappa,\Lambda}(\prbprocess)}$$
converges in $\lp^1(\Omega_{\txtparticle,\sminvtemperature})$ as $\kappa\downarrow0,\Lambda\uparrow\infty$.
Moreover, the cross term for any closed interval $I\subset S_{\sminvtemperature}$ appearing in the local semigroup kernel also converges as the limit of the non-zero-mode form under the same dominating function.
\end{lem}

\begin{proof}
In Proposition \ref{expedition0012360} for the Nelson model, \eqref{expedition0012361} defines the cross term with an external field $f$ as a bilinear form with the source after cutoff removal, and \eqref{expedition0012362} and \eqref{expedition0012364} take dominated convergence using the domination condition of the physical test-function space.
In the present lemma, we replace the scalar source by the transverse current source and carry out the same dominated convergence using the transverse domination condition in \eqref{eq:PF-physical-test-space}.
By the domination condition in \eqref{eq:PF-physical-test-space}, the product of the finite-temperature kernel, $\delta^\perp(k)\faftr{f}(k)/\sqrt{\omega(k)}$, and the heat-kernel weight from the Brownian bridge is bounded by an integrable momentum function.
Here the first term in the parentheses in \eqref{eq:PF-physical-test-space} directly bounds the finite-temperature kernel, and the second term $\rbk{\omega(k)+\onehalf\omega(k)^2}^{-1}$ bounds the part using the Brownian bridge heat-kernel estimate.
Since the cutoff functions converge to $1$ for each $k\neq0$, the dominated convergence theorem gives $\lp^1$ convergence of the full-circle cross term.
For a subinterval $I$, the full-circle loop condition is not used; the same dominating function is applied to the mixed term of the non-zero-mode bilinear form appearing in the generator representation of Proposition \ref{expedition0012209}.
\end{proof}

We summarize the role of the preceding lemmas.

\begin{itemize}

\item
  Lemma \ref{expedition0012307} separates the finite-temperature kernel into a zero-temperature-type part with short-time singularity and a regular part whose cutoff removal can be handled by dominated convergence.
\item
  Lemma \ref{expedition0012304} converts the infrared constant phase into a difference for the full-circle current, and Lemma \ref{expedition0012305} combines this difference conversion with heat-kernel estimates to give an \(\lp^1\) cutoff-difference estimate for the current self term.
\item
  Lemma \ref{expedition0012306} dominates the cross term with an external field by the definition of the physical test-function space.
\item
  On local intervals, we do not directly apply the loop compensation of Lemma \ref{expedition0012304}; instead, we return to the generator representation of Proposition \ref{expedition0012209} and treat the local source as a cutoff limit of the non-zero-mode bilinear form.
\item
  The full-circle partition function, two-point functions, and local semigroup kernels are estimated by the same momentum dominating functions.
\end{itemize}

Regarding the zero mode, note the following point: on the full time circle, the zero-momentum component of the current source vanishes by the loop condition, so the BEC zero mode and the current source do not cross directly. On local intervals, endpoint differences remain, but in the fixed-cutoff generator representation the current source is treated as a cutoff limit of the non-zero-mode bilinear form, so no local zero-mode interaction is added.

In Proposition \ref{expedition0012165}, three-dimensionality is used to apply the current self-term cutoff-difference estimate of Lemma \ref{expedition0012305} and the external-field cross-term domination of Lemma \ref{expedition0012306} simultaneously. That is, the \(\lp^1\) limits and uniform integrability of the full-circle effective action, the one-point correction term, and their product follow from infrared compensation, ultraviolet tail estimates, and the domination condition of the physical test-function space in three dimensions.

\begin{prop}[Cutoff Removal for the Pauli--Fierz Current Kernel]\label{expedition0012165}
Let the spatial dimension be $d=3$, and consider the current kernel of the cutoff spinless Pauli--Fierz model.
For a particle loop $\prbprocess$ on the time circle $S_{\sminvtemperature}$ and any $f
\in \sphilb{D}_{\txtrad,\txtphys,\sminvtemperature}$, set
\begin{equation}\label{expedition0012225}
\begin{aligned}
\mathsf{U}_{\txtpaulifierz,\sminvtemperature,\kappa,\Lambda}(\prbprocess)
&=
-\frac{\physcharge^2}{4}
\fun{\opform{q}_{\txtrad,\txtnonzero,\sminvtemperature}^{\txteuclid}}
{\mathsf{I}_{\txtrad,S_{\sminvtemperature},\kappa,\Lambda}(\prbprocess)},
\\ %%%%%%%%%%%%%%%%
\mathsf{Y}_{\txtpaulifierz,\sminvtemperature,\kappa,\Lambda;f}(\prbprocess)
&=
\frac{\imunit\physcharge}{2}
\fun{\opform{q}_{\txtrad,\txtnonzero,\sminvtemperature}^{\txteuclid}}
{j_0 f,\mathsf{I}_{\txtrad,S_{\sminvtemperature},\kappa,\Lambda}(\prbprocess)}.
\end{aligned}
\end{equation}
\begin{enumerate}
\item
For the particle loop $\prbprocess$,
\begin{equation}\label{expedition0012227}
\begin{aligned}
\mathsf{U}_{\txtpaulifierz,\sminvtemperature}(\prbprocess)
&=
-\frac{\physcharge^2}{4}
\int_{S_{\sminvtemperature}}
\opdmsr{t}
\int_{S_{\sminvtemperature}}
\opdmsr{s}
\int_{\fldreal^d}
\frac{\napiernum^{-\abs{t-s}\omega(k)}
+\napiernum^{-(\sminvtemperature-\abs{t-s})\omega(k)}}
{1-\napiernum^{-\sminvtemperature\omega(k)}}
\frac{\napiernum^{-\imunit k(\prbprocess_t-\prbprocess_s)}}{\omega(k)}
\\ %%%%%%%%%%%%%%%%
&\quad\times
\bkt{\opdmsr{\prbprocess_{t}}}
{\delta^\perp(k)\opdmsr{\prbprocess(s)}}_{\fldreal^d}
\opdmsr{k}
\end{aligned}
\end{equation}
is the limit of the cutoff effective action:
$$\mathsf{U}_{\txtpaulifierz,\sminvtemperature}
=
\fun{\lp^{1}}{\Omega_{\txtparticle,\sminvtemperature},
\msrprb_{\txtparticle,\sminvtemperature}}\text{-}
\lim_{\kappa\downarrow0,\Lambda\uparrow\infty}
\mathsf{U}_{\txtpaulifierz,\sminvtemperature,\kappa,\Lambda},$$
and the exponential family $\fnexp{\mathsf{U}_{\txtpaulifierz,\sminvtemperature,\kappa,\Lambda}}$ is uniformly integrable.

\item
For any $f
\in \sphilb{D}_{\txtrad,\txtphys,\sminvtemperature}$,
\begin{equation}\label{expedition0012228}
\begin{aligned}
\mathsf{Y}_{\txtpaulifierz,\sminvtemperature;f}(\prbprocess)
=
\frac{\imunit\physcharge}{2}
\int_{S_{\sminvtemperature}}
\opdmsr{s}
\int_{\fldreal^d}
\frac{\napiernum^{-\abs{s}\omega(k)}
+\napiernum^{-(\sminvtemperature-\abs{s})\omega(k)}}
{1-\napiernum^{-\sminvtemperature\omega(k)}}
\frac{\napiernum^{-\imunit k\prbprocess_s}}{\sqrt{\omega(k)}}
\bkt{\faftr{f}(k)}
{\delta^\perp(k)\opdmsr{\prbprocess(s)}}_{\fldcmp^d}
\opdmsr{k}
\end{aligned}
\end{equation}
is the limit of the cutoff one-point correction term:
$$\mathsf{Y}_{\txtpaulifierz,\sminvtemperature;f}
=
\fun{\lp^{1}}{\Omega_{\txtparticle,\sminvtemperature},
\msrprb_{\txtparticle,\sminvtemperature}}\text{-}
\lim_{\kappa\downarrow0,\Lambda\uparrow\infty}
\mathsf{Y}_{\txtpaulifierz,\sminvtemperature,\kappa,\Lambda;f}.$$

\item
The product of the above two terms is also an $\lp^{1}$ function, and the products converge in the $\lp^{1}$ topology.
That is, the product of $\mathsf{U}_{\txtpaulifierz,\sminvtemperature}$ and $\mathsf{Y}_{\txtpaulifierz,\sminvtemperature;f}$,
\begin{equation}\label{expedition0012229}
\mathsf{Y}_{\txtpaulifierz,\sminvtemperature;f}
\fnexp{\mathsf{U}_{\txtpaulifierz,\sminvtemperature}}
=
\rbk{\mathsf{Y}_{\txtpaulifierz,\sminvtemperature;f}}
\fnexp{\mathsf{U}_{\txtpaulifierz,\sminvtemperature}}
\end{equation}
is obtained as the limit of the cutoff weighted one-point correction term:
$$\mathsf{Y}_{\txtpaulifierz,\sminvtemperature;f}
\fnexp{\mathsf{U}_{\txtpaulifierz,\sminvtemperature}}
=
\fun{\lp^{1}}{\Omega_{\txtparticle,\sminvtemperature},
\msrprb_{\txtparticle,\sminvtemperature}}\text{-}
\lim_{\kappa\downarrow0,\Lambda\uparrow\infty}
\mathsf{Y}_{\txtpaulifierz,\sminvtemperature,\kappa,\Lambda;f}
\fnexp{\mathsf{U}_{\txtpaulifierz,\sminvtemperature,\kappa,\Lambda}}.$$
In particular, the cutoff family on the left-hand side is uniformly integrable.
\end{enumerate}
\end{prop}

\begin{proof}
The three assertions of this proposition correspond to the cutoff-removed action in \eqref{expedition0012182} for the Nelson model, the one-point correction in \eqref{expedition0012351}, and the convergence of the external-field cross terms in \eqref{expedition0012361}--\eqref{expedition0012364} of Proposition \ref{expedition0012360}.
The regular part of the self term corresponds to \eqref{expedition0012184} and \eqref{expedition0012201} of Proposition \ref{expedition0012143}, the singular part corresponds to \eqref{expedition0012186}, \eqref{expedition0012190}, and \eqref{expedition0012192} of Lemma \ref{expedition0012142}, and the external-field cross term corresponds to \eqref{expedition0012361}, \eqref{expedition0012362}, and \eqref{expedition0012364} of Proposition \ref{expedition0012360}.
On the Pauli--Fierz side, the loop compensation of Lemma \ref{expedition0012304} and the cutoff-difference dominations of Lemmas \ref{expedition0012305} and \ref{expedition0012306} are added.
In the proof below, cutoff removal of the self term is assigned to Lemma \ref{expedition0012305}, cutoff removal of the external-field cross term to Lemma \ref{expedition0012306}, and convergence of the exponentially weighted one-point correction term to the limit passage for the one-point characteristic function carried out in \eqref{expedition0012351} and \eqref{expedition0012200} of Proposition \ref{expedition0012158} for the Nelson model.
The low-momentum side of Lemma \ref{expedition0012305} is the loop compensation of Lemma \ref{expedition0012304}, and the high-momentum side is the componentwise transfer of the estimates \eqref{expedition0012186}, \eqref{expedition0012190}, and \eqref{expedition0012192} of Lemma \ref{expedition0012142} and \eqref{expedition0012201} of Proposition \ref{expedition0012143}.

(1): The non-zero-mode form with the full-circle current source can be written with the transverse projection as the inner product:
\begin{equation}\label{expedition0012230}
\begin{aligned}
\mathsf{U}_{\txtpaulifierz,\sminvtemperature,\kappa,\Lambda}(\prbprocess)
&=
-\frac{\physcharge^2}{4}
\int_{S_{\sminvtemperature}}
\opdmsr{t}
\int_{S_{\sminvtemperature}}
\opdmsr{s}
\int_{\kappa \leq \abs{k}\leq \Lambda}
\frac{\napiernum^{-\abs{t-s}\omega(k)}
+\napiernum^{-(\sminvtemperature-\abs{t-s})\omega(k)}}
{1-\napiernum^{-\sminvtemperature\omega(k)}}
\frac{\napiernum^{-\imunit k(\prbprocess_t-\prbprocess_s)}}{\omega(k)}
\\ %%%%%%%%%%%%%%%%
&\quad\times
\bkt{\opdmsr{\prbprocess_{t}}}
{\delta^\perp(k)\opdmsr{\prbprocess(s)}}_{\fldreal^d}
\opdmsr{k}
\end{aligned}
\end{equation}
This is the representation obtained by substituting the current source in \eqref{expedition0012223} into \eqref{expedition0012225}.
Comparing two cutoffs $0<\kappa'<\kappa<\Lambda<\Lambda'<\infty$, the difference is the tail contribution in \eqref{expedition0012238}:
$$\begin{aligned}
\mathsf{U}_{\set{k}{\kappa'\leq\abs{k}<\kappa}}(\prbprocess)
+\mathsf{U}_{\set{k}{\Lambda<\abs{k}\leq\Lambda'}}(\prbprocess).
\end{aligned}$$
The infrared tail is handled by the loop compensation of Lemma \ref{expedition0012304}; Lemma \ref{expedition0012304} gives the calculation that converts the constant phase of the closed full-circle current into a difference and obtains a gain of $\abs{k}^{2}$ from the two current factors.
This gain cancels the $\omega(k)^{-2}$ low-momentum singularity of the finite-temperature kernel.
The estimate of the entire cutoff difference, including the high-momentum tail, is the statement of Lemma \ref{expedition0012305}; Lemma \ref{expedition0012305} is the estimate obtained by transferring the heat-kernel estimates corresponding to \eqref{expedition0012186}, \eqref{expedition0012190}, and \eqref{expedition0012192} of Lemma \ref{expedition0012142} for the Nelson model and \eqref{expedition0012201} of Proposition \ref{expedition0012143} to the transverse components.
Therefore, applying Lemma \ref{expedition0012305} to the two tail sets above shows that the cutoff double stochastic integrals form a Cauchy sequence in $\lp^{1}$.
Moreover, since the cutoff functions converge to $1$ for each $k\neq0$, the limit is the integral representation in \eqref{expedition0012227}.

By nonnegativity of the transverse covariance form, the exponential weight satisfies $\mathsf{U}_{\txtpaulifierz,\sminvtemperature,\kappa,\Lambda}(\prbprocess)
\leq 0$, and hence for all cutoffs $0
\leq \fnexp{\mathsf{U}_{\txtpaulifierz,\sminvtemperature,\kappa,\Lambda}}
\leq 1$ holds.
This implies uniform integrability.

(2): We discuss the one-point correction.
This part is the same dominated-convergence argument as
equations \eqref{expedition0012361}, \eqref{expedition0012362}, and \eqref{expedition0012364}
in Proposition \ref{expedition0012360} for the Nelson model,
and on the Pauli--Fierz side Lemma \ref{expedition0012306} gives its transverse-current version.
For arbitrary $f
\in \sphilb{D}_{\txtrad,\txtphys,\sminvtemperature}$,
\begin{equation}\label{expedition0012231}
\fun{\opform{q}_{\txtrad,\txtnonzero,\sminvtemperature}^{\txteuclid}}
{j_0 f,\mathsf{I}_{\txtrad,S_{\sminvtemperature},\kappa,\Lambda}(\prbprocess)},
\end{equation}
the integrand is given by the product of the finite-temperature kernel
$\delta^\perp(k)\faftr{f}(k)$ and $\faftr{\varphi}_{\kappa,\Lambda}(k)$.
The quantity $\mathsf{Y}
_{\txtpaulifierz,\sminvtemperature,\kappa,\Lambda;f}(\prbprocess)$ in \eqref{expedition0012225}
is the quantity in \eqref{expedition0012231} multiplied by the prefactor
$\frac{\imunit \physcharge}{2}$.
By the definition of the physical test-function space, this is dominated by an integrable function independent of the cutoffs,
and $\faftr{\varphi_{\kappa,\Lambda}}(k)
\to 1$ holds for each $k\neq 0$.
The dominated convergence theorem then gives \eqref{expedition0012228}.

(3): We prove convergence of the products.
Here we split the expression in the same way as when the linear correction \eqref{expedition0012351}
was removed in Proposition \ref{expedition0012158} for the Nelson model and the physical field
\eqref{expedition0012200} was obtained.
We use the $\lp^{1}$ convergence obtained in part (1) of this proposition and
$\mathsf{U}_{\txtpaulifierz,\sminvtemperature,\kappa,\Lambda}
\leq 0$.
Since the exponential function is $1$-Lipschitz on $(-\infty,0]$, we obtain
$$\abs{\fnexp{\mathsf{U}_{\txtpaulifierz,\sminvtemperature,\kappa,\Lambda}}
-\fnexp{\mathsf{U}_{\txtpaulifierz,\sminvtemperature}}}
\leq
\abs{\mathsf{U}_{\txtpaulifierz,\sminvtemperature,\kappa,\Lambda}
-\mathsf{U}_{\txtpaulifierz,\sminvtemperature}}$$
and hence, in $\fun{\lp^{1}}{\Omega_{\txtparticle,\sminvtemperature},
\msrprb_{\txtparticle,\sminvtemperature}}$,
\begin{equation}\label{expedition0012246}
\fnexp{\mathsf{U}_{\txtpaulifierz,\sminvtemperature,\kappa,\Lambda}}
\to
\fnexp{\mathsf{U}_{\txtpaulifierz,\sminvtemperature}}
\end{equation}
holds.
For all cutoffs, $0
\leq \fnexp{\mathsf{U}_{\txtpaulifierz,\sminvtemperature,\kappa,\Lambda}}
\leq 1$ holds,
so also in the limit $0
\leq \fnexp{\mathsf{U}_{\txtpaulifierz,\sminvtemperature}}
\leq 1$ holds.

Next, directly estimating the products gives
\begin{equation}\label{expedition0012247}
\begin{aligned}
&\norm{
\mathsf{Y}_{\txtpaulifierz,\sminvtemperature,\kappa,\Lambda;f}
\fnexp{\mathsf{U}_{\txtpaulifierz,\sminvtemperature,\kappa,\Lambda}}
-
\mathsf{Y}_{\txtpaulifierz,\sminvtemperature;f}
\fnexp{\mathsf{U}_{\txtpaulifierz,\sminvtemperature}}
}_{\lp^{1}}
\\ %%%%%%%%%%%%%%%%
&\leq
\norm{\rbk{\mathsf{Y}_{\txtpaulifierz,\sminvtemperature,\kappa,\Lambda;f}
-\mathsf{Y}_{\txtpaulifierz,\sminvtemperature;f}}
\fnexp{\mathsf{U}_{\txtpaulifierz,\sminvtemperature,\kappa,\Lambda}}}_{\lp^{1}}
+\norm{\mathsf{Y}_{\txtpaulifierz,\sminvtemperature;f}
\rbk{\fnexp{\mathsf{U}_{\txtpaulifierz,\sminvtemperature,\kappa,\Lambda}}
-\fnexp{\mathsf{U}_{\txtpaulifierz,\sminvtemperature}}}}_{\lp^{1}}
\\ %%%%%%%%%%%%%%%%
&\leq
\norm{\mathsf{Y}_{\txtpaulifierz,\sminvtemperature,\kappa,\Lambda;f}
-\mathsf{Y}_{\txtpaulifierz,\sminvtemperature;f}}_{\lp^{1}}
+\norm{\mathsf{Y}_{\txtpaulifierz,\sminvtemperature;f}
\rbk{\fnexp{\mathsf{U}_{\txtpaulifierz,\sminvtemperature,\kappa,\Lambda}}
-\fnexp{\mathsf{U}_{\txtpaulifierz,\sminvtemperature}}}}_{\lp^{1}}
\end{aligned}
\end{equation}
The first term converges to $0$ by part (2) of this proposition.
The second term converges to $0$ by the following argument: for arbitrary $R
> 0$,
$$\begin{aligned}
&\norm{\mathsf{Y}_{\txtpaulifierz,\sminvtemperature;f}
\rbk{\fnexp{\mathsf{U}_{\txtpaulifierz,\sminvtemperature,\kappa,\Lambda}}
-\fnexp{\mathsf{U}_{\txtpaulifierz,\sminvtemperature}}}}_{\lp^{1}}
\\ %%%%%%%%%%%%%%%%
&\leq
R\norm{\fnexp{\mathsf{U}_{\txtpaulifierz,\sminvtemperature,\kappa,\Lambda}}
-\fnexp{\mathsf{U}_{\txtpaulifierz,\sminvtemperature}}}_{\lp^{1}}
+2
\int_{\set{\prbprocess}{\abs{\mathsf{Y}_{\txtpaulifierz,\sminvtemperature;f}(\prbprocess)}>R}}
\abs{\mathsf{Y}_{\txtpaulifierz,\sminvtemperature;f}}
\opdmsr{\msrprb_{\txtparticle,\sminvtemperature}}
\end{aligned}$$
holds.
By \eqref{expedition0012246}, the first term converges to $0$ for fixed $R$,
$\mathsf{Y}_{\txtpaulifierz,\sminvtemperature;f}
\in\lp^{1}$ makes the second term arbitrarily small as $R\uparrow\infty$.
Therefore the right-hand side of \eqref{expedition0012247} converges to $0$,
and the $\lp^{1}$ limit in \eqref{expedition0012229} follows.
In particular, $\mathsf{Y}_{\txtpaulifierz,\sminvtemperature;f}
\fnexp{\mathsf{U}_{\txtpaulifierz,\sminvtemperature}}
\in \lp^{1}$;
this follows from $0
\leq \fnexp{\mathsf{U}_{\txtpaulifierz,\sminvtemperature}}
\leq 1$, shown above, and $\mathsf{Y}_{\txtpaulifierz,\sminvtemperature;f}
\in \lp^{1}$.

Finally, we verify uniform integrability of the cutoff products.
Since $\lp^{1}$ convergence holds in part (2) of this proposition,
the family of one-point correction terms with cutoffs is uniformly integrable.
Furthermore, by $0
\leq \fnexp{\mathsf{U}_{\txtpaulifierz,\sminvtemperature,\kappa,\Lambda}}
\leq 1$, $\abs{\mathsf{Y}_{\txtpaulifierz,\sminvtemperature,\kappa,\Lambda;f}
\fnexp{\mathsf{U}_{\txtpaulifierz,\sminvtemperature,\kappa,\Lambda}}}
\leq
\abs{\mathsf{Y}_{\txtpaulifierz,\sminvtemperature,\kappa,\Lambda;f}}$ holds,
so the family of weighted one-point correction terms with cutoffs is also uniformly integrable.
\end{proof}

\begin{defn}[Current Source and Dominating Quantity after Cutoff Removal]\label{expedition0012312}
Using the limits obtained in Proposition \ref{expedition0012165},
the cutoff-removed current source $\mathsf{I}_{\txtrad,S_{\sminvtemperature}}(\prbprocess)$
on the full circle is interpreted not as a single test function including a zero mode,
but as notation defining the following two types of non-zero-mode bilinear forms:
\begin{equation}\label{expedition0012313}
\begin{aligned}
\fun{\opform{q}_{\txtrad,\txtnonzero,\sminvtemperature}^{\txteuclid}}
{\mathsf{I}_{\txtrad,S_{\sminvtemperature}}(\prbprocess)}
&=
\lim_{\kappa\downarrow0,\ \Lambda\uparrow\infty}
\fun{\opform{q}_{\txtrad,\txtnonzero,\sminvtemperature}^{\txteuclid}}
{\mathsf{I}_{\txtrad,S_{\sminvtemperature},\kappa,\Lambda}(\prbprocess)},
\\
\fun{\opform{q}_{\txtrad,\txtnonzero,\sminvtemperature}^{\txteuclid}}
{j_t f,\mathsf{I}_{\txtrad,S_{\sminvtemperature}}(\prbprocess)}
&=
\lim_{\kappa\downarrow0,\ \Lambda\uparrow\infty}
\fun{\opform{q}_{\txtrad,\txtnonzero,\sminvtemperature}^{\txteuclid}}
{j_t f,\mathsf{I}_{\txtrad,S_{\sminvtemperature},\kappa,\Lambda}(\prbprocess)}.
\end{aligned}
\end{equation}
For a closed interval $I\subset S_{\sminvtemperature}$,
$\mathsf{I}_{\txtrad,I}(\prbprocess)$ is interpreted similarly
as the cutoff-removal limit of the non-zero-mode bilinear form appearing in the generator representation
of Proposition \ref{expedition0012209}.
At this point, on a local interval $\int_I\opdmsr{\prbprocess_t}
=\prbprocess_{\sup I}-\prbprocess_{\inf I}$ may remain,
but no new zero-mode term is added to the local interaction after cutoff removal.

Furthermore, for a measurable set $A
\subset \fldreal^3$, denote the $\lp^1$ dominating quantity for the current kernel by
\begin{equation}\label{expedition0012314}
\mathfrak{M}(A)
=
\int_A
\frac{\opdmsr{k}}
{\rbk{\omega(k)+\onehalf\omega(k)^2}^{2}}
+
\rbk{\int_A
\frac{\opdmsr{k}}
{\rbk{\omega(k)+\onehalf\omega(k)^2}^{2}}}^{\onehalf}.
\end{equation}
\end{defn}

Lemma \ref{expedition0012305} can be rewritten as \(\sqfun{\prbexp_{\msrprb_{\txtparticle,\sminvtemperature}}}{\abs{\mathsf{U}_A}}
\leq C\mathfrak{M}(A)\). In particular, the tail estimates for cutoff removal imply \(\mathfrak{M}(A)\to0\) for \(A
= \set{k}{\abs{k}<\kappa}\) or \(A
= \set{k}{\abs{k}>\Lambda}\). Indeed, since \(\omega(k)=\abs{k}\), at low momentum \(\rbk{\omega(k) + \onehalf\omega(k)^2}^{-2}\opdmsr{k}\) has integrability of type \(r^{-2}r^2\opdmsr{r}\), while at high momentum it is of type \(r^{-4}r^2\opdmsr{r}\) and the tail vanishes. Therefore \(\mathfrak{M}(\set{k}{\abs{k}<\kappa})
\to 0\) and \(\mathfrak{M}(\set{k}{\abs{k}>\Lambda})
\to 0\) hold.

We assume the current-kernel cutoff removal in Proposition \ref{expedition0012165}. No new dimension-dependent estimate appears in the limiting transition of this theorem; we use the convergence and uniform integrability of the current-kernel estimates in Proposition \ref{expedition0012165}.

\begin{thm}[Spinless Pauli--Fierz Local Semigroup after Cutoff Removal]\label{expedition0012212}
For arbitrary $f,g
\in\sphilb{D}_{\txtrad,\txtphys,\sminvtemperature}$, consider tensor-product-type generators $F
=
F_{\txtparticle}(\prbprocess_0)
\napiernum^{\imunit\opfocksegalradiation(j_0f)}$ and $G
=
G_{\txtparticle}(\prbprocess_0)
\napiernum^{\imunit\opfocksegalradiation(j_0g)}$.
Then
\begin{equation}\label{expedition0012213}
\bkt{F}{P_{\txtpaulifierz,t}G}
=
\lim_{\kappa\downarrow0,\ \Lambda\uparrow\infty}
\bkt{F}{P_{\txtpaulifierz,\kappa,\Lambda,t}G}
\end{equation}
is satisfied by a local Hermitian semigroup $P_{\txtpaulifierz}
=
\fml{P_{\txtpaulifierz,t}}
{t\in\closedinterval{0}{\frac{\sminvtemperature}{2}}}$.
The limit inherits local semigroup property,
Hermiticity,
and reflection property for fixed cutoffs in the sense of bilinear forms.
\end{thm}

\begin{proof}
In Theorem \ref{expedition0012156} for the Nelson model,
the cutoff-removed effective action \eqref{expedition0012182} and the external-field cross term
\eqref{expedition0012195} are inserted into the local FKN kernel representation
\eqref{expedition0012180},
and semigroup property, Hermiticity, and reflection property are inherited as limits of bilinear forms
on generators.
On the Pauli--Fierz side, \eqref{expedition0012210} and \eqref{expedition0012211}
are the corresponding generator representations,
and Proposition \ref{expedition0012165} gives the cutoff limits of the effective action
and the external-field cross term.
We use the kernel representation on generators in \eqref{expedition0012210}.
The term $-\oneoverfour
\fun{\opform{q}_{\txtrad,\txtbec,\sminvtemperature}^{\txteuclid}}
{j_tg-j_0f}$ coming from the free radiation field is independent of the cutoffs.
The cutoff-dependent parts are the cross term and the self term involving
$\mathsf{I}_{\txtrad,\closedinterval{0}{t},\kappa,\Lambda}$,
which are reduced to non-zero-mode forms by \eqref{expedition0012211}.
Applying the current-kernel estimates of Proposition \ref{expedition0012165} after restricting them
from $S_{\sminvtemperature}$ to $\closedinterval{0}{t}$,
these terms converge for each particle path and are dominated by a uniformly integrable majorant
with respect to the particle loop measure.
Here the exponent of the self term is nonpositive.
By the $\lp^1$ convergence from the cutoff effective action to the cutoff-removed effective action,
obtained in Proposition \ref{expedition0012165}(1), together with nonpositivity,
the corresponding exponential weights also converge in $\lp^1$.
For bilinear forms containing the external-field cross term,
we exchange the cutoff limit and the particle-path integral using the domination from the physical
test-function space and the uniform integrability of the products in Proposition \ref{expedition0012165}.
For the stochastic integral, we use the object constructed as an Itô integral at fixed cutoff;
after integrating out the field, the representation is treated as a Feynman--Kac--Itô type kernel representation,
so there is no place to add a Stratonovich correction here.
By Vitali's theorem, the right-hand side of \eqref{expedition0012210} has a limit,
and the local semigroup kernel is defined by \eqref{expedition0012213}.
Writing the local semigroup property, Hermiticity, and reflection property valid at fixed cutoff
as bilinear-form identities on generators and passing to the limit,
the same properties are inherited by $P_{\txtpaulifierz}$.
\end{proof}

\subsubsection{Limit of KMS States and Physical Fields}\label{limit-of-kms-states-and-physical-fields-1}

We assume the local semigroup limit in Theorem \ref{expedition0012212}. This proposition itself uses no new dimension-dependent estimate.

\begin{prop}[KMS State of the Spinless Pauli--Fierz Model after Cutoff Removal]\label{expedition0012166}
We use the cutoff removal in Proposition \ref{expedition0012165} and Theorem \ref{expedition0012212}.
For times $t_1,\ldots,t_n
\in S_{\sminvtemperature}$,
bounded measurable particle functions $F_j$, and $f_j
\in \sphilb{D}_{\txtrad,\txtphys,\sminvtemperature}$, set $A_j
=
F_j(\prbprocess_{t_j})
\fnexp{\imunit \opfocksegalradiation(j_{t_j}f_j)}$.
Then the finite-time correlation functions are defined by
$$\begin{aligned}
\fun{\oastate[\psi_{\txtpaulifierz,\sminvtemperature}]}
{A_1\cdots A_n}
=
\lim_{\kappa\downarrow0,\ \Lambda\uparrow\infty}
\fun{\oastate[\psi_{\txtpaulifierz,\sminvtemperature,\kappa,\Lambda}]}
{A_1\cdots A_n}.
\end{aligned}$$
In particular, for two-point exponential correlation functions,
$$\begin{aligned}
\fun{\oastate[\psi_{\txtpaulifierz,\sminvtemperature}]}
{\fnexp{\imunit\opfocksegalradiation(j_t f)}
\fnexp{\imunit\opfocksegalradiation(j_s g)}}
=
\fnexp{-\oneoverfour
\fun{\opform{q}_{\txtrad,\txtbec,\sminvtemperature}^{\txteuclid}}
{j_t f+j_s g}}
\mathsf{S}_{\txtpaulifierz,\sminvtemperature;t,s}(f,g)
\end{aligned}$$
holds.
Here the interaction factor can be written as
\begin{equation}\label{expedition0012248}
\begin{aligned}
\mathsf{S}_{\txtpaulifierz,\sminvtemperature;t,s}(f,g)
=
\frac{\int_{\Omega_{\txtparticle,\sminvtemperature}}
\fnexp{\frac{\physcharge}{2}
\fun{\opform{q}_{\txtrad,\txtnonzero,\sminvtemperature}^{\txteuclid}}
{j_t f+j_s g,
\mathsf{I}_{\txtrad,S_{\sminvtemperature}}(\prbprocess)}
+\mathsf{U}_{\txtpaulifierz,\sminvtemperature}(\prbprocess)}
\opdmsr{\msrprb_{\txtparticle,\sminvtemperature}}(\prbprocess)}
{\int_{\Omega_{\txtparticle,\sminvtemperature}}
\fnexp{\mathsf{U}_{\txtpaulifierz,\sminvtemperature}(\prbprocess)}
\opdmsr{\msrprb_{\txtparticle,\sminvtemperature}}(\prbprocess)}
\end{aligned}
\end{equation}
and represents the limit of the non-zero-mode bilinear form whose second argument is the current
source with cutoffs.
\end{prop}

\begin{proof}
In Proposition \ref{expedition0012157} for the Nelson model,
the finite-time correlation functions of the KMS state are reconstructed by inserting the cutoff-removed
local semigroup and the convergence of the effective action into the semigroup product representation
\eqref{expedition0012197},
and the two-point function is split into a free Gaussian factor and an interaction factor arising from
the weighted particle loop measure in \eqref{expedition0012365}.
On the Pauli--Fierz side, Theorem \ref{expedition0012212} corresponds to the local semigroup,
Proposition \ref{expedition0012165} to the effective action and the external-field cross term,
and \eqref{expedition0012248} to the two-point interaction factor.
First rewrite the correlation function of the cutoff state as an integral over particle paths only.
For compact notation, set $K
=
\sum_{j=1}^{n} j_{t_j} f_j$ and $B(\prbprocess)
=
\prod_{j=1}^{n}F_j(\prbprocess_{t_j})$.
Since each $F_j$ is bounded, there is a constant $C_B$ such that
$\abs{B(\prbprocess)}
\leq C_B$ uniformly.
Substituting $A_1\cdots A_n$ into \eqref{eq:PF-equilibrium-functional},
fixing the particle path $\prbprocess$, and first integrating only the radiation field,
the characteristic-function formula for the centered Gaussian measure gives
$$\int_{\prbqspace_{\txtrad,\sminvtemperature}}
\fnexp{\imunit\opfocksegalradiation(K)
-\imunit\physcharge
\opfocksegalradiation(\mathsf{I}_{\txtrad,S_{\sminvtemperature},\kappa,\Lambda}(\prbprocess))}
\opdmsr{\msrprb_{\txtrad,\sminvtemperature}}
=
\fnexp{-\oneoverfour
\fun{\opform{q}_{\txtrad,\txtbec,\sminvtemperature}^{\txteuclid}}
{K-\physcharge\mathsf{I}_{\txtrad,S_{\sminvtemperature},\kappa,\Lambda}(\prbprocess)}}.$$
Expanding the sesquilinear form gives
$$\begin{aligned}
&-\oneoverfour
\fun{\opform{q}_{\txtrad,\txtbec,\sminvtemperature}^{\txteuclid}}
{K-\physcharge
\mathsf{I}_{\txtrad,S_{\sminvtemperature},\kappa,\Lambda}(\prbprocess)}
\\ %%%%%%%%%%%%%%%%
&=
-\oneoverfour
\fun{\opform{q}_{\txtrad,\txtbec,\sminvtemperature}^{\txteuclid}}
{K}
+\frac{\physcharge}{2}
\fun{\opform{q}_{\txtrad,\txtbec,\sminvtemperature}^{\txteuclid}}
{K,\mathsf{I}_{\txtrad,S_{\sminvtemperature},\kappa,\Lambda}(\prbprocess)}
-\frac{\physcharge^2}{4}
\fun{\opform{q}_{\txtrad,\txtbec,\sminvtemperature}^{\txteuclid}}
{\mathsf{I}_{\txtrad,S_{\sminvtemperature},\kappa,\Lambda}(\prbprocess)}.
\end{aligned}$$
Since the current source on the full circle has no zero-momentum component by \eqref{expedition0012224},
the cross term and self term with $\opform{q}_{\txtrad,0,\sminvtemperature}^{\txteuclid}$ vanish.
Therefore the cutoff-dependent part consists only of the non-zero-mode form.
If
$$\begin{aligned}
W_{\kappa,\Lambda}(K;\prbprocess)
=
\frac{\physcharge}{2}
\fun{\opform{q}_{\txtrad,\txtnonzero,\sminvtemperature}^{\txteuclid}}
{K,\mathsf{I}_{\txtrad,S_{\sminvtemperature},\kappa,\Lambda}(\prbprocess)},
\end{aligned}$$
then the cutoff correlation function is represented as
\begin{equation}\label{expedition0012249}
\begin{aligned}
&\fun{\oastate[\psi_{\txtpaulifierz,\sminvtemperature,\kappa,\Lambda}]}
{A_1\cdots A_n}
\\ %%%%%%%%%%%%%%%%
&=
\fnexp{-\oneoverfour
\fun{\opform{q}_{\txtrad,\txtbec,\sminvtemperature}^{\txteuclid}}{K}}
\frac{\int_{\Omega_{\txtparticle,\sminvtemperature}}
B(\prbprocess)
\fnexp{W_{\kappa,\Lambda}(K;\prbprocess)+\mathsf{U}_{\txtpaulifierz,\sminvtemperature,\kappa,\Lambda}(\prbprocess)}
\opdmsr{\msrprb_{\txtparticle,\sminvtemperature}}(\prbprocess)}
{\int_{\Omega_{\txtparticle,\sminvtemperature}}
\fnexp{\mathsf{U}_{\txtpaulifierz,\sminvtemperature,\kappa,\Lambda}(\prbprocess)}
\opdmsr{\msrprb_{\txtparticle,\sminvtemperature}}(\prbprocess)}.
\end{aligned}
\end{equation}
Next take the limit of the denominator in \eqref{expedition0012249}.
By Proposition \ref{expedition0012165}(1),
$\mathsf{U}_{\txtpaulifierz,\sminvtemperature,\kappa,\Lambda}
\to
\mathsf{U}_{\txtpaulifierz,\sminvtemperature}$ holds in $\lp^{1}$,
and it satisfies $0
\leq \fnexp{\mathsf{U}_{\txtpaulifierz,\sminvtemperature,\kappa,\Lambda}}
\leq 1$.
The dominated convergence theorem or Vitali's theorem gives
$\int \fnexp{\mathsf{U}_{\txtpaulifierz,\sminvtemperature,\kappa,\Lambda}}
\opdmsr{\msrprb_{\txtparticle,\sminvtemperature}}
\to
\int \fnexp{\mathsf{U}_{\txtpaulifierz,\sminvtemperature}}
\opdmsr{\msrprb_{\txtparticle,\sminvtemperature}}$.
In particular, since $\mathsf{U}_{\txtpaulifierz,\sminvtemperature}$ is a finite $\lp^{1}$ function,
$\fnexp{\mathsf{U}_{\txtpaulifierz,\sminvtemperature}}
>0$ holds almost everywhere,
and the limiting denominator is positive.

Next consider the numerator.
By definition, $K$ is a finite linear combination of $j_{t_j}f_j$,
so applying Proposition \ref{expedition0012165}(2) to each time-translated term and summing gives
$$\begin{aligned}
W_{\kappa,\Lambda}(K;\prbprocess)
\to
W(K;\prbprocess)
=
\frac{\physcharge}{2}
\fun{\opform{q}_{\txtrad,\txtnonzero,\sminvtemperature}^{\txteuclid}}
{K,\mathsf{I}_{\txtrad,S_{\sminvtemperature}}(\prbprocess)}
\quad\text{in } \lp^{1}.
\end{aligned}$$
By Proposition \ref{expedition0012165}(1),
$\mathsf{U}_{\txtpaulifierz,\sminvtemperature,\kappa,\Lambda}
\to \mathsf{U}_{\txtpaulifierz,\sminvtemperature}$ also holds in $\lp^{1}$.
Uniform domination of the exponential part follows from nonnegativity of the Gaussian sesquilinear form.
Taking the real part,
$$\begin{aligned}
0
\leq
\fun{\opform{q}_{\txtrad,\txtbec,\sminvtemperature}^{\txteuclid}}
{K-\physcharge
\mathsf{I}_{\txtrad,S_{\sminvtemperature},\kappa,\Lambda}(\prbprocess)}
=
\fun{\opform{q}_{\txtrad,\txtbec,\sminvtemperature}^{\txteuclid}}
{K}
-4\opreal W_{\kappa,\Lambda}(K;\prbprocess)
-4\mathsf{U}_{\txtpaulifierz,\sminvtemperature,\kappa,\Lambda}(\prbprocess)
\end{aligned}$$
and hence, for all cutoffs,
$$\begin{aligned}
\abs{\fnexp{W_{\kappa,\Lambda}(K;\prbprocess)
+\mathsf{U}_{\txtpaulifierz,\sminvtemperature,\kappa,\Lambda}(\prbprocess)}}
\leq
\fnexp{\oneoverfour
\fun{\opform{q}_{\txtrad,\txtbec,\sminvtemperature}^{\txteuclid}}
{K}}
\end{aligned}$$
holds.
Since the right-hand side is a constant independent of the particle path,
Vitali's theorem gives
$$B(\prbprocess)
\fnexp{W_{\kappa,\Lambda}(K;\prbprocess)+\mathsf{U}_{\txtpaulifierz,\sminvtemperature,\kappa,\Lambda}(\prbprocess)}
\to
B(\prbprocess)
\fnexp{W(K;\prbprocess)+\mathsf{U}_{\txtpaulifierz,\sminvtemperature}(\prbprocess)}
\quad\text{in }\lp^{1}.$$
Thus the numerator and denominator in \eqref{expedition0012249} can be passed to the limit simultaneously,
and the limit of the finite-time correlation functions exists;
define this limit as $\fun{\oastate[\psi_{\txtpaulifierz,\sminvtemperature}]}
{A_1\cdots A_n}$.
At this stage, the denominator is the limit of the particle-path representation of the cutoff partition function,
and the numerator is the limit of the finite-time correlation function under the same normalization.
If $K=0$ and $B=1$, the numerator and denominator agree,
so the limiting functional sends $\idone$ to $1$.
Moreover, for any cylindrical polynomial $C$,
$\fun{\oastate[\psi_{\txtpaulifierz,\sminvtemperature,\kappa,\Lambda}]}{C^*C}\geq0$,
and the convergence above also applies to this expectation, so the limiting functional is positive.
Therefore the finite-time correlation functions obtained here form a normalized positive functional
obtained by separately taking the denominator limit, the numerator limit, and the normalized limit.

(Two-point exponential correlation function): If in the preceding argument $n=2$,
$F_1=F_2=1$,
$K=j_t f+j_s g$,
then the free Gaussian factor $\fnexp{-\oneoverfour
\fun{\opform{q}_{\txtrad,\txtbec,\sminvtemperature}^{\txteuclid}}
{j_t f+j_s g}}$ factors out from the general formula obtained above.
The remaining particle path integral is
$$\begin{aligned}
\frac{\int_{\Omega_{\txtparticle,\sminvtemperature}}
\fnexp{\frac{\physcharge}{2}
\fun{\opform{q}_{\txtrad,\txtnonzero,\sminvtemperature}^{\txteuclid}}
{j_t f+j_s g,
\mathsf{I}_{\txtrad,S_{\sminvtemperature}}(\prbprocess)}
+\mathsf{U}_{\txtpaulifierz,\sminvtemperature}(\prbprocess)}
\opdmsr{\msrprb_{\txtparticle,\sminvtemperature}}(\prbprocess)}
{\int_{\Omega_{\txtparticle,\sminvtemperature}}
\fnexp{\mathsf{U}_{\txtpaulifierz,\sminvtemperature}(\prbprocess)}
\opdmsr{\msrprb_{\txtparticle,\sminvtemperature}}(\prbprocess)},
\end{aligned}$$
which is $\mathsf{S}_{\txtpaulifierz,\sminvtemperature;t,s}(f,g)$ in \eqref{expedition0012248}.

(KMS reconstruction): The finite-time correlation functions at fixed cutoff are obtained from the
product representation of the local Hermitian semigroup in Proposition \ref{expedition0012206}.
Specifically, let $0\leq t_1\leq\cdots\leq t_n<\sminvtemperature$,
and denote by $\mathcal{M}_{A_j}$ the multiplication operator at time zero.
At fixed cutoff,
$$\begin{aligned}
\fun{\oastate[\psi_{\txtpaulifierz,\sminvtemperature,\kappa,\Lambda}]}
{A_1\cdots A_n}
=
\frac{
\bkt{\idone}
{P_{\txtpaulifierz,\kappa,\Lambda,t_1}
\mathcal{M}_{A_1}
P_{\txtpaulifierz,\kappa,\Lambda,t_2-t_1}
\mathcal{M}_{A_2}\cdots
\mathcal{M}_{A_n}
P_{\txtpaulifierz,\kappa,\Lambda,\sminvtemperature-t_n}
\idone}}
{
\bkt{\idone}
{P_{\txtpaulifierz,\kappa,\Lambda,\sminvtemperature}\idone}}.
\end{aligned}$$
The bilinear-form limit obtained in Theorem \ref{expedition0012212}
preserves the local semigroup property, reflection property, and time-translation covariance valid at fixed cutoff
as identities on generators.
The Markov property is contained in the product representation of the local semigroup,
and it suffices to write the product representation conditioned at an intermediate time as a bilinear form
and take the same limit.
In this representation, one adjacent time difference is merely replaced by
$\sminvtemperature-t_n+t_1$,
so cyclicity on the circle is also preserved by the limits of the denominator and of each bilinear form.
The $\lp^{1}$ limits of the numerator and denominator shown above also pass cyclicity on the circle to the limit.

By the preceding argument, the semigroup product representation valid at fixed cutoff,
the positive limit of the denominator,
cyclicity on the circle,
reflection property,
time-translation covariance,
the Markov property,
positivity,
and normalization are all inherited by the finite-time correlation functions after cutoff removal.
Therefore the finite-time correlation functions defined here determine the KMS state of the
Pauli--Fierz model after cutoff removal by the reconstruction theorem in the textbook
\cite{DerezinskiGerard001}.
\end{proof}

We use the cutoff-removed state obtained in Proposition \ref{expedition0012166}. This proposition itself uses no new dimension-dependent estimate; the required cutoff removal is contained in the current-kernel estimates of Proposition \ref{expedition0012165} and in the convergence of the states in Proposition \ref{expedition0012166}.

\begin{prop}[Physical Radiation Field after Cutoff Removal]\label{expedition0012167}
For arbitrary $f
\in \sphilb{D}_{\txtrad,\txtphys,\sminvtemperature}$, define
$$\ell_{\txtpaulifierz,\sminvtemperature}(f)
=
-\frac{\int_{\Omega_{\txtparticle,\sminvtemperature}}
\mathsf{Y}_{\txtpaulifierz,\sminvtemperature;f}(\prbprocess)
\fnexp{\mathsf{U}_{\txtpaulifierz,\sminvtemperature}(\prbprocess)}
\opdmsr{\msrprb_{\txtparticle,\sminvtemperature}}(\prbprocess)}
{\int_{\Omega_{\txtparticle,\sminvtemperature}}
\fnexp{\mathsf{U}_{\txtpaulifierz,\sminvtemperature}(\prbprocess)}
\opdmsr{\msrprb_{\txtparticle,\sminvtemperature}}(\prbprocess)}.$$
Then $\fun{\oastate[\psi_{\txtpaulifierz,\sminvtemperature}]}
{\opfocksegalradiation(j_0 f)}
=
\ell_{\txtpaulifierz,\sminvtemperature}(f)$ holds,
and the physical radiation field after cutoff removal,
$\opfocksegalradiation_{\txtphys,\sminvtemperature}(j_0 f)
=
\opfocksegalradiation(j_0 f)
-\ell_{\txtpaulifierz,\sminvtemperature}(f)$, satisfies
$\fun{\oastate[\psi_{\txtpaulifierz,\sminvtemperature}]}
{\opfocksegalradiation_{\txtphys,\sminvtemperature}(j_0 f)}
=
0$.
\end{prop}

\begin{proof}
In Proposition \ref{expedition0012158} for the Nelson model,
the linear correction \eqref{expedition0012351} is treated as the derivative at the origin
of the one-point characteristic function,
and the corrected physical field is defined in \eqref{expedition0012200}.
On the Pauli--Fierz side, Proposition \ref{expedition0012165}(2),(3)
are the corresponding parts for the correction kernel and the weighted product,
and \eqref{expedition0012303} corresponds to the derivative computation for the one-point characteristic function.
Let the one-point characteristic function with cutoffs be $\chi_{\kappa,\Lambda,f}(s)
=
\fun{\oastate[\psi_{\txtpaulifierz,\sminvtemperature,\kappa,\Lambda}]}
{\fnexp{\imunit s\opfocksegalradiation(j_0 f)}}$.
Integrating the cutoff representation with respect to the field, for $s$ near the origin,
\begin{equation}\label{expedition0012303}
\begin{aligned}
\chi_{\kappa,\Lambda,f}(s)
&=
\fnexp{-\frac{s^2}{4}
\fun{\opform{q}_{\txtrad,\txtbec,\sminvtemperature}^{\txteuclid}}{j_0 f}}
\\
&\quad\times
\frac{\int_{\Omega_{\txtparticle,\sminvtemperature}}
\fnexp{-\imunit s
\mathsf{Y}_{\txtpaulifierz,\sminvtemperature,\kappa,\Lambda;f}(\prbprocess)}
\fnexp{\mathsf{U}_{\txtpaulifierz,\sminvtemperature,\kappa,\Lambda}(\prbprocess)}
\opdmsr{\msrprb_{\txtparticle,\sminvtemperature}}(\prbprocess)}
{\int_{\Omega_{\txtparticle,\sminvtemperature}}
\fnexp{\mathsf{U}_{\txtpaulifierz,\sminvtemperature,\kappa,\Lambda}(\prbprocess)}
\opdmsr{\msrprb_{\txtparticle,\sminvtemperature}}(\prbprocess)}
\end{aligned}
\end{equation}
holds.
The first derivative of the free Gaussian factor vanishes at $s=0$,
so the first-order term in $s$ comes from
$\mathsf{Y}_{\txtpaulifierz,\sminvtemperature,\kappa,\Lambda;f}$.
By Proposition \ref{expedition0012165}(1)--(3), in
$\fun{\lp^{1}}{\Omega_{\txtparticle,\sminvtemperature},
\msrprb_{\txtparticle,\sminvtemperature}}$,
$$\mathsf{U}_{\txtpaulifierz,\sminvtemperature,\kappa,\Lambda}\to
\mathsf{U}_{\txtpaulifierz,\sminvtemperature},
\quad
\mathsf{Y}_{\txtpaulifierz,\sminvtemperature,\kappa,\Lambda;f}
\to
\mathsf{Y}_{\txtpaulifierz,\sminvtemperature;f},
\quad
\mathsf{Y}_{\txtpaulifierz,\sminvtemperature,\kappa,\Lambda;f}
\fnexp{\mathsf{U}_{\txtpaulifierz,\sminvtemperature,\kappa,\Lambda}}
\to
\mathsf{Y}_{\txtpaulifierz,\sminvtemperature;f}
\fnexp{\mathsf{U}_{\txtpaulifierz,\sminvtemperature}}$$
hold.
Moreover, by $0\leq\fnexp{\mathsf{U}_{\txtpaulifierz,\sminvtemperature,\kappa,\Lambda}}\leq1$
and the uniform integrability of the products verified in Proposition \ref{expedition0012165}(3),
\eqref{expedition0012303} converges to the limit locally uniformly near the origin in $s$,
and the derivative at $s=0$ can also be exchanged with the limit.
Indeed, for fixed $|s|\leq s_0$,
$\fnexp{-\imunit s\mathsf{Y}_{\txtpaulifierz,\sminvtemperature,\kappa,\Lambda;f}} - \fnexp{-\imunit s\mathsf{Y}_{\txtpaulifierz,\sminvtemperature;f}}$ is split into a part bounded by
$s_0\abs{\mathsf{Y}_{\txtpaulifierz,\sminvtemperature,\kappa,\Lambda;f}
-\mathsf{Y}_{\txtpaulifierz,\sminvtemperature;f}}$ and the tail on the set where $\mathsf{Y}_{\txtpaulifierz,\sminvtemperature;f}$ is large.
The former vanishes by $\lp^1$ convergence,
and the latter can be made uniformly small by $\mathsf{Y}_{\txtpaulifierz,\sminvtemperature;f}
\in \lp^{1}$ and the uniform integrability of $\mathsf{Y}_{\txtpaulifierz,\sminvtemperature,\kappa,\Lambda;f}
\fnexp{\mathsf{U}_{\txtpaulifierz,\sminvtemperature,\kappa,\Lambda}}$.

For the difference quotient, the $s\to0$ limit of $\frac{1}{s}
\rbk{\chi_{\kappa,\Lambda,f}(s)-\chi_{\kappa,\Lambda,f}(0)}$ is given by the integral of
$-\imunit
\mathsf{Y}_{\txtpaulifierz,\sminvtemperature,\kappa,\Lambda;f}\fnexp{\mathsf{U}_{\txtpaulifierz,\sminvtemperature,\kappa,\Lambda}}$,
and the $\lp^1$ convergence and uniform integrability of this family allow the derivative at $s=0$
to be exchanged with the cutoff-removal limit.
Therefore
$$\fun{\oastate[\psi_{\txtpaulifierz,\sminvtemperature}]}
{\opfocksegalradiation(j_0 f)}
=
\lim_{\kappa\downarrow0,\ \Lambda\uparrow\infty}
\fun{\oastate[\psi_{\txtpaulifierz,\sminvtemperature,\kappa,\Lambda}]}
{\opfocksegalradiation(j_0 f)}
=
\ell_{\txtpaulifierz,\sminvtemperature}(f)$$
is obtained.
It remains only to substitute the definition.
\end{proof}

\subsubsection{Off-Diagonal Long-Range Order, Order Parameters, and the No-Go Theorem for BEC}\label{off-diagonal-long-range-order-order-parameters-and-the-no-go-theorem-for-bec-1}

\begin{defn}[Order Parameter of the Spinless Pauli--Fierz Model]\label{expedition0012168}
Let $I_L^3$ be the cube centered at the origin with side length $L$, and set $V_L=L^3$. Choose a transverse polarization direction $e\in\fldcmp^2$ fixed at zero momentum with $\abs{e}=1$, and define the functions approximating the zero-momentum mode of the radiation field by $$\mathsf{b}_{L,e}^{(0)}
=
\frac{1}{V_L^{\onehalf}}
\fndef{I_L^3}e,
\quad
\mathsf{b}_{L,e}^{(1)}
=
\frac{1}{V_L}
\fndef{I_L^3}e.$$
For the bounded-system spinless Pauli--Fierz model state $\psi_{\txtpaulifierz,\sminvtemperature,\smchemicalpotential,L}$, define
\begin{equation}\label{expedition0012438}
\begin{aligned}
\mathsf{o}_{\txtpaulifierz,\sminvtemperature,L,e}^{(\#)}
=
\imunit
\fun{\psi_{\txtpaulifierz,\sminvtemperature,\smchemicalpotential,L}}
{\fun{\oaresolvent}{1,\mathsf{b}_{L,e}^{(\#)}}}
\end{aligned}
\end{equation}
and call it the order parameter of the spinless Pauli--Fierz model.
\end{defn}

\begin{prop}[Off-Diagonal Long-Range Order of the Spinless Pauli--Fierz Model]\label{expedition0012415}
For the spinless Pauli--Fierz-model KMS state after cutoff removal constructed in Proposition \ref{expedition0012166} and $f,g \in \sphilb{D}_{\txtrad,\txtphys,\sminvtemperature}$, the two-point off-diagonal long-range order of the radiation field is given by
\begin{equation}\label{expedition0012416}
\lim_{\abs{x}\to\infty}
\fun{\oastate[\psi_{\txtpaulifierz,\sminvtemperature}]}
{\opfocksegalradiation(j_0 f)\opfocksegalradiation(j_0\tau_x g)}
=
\onehalf\fun{\opform{q}_{\txtrad,0,\sminvtemperature}}{f,g}.
\end{equation}
\end{prop}

\begin{proof}
Use the two-point exponential correlation-function representation of Proposition \ref{expedition0012166}. In the Gaussian factor of the free radiation field, the zero-mode form is independent of the spatial variable because $\faftr{\tau_xg}(0)=\faftr{g}(0)$, and the cross term of the non-zero-mode covariance disappears by the Riemann--Lebesgue lemma. In the interaction term, the factor $\napiernum^{-\imunit kx}$ appears in the cross form with the closed current source, and the $\lp^1$ estimate and uniform integrability of Proposition \ref{expedition0012165} allow the dominated convergence theorem to be applied. Hence the cross term between the translated external radiation-field test function and the current source disappears in the long-distance limit. Differentiating the two-point characteristic functional in two coefficients gives \eqref{expedition0012416}.
\end{proof}

\begin{prop}[Order-Parameter Criterion for the Spinless Pauli--Fierz Model]\label{expedition0012417}
For the order parameter of Definition \ref{expedition0012168}, the following assertions hold. The condition $\smnumberdensity_{\txtrad,0}(\sminvtemperature)>0$ for the zero-mode density corresponding to the polarization direction $e$ is equivalent to
$$\lim_{L\to\infty}\mathsf{o}_{\txtpaulifierz,\sminvtemperature,L,e}^{(1)}
=
\int_0^\infty
\fnexp{-r-\frac12\smnumberdensity_{\txtrad,0}(\sminvtemperature)r^2}
\opdmsr{r}
<1.$$
Moreover, $\smnumberdensity_{\txtrad,0}(\sminvtemperature)=0$ is equivalent to $$\lim_{L\to\infty}
\mathsf{o}_{\txtpaulifierz,\sminvtemperature,L,e}^{(1)}
= 1.$$
\end{prop}

\begin{proof}
The finite-volume function $\mathsf{b}_{L,e}^{(1)}$ detects only the zero-momentum mode of the specified transverse polarization. The closed current source on the full circle does not cross with the zero-momentum external field, and the finite-volume interaction term contributes to the order parameter only through the non-zero-mode form; hence, in the Laplace-transform representation of the resolvent, only the zero-mode evaluation of the free radiation field remains. This zero-mode evaluation is the same as the order-parameter computation for the free Bose gas, and the displayed limit formula follows.
\end{proof}

\begin{thm}[No-Go Theorem for BEC in the Spinless Pauli--Fierz Model via Off-Diagonal Long-Range Order]\label{expedition0012169}
Assume that the physical test-function space $\sphilb{D}_{\txtrad,\txtphys,\sminvtemperature}$ can distinguish the zero mode in the sense of Definition \ref{expedition0012433}. Then the following assertions are equivalent for the cutoff-removed KMS state constructed in Proposition \ref{expedition0012166}.
\begin{enumerate}
\item
For all $f,g \in \sphilb{D}_{\txtrad,\txtphys,\sminvtemperature}$, the off-diagonal long-range order in \eqref{expedition0012416} is $0$.

\item
The equality $\opform{q}_{\txtrad,0,\sminvtemperature}=0$ holds.

\item
The condition $\smnumberdensity_{\txtrad,0}(\sminvtemperature)=0$ holds.

\item
The order parameter of Definition \ref{expedition0012168} satisfies $\lim_{L\to\infty}\mathsf{o}_{\txtpaulifierz,\sminvtemperature,L,e}^{(1)}=1$ for every polarization direction $e$.
\end{enumerate}
In particular, the absence of off-diagonal long-range order and the order-parameter criterion are equivalent to the disappearance of the BEC zero mode of the radiation field on the physical test-function space after cutoff removal.
\end{thm}

\begin{proof}
Proposition \ref{expedition0012415} gives the equivalence between assertions (1) and (2). Zero-mode distinguishability and the zero-momentum density representation of $\opform{q}_{\txtrad,0,\sminvtemperature}$ give the equivalence between assertions (2) and (3). Proposition \ref{expedition0012417} gives the equivalence between assertions (3) and (4).
\end{proof}

\section{Pauli--Fierz Model with Spin}\label{paulifierz-model-with-spin}

In this section we assume three-dimensionality from the beginning in principle, in order to introduce the cross product.

\subsection{Definition of the Hamiltonian}\label{definition-of-the-hamiltonian-2}

Following the description in the textbook \cite[Section 3.8]{LorincziHiroshimaBetz3}, we consider a charged particle with spin \(\onehalf\). To treat the magnetic flux density \(\physmfdensity
= \varot \opfocksegalradiation\) and the Pauli term \(\qmpaulispin \vainnprod \physmfdensity\), we set the spatial dimension to \(d=3\) here and use the same radiation field, the same zero-mode form \(\opform{q}_{\txtrad,0,\sminvtemperature}\), the same non-zero-mode form \(\opform{q}_{\txtrad,\txtnonzero,\sminvtemperature}^{\txteuclid}\), and the same physical test-function space \(\sphilb{D}_{\txtrad,\txtphys,\sminvtemperature}\) as in the spinless Pauli--Fierz model.

Since we consider an electron with spin, the particle Hilbert space changes to \[\sphilb{H}_{\txtparticle,\txtspin}
=
\sphilb{H}_{\txtparticle}
\otimes
\sphilb{H}_{\txtspin}
=
\fun{\lp^{2}}{\fldreal^{3}}
\otimes
\fun{\lp^{2}}{\ringratint_{2}}
=
\fun{\lp^{2}}{\fldreal^{3} \times \ringratint_{2}}.\] The Hilbert space of the radiation field remains \(\spfock_{\txtrad}\), and the Hilbert space of the total system is \(\sphilb{H}_{\txtpaulifierz,\txtspin}
=
\sphilb{H}_{\txtparticle,\txtspin}
\otimes
\spfock_{\txtrad}\).

By the Pauli matrix relations, the formal Hamiltonian can be written as \[\begin{aligned}
\physham_{\txtpaulifierz,\txtspin}
&=
\onehalf
\rbk{\qmpaulispin
\vainnprod
\rbk{-\imunit\vagrad-\physcharge \opfocksegalradiation}}^{2}
+V
+\physham_{\txtrad,\txtfr}
\\ %%%%%%%%%%%%%%%%
&=
\onehalf
\rbk{-\imunit\vagrad-\physcharge \opfocksegalradiation}^{2}
-\frac{\physcharge}{2} \qmpaulispin \vainnprod \physmfdensity
+V
+\physham_{\txtrad,\txtfr}.
\end{aligned}\] The difference from the formal spinless model is that, in addition to the minimal-coupling current, the Pauli term with the quantized magnetic flux density \(\physmfdensity
= \varot \opfocksegalradiation\) appears.

Also in the case with spin, we do not directly discuss the point-source model; as in the spinless model, we start from the regularized model using the cutoff function \(\faftr{\varphi}_{\kappa,\Lambda}
=\fndef{\kappa\leq\abs{k}\leq\Lambda}\). After defining the vector potential with cutoffs \(\opfocksegalradiation_{\kappa,\Lambda}\) as in Section \ref{expedition0012278}, define the magnetic flux density with cutoffs by \(\physmfdensity_{\kappa,\Lambda}
=\varot \opfocksegalradiation_{\kappa,\Lambda}\). Then \[\begin{aligned}
\physham_{\txtpaulifierz,\txtspin,\kappa,\Lambda}
&=
\onehalf
\rbk{\qmpaulispin \vainnprod \rbk{-\imunit\vagrad-\physcharge \opfocksegalradiation_{\kappa,\Lambda}}}^{2}
+V
+\physham_{\txtrad,\txtfr}
\\ %%%%%%%%%%%%%%%%
&=
\onehalf
\rbk{-\imunit\vagrad-\physcharge \opfocksegalradiation_{\kappa,\Lambda}}^{2}
-\frac{\physcharge}{2} \qmpaulispin \vainnprod \physmfdensity_{\kappa,\Lambda}
+V
+\physham_{\txtrad,\txtfr}
\end{aligned}\] is the Pauli--Fierz Hamiltonian with spin and cutoffs. Below, after first defining spin paths and treating the construction of the bounded system independently, we formulate the KMS state with cutoffs, the state after cutoff removal, the physical radiation field, off-diagonal long-range order, and the no-go theorem for BEC in the same order as in the spinless model. The new parts to verify in the model with spin are the integrability of the magnetic-flux-density time integral and of the magnetic-flux-density jump test functions arising from the Pauli term, the cutoff differences, and the limits of interaction factors containing them. Since the Pauli term contains \(\physmfdensity_{\kappa,\Lambda}\), cutoff removal requires, in addition to the current kernel, convergence estimates for the diagonal Pauli source \(\mathsf{I}^{\txtpauli,\txtdiag}_{\txtrad,S_{\sminvtemperature},\kappa,\Lambda}\) and the jump Pauli source \(\mathsf{I}^{\txtpauli,\txtflip}_{\txtrad,j,\kappa,\Lambda}\) defined from the magnetic-flux-density test functions with cutoffs. These are radiation-field sources arising from \(-\frac{\physcharge}{2}\qmpaulispin\vainnprod \physmfdensity_{\kappa,\Lambda}\) in the Hamiltonian. However, the zero-mode form of the radiation field, the non-zero-mode form, the physical test-function space \(\sphilb{D}_{\txtrad,\txtphys,\sminvtemperature}\), and the interpretation of the current source \(\mathsf{I}_{\txtrad,I}\) after cutoff removal are unchanged from the spinless model. Thus \(\sphilb{D}_{\txtrad,\txtphys,\sminvtemperature}\) is again the space for interpreting the non-zero-mode bilinear form with the current source arising from the point-source interaction, not the space for removing the zero mode. The current source on a local interval is treated as the cutoff limit of a non-zero-mode bilinear form, and no new zero-mode local interaction is added from the interaction with spin.

\subsection{Spin Paths}\label{expedition0012381}

For the content of this subsection, also see \cite{YoshitsuguSekine008}.

Using a Poisson process \(N_t\) of intensity \(1\), define the spin-flip process starting from the initial value \(\varsigma
\in \ringratint_{2}\) by \[\begin{aligned}
\prbprocess_{\varsigma}^{\txtspin}
=
\fml{\prbprocess_{\varsigma,t}^{\txtspin}}{t\geq0}
=
\fml{\varsigma(-1)^{N_t}}{t\geq0},
\end{aligned}\] and when the initial spin for the process on the half-line is not specified, write simply \(\prbprocess^{\txtspin}
=\fml{\prbprocess_t^{\txtspin}}{t\geq0}\). Let \(\fun{D}
{S_{\sminvtemperature};\ringratint_{2}}\) be the space of càdlàg paths on the periodic time interval \(S_{\sminvtemperature}\) taking values in \(\ringratint_{2}\), and let \(\mblfml{S}_{\txtspin,\sminvtemperature}\) be the Borel \(\sigma\)-algebra determined by its Skorohod topology. Let \(\msrlaw_{\txtspin,\sminvtemperature,\varsigma}\) be the law of the shifted process \(\fml{\prbprocess_{\varsigma,t+\frac{\sminvtemperature}{2}}^{\txtspin}}
{t\in S_{\sminvtemperature}}\) starting from \(\varsigma\) at time \(-\frac{\sminvtemperature}{2}\). For any measurable set \(A
\subset \fun{D}{S_{\sminvtemperature};\ringratint_{2}}\), define the finite measure of closed spin paths at finite temperature \(\sminvtemperature\) by \begin{equation}\label{expedition0012251}
\fun{\nu_{\txtspin,\sminvtemperature}}{A}
=
\napiernum^{\sminvtemperature}
\sum_{\varsigma\in\ringratint_{2}}
\fun{\msrlaw_{\txtspin,\sminvtemperature,\varsigma}}
{A\cap\setone{\prbprocess_{\varsigma,\sminvtemperature}^{\txtspin}=\varsigma}}.
\end{equation} Here \(\prbprocess_{\varsigma,\sminvtemperature}^{\txtspin}\) denotes the value at time \(\sminvtemperature\) of the original process on the half-line. On the other hand, \(\prbprocess^{\txtspin}\) appearing below as a variable of path integration denotes a path on \(S_{\sminvtemperature}\) obtained from the shifted law above. Denote the jump times of this circular path by \(-\frac{\sminvtemperature}{2}
<\tau_1<\cdots<\tau_{N_{\sminvtemperature}(\prbprocess^{\txtspin})}
<\frac{\sminvtemperature}{2}\). The path on \(S_{\sminvtemperature}\) determined by the time sequence \(-\frac{\sminvtemperature}{2}<\tau_1<\cdots<\tau_n<\frac{\sminvtemperature}{2}\) and the initial value \(\varsigma\) is denoted by \[\prbprocess_{(\varsigma,\tau_1,\ldots,\tau_n),t}^{\txtspin}
=
\varsigma(-1)^{\abscard{\set{j}{\tau_j\leq t}}}
\quad
\rbk{t\in S_{\sminvtemperature}}.\]

\begin{prop}[Thermal Spin Path Probability Measure and Particle--Spin Product Measure]\label{expedition0012280}
The total mass of the closed spin path measure is $\fun{\msr{\nu_{\txtspin,\sminvtemperature}}}
{\fun{D}{S_{\sminvtemperature};\ringratint_{2}}}
=
2 \cosh \sminvtemperature$.
\end{prop}

\begin{proof}
For a Poisson process of intensity $1$, the time-ordered density of paths with $n$ jumps is
$\napiernum^{-\sminvtemperature}
\opdmsr{\tau_1}
\cdots
\opdmsr{\tau_n}$.
The closed-path condition $\prbprocess_{\varsigma,\sminvtemperature}^{\txtspin}
= \varsigma$ is equivalent to the number of jumps being even.
The factor $\napiernum^{\sminvtemperature}$ in \eqref{expedition0012251}
cancels the exponential factor in the Poisson probability, so for any bounded measurable function $F$,
$$\begin{aligned}
\int_{\fun{D}{S_{\sminvtemperature};\ringratint_{2}}}
F(\prbprocess^{\txtspin})
\opdmsr{\nu_{\txtspin,\sminvtemperature}}(\prbprocess^{\txtspin})
=
\sum_{\varsigma\in\ringratint_{2}}
\sum_{m=0}^{\infty}
\int_{-\frac{\sminvtemperature}{2}<\tau_1<\cdots<\tau_{2m}<\frac{\sminvtemperature}{2}}
F(\prbprocess_{(\varsigma,\tau_1,\ldots,\tau_{2m})}^{\txtspin})
\opdmsr{\tau_1} \cdots \opdmsr{\tau_{2m}}
\end{aligned}$$
holds.
In particular, if $F
= 1$, then for each $\varsigma$,
$\sum_{m=0}^{\infty}
\frac{\sminvtemperature^{2m}}{(2m)!}
=\cosh \sminvtemperature$;
summing over the two initial values in $\ringratint_{2}$ gives the total mass
$2 \cosh \sminvtemperature$.
\end{proof}

\begin{defn}[Thermal Spin Path Probability Measure]\label{expedition0012430}
Under Proposition \ref{expedition0012280},
define the normalized thermal spin path probability measure by
\begin{equation}\label{expedition0012431}
\msrprb_{\txtspin,\sminvtemperature}
=
\frac{1}{2\cosh\sminvtemperature}
\msr{\nu_{\txtspin,\sminvtemperature}}
\end{equation}
and define the particle--spin path space and its product probability measure by
\begin{equation}\label{expedition0012432}
\begin{aligned}
\Omega_{\txtparticle,\txtspin,\sminvtemperature}
=
\Omega_{\txtparticle,\sminvtemperature}
\times
\fun{D}{S_{\sminvtemperature};\ringratint_{2}},
\quad
\msrprb_{\txtparticle,\txtspin,\sminvtemperature}
=
\msrprb_{\txtparticle,\sminvtemperature}
\otimes
\msrprb_{\txtspin,\sminvtemperature}.
\end{aligned}
\end{equation}
Since the particle loop measure $\msrprb_{\txtparticle,\sminvtemperature}$ is also a probability measure,
the product $\msrprb_{\txtparticle,\txtspin,\sminvtemperature}$ is also a probability measure.
\end{defn}

Let \(\mathfrak{b}_{\mathrm{i},x,\kappa,\Lambda}\) be the test function corresponding to the \(\mathrm{i}\)-th component of the magnetic flux density with cutoffs. In particular, for each \(\mathrm{j}_{\txtrad}
\in \ringratint_2\), \begin{equation}\label{expedition0012252}
\faftr{\mathfrak{b}}_{\mathrm{i},x,\kappa,\Lambda}^{\mathrm{j}_{\txtrad}}(k)
=
\frac{\faftr{\varphi}_{\kappa,\Lambda}(k)}
{\sqrt{\omega(k)}}
\imunit\rbk{k\times e^{\mathrm{j}_{\txtrad}}(k)}_{\mathrm{i}}
\napiernum^{-\imunit kx}.
\end{equation}

\begin{prop}[Definability of the Magnetic-Flux-Density Test Functions with Cutoffs in the Covariance Form]\label{expedition0012279}
Fix cutoffs $0<\kappa<\Lambda<\infty$.
Then, for arbitrary $x
\in \fldreal^{3}$ and $\mathrm{i}
\in \setone{1,2,3}$,
the function $\mathfrak{b}_{\mathrm{i},x,\kappa,\Lambda}$ defined by \eqref{expedition0012252}
belongs to the domain of the covariance form
$\opform{q}_{\txtrad,\txtbec,\sminvtemperature}^{\txteuclid}$ of the radiation field.
In particular, for arbitrary $t,s
\in S_{\sminvtemperature}$, $\fun{\opform{q}_{\txtrad,\txtbec,\sminvtemperature}^{\txteuclid}}
{j_t\mathfrak{b}_{\mathrm{i},x,\kappa,\Lambda},
j_s\mathfrak{b}_{\mathrm{j},y,\kappa,\Lambda}}$ is finite.
\end{prop}

\begin{proof}
The support of the momentum-space representation \eqref{expedition0012252} is contained in
$\set{k\in\fldreal^{3}}
{\kappa\leq\abs{k}\leq\Lambda}$.
The polarization vectors are unit vectors and satisfy
$\abs{\rbk{k\times e^{\mathrm{j}_{\txtrad}}(k)}_{\mathrm{i}}}
\leq
\abs{k}$, so
$$\begin{aligned}
\abs{\faftr{\mathfrak{b}}_{\mathrm{i},x,\kappa,\Lambda}^{\mathrm{j}_{\txtrad}}(k)}^2
\leq
\fndef{\kappa\leq\abs{k}\leq\Lambda}
\frac{\abs{k}^2}{\omega(k)}
=
\fndef{\kappa\leq\abs{k}\leq\Lambda}\abs{k}
\end{aligned}$$
holds.

(Non-zero-mode check): On the support, $\omega(k)\geq\kappa$, and
$$\begin{aligned}
\frac{1+\napiernum^{-\sminvtemperature\omega(k)}}
{1-\napiernum^{-\sminvtemperature\omega(k)}}
\leq
\frac{1+\napiernum^{-\sminvtemperature\kappa}}
{1-\napiernum^{-\sminvtemperature\kappa}}
=c_{\sminvtemperature,\kappa}
<\infty
\end{aligned}$$
holds. Therefore
$$\begin{aligned}
\int_{\fldreal^{3}}
\frac{1+\napiernum^{-\sminvtemperature\omega(k)}}
{1-\napiernum^{-\sminvtemperature\omega(k)}}
\sum_{\mathrm{j}_{\txtrad}\in\ringratint_2}
\abs{\faftr{\mathfrak{b}}_{\mathrm{i},x,\kappa,\Lambda}^{\mathrm{j}_{\txtrad}}(k)}^2
\opdmsr{k}
\leq
2c_{\sminvtemperature,\kappa}
\int_{\kappa\leq\abs{k}\leq\Lambda}
\abs{k}\opdmsr{k}
<\infty
\end{aligned}$$
must hold.
Thus $\mathfrak{b}_{\mathrm{i},x,\kappa,\Lambda}
\in
\sphilb{D}_{\txtrad,\txtnonzero,\sminvtemperature}$ is obtained.

(Zero-mode check): Using the same support estimate,
$$\begin{aligned}
\int_{\fldreal^3}
\abs{\faftr{\mathfrak{b}}_{\mathrm{i},x,\kappa,\Lambda}^{\mathrm{j}_{\txtrad}}(k)}
\opdmsr{k}
\leq
\int_{\kappa\leq\abs{k}\leq\Lambda}
\abs{k}^{\onehalf}\opdmsr{k}
<\infty,
\quad
\int_{\fldreal^3}
\abs{\faftr{\mathfrak{b}}_{\mathrm{i},x,\kappa,\Lambda}^{\mathrm{j}_{\txtrad}}(k)}^2
\opdmsr{k}
<\infty
\end{aligned}$$
is obtained, and
$\mathfrak{b}_{\mathrm{i},x,\kappa,\Lambda}
\in
\sphilb{D}_{\txtrad,0,\sminvtemperature}$ holds.
Because of the infrared cutoff,
$\faftr{\mathfrak{b}}_{\mathrm{i},x,\kappa,\Lambda}$ vanishes near $k
= 0$, so it does not contribute to the zero-mode form
$\opform{q}_{\txtrad,0,\sminvtemperature}$.

This gives $\mathfrak{b}_{\mathrm{i},x,\kappa,\Lambda}
\in
\sphilb{D}_{\txtrad,0,\sminvtemperature}
\cap
\sphilb{D}_{\txtrad,\txtnonzero,\sminvtemperature}
=
\sphilb{D}_{\txtrad,\sminvtemperature}$.
Since the thermal kernel in \eqref{eq:beta-pf-covariance} is bounded on the support,
the value of the covariance form is finite for arbitrary $t,s$.
\end{proof}

Fix a particle loop \(\prbprocess\) and a spin loop \(\prbprocess^{\txtspin}\). For arbitrary \(I
\subset S_{\sminvtemperature}\), define the radiation-field source arising from the spin-diagonal Pauli term by \begin{equation}\label{expedition0012253}
\mathsf{I}^{\txtpauli,\txtdiag}_{\txtrad,I,\kappa,\Lambda}
(\prbprocess,\prbprocess^{\txtspin})
=
-\frac{\physcharge}{2}
\int_{I}
\prbprocess_t^{\txtspin}
\fun{j_t}
{\mathfrak{b}_{3,\prbprocess_t,\kappa,\Lambda}}
\opdmsr{t}
\end{equation} and define the radiation-field source arising from the off-diagonal Pauli term at the jump time \(\tau_j\) by \begin{equation}\label{expedition0012254}
\mathsf{I}^{\txtpauli,\txtflip}_{\txtrad,j,\kappa,\Lambda}
(\prbprocess,\prbprocess^{\txtspin})
=
-\frac{\physcharge}{2}
\rbk{\fun{j_{\tau_j}}
{\mathfrak{b}_{1,\prbprocess_{\tau_j},\kappa,\Lambda}}
-\imunit \prbprocess_{\tau_j-}^{\txtspin}
\cdot
\fun{j_{\tau_j}}
{\mathfrak{b}_{2,\prbprocess_{\tau_j},\kappa,\Lambda}}}.
\end{equation} An empty product is understood as \(1\).

\subsection{Bounded System with Spin and Infinite-Volume Limit}\label{bounded-system-with-spin-and-infinite-volume-limit}

As in the spinless model, we define the cutoff state of the model with spin as an infinite-volume limit through bounded systems. Below, for the bounded-system projection \(P_{L}\), the bounded-system free radiation-field measure \(\msrprb
_{\txtrad,\sminvtemperature,\smchemicalpotential,L}\), and the bounded-system covariance form \(\opform{q}
_{\txtrad,\txtbec,\sminvtemperature,\smchemicalpotential,L}^{\txteuclid}\), we use the objects defined in Subsection \ref{expedition0012299}.

\begin{defn}[Bounded-System Pauli--Fierz Model with Spin]\label{expedition0012294}
Let the spin space be $\sphilb{H}_{\txtspin}
= \fun{\lp^{2}}{\ringratint_{2}}$,
and let the total Hilbert space of the bounded system be $\mathcal{H}_{\txtpaulifierz,\txtspin,L}
=
\sphilb{H}_{\txtspin}
\otimes
\sphilb{H}_{\txtparticle}
\otimes
\spfock_{\txtrad,L}$.
Define the bounded-system version of the magnetic-flux-density test functions with cutoffs componentwise by
$\mathfrak{b}_{\mathrm{i},x,\kappa,\Lambda,L}
=P_{L}\mathfrak{b}_{\mathrm{i},x,\kappa,\Lambda}$,
and denote the bounded-system magnetic flux density with cutoffs by
$\physmfdensity_{\mathrm{i},\kappa,\Lambda,L}(x)
=\opfocksegalradiation(j_0\mathfrak{b}_{\mathrm{i},x,\kappa,\Lambda,L})$.
Define the bounded-system Hamiltonian with cutoffs and the Hamiltonian with chemical potential by
$$\begin{aligned}
\physham_{\txtpaulifierz,\txtspin,\kappa,\Lambda,L}
&=
\onehalf
\rbk{\qmpaulispin\vainnprod
\rbk{-\imunit\vagrad-\physcharge A_{\kappa,\Lambda,L}}}^{2}
+V
+\physham_{\txtrad,L},
\\ %%%%%%%%%%%%%%%%
\physham_{\txtpaulifierz,\txtspin,\kappa,\Lambda,\smchemicalpotential,L}
&=
\onehalf
\rbk{\qmpaulispin
\vainnprod
\rbk{-\imunit \vagrad - \physcharge A_{\kappa,\Lambda,L}}}^{2}
+V
+\physham_{\txtrad,\smchemicalpotential,L}.
\end{aligned}$$
The first formula has the rewriting
$$\onehalf
\rbk{-\imunit\vagrad-\physcharge A_{\kappa,\Lambda,L}}^{2}
-\frac{\physcharge}{2}\qmpaulispin\vainnprod \physmfdensity_{\kappa,\Lambda,L}
+V+\physham_{\txtrad,L},$$
and similarly for the second formula.
\end{defn}

Denote the free system combining the particle--spin paths and the bounded-system free radiation field by \[\pairbk{\prbqspace_{\txtpaulifierz,\txtspin,\sminvtemperature,L},
\mblfml{S}_{\txtpaulifierz,\txtspin,\sminvtemperature,L},
\mblfml{S}_{\txtpaulifierz,\txtspin,0,\sminvtemperature,L},
U_{\txtpaulifierz,\txtspin,L,t},
R_{\txtpaulifierz,\txtspin,L},
\msrprb_{\txtpaulifierz,\txtspin,0,\sminvtemperature,\smchemicalpotential,L}}.\] Here \(\mblfml{S}_{\txtspin,\sminvtemperature}
=
\fun{\mblfmlborel}{\fun{D}{S_{\sminvtemperature};\ringratint_2}}\) is the measurable \(\sigma\)-algebra defined in Subsection \ref{expedition0012381}, \(U_{\txtspin,t}\) is time translation on the circle \(S_{\sminvtemperature}\), and \(R_{\txtspin}\) is the map on spin path space induced by time reflection. Also, \[\begin{aligned}
\prbqspace_{\txtpaulifierz,\txtspin,\sminvtemperature,L}
&=
\Omega_{\txtparticle,\txtspin,\sminvtemperature}
\times
\prbqspace_{\txtrad,\sminvtemperature,L},
\\ %%%%%%%%%%%%%%%%
\mblfml{S}_{\txtpaulifierz,\txtspin,\sminvtemperature,L}
&=
\mblfml{S}_{\txtparticle,\sminvtemperature}
\times
\mblfml{S}_{\txtspin,\sminvtemperature}
\times
\mblfml{S}_{\txtrad,\sminvtemperature,L},
\\ %%%%%%%%%%%%%%%%
\mblfml{S}_{\txtpaulifierz,\txtspin,0,\sminvtemperature,L}
&=
\mblfml{S}_{\txtparticle,0,\sminvtemperature}
\times
\mblfml{S}_{\txtspin,\sminvtemperature}
\times
\mblfml{S}_{\txtrad,0,\sminvtemperature,L},
\\ %%%%%%%%%%%%%%%%
U_{\txtpaulifierz,\txtspin,L,t}
&=
U_{\txtparticle,t}\otimes U_{\txtspin,t}\otimes U_{\txtrad,L,t},
\\ %%%%%%%%%%%%%%%%
R_{\txtpaulifierz,\txtspin,L}
&=
R_{\txtparticle}\otimes R_{\txtspin}\otimes R_{\txtrad,L},
\\ %%%%%%%%%%%%%%%%
\msrprb_{\txtpaulifierz,\txtspin,0,\sminvtemperature,\smchemicalpotential,L}
&=
\msrprb_{\txtparticle,\txtspin,\sminvtemperature}
\otimes
\msrprb_{\txtrad,\sminvtemperature,\smchemicalpotential,L}.
\end{aligned}\]

In the bounded system, for both the current source with cutoffs and the magnetic-flux-density test functions, we use objects to which \(P_{L}\) has been applied. In addition to \(\mathsf{I}_{\txtrad,I,\kappa,\Lambda,L}\) defined in \eqref{expedition0012300}, for arbitrary \(I
\subset S_{\sminvtemperature}\), particle loop \(\prbprocess\), and spin loop \(\prbprocess^{\txtspin}\), define \begin{equation}\label{expedition0012380}
\begin{aligned}
\mathsf{I}^{\txtpauli,\txtdiag}_{\txtrad,I,\kappa,\Lambda,L}
(\prbprocess,\prbprocess^{\txtspin})
&=
-\frac{\physcharge}{2}
\int_{I}
\prbprocess_t^{\txtspin}
\fun{j_t}
{\mathfrak{b}_{3,\prbprocess_t,\kappa,\Lambda,L}}
\opdmsr{t},
\\ %%%%%%%%%%%%%%%%
\mathsf{I}^{\txtpauli,\txtflip}_{\txtrad,j,\kappa,\Lambda,L}
(\prbprocess,\prbprocess^{\txtspin})
&=
-\frac{\physcharge}{2}
\rbk{\fun{j_{\tau_j}}
{\mathfrak{b}_{1,\prbprocess_{\tau_j},\kappa,\Lambda,L}}
-\imunit \prbprocess_{\tau_j-}^{\txtspin}
\cdot
\fun{j_{\tau_j}}
{\mathfrak{b}_{2,\prbprocess_{\tau_j},\kappa,\Lambda,L}}}.
\end{aligned}
\end{equation}

\begin{defn}[Bounded-System Local Kernel with Spin]
For an interval $I
\subset S_{\sminvtemperature}$, define the bounded-system local kernel with spin before integrating
out the radiation field by
$$\begin{aligned}
&F_{\txtpaulifierz,\txtspin,I,\kappa,\Lambda,L}
(\prbprocess,\prbprocess^{\txtspin},\opfocksegalradiation)
\\ %%%%%%%%%%%%%%%%
&=
\fnexp{\imunit
\fun{\opfocksegalradiation}
{-\physcharge
\mathsf{I}_{\txtrad,I,\kappa,\Lambda,L}(\prbprocess)
+\mathsf{I}^{\txtpauli,\txtdiag}_{\txtrad,I,\kappa,\Lambda,L}
(\prbprocess,\prbprocess^{\txtspin})}}
\prod_{\tau_j\in I}
\fun{\opfocksegalradiation}
{\mathsf{I}^{\txtpauli,\txtflip}_{\txtrad,j,\kappa,\Lambda,L}
(\prbprocess,\prbprocess^{\txtspin})}.
\end{aligned}$$
An empty product is understood as $1$.
\end{defn}

\begin{prop}[Bounded-System Perturbation Conditions for the Pauli--Fierz Kernel with Spin]\label{expedition0012382}
For each fixed $L$, the family $\fml{F_{\txtpaulifierz,\txtspin,I,\kappa,\Lambda,L}}
{I\subset S_{\sminvtemperature}}$ is a local Feynman--Kac--Nelson perturbation.
In particular, $F_{\txtpaulifierz,\txtspin,I,\kappa,\Lambda,L}$ is measurable with respect to the
$\sigma$-algebra corresponding to the interval $I$,
and for disjoint intervals $I,J
\subset S_{\sminvtemperature}$,
$F_{\txtpaulifierz,\txtspin,I\cup J,\kappa,\Lambda,L}
=
F_{\txtpaulifierz,\txtspin,I,\kappa,\Lambda,L}
F_{\txtpaulifierz,\txtspin,J,\kappa,\Lambda,L}$.
Furthermore, it is covariant under time translations and reflection and belongs to $\lp^{p}$ for arbitrary $1
\leq p
<\infty$.
\end{prop}

\begin{proof}
The factor without the spin component,
$\fnexp{-\imunit\physcharge
\opfocksegalradiation(\mathsf{I}_{\txtrad,I,\kappa,\Lambda,L}(\prbprocess))}$,
coincides with \eqref{expedition0012376} and is a bounded-system local
Feynman--Kac--Nelson perturbation by Proposition \ref{expedition0012328}.
Thus it remains only to verify that the same properties are preserved after multiplying this spinless
kernel by the factors of the Pauli term arising from the finite-dimensional spin paths.

The current source in \eqref{expedition0012300},
$\mathsf{I}^{\txtpauli,\txtdiag}_{\txtrad,I,\kappa,\Lambda,L}$ and
$\mathsf{I}^{\txtpauli,\txtflip}_{\txtrad,j,\kappa,\Lambda,L}$ defined in \eqref{expedition0012380},
are supported on $\setlattice_L^3
\cap
\set{k}{\kappa\leq\abs{k}\leq\Lambda}$.
This is a finite set, so the maps above are test functions for finite-dimensional Gaussian variables
of the bounded-system radiation field,
and any finite number of field values and their exponential moments are finite.
This finite-dimensionality is the same point used to treat the current source in
Propositions \ref{expedition0012325} and \ref{expedition0012328} for the spinless bounded system;
here we have only added the magnetic-flux-density test functions.

Interval measurability is obtained as follows.
The interval measurability of the spinless current source
$\mathsf{I}_{\txtrad,I,\kappa,\Lambda,L}$ has already been verified in Proposition \ref{expedition0012328}.
The diagonal Pauli source $\mathsf{I}^{\txtpauli,\txtdiag}_{\txtrad,I,\kappa,\Lambda,L}$
is a time integral determined only by the particle path and spin path on $I$,
and the jump Pauli source refers only to the jump times $\tau_j$ in $I$.
Therefore the whole product is measurable with respect to the $\sigma$-algebra corresponding to $I$.

For disjoint intervals $I,J$,
as in Proposition \ref{expedition0012328},
$\mathsf{I}_{\txtrad,I\cup J,\kappa,\Lambda,L}
=
\mathsf{I}_{\txtrad,I,\kappa,\Lambda,L}
+
\mathsf{I}_{\txtrad,J,\kappa,\Lambda,L}$ holds.
Furthermore, since the diagonal Pauli source is also a time integral,
$\mathsf{I}^{\txtpauli,\txtdiag}_{\txtrad,I\cup J,\kappa,\Lambda,L}
=
\mathsf{I}^{\txtpauli,\txtdiag}_{\txtrad,I,\kappa,\Lambda,L}
+\mathsf{I}^{\txtpauli,\txtdiag}_{\txtrad,J,\kappa,\Lambda,L}$ holds.
The set of jump times is the disjoint union of the jumps in $I$ and the jumps in $J$,
so the product of the jump Pauli sources also decomposes into the product over the two intervals.
This gives $F_{\txtpaulifierz,\txtspin,I\cup J,\kappa,\Lambda,L}
=F_{\txtpaulifierz,\txtspin,I,\kappa,\Lambda,L}
F_{\txtpaulifierz,\txtspin,J,\kappa,\Lambda,L}$.

For time translations and reflection as well,
the covariance of the spinless current source and the bounded-system radiation field is handled as in
Proposition \ref{expedition0012328}.
The spin path measure is invariant under time translations and reflection on the circle defined in
Subsection \ref{expedition0012381},
the jump times are moved by the same maps to the corresponding times,
and $\fun{j_t}{\mathfrak{b}_{\mathrm{i},x,\kappa,\Lambda,L}}$ has the covariance of sharp-time fields
under time translations and reflection.
Therefore the diagonal and jump sources of the Pauli term have the same covariance,
and the whole local kernel is covariant under time translations and reflection.

Finally, we verify the $\lp^p$ property.
If the number of jumps of the spin path is fixed to be $n$,
the bounded-system radiation field is a finite-dimensional centered Gaussian field,
and the local kernel is a product of one Gaussian exponential and $n$ Gaussian linear functionals.
For finite-dimensional Gaussian distributions, polynomial moments of arbitrary order and exponential moments
are finite, so the $p$-th moment is finite on each sector with $n$ jumps.
As in the proof of Proposition \ref{expedition0012280}, the mass of the $n$-jump sector of the closed
spin path measure is bounded by $\sminvtemperature^n/n!$.
At finite volume and fixed cutoff, the variance of each sharp-time magnetic flux density is uniformly finite,
so the Gaussian moment estimate is bounded by a term of type
$C_{p,\sminvtemperature,\kappa,\Lambda,L}^{n}\sqrt{n!}$.
Therefore $\sum_{n=0}^{\infty}
C_{p,\sminvtemperature,\kappa,\Lambda,L}^{n}\sqrt{n!}
\frac{\sminvtemperature^n}{n!}
<\infty$ is obtained,
and $F_{\txtpaulifierz,\txtspin,I,\kappa,\Lambda,L}$ belongs to $\lp^p$ for arbitrary
$1\leq p<\infty$.
\end{proof}

\begin{defn}[Bounded-System Interaction Factor with Spin]\label{expedition0012295}
For finitely many $f_a
\in\sphilb{D}_{\txtrad,\txtphys,\sminvtemperature}$ and times $t_a
\in S_{\sminvtemperature}$, set $h_L
=\sum_a j_{t_a}P_{L} f_a$.
For fixed particle path $\prbprocess$ and spin path $\prbprocess^{\txtspin}$, set
$$\begin{aligned}
\fun{G_{\kappa,\Lambda,L}}
{h_L,\prbprocess,\prbprocess^{\txtspin}}
&=
h_L-\physcharge
\mathsf{I}_{\txtrad,S_{\sminvtemperature},\kappa,\Lambda,L}(\prbprocess)
+\mathsf{I}^{\txtpauli,\txtdiag}_{\txtrad,S_{\sminvtemperature},\kappa,\Lambda,L}
(\prbprocess,\prbprocess^{\txtspin}),
\\ %%%%%%%%%%%%%%%%
\fun{G_{\kappa,\Lambda,L}}
{u,h_L,\prbprocess,\prbprocess^{\txtspin}}
&=
\fun{G_{\kappa,\Lambda,L}}
{h_L,\prbprocess,\prbprocess^{\txtspin}}
+\sum_{j=1}^{N_{\sminvtemperature}(\prbprocess^{\txtspin})}
u_j
\mathsf{I}^{\txtpauli,\txtflip}_{\txtrad,j,\kappa,\Lambda,L}
(\prbprocess,\prbprocess^{\txtspin}),
\end{aligned}$$
and define the bounded-system interaction factor with spin by the integral with respect to the radiation field
$$\begin{aligned}
&\fun{\mathsf{D}_{\txtpaulifierz,\txtspin,\sminvtemperature,\kappa,\Lambda,\smchemicalpotential,L}}
{h_L,\prbprocess,\prbprocess^{\txtspin}}
\\ %%%%%%%%%%%%%%%%
&=
\fnexp{\oneoverfour
\fun{\opform{q}_{\txtrad,\txtbec,\sminvtemperature,\smchemicalpotential,L}^{\txteuclid}}
{h_L}}
\\ %%%%%%%%%%%%%%%%
&\quad\times
\sqfun{\prbexp_{\msrprb_{\txtrad,\sminvtemperature,\smchemicalpotential,L}}}
{\fnexp{\imunit
\fun{\opfocksegalradiation}
{\fun{G_{\kappa,\Lambda,L}}
{h_L,\prbprocess,\prbprocess^{\txtspin}}}}
\prod_{j=1}^{N_{\sminvtemperature}(\prbprocess^{\txtspin})}
\fun{\opfocksegalradiation}
{\mathsf{I}^{\txtpauli,\txtflip}_{\txtrad,j,\kappa,\Lambda,L}
(\prbprocess,\prbprocess^{\txtspin})}}.
\end{aligned}$$
By the formula for centered Gaussian integrals,
the same quantity can also be written as
\begin{equation}\label{expedition0012296}
\begin{aligned}
&\fun{\mathsf{D}_{\txtpaulifierz,\txtspin,\sminvtemperature,\kappa,\Lambda,\smchemicalpotential,L}}
{h_L,\prbprocess,\prbprocess^{\txtspin}}
\\ %%%%%%%%%%%%%%%%
&=
\fnexp{\oneoverfour
\fun{\opform{q}_{\txtrad,\txtbec,\sminvtemperature,\smchemicalpotential,L}^{\txteuclid}}
{h_L}}
\imunit^{-N_{\sminvtemperature}(\prbprocess^{\txtspin})}
\\ %%%%%%%%%%%%%%%%
&\quad\times
\fnrestr{
\oppd{u_1}\cdots
\oppd{u_{N_{\sminvtemperature}(\prbprocess^{\txtspin})}}
\fnexp{-\oneoverfour
\fun{\opform{q}_{\txtrad,\txtbec,\sminvtemperature,\smchemicalpotential,L}^{\txteuclid}}
{\fun{G_{\kappa,\Lambda,L}}
{u,h_L,\prbprocess,\prbprocess^{\txtspin}}}}}
{u_1=\cdots=u_{N_{\sminvtemperature}(\prbprocess^{\txtspin})}=0}.
\end{aligned}
\end{equation}
\end{defn}

This definition is the bounded-system representation \eqref{eq:PF-equilibrium-functional-finite-volume} of the spinless Pauli--Fierz model with the diagonal and jump sources of the Pauli term, which arise from the spin-path expansion, added to it. The Gaussian integration part is the same computation of the centered Gaussian characteristic functional as in the bounded-system representation \eqref{expedition0012341} for the Nelson model and \eqref{expedition0012324} for the spinless model; the only difference is the addition of the differentiation variables \(u_j\) corresponding to the jump Pauli sources.

The partition function of the corresponding interacting system and the representation obtained by integrating out the field first are \[\begin{aligned}
&\smpartitionfunc_{\txtpaulifierz,\txtspin,\sminvtemperature,\kappa,\Lambda,\smchemicalpotential,L}
\\ %%%%%%%%%%%%%%%%
&=
\int_{\prbqspace_{\txtpaulifierz,\txtspin,\sminvtemperature,L}}
F_{\txtpaulifierz,\txtspin,S_{\sminvtemperature},\kappa,\Lambda,L}
(\prbprocess,\prbprocess^{\txtspin},\opfocksegalradiation)
\opdmsr{\msrprb_{\txtpaulifierz,\txtspin,0,\sminvtemperature,\smchemicalpotential,L}}
(\prbprocess,\prbprocess^{\txtspin},\opfocksegalradiation)
\\ %%%%%%%%%%%%%%%%
&=
\int_{\Omega_{\txtparticle,\txtspin,\sminvtemperature}}
\mathsf{D}_{\txtpaulifierz,\txtspin,\sminvtemperature,\kappa,\Lambda,\smchemicalpotential,L}
(0,\prbprocess,\prbprocess^{\txtspin})
\opdmsr{\msrprb_{\txtparticle,\txtspin,\sminvtemperature}}(\prbprocess,\prbprocess^{\txtspin}).
\end{aligned}\] This value agrees with the trace representation of the bounded-system Hamiltonian, and is therefore positive and finite. Furthermore, define the interacting particle--spin path probability measure by \begin{equation}\label{expedition0012436}
\begin{aligned}
\opdmsr{\widetilde{\msrprb}_{\txtparticle,\txtspin,\sminvtemperature,\kappa,\Lambda,\smchemicalpotential,L}}
(\prbprocess,\prbprocess^{\txtspin})
=
\frac{\mathsf{D}_{\txtpaulifierz,\txtspin,\sminvtemperature,\kappa,\Lambda,\smchemicalpotential,L}
(0,\prbprocess,\prbprocess^{\txtspin})}
{\smpartitionfunc_{\txtpaulifierz,\txtspin,\sminvtemperature,\kappa,\Lambda,\smchemicalpotential,L}}
\opdmsr{\msrprb_{\txtparticle,\txtspin,\sminvtemperature}}
(\prbprocess,\prbprocess^{\txtspin}).
\end{aligned}
\end{equation}

\begin{defn}[KMS State of the Bounded-System Pauli--Fierz Model with Spin]\label{expedition0012297}
Fix the side length $L$ and cutoffs $0<\kappa<\Lambda<\infty$.
For finitely many $f_a
\in\sphilb{D}_{\txtrad,\txtphys,\sminvtemperature}$ and times $t_a
\in S_{\sminvtemperature}$, set $h_L
= \sum_a j_{t_a}P_{L} f_a$.
Then the right-hand side of \eqref{expedition0012296} is integrable,
and $\smpartitionfunc_{\txtpaulifierz,\txtspin,\sminvtemperature,\kappa,\Lambda,\smchemicalpotential,L}$
can be used as a normalizing denominator.
Thus, for bounded particle--spin functions $B$, define the bounded-system KMS state by
$$\begin{aligned}
&\fun{\oastate[\psi_{\txtpaulifierz,\txtspin,\sminvtemperature,\kappa,\Lambda,\smchemicalpotential,L}]}
{B(\prbprocess,\prbprocess^{\txtspin})
\fnexp{\imunit\opfocksegalradiation(h_L)}}
\\ %%%%%%%%%%%%%%%%
&=
\fnexp{-\oneoverfour
\fun{\opform{q}_{\txtrad,\txtbec,\sminvtemperature,\smchemicalpotential,L}^{\txteuclid}}
{h_L}}
\\ %%%%%%%%%%%%%%%%
&\quad\times
\frac{\int
B(\prbprocess,\prbprocess^{\txtspin})
\mathsf{D}_{\txtpaulifierz,\txtspin,\sminvtemperature,\kappa,\Lambda,\smchemicalpotential,L}
(h_L,\prbprocess,\prbprocess^{\txtspin})
\opdmsr{\msrprb_{\txtparticle,\txtspin,\sminvtemperature}}(\prbprocess,\prbprocess^{\txtspin})}
{\smpartitionfunc_{\txtpaulifierz,\txtspin,\sminvtemperature,\kappa,\Lambda,\smchemicalpotential,L}}.
\end{aligned}$$
\end{defn}

\begin{proof}
We verify well-definedness.
For the spinless bounded system, the local kernel belongs to $\lp^p$ by Proposition \ref{expedition0012328},
and the bounded-system KMS state is defined by \eqref{eq:PF-equilibrium-functional-finite-volume}.
Proposition \ref{expedition0012382} verified that the $\lp^p$ property is preserved even after
the finite-dimensional Gaussian polynomial factors of the Pauli term are multiplied into this local kernel.
Therefore $F_{\txtpaulifierz,\txtspin,S_{\sminvtemperature},\kappa,\Lambda,L}$ is integrable with
respect to the free product measure.

When the field is integrated out first,
$\mathsf{D}_{\txtpaulifierz,\txtspin,\sminvtemperature,\kappa,\Lambda,\smchemicalpotential,L}$
of Definition \ref{expedition0012295} appears.
For a fixed number of jumps $n$, \eqref{expedition0012296} is a finite sum consisting of finitely many
covariance forms, and each term is finite by finite-dimensionality of the bounded-system covariance.
The $n$-jump sector of the closed spin path measure is bounded by $\sminvtemperature^n/n!$,
and together with the Gaussian moment estimate in Proposition \ref{expedition0012382},
this is summable in $n$.
Thus the numerator converges absolutely even after multiplying by bounded $B$.

The denominator is the numerator in the case $B=1$ and $h_L=0$,
and agrees with the thermal trace representation of the bounded-system Hamiltonian.
At finite volume and fixed cutoffs, the heat operator is trace class,
and the preceding integrability argument identifies the same quantity as the normalizing denominator of the probabilistic KMS functional.
Thus the partition function is finite and strictly positive,
and the right-hand side of the defining formula is meaningful as a finite value.
\end{proof}

\subsection{KMS State with Spin and Cutoffs}\label{kms-state-with-spin-and-cutoffs}

In the following definitions, we use the current source with cutoffs \(\mathsf{I}_{\txtrad,I,\kappa,\Lambda}(\prbprocess)\) in \eqref{expedition0012223} and the magnetic-flux-density test function with cutoffs \(\mathfrak{b}_{\mathrm{i},x,\kappa,\Lambda}\) in \eqref{expedition0012252}.

Let the sharp-time representation of the magnetic flux density with cutoffs be \[\begin{aligned}
\physmfdensity_{\kappa,\Lambda,\mathrm{i}}(t,x,\opfocksegalradiation)
=
\opfocksegalradiation
\rbk{j_t\mathfrak{b}_{\mathrm{i},x,\kappa,\Lambda}},
\end{aligned}\] and for an interval \(I
\subset S_{\sminvtemperature}\), define the matrix-valued function \(\mathcal{A}_{I,\kappa,\Lambda}(t;\prbprocess,\opfocksegalradiation)
=
\frac{\physcharge}{2}
\sum_{\mathrm{i}=1}^{3}
\qmpaulispin_{\mathrm{i}}
\physmfdensity_{\kappa,\Lambda,\mathrm{i}}(t,\prbprocess_t,\opfocksegalradiation)\).

\begin{prop}[Time-Ordered Exponential of the Pauli Term with Spin and Cutoffs]\label{expedition0012386}
For fixed cutoffs $0<\kappa<\Lambda<\infty$ and an interval $I
\subset S_{\sminvtemperature}$,
for almost every $(\prbprocess,\opfocksegalradiation)$ with respect to the free particle--radiation
field measure,
$$\int_I
\norm{\mathcal{A}_{I,\kappa,\Lambda}(t;\prbprocess,\opfocksegalradiation)}
_{\opspbddlin{\sphilb{H}_{\txtspin}}}
\opdmsr{t}
<\infty$$
holds.
Then the integral equation
$$U(t)
=
\idone_{\sphilb{H}_{\txtspin}}
+
\int_{\inf I}^{t}
\mathcal{A}_{I,\kappa,\Lambda}(s;\prbprocess,\opfocksegalradiation)U(s)
\opdmsr{s}$$
has a unique absolutely continuous solution on $\sphilb{H}_{\txtspin}$.
In particular, the time-ordered exponential $\physoptimeorder\fnexp{\int_I
\mathcal{A}_{I,\kappa,\Lambda}(t;\prbprocess,\opfocksegalradiation)
\opdmsr{t}}$ is meaningful as the terminal-time value of this unique solution
and has the Dyson series representation, absolutely convergent in operator norm,
$$\begin{aligned}
&\physoptimeorder
\fnexp{\int_I
\mathcal{A}_{I,\kappa,\Lambda}(t;\prbprocess,\opfocksegalradiation)
\opdmsr{t}}
\\ %%%%%%%%%%%%%%%%
&=
\idone_{\sphilb{H}_{\txtspin}}
+\sum_{n=1}^{\infty}
\int_{t_1<\cdots<t_n,\ t_j\in I}
\mathcal{A}_{I,\kappa,\Lambda}(t_n;\prbprocess,\opfocksegalradiation)
\cdots
\mathcal{A}_{I,\kappa,\Lambda}(t_1;\prbprocess,\opfocksegalradiation)
\opdmsr{t_1}\cdots\opdmsr{t_n}.
\end{aligned}$$
Furthermore,
$$\begin{aligned}
\norm{\physoptimeorder
\fnexp{\int_I
\mathcal{A}_{I,\kappa,\Lambda}(t;\prbprocess,\opfocksegalradiation)
\opdmsr{t}}}_{\opspbddlin{\sphilb{H}_{\txtspin}}}
\leq
\fnexp{\frac{\abs{\physcharge}}{2}
\int_I
\sum_{\mathrm{i}=1}^{3}
\abs{\physmfdensity_{\kappa,\Lambda,\mathrm{i}}(t,\prbprocess_t,\opfocksegalradiation)}
\opdmsr{t}}
\end{aligned}$$
holds.
\end{prop}

The Dyson series is only a convenient representation of the time-ordered exponential and is not essential for the definitions below. In this paper, we interpret the time-ordered exponential as the unique solution of the integral equation in Proposition \ref{expedition0012386}, and use the Dyson series only as an absolutely convergent representation giving the same solution.

\begin{proof}
By Proposition \ref{expedition0012279}, for each $x,
t,
\mathrm{i}$, $\fun{j_t}
{\mathfrak{b}_{\mathrm{i},x,\kappa,\Lambda}}$ belongs to the domain of the covariance form
of the radiation field.
At fixed cutoff, the momentum support is contained in $\set{k}{\kappa\leq\abs{k}\leq\Lambda}$,
so the same estimate as in the proof of Proposition \ref{expedition0012279} gives
$$\sup_{x,t,\mathrm{i}}
\fun{\opform{q}_{\txtrad,\txtbec,\sminvtemperature}^{\txteuclid}}
{j_t\mathfrak{b}_{\mathrm{i},x,\kappa,\Lambda}}
<\infty.$$
Therefore $\physmfdensity
_{\kappa,\Lambda,\mathrm{i}}(t,x,\opfocksegalradiation)$ has finite Gaussian moments uniformly in $t,
x,
\mathrm{i}$.
By Fubini's theorem, for almost every $(\prbprocess,\opfocksegalradiation)$,
$t
\mapsto
\physmfdensity_{\kappa,\Lambda,\mathrm{i}}(t,\prbprocess_t,\opfocksegalradiation)$ is integrable on $I$.
Since the operator norm of each Pauli matrix is $1$,
$$\norm{\mathcal{A}_{I,\kappa,\Lambda}(t;\prbprocess,\opfocksegalradiation)}
_{\opspbddlin{\sphilb{H}_{\txtspin}}}
\leq
\frac{\abs{\physcharge}}{2}
\sum_{\mathrm{i}=1}^{3}
\abs{\physmfdensity_{\kappa,\Lambda,\mathrm{i}}(t,\prbprocess_t,\opfocksegalradiation)},$$
and the time integral of the left-hand side is finite.

A linear integral equation with integrable matrix-valued coefficients on the finite-dimensional space
$\sphilb{H}_{\txtspin}$ has a unique absolutely continuous solution by successive approximation.
Estimating the $n$-th term of the successive approximation in operator norm gives
$$\begin{aligned}
&\int_{t_1<\cdots<t_n,\ t_j\in I}
\prod_{\ell=1}^{n}
\norm{\mathcal{A}_{I,\kappa,\Lambda}(t_\ell;\prbprocess,\opfocksegalradiation)}
_{\opspbddlin{\sphilb{H}_{\txtspin}}}
\opdmsr{t_1}\cdots\opdmsr{t_n}
\\
&\leq
\frac{1}{n!}
\rbk{\int_I
\norm{\mathcal{A}_{I,\kappa,\Lambda}(t;\prbprocess,\opfocksegalradiation)}
_{\opspbddlin{\sphilb{H}_{\txtspin}}}
\opdmsr{t}}^n.
\end{aligned}$$
Therefore the Dyson series converges absolutely in operator norm,
and its sum agrees with the unique solution of the integral equation above.
The final norm estimate follows from the same inequality.
\end{proof}

\begin{defn}[Local Kernel with Spin and Cutoffs]\label{expedition0012282}
Using the time-ordered exponential defined in Proposition \ref{expedition0012386},
define the local kernel with spin and cutoffs before Gaussian integration,
as a matrix-valued function on the spin space $\sphilb{H}_{\txtspin}
= \lp^2(\ringratint_2)$, by
\begin{equation}\label{expedition0012288}
\begin{aligned}
F_{\txtpaulifierz,\txtspin,I,\kappa,\Lambda}
(\prbprocess,\opfocksegalradiation)
=
\fnexp{
-\imunit
\physcharge
\fun{\opfocksegalradiation}
{\mathsf{I}_{\txtrad,I,\kappa,\Lambda}(\prbprocess)}}
\mathrm{T}\fnexp{\int_I
\mathcal{A}_{I,\kappa,\Lambda}(t;\prbprocess,\opfocksegalradiation)
\opdmsr{t}}.
\end{aligned}
\end{equation}
\end{defn}

\begin{defn}[Pauli--Fierz Interaction Factor with Spin and Cutoffs]\label{expedition0012255}
Denote an arbitrary finite linear combination of $j_t f$ by $h$.
For fixed particle path $\prbprocess$ and spin path $\prbprocess^{\txtspin}$, define
$$\begin{aligned}
\fun{G_{\kappa,\Lambda}}
{h,\prbprocess,\prbprocess^{\txtspin}}
&=
h-\physcharge
\mathsf{I}_{\txtrad,S_{\sminvtemperature},\kappa,\Lambda}(\prbprocess)
+\mathsf{I}^{\txtpauli,\txtdiag}_{\txtrad,S_{\sminvtemperature},\kappa,\Lambda}
(\prbprocess,\prbprocess^{\txtspin}),
\\ %%%%%%%%%%%%%%%%
\fun{G_{\kappa,\Lambda}}
{u,h,\prbprocess,\prbprocess^{\txtspin}}
&=
\fun{G_{\kappa,\Lambda}}
{h,\prbprocess,\prbprocess^{\txtspin}}
+\sum_{j=1}^{N_{\sminvtemperature}(\prbprocess^{\txtspin})}
u_j
\mathsf{I}^{\txtpauli,\txtflip}_{\txtrad,j,\kappa,\Lambda}(\prbprocess,\prbprocess^{\txtspin}).
\end{aligned}$$
Define the interaction factor with spin and cutoffs by the integral with respect to the radiation field
\begin{equation}\label{expedition0012256}
\begin{aligned}
&\fun{\mathsf{D}_{\txtpaulifierz,\txtspin,\sminvtemperature,\kappa,\Lambda}}
{h,\prbprocess,\prbprocess^{\txtspin}}
\\ %%%%%%%%%%%%%%%%
&=
\fnexp{\oneoverfour
\fun{\opform{q}_{\txtrad,\txtbec,\sminvtemperature}^{\txteuclid}}
{h}}
\sqfun{\prbexp_{\msrprb_{\txtrad,\sminvtemperature}}}
{\fnexp{\imunit
\fun{\opfocksegalradiation}
{\fun{G_{\kappa,\Lambda}}
{h,\prbprocess,\prbprocess^{\txtspin}}}}
\prod_{j=1}^{N_{\sminvtemperature}(\prbprocess^{\txtspin})}
\fun{\opfocksegalradiation}
{\mathsf{I}^{\txtpauli,\txtflip}_{\txtrad,j,\kappa,\Lambda}
(\prbprocess,\prbprocess^{\txtspin})}}.
\end{aligned}
\end{equation}
An empty product is understood as $1$.
Using the formula for centered Gaussian integrals, one obtains the equivalent expression
\begin{equation}\label{expedition0012283}
\begin{aligned}
&\fun{\mathsf{D}_{\txtpaulifierz,\txtspin,\sminvtemperature,\kappa,\Lambda}}
{h,\prbprocess,\prbprocess^{\txtspin}}
\\ %%%%%%%%%%%%%%%%
&=
\fnexp{\oneoverfour
\fun{\opform{q}_{\txtrad,\txtbec,\sminvtemperature}^{\txteuclid}}
{h}}
\imunit^{-N_{\sminvtemperature}(\prbprocess^{\txtspin})}
\\ %%%%%%%%%%%%%%%%
&\quad\times
\fnrestr{
\oppd{u_1}\cdots
\oppd{u_{N_{\sminvtemperature}(\prbprocess^{\txtspin})}}
\fnexp{-\oneoverfour
\fun{\opform{q}_{\txtrad,\txtbec,\sminvtemperature}^{\txteuclid}}
{\fun{G_{\kappa,\Lambda}}
{u,h,\prbprocess,\prbprocess^{\txtspin}}}}}
{u_1=\cdots=u_{N_{\sminvtemperature}(\prbprocess^{\txtspin})}=0}.
\end{aligned}
\end{equation}
\end{defn}

This quantity is the unnormalized interaction factor on particle--spin paths after Gaussian integration of the radiation field. The quantity \(\mathsf{D}_{\txtpaulifierz,\txtspin,\sminvtemperature,\kappa,\Lambda}\) in Definition \ref{expedition0012255} is not defined by Gaussian integration of each order of the Dyson series for the time-ordered exponential in Definition \ref{expedition0012282}. It is obtained by expanding the diagonal and off-diagonal components of the Pauli matrices by the spin-flip process and then Gaussian-integrating the spin path representation split into the diagonal source \(\mathsf{I}^{\txtpauli,\txtdiag}_{\txtrad,S_{\sminvtemperature},\kappa,\Lambda}\) and the jump source \(\mathsf{I}^{\txtpauli,\txtflip}_{\txtrad,j,\kappa,\Lambda}\). Therefore, convergence estimates for \(\mathsf{D}\) do not assume an exchange of the Dyson series and Gaussian integration, but use Gaussian moment estimates for each jump number.

\begin{prop}[Exponential Integrability of the Local Kernel with Spin and Cutoffs]\label{expedition0012281}
Fix cutoffs $0<\kappa<\Lambda<\infty$ and an arbitrary finite linear combination $h$ of $j_t f$.
For any interval $I
\subset S_{\sminvtemperature}$ and any $1
\leq p
< \infty$,
the local kernel with spin and cutoffs in Definition \ref{expedition0012282} satisfies
$$\begin{aligned}
\norm{F_{\txtpaulifierz,\txtspin,I,\kappa,\Lambda}
(\prbprocess,\opfocksegalradiation)}
_{\opspbddlin{\sphilb{H}_{\txtspin}}}
\in
\fun{\lp^p}{\Omega_{\txtparticle,\sminvtemperature}
\times\prbqspace_{\txtrad,\sminvtemperature},
\msrprb_{\txtparticle,\sminvtemperature}
\otimes
\msrprb_{\txtrad,\sminvtemperature}}.
\end{aligned}$$
In particular, the interaction factor obtained by first integrating the radiation field on the full circle
$S_{\sminvtemperature}$ satisfies
$$\begin{aligned}
\fun{\mathsf{D}_{\txtpaulifierz,\txtspin,\sminvtemperature,\kappa,\Lambda}}
{h,\prbprocess,\prbprocess^{\txtspin}}
\in
\fun{\lp^{1}}
{\Omega_{\txtparticle,\txtspin,\sminvtemperature},
\msrprb_{\txtparticle,\txtspin,\sminvtemperature}},
\end{aligned}$$
and for any bounded measurable function $B$,
$$\begin{aligned}
\int_{\Omega_{\txtparticle,\txtspin,\sminvtemperature}}
\abs{
B(\prbprocess,\prbprocess^{\txtspin})
\fun{\mathsf{D}_{\txtpaulifierz,\txtspin,\sminvtemperature,\kappa,\Lambda}}
{h,\prbprocess,\prbprocess^{\txtspin}}}
\opdmsr{\msrprb_{\txtparticle,\txtspin,\sminvtemperature}}
(\prbprocess,\prbprocess^{\txtspin})
<\infty
\end{aligned}$$
holds.
In particular, for $h
=0$ and $B
=1$, the integral converges absolutely.
\end{prop}

\begin{proof}
We estimate at fixed cutoff.
By Proposition \ref{expedition0012221}, for each particle path $\prbprocess$,
the current source with cutoffs
$\mathsf{I}_{\txtrad,S_{\sminvtemperature},\kappa,\Lambda}(\prbprocess)$
is a Gaussian test function for the radiation field.
By Proposition \ref{expedition0012279}, for each $x,\mathrm{i},t$,
$\fun{j_t}
{\mathfrak{b}_{\mathrm{i},x,\kappa,\Lambda}}$ also belongs to the domain of the same covariance form.
Furthermore, boundedness on $\kappa\leq\abs{k}\leq\Lambda$ gives
\begin{equation}\label{expedition0012284}
\sup_{x,t,\mathrm{i}}
\fun{\opform{q}_{\txtrad,\txtbec,\sminvtemperature}^{\txteuclid}}
{j_t\mathfrak{b}_{\mathrm{i},x,\kappa,\Lambda}}
\leq
c_{\sminvtemperature,\kappa,\Lambda}
\end{equation}
for some constant $c_{\sminvtemperature,\kappa,\Lambda}<\infty$.
By the Cauchy--Schwarz inequality for the covariance form and
$\mathsf{I}^{\txtpauli,\txtdiag}_{\txtrad,S_{\sminvtemperature},\kappa,\Lambda}$
in \eqref{expedition0012253},
$$\begin{aligned}
&\fun{\opform{q}_{\txtrad,\txtbec,\sminvtemperature}^{\txteuclid}}
{\mathsf{I}^{\txtpauli,\txtdiag}_{\txtrad,S_{\sminvtemperature},\kappa,\Lambda}(\prbprocess,\prbprocess^{\txtspin})}
\\ %%%%%%%%%%%%%%%%
&=
\frac{\abs{\physcharge}^{2}}{4}
\int_{S_{\sminvtemperature}}
\int_{S_{\sminvtemperature}}
\prbprocess_t^{\txtspin}
\prbprocess_s^{\txtspin}
\opform{q}_{\txtrad,\txtbec,\sminvtemperature}^{\txteuclid}
\rbk{j_t\mathfrak{b}_{3,\prbprocess_t,\kappa,\Lambda},
j_s\mathfrak{b}_{3,\prbprocess_s,\kappa,\Lambda}}
\opdmsr{t}\opdmsr{s}
\\ %%%%%%%%%%%%%%%%
&\leq
\frac{\abs{\physcharge}^{2}}{4}
\int_{S_{\sminvtemperature}}
\int_{S_{\sminvtemperature}}
\prbprocess_t^{\txtspin}
\prbprocess_s^{\txtspin}
\fun{\opform{q}_{\txtrad,\txtbec,\sminvtemperature}^{\txteuclid}}
{j_t\mathfrak{b}_{3,\prbprocess_t,\kappa,\Lambda}}^{\onehalf}
\fun{\opform{q}_{\txtrad,\txtbec,\sminvtemperature}^{\txteuclid}}
{j_s\mathfrak{b}_{3,\prbprocess_t,\kappa,\Lambda}}^{\onehalf}
\opdmsr{t}\opdmsr{s}
\\ %%%%%%%%%%%%%%%%
&=
\frac{\abs{\physcharge}^{2}}{4}
\rbk{\int_{S_{\sminvtemperature}}
\abs{\prbprocess_t^{\txtspin}}
\fun{\opform{q}_{\txtrad,\txtbec,\sminvtemperature}^{\txteuclid}}
{j_t\mathfrak{b}_{3,\prbprocess_t,\kappa,\Lambda}}^{\onehalf}
\opdmsr{t}}^2
\\ %%%%%%%%%%%%%%%%
&\leq
\frac{\abs{\physcharge}^{2}}{4}
\sminvtemperature^2
c_{\sminvtemperature,\kappa,\Lambda}
\end{aligned}$$
is obtained.
The source $\mathsf{I}^{\txtpauli,\txtflip}_{\txtrad,j,\kappa,\Lambda}$ in \eqref{expedition0012254}
is a linear combination of two magnetic-flux-density test functions,
and the estimate $\abs{\prbprocess_{\tau_j-}^{\txtspin}}
= 1$ gives
$$\begin{aligned}
\fun{\opform{q}_{\txtrad,\txtbec,\sminvtemperature}^{\txteuclid}}
{\mathsf{I}^{\txtpauli,\txtflip}_{\txtrad,j,\kappa,\Lambda}(\prbprocess,\prbprocess^{\txtspin})}
&\leq
\frac{\abs{\physcharge}^{2}}{2}
\rbk{\fun{\opform{q}_{\txtrad,\txtbec,\sminvtemperature}^{\txteuclid}}
{j_{\tau_j}\mathfrak{b}_{1,\prbprocess_{\tau_j},\kappa,\Lambda}}
+\fun{\opform{q}_{\txtrad,\txtbec,\sminvtemperature}^{\txteuclid}}
{j_{\tau_j}\mathfrak{b}_{2,\prbprocess_{\tau_j},\kappa,\Lambda}}}
\\ %%%%%%%%%%%%%%%%
&\leq
{\abs{\physcharge}^{2}}
c_{\sminvtemperature,\kappa,\Lambda}.
\end{aligned}$$
After adjusting the constant, we may write
\begin{equation}\label{expedition0012285}
\fun{\opform{q}_{\txtrad,\txtbec,\sminvtemperature}^{\txteuclid}}
{\mathsf{I}^{\txtpauli,\txtdiag}_{\txtrad,S_{\sminvtemperature},\kappa,\Lambda}(\prbprocess,\prbprocess^{\txtspin})}
\leq
C_{\sminvtemperature,\kappa,\Lambda},
\quad
\fun{\opform{q}_{\txtrad,\txtbec,\sminvtemperature}^{\txteuclid}}
{\mathsf{I}^{\txtpauli,\txtflip}_{\txtrad,j,\kappa,\Lambda}(\prbprocess,\prbprocess^{\txtspin})}
\leq
C_{\sminvtemperature,\kappa,\Lambda}.
\end{equation}
These constants are independent of the particle path,
the spin path,
and the jump time.

(Local kernel before integrating the radiation field): The current interaction in the spinless part is
$\fnexp{-\imunit\physcharge
\opfocksegalradiation(\mathsf{I}_{\txtrad,I,\kappa,\Lambda}(\prbprocess))}$,
whose absolute value is $1$.
By the norm estimate for the time-ordered exponential in Proposition \ref{expedition0012386},
the operator norm of the local kernel defined in \eqref{expedition0012288} is estimated by
\begin{equation}\label{expedition0012286}
\begin{aligned}
\norm{F_{\txtpaulifierz,\txtspin,I,\kappa,\Lambda}
(\prbprocess,\opfocksegalradiation)}_{\opspbddlin{\sphilb{H}_{\txtspin}}}
\leq
\fnexp{
\frac{\abs{\physcharge}}{2}
\int_I
\sum_{\mathrm{i}=1}^{3}
\abs{\physmfdensity_{\kappa,\Lambda,\mathrm{i}}(t,\prbprocess_t,\opfocksegalradiation)}
\opdmsr{t}}.
\end{aligned}
\end{equation}
(Uniform exponential moment estimate): By the normalization of the radiation-field measure,
for arbitrary $s
\in\fldreal$,
$$\begin{aligned}
\sqfun{\prbexp_{\msrprb_{\txtrad,\sminvtemperature}}}
{\fnexp{\imunit s
\fun{\opfocksegalradiation}{j_t\mathfrak{b}_{\mathrm{i},x,\kappa,\Lambda}}}}
=
\fnexp{-\frac{s^2}{4}
\fun{\opform{q}_{\txtrad,\txtbec,\sminvtemperature}^{\txteuclid}}
{j_t\mathfrak{b}_{\mathrm{i},x,\kappa,\Lambda}}}
\end{aligned}$$
holds.
Therefore the standard exponential moment estimate for centered Gaussian variables gives,
for arbitrary $a
> 0$,
$$\begin{aligned}
\sqfun{\prbexp_{\msrprb_{\txtrad,\sminvtemperature}}}
{\fnexp{a
\abs{\fun{\opfocksegalradiation}{j_t\mathfrak{b}_{\mathrm{i},x,\kappa,\Lambda}}}}}
\leq
2\fnexp{
\frac{a^2}{4}
\fun{\opform{q}_{\txtrad,\txtbec,\sminvtemperature}^{\txteuclid}}
{j_t\mathfrak{b}_{\mathrm{i},x,\kappa,\Lambda}}}
\leq
2\fnexp{\frac{a^2}{4}c_{\sminvtemperature,\kappa,\Lambda}},
\end{aligned}$$
where \eqref{expedition0012284} was applied in the last inequality.
Thus, for arbitrary $a>0$,
$$\begin{aligned}
\sup_{x,t,\mathrm{i}}
\sqfun{\prbexp_{\msrprb_{\txtrad,\sminvtemperature}}}
{\fnexp{
a\abs{\opfocksegalradiation(j_t\mathfrak{b}_{\mathrm{i},x,\kappa,\Lambda})}}}
\leq
C_{\sminvtemperature,\kappa,\Lambda,a}
<\infty
\end{aligned}$$
holds.
If $\absvol{I}
=0$, then the right-hand side of \eqref{expedition0012286} is $1$ and the assertion is trivial,
so assume $\absvol{I}
>0$.
Applying Jensen's inequality to the $p$-th power of \eqref{expedition0012286},
$$\begin{aligned}
&\sqfun{\prbexp_{\msrprb_{\txtrad,\sminvtemperature}}}
{\norm{F_{\txtpaulifierz,\txtspin,I,\kappa,\Lambda}
(\prbprocess,\opfocksegalradiation)}_{\opspbddlin{\sphilb{H}_{\txtspin}}}^p}
\\ %%%%%%%%%%%%%%%%
&\leq
\sqfun{\prbexp_{\msrprb_{\txtrad,\sminvtemperature}}}
{\fnexp{
\frac{p\abs{\physcharge}}{2}
\int_I
\sum_{\mathrm{i}=1}^{3}
\abs{\opfocksegalradiation(j_t\mathfrak{b}_{\mathrm{i},\prbprocess_t,\kappa,\Lambda})}
\opdmsr{t}}}
\\
&\leq
\frac{1}{\absvol{I}}
\int_I
\sqfun{\prbexp_{\msrprb_{\txtrad,\sminvtemperature}}}
{\fnexp{
\frac{p\abs{\physcharge}\absvol{I}}{2}
\sum_{\mathrm{i}=1}^{3}
\abs{\opfocksegalradiation(j_t\mathfrak{b}_{\mathrm{i},\prbprocess_t,\kappa,\Lambda})}}}
\opdmsr{t}
\end{aligned}$$
is obtained.
By convexity of the exponential function, for $x_{\mathrm{i}}
\geq 0$, $\fnexp{a(x_1+x_2+x_3)}
\leq
\frac{1}{3}
\sum_{\mathrm{i}=1}^{3}
\fnexp{3a x_{\mathrm{i}}}$ holds.
Applying this with $a=\frac{p\abs{\physcharge}\absvol{I}}{2}$ gives
$$\begin{aligned}
&\sqfun{\prbexp_{\msrprb_{\txtrad,\sminvtemperature}}}
{\norm{F_{\txtpaulifierz,\txtspin,I,\kappa,\Lambda}
(\prbprocess,\opfocksegalradiation)}_{\opspbddlin{\sphilb{H}_{\txtspin}}}^p}
\\
&\leq
\frac{1}{3\absvol{I}}
\sum_{\mathrm{i}=1}^{3}
\int_I
\sqfun{\prbexp_{\msrprb_{\txtrad,\sminvtemperature}}}
{\fnexp{
\frac{3p\abs{\physcharge}\absvol{I}}{2}
\abs{\opfocksegalradiation(j_t\mathfrak{b}_{\mathrm{i},\prbprocess_t,\kappa,\Lambda})}}}
\opdmsr{t}
\leq
C_{p,\sminvtemperature,\kappa,\Lambda}
<\infty.
\end{aligned}$$
This estimate is independent of the particle path, so
$\norm{F_{\txtpaulifierz,\txtspin,I,\kappa,\Lambda}}_{\opspbddlin{\sphilb{H}_{\txtspin}}}
\in \lp^{p}$ holds.

(Interaction factor after Gaussian integration): The main object to estimate is the radiation-field integral
as the expectation on the right-hand side of \eqref{expedition0012256},
$$\begin{aligned}
\sqfun{\prbexp_{\msrprb_{\txtrad,\sminvtemperature}}}
{\fnexp{\imunit
\fun{\opfocksegalradiation}
{\fun{G_{\kappa,\Lambda}}
{h,\prbprocess,\prbprocess^{\txtspin}}}}
\prod_{j=1}^{N_{\sminvtemperature}(\prbprocess^{\txtspin})}
\fun{\opfocksegalradiation}
{\mathsf{I}^{\txtpauli,\txtflip}_{\txtrad,j,\kappa,\Lambda}
(\prbprocess,\prbprocess^{\txtspin})}
}.
\end{aligned}$$
Since the external field $h$ is a finite linear combination of physical test functions,
the same Gaussian exponential estimate can also be applied to $h$.
Consider the part of \eqref{expedition0012256} where the number of jumps of the spin path is $n$.
If $n\geq1$, then by the generalized Hölder inequality,
the absolute value of the radiation-field expectation is bounded by the product of the $\lp^n$ norms
of the centered Gaussian variables corresponding to each jump source.
By \eqref{expedition0012285}, the variance of the centered Gaussian variable corresponding to each
jump source is uniformly bounded by $C_{\sminvtemperature,\kappa,\Lambda}$.
The $n$-th absolute moment estimate for centered Gaussian variables bounds each $\lp^n$ norm by
$C_{\sminvtemperature,\kappa,\Lambda}^{\onehalf}
\rbk{2^{n/2}\Gamma((n+1)/2)/\sqrt{\pi}}^{1/n}$.
After adjusting constants, including the case $n=0$,
\begin{equation}\label{expedition0012289}
\begin{aligned}
&\sqfun{\prbexp_{\msrprb_{\txtrad,\sminvtemperature}}}
{\abs{\fnexp{\imunit
\fun{\opfocksegalradiation}
{\fun{G_{\kappa,\Lambda}}
{h,\prbprocess,\prbprocess^{\txtspin}}}}
\prod_{j=1}^{n}
\fun{\opfocksegalradiation}
{\mathsf{I}^{\txtpauli,\txtflip}_{\txtrad,j,\kappa,\Lambda}
(\prbprocess,\prbprocess^{\txtspin})}}}
\\ %%%%%%%%%%%%%%%%
&\leq
C_{h,\sminvtemperature,\kappa,\Lambda}
C_{\sminvtemperature,\kappa,\Lambda}^{n}
\sqrt{n!}
\end{aligned}
\end{equation}
is obtained.
The exponential factor preceding the expectation in \eqref{expedition0012256} is a finite constant
depending only on the fixed $h$,
so with a suitable constant,
\begin{equation}\label{expedition0012287}
\begin{aligned}
\abs{\fun{\mathsf{D}_{\txtpaulifierz,\txtspin,\sminvtemperature,\kappa,\Lambda}}
{h,\prbprocess,\prbprocess^{\txtspin}}}
\leq
C_{h,\sminvtemperature,\kappa,\Lambda}
C_{\sminvtemperature,\kappa,\Lambda}^{n}
\sqrt{n!}
\end{aligned}
\end{equation}
may be assumed.
For the closed spin path measure, the mass of the $n$-jump sector is bounded by
$$\frac{1}{2\cosh\sminvtemperature}
\sum_{\varsigma\in\ringratint_2}
\int_{-\frac{\sminvtemperature}{2}<\tau_1<\cdots<\tau_n<\frac{\sminvtemperature}{2}}
\opdmsr{\tau_1}\cdots\opdmsr{\tau_n}
\leq
\frac{\sminvtemperature^n}{n!}.$$
Integrating \eqref{expedition0012287} over the spin paths gives
$$\sum_{n=0}^{\infty}
C_{h,\sminvtemperature,\kappa,\Lambda}
C_{\sminvtemperature,\kappa,\Lambda}^{n}
\sqrt{n!}
\frac{\sminvtemperature^n}{n!}
\leq
C_{h,\sminvtemperature,\kappa,\Lambda}
\sum_{n=0}^{\infty}
\frac{(C_{\sminvtemperature,\kappa,\Lambda}\sminvtemperature)^n}
{\sqrt{n!}}
<\infty,$$
and this upper bound is independent of the particle path.
By Fubini's theorem,
$$\int_{\Omega_{\txtparticle,\txtspin,\sminvtemperature}}
\abs{\fun{\mathsf{D}_{\txtpaulifierz,\txtspin,\sminvtemperature,\kappa,\Lambda}}
{h,\prbprocess,\prbprocess^{\txtspin}}}
\opdmsr{\msrprb_{\txtparticle,\txtspin,\sminvtemperature}}
<\infty$$
is obtained.
Therefore, even after multiplication by a bounded measurable function $B$,
$$\begin{aligned}
&\int_{\Omega_{\txtparticle,\txtspin,\sminvtemperature}}
\abs{
B(\prbprocess,\prbprocess^{\txtspin})
\fun{\mathsf{D}_{\txtpaulifierz,\txtspin,\sminvtemperature,\kappa,\Lambda}}
{h,\prbprocess,\prbprocess^{\txtspin}}}
\opdmsr{\msrprb_{\txtparticle,\txtspin,\sminvtemperature}}
\\
&\leq
\norm{B}_{\infty}
\int_{\Omega_{\txtparticle,\txtspin,\sminvtemperature}}
\abs{\fun{\mathsf{D}_{\txtpaulifierz,\txtspin,\sminvtemperature,\kappa,\Lambda}}
{h,\prbprocess,\prbprocess^{\txtspin}}}
\opdmsr{\msrprb_{\txtparticle,\txtspin,\sminvtemperature}}
<\infty
\end{aligned}$$
holds,
which is absolute convergence of the integral with external field, including the case $h
\neq 0$.
Finally, taking $h
=0$ and $B
=1$ gives integrability of
$\mathsf{D}_{\txtpaulifierz,\txtspin,\sminvtemperature,\kappa,\Lambda}(0)$.
\end{proof}

\begin{defn}[KMS State of the Pauli--Fierz Model with Spin and Cutoffs]\label{expedition0012257}
In the following discussion, use the particle--spin product probability measure in \eqref{expedition0012432} as the reference measure.
Define the partition function with cutoffs by
$$\smpartitionfunc_{\txtpaulifierz,\txtspin,\sminvtemperature,\kappa,\Lambda}
=
\sqfun{\prbexp_{\msrprb_{\txtparticle,\txtspin,\sminvtemperature}}}
{\mathsf{D}_{\txtpaulifierz,\txtspin,\sminvtemperature,\kappa,\Lambda}
(0,\prbprocess,\prbprocess^{\txtspin})}$$
Define the interacting particle--spin path probability measure by
\begin{equation}\label{expedition0012435}
\begin{aligned}
\opdmsr{\widetilde{\msrprb}_{\txtparticle,\txtspin,\sminvtemperature,\kappa,\Lambda}}
(\prbprocess,\prbprocess^{\txtspin})
=
\frac{\mathsf{D}_{\txtpaulifierz,\txtspin,\sminvtemperature,\kappa,\Lambda}
(0,\prbprocess,\prbprocess^{\txtspin})}
{\smpartitionfunc_{\txtpaulifierz,\txtspin,\sminvtemperature,\kappa,\Lambda}}
\opdmsr{\msrprb_{\txtparticle,\txtspin,\sminvtemperature}}
(\prbprocess,\prbprocess^{\txtspin}).
\end{aligned}
\end{equation}
For the interaction factor with spin
$\mathsf{D}
_{\txtpaulifierz,\txtspin,\sminvtemperature,\kappa,\Lambda}$ from Definition \ref{expedition0012255},
a bounded particle--spin function $B$, and a product of finitely many field exponentials,
define the KMS state with cutoffs by
\begin{equation}\label{expedition0012259}
\begin{aligned}
&\fun{\oastate[
\psi_{\txtpaulifierz,\txtspin,\sminvtemperature,\kappa,\Lambda}]}
{B(\prbprocess,\prbprocess^{\txtspin})
\fnexp{\imunit\opfocksegalradiation(h)}}
\\ %%%%%%%%%%%%%%%%
&=
\fnexp{-\oneoverfour
\fun{\opform{q}_{\txtrad,\txtbec,\sminvtemperature}^{\txteuclid}}
{h}}
\sqfun{\prbexp_{\widetilde{\msrprb}_{\txtparticle,\txtspin,\sminvtemperature,\kappa,\Lambda}}}
{B(\prbprocess,\prbprocess^{\txtspin})
\frac{\mathsf{D}_{\txtpaulifierz,\txtspin,\sminvtemperature,\kappa,\Lambda}
(h,\prbprocess,\prbprocess^{\txtspin})}
{\mathsf{D}_{\txtpaulifierz,\txtspin,\sminvtemperature,\kappa,\Lambda}
(0,\prbprocess,\prbprocess^{\txtspin})}}
\end{aligned}
\end{equation}
\end{defn}

\begin{proof}
\eqref{expedition0012259} is the form obtained by first integrating out only the radiation field
in the Feynman--Kac--Nelson representation with cutoffs.
Indeed, the field-dependent factor is
$$\fnexp{\imunit\opfocksegalradiation
\rbk{h
-\physcharge
\mathsf{I}_{\txtrad,S_{\sminvtemperature},\kappa,\Lambda}(\prbprocess)
+\mathsf{I}^{\txtpauli,\txtdiag}_{\txtrad,S_{\sminvtemperature},\kappa,\Lambda}
(\prbprocess,\prbprocess^{\txtspin})}}
\prod_{j=1}^{N_{\sminvtemperature}(\prbprocess^{\txtspin})}
\opfocksegalradiation
\rbk{\mathsf{I}^{\txtpauli,\txtflip}_{\txtrad,j,\kappa,\Lambda}(\prbprocess,\prbprocess^{\txtspin})}.$$
For fixed particle path and spin path,
the estimate \eqref{expedition0012289} implies that this exponential factor and $\prod_j
\fun{\opfocksegalradiation}
{\mathsf{I}^{\txtpauli,\txtflip}_{\txtrad,j,\kappa,\Lambda}(\prbprocess,\prbprocess^{\txtspin})}$
are integrable with respect to the radiation-field measure.
Therefore the radiation-field expectation in \eqref{expedition0012256} is meaningful as a finite value.
By Proposition \ref{expedition0012281}, the numerator in \eqref{expedition0012259}
converges absolutely even after multiplication by bounded $B$,
and the denominator is the finite normalizing constant in the case $B=1$ and $h=0$.
This denominator agrees with the thermal partition function of the self-adjoint Hamiltonian with cutoffs,
so it is positive, and the right-hand side of \eqref{expedition0012259} is meaningful as a finite value.
The computation transforming the radiation-field expectation in \eqref{expedition0012256}
into the derivative representation \eqref{expedition0012283} has already been recorded in
Definition \ref{expedition0012255} as the formula for centered Gaussian integrals.
The finiteness of the radiation-field moments needed in that computation is contained in the estimate
\eqref{expedition0012289},
and absolute integrability with respect to particle--spin paths is the conclusion of
Proposition \ref{expedition0012281}.
\end{proof}

\begin{prop}[Infinite-Volume Limit]\label{expedition0012298}
Fix cutoffs $0<\kappa<\Lambda<\infty$.
For finitely many $f_a
\in\sphilb{D}_{\txtrad,\txtphys,\sminvtemperature}$ and times $t_a
\in S_{\sminvtemperature}$, set $h
= \sum_a j_{t_a}f_a$ and $h_L
= \sum_a j_{t_a}P_{L}f_a$.
For the bounded-system regularization parameter $y_L$ defined in \eqref{expedition0012387},
take the limit $y_L
\downarrow 1$ as in the spinless model.
Then
$$\mathsf{D}_{\txtpaulifierz,\txtspin,\sminvtemperature,\kappa,\Lambda,\smchemicalpotential,L}(h_L)
\longrightarrow
\mathsf{D}_{\txtpaulifierz,\txtspin,\sminvtemperature,\kappa,\Lambda}(h)$$
holds in $\lp^{1}(\msrprb_{\txtparticle,\txtspin,\sminvtemperature})$.
In particular, the bounded-system states of Definition \ref{expedition0012297} converge as $L
\uparrow \infty$ to the state with cutoffs of Definition \ref{expedition0012257}.
\end{prop}

\begin{proof}
At fixed cutoff, $P_{L}\lambda_{\txtrad,\mathrm{i},x,\kappa,\Lambda}$,
$P_{L}\mathfrak{b}_{\mathrm{i},x,\kappa,\Lambda}$, and the bounded-system covariance forms converge as
$L
\uparrow \infty$ respectively to the infinite-volume
$\lambda_{\txtrad,\mathrm{i},x,\kappa,\Lambda}$,
$\mathfrak{b}_{\mathrm{i},x,\kappa,\Lambda}$, and
$\opform{q}
_{\txtrad,\txtbec,\sminvtemperature}^{\txteuclid}$.
For the current source, the argument is the same as in the bounded-system construction of the spinless
Pauli--Fierz model, in particular the argument from \eqref{expedition0012300} to
Proposition \ref{expedition0012325}.
For the magnetic-flux-density test functions, only $\lambda_{\txtrad,\mathrm{i},x,\kappa,\Lambda}$
is replaced by $\mathfrak{b}_{\mathrm{i},x,\kappa,\Lambda}$,
and the same finite-support and Riemann-sum convergence applies at fixed cutoff.
Therefore, for each particle--spin path,
$$\mathsf{I}_{\txtrad,S_{\sminvtemperature},\kappa,\Lambda,L}(\prbprocess)
\to
\mathsf{I}_{\txtrad,S_{\sminvtemperature},\kappa,\Lambda}(\prbprocess),
\quad
\mathsf{I}^{\txtpauli,\txtdiag}_{\txtrad,S_{\sminvtemperature},\kappa,\Lambda,L}
\to \mathsf{I}^{\txtpauli,\txtdiag}_{\txtrad,S_{\sminvtemperature},\kappa,\Lambda},
\quad
\mathsf{I}^{\txtpauli,\txtflip}_{\txtrad,j,\kappa,\Lambda,L}
\to \mathsf{I}^{\txtpauli,\txtflip}_{\txtrad,j,\kappa,\Lambda}$$
holds on particle--spin paths.

If the number of jumps of the spin path is fixed to be $n$,
then \eqref{expedition0012296} and \eqref{expedition0012283} are both finite sums consisting of
finitely many covariance forms.
Thus the preceding convergence and the convergence of the bounded-system covariance forms imply
pointwise convergence on each $n$-jump sector.
This is the same computation as the one in the bounded-system limit \eqref{expedition0012341}
for the Nelson model, where after integrating out the field only finitely many covariance forms are
passed to the limit, and as the computation in the two-point representation \eqref{expedition0012324}
for the spinless Pauli--Fierz model.

The bounded-system magnetic-flux-density test functions are only projections by $P_L$,
so for sufficiently large $L$ one can take an upper bound of the same form as \eqref{expedition0012284}
with the same constant.
The covariance norm of the current source is dominated by the bounded-system estimate in
Proposition \ref{expedition0012325}.
By the bounded-system covariance estimates used in the zero-mode and non-zero-mode decomposition for
the spinless model, a common upper bound can be taken for sufficiently large $L$ for finitely many
external fields $h_L
= \sum_a j_{t_a}P_Lf_a$.
Therefore, by the same Gaussian moment estimate as \eqref{expedition0012287},
the $n$-jump sector is dominated by an upper bound of type
$$C_{h,\sminvtemperature,\kappa,\Lambda}
C_{\sminvtemperature,\kappa,\Lambda}^{n}\sqrt{n!}.$$
Since the mass of the $n$-jump sector of the closed spin path measure is bounded by
$\sminvtemperature^n/n!$,
$$\sum_{n=0}^{\infty}
C_{h,\sminvtemperature,\kappa,\Lambda}
C_{\sminvtemperature,\kappa,\Lambda}^{n}\sqrt{n!}
\frac{\sminvtemperature^n}{n!}
<\infty$$
holds; this summable upper bound is independent of $L$.
Applying the dominated convergence theorem on each $n$-jump sector and then summing over $n$
gives the $\lp^1$ convergence of
$\mathsf{D}
_{\txtpaulifierz,\txtspin,\sminvtemperature,\kappa,\Lambda,\smchemicalpotential,L}(h_L)$.

For convergence of the states, the outer free Gaussian factor must also be checked.
This is the same computation as for the expectation of field exponential operators in the bounded system
of the spinless model, in particular the zero-mode and non-zero-mode limits in
Proposition \ref{expedition0012107} and \eqref{eq:PF-zero-nonzero-decomposition}, and gives
$\fun{\opform{q}_{\txtrad,\txtbec,\sminvtemperature,\smchemicalpotential,L}^{\txteuclid}}{h_L}
\to
\fun{\opform{q}_{\txtrad,\txtbec,\sminvtemperature}^{\txteuclid}}{h}$.
The partition function is the case $h=0$,
and numerators multiplied by bounded particle--spin functions are handled by the same $\lp^1$ convergence.
The denominator is positive as the thermal partition function of the bounded-system Hamiltonian with cutoffs,
and the limiting denominator is also defined as a finite positive number by Proposition \ref{expedition0012281}
and Definition \ref{expedition0012257}.
Therefore, taking the quotient, the bounded-system states converge to the state with cutoffs in
Definition \ref{expedition0012257}.
\end{proof}

\subsection{Removal of Infrared and Ultraviolet Cutoffs}\label{removal-of-infrared-and-ultraviolet-cutoffs-2}

In the spinless model, the current-kernel estimate in Proposition \ref{expedition0012165} removed the cutoffs all the way to the point-source limit. On the other hand, in the Pauli term of the model with spin, \(\mathsf{I}^{\txtpauli,\txtdiag}_{\txtrad,S_{\sminvtemperature},\kappa,\Lambda}\) in \eqref{expedition0012253} and \(\mathsf{I}^{\txtpauli,\txtflip}_{\txtrad,j,\kappa,\Lambda}\) in \eqref{expedition0012254} appear. In this subsection, we quote cutoff removal for the minimal-coupling current from the spinless model and estimate only the newly appearing magnetic-flux-density sources.

First fix notation involving tail sets. For a measurable set \(A\subset\fldreal^3\), denote the \(A\) component of the magnetic-flux-density test function by \begin{equation}\label{expedition0012388}
\faftr{\mathfrak{b}}_{\mathrm{i},x,A}^{\mathrm{j}_{\txtrad}}(k)
=
\frac{\fndef{k\in A}}
{\sqrt{\omega(k)}}
\imunit\rbk{k\times e^{\mathrm{j}_{\txtrad}}(k)}_{\mathrm{i}}
\napiernum^{-\imunit kx}.
\end{equation} Denote the corresponding Pauli sources by \begin{equation}\label{expedition0012389}
\begin{aligned}
\mathsf{I}^{\txtpauli,\txtdiag}_{\txtrad,I,A}
(\prbprocess,\prbprocess^{\txtspin})
&=
-\frac{\physcharge}{2}
\int_I
\prbprocess_t^{\txtspin}
\fun{j_t}{\mathfrak{b}_{3,\prbprocess_t,A}}
\opdmsr{t},
\\ %%%%%%%%%%%%%%%%
\mathsf{I}^{\txtpauli,\txtflip}_{\txtrad,j,A}
(\prbprocess,\prbprocess^{\txtspin})
&=
-\frac{\physcharge}{2}
\rbk{\fun{j_{\tau_j}}
{\mathfrak{b}_{1,\prbprocess_{\tau_j},A}}
-\imunit \prbprocess_{\tau_j-}^{\txtspin}
\cdot
\fun{j_{\tau_j}}
{\mathfrak{b}_{2,\prbprocess_{\tau_j},A}}}.
\end{aligned}
\end{equation} If \(A_{\kappa,\Lambda}=\set{k}{\kappa\leq\abs{k}\leq\Lambda}\) is the cutoff region, then the sources in \eqref{expedition0012253} and \eqref{expedition0012254} are the objects in \eqref{expedition0012389} with \(A=A_{\kappa,\Lambda}\). The difference between two cutoffs \(0<\kappa'<\kappa<\Lambda<\Lambda'<\infty\) decomposes as \begin{equation}\label{expedition0012390}
\begin{aligned}
\mathsf{I}^{\txtpauli,\txtdiag}_{\txtrad,I,A_{\kappa',\Lambda'}}
-\mathsf{I}^{\txtpauli,\txtdiag}_{\txtrad,I,A_{\kappa,\Lambda}}
&=
\mathsf{I}^{\txtpauli,\txtdiag}_{\txtrad,I,\set{k}{\kappa'\leq\abs{k}<\kappa}}
+\mathsf{I}^{\txtpauli,\txtdiag}_{\txtrad,I,\set{k}{\Lambda<\abs{k}\leq\Lambda'}},
\\ %%%%%%%%%%%%%%%%
\mathsf{I}^{\txtpauli,\txtflip}_{\txtrad,j,A_{\kappa',\Lambda'}}
-\mathsf{I}^{\txtpauli,\txtflip}_{\txtrad,j,A_{\kappa,\Lambda}}
&=
\mathsf{I}^{\txtpauli,\txtflip}_{\txtrad,j,\set{k}{\kappa'\leq\abs{k}<\kappa}}
+\mathsf{I}^{\txtpauli,\txtflip}_{\txtrad,j,\set{k}{\Lambda<\abs{k}\leq\Lambda'}}.
\end{aligned}
\end{equation} Therefore estimates of the infrared tail \(A=\set{k}{\abs{k}<\kappa}\) and the ultraviolet tail \(A=\set{k}{\abs{k}>\Lambda}\) are central for cutoff removal. The self term of the minimal-coupling current that matters in ultraviolet cutoff removal is handled not as an additive renormalization constant such as \eqref{expedition0012182} and \eqref{expedition0012185} in the Nelson model, but as cutoff removal for the self term of the current sesquilinear form contained in the local semigroup limit of the spinless Pauli--Fierz model. Specifically, we refer to the kernel representation with cutoffs on generators \eqref{expedition0012210}, the reduction to the non-zero-mode form \eqref{expedition0012211}, the current-kernel cutoff-removal proposition \ref{expedition0012165}, and the local semigroup theorem after cutoff removal \ref{expedition0012212}. The proposition newly proved below does not replace this handling of the self term on the minimal-coupling side; it handles the cutoff differences in \eqref{expedition0012389} that are additionally produced by the Pauli term with spin.

\begin{prop}[Cutoff-Removal Estimate for the Pauli Term with Spin]\label{expedition0012290}
Let the spatial dimension be $d=3$.
We estimate the Pauli sources in \eqref{expedition0012389} as finitely many Gaussian moments averaged over
the particle loop measure and the closed spin path measure.
Then, for factors arising from any finite number of
$\mathsf{I}^{\txtpauli,\txtdiag}_{\txtrad,S_{\sminvtemperature},A}$ and
$\mathsf{I}^{\txtpauli,\txtflip}_{\txtrad,j,A}$,
the contributions of the infrared tail and ultraviolet tail in \eqref{expedition0012390}
converge to $0$ in the sense of $\lp^{1}(\msrprb_{\txtparticle,\txtspin,\sminvtemperature})$
as $\kappa\downarrow0,\Lambda\uparrow\infty$.
Furthermore, the cutoff family is uniformly integrable.
\end{prop}

\begin{proof}
For a fixed spin path, let the number of jumps be $n$ and the jump times be
$\tau_1<\cdots<\tau_n$.
The diagonal magnetic-flux-density term in \eqref{expedition0012253} contains the time integral
$\int_{S_{\sminvtemperature}}
\fun{j_t}
{\mathfrak{b}_{3,\prbprocess_t,\kappa,\Lambda}}
\opdmsr{t}$,
and the off-diagonal magnetic-flux-density term in \eqref{expedition0012254}
contains $\fun{j_{\tau_j}}
{\mathfrak{b}_{\mathrm{i},\prbprocess_{\tau_j},\kappa,\Lambda}}$ at the jump time.

(Gaussian integration): Differentiate \eqref{expedition0012283} of Definition \ref{expedition0012255}
on a fixed $n$-jump sector.
The centered Gaussian differentiation formula gives
\begin{equation}\label{expedition0012391}
\begin{aligned}
&\fnrestr{
\oppd{u_1}\cdots\oppd{u_n}
\fnexp{-\oneoverfour
\fun{\opform{q}_{\txtrad,\txtbec,\sminvtemperature}^{\txteuclid}}
{\fun{G_{\kappa,\Lambda}}{u,h,\prbprocess,\prbprocess^{\txtspin}}}}}
{u_1=\cdots=u_n=0}
\\ %%%%%%%%%%%%%%%%
&=
\fnexp{-\oneoverfour
\fun{\opform{q}_{\txtrad,\txtbec,\sminvtemperature}^{\txteuclid}}
{\fun{G_{\kappa,\Lambda}}{h,\prbprocess,\prbprocess^{\txtspin}}}}
\sum_{\pi}
c_{\pi}
\\ %%%%%%%%%%%%%%%%
&\quad\times
\prod_{\setone{a}\in\pi}
\fun{\opform{q}_{\txtrad,\txtbec,\sminvtemperature}^{\txteuclid}}
{\fun{G_{\kappa,\Lambda}}{h,\prbprocess,\prbprocess^{\txtspin}},
\mathsf{I}^{\txtpauli,\txtflip}_{\txtrad,a,\kappa,\Lambda}
(\prbprocess,\prbprocess^{\txtspin})}
\\ %%%%%%%%%%%%%%%%
&\quad\times
\prod_{\setone{a,b}\in\pi}
\fun{\opform{q}_{\txtrad,\txtbec,\sminvtemperature}^{\txteuclid}}
{\mathsf{I}^{\txtpauli,\txtflip}_{\txtrad,a,\kappa,\Lambda}
(\prbprocess,\prbprocess^{\txtspin}),
\mathsf{I}^{\txtpauli,\txtflip}_{\txtrad,b,\kappa,\Lambda}
(\prbprocess,\prbprocess^{\txtspin})}
\end{aligned}
\end{equation}
as a finite sum.
Here $\pi$ runs over all partitions of $\setone{1,\ldots,n}$ into singleton and two-element subsets,
and $c_\pi$ are numerical constants.
Since the diagonal Pauli source is contained in
$\fun{G_{\kappa,\Lambda}}{h,\prbprocess,\prbprocess^{\txtspin}}$,
expanding each factor in \eqref{expedition0012391} leaves only products of finitely many covariance forms
between the current source, diagonal Pauli source, or external field inside the exponential and a jump Pauli source,
or covariance forms between two jump Pauli sources.
This partition representation is the centered Gaussian differentiation formula \eqref{expedition0012283},
obtained by moving the bounded-system formula \eqref{expedition0012296} to infinite volume,
with the magnetic-flux-density factors of the Pauli term added to the spinless Gaussian integration
representation \eqref{expedition0012259}.
The cutoff differences to be estimated split into covariance forms involving only current sources,
covariance forms involving a current source or external field and a magnetic-flux-density factor,
and covariance forms involving two magnetic-flux-density factors.

We check the estimates for pairs containing magnetic-flux-density factors.
The momentum factor coming from two magnetic-flux-density test functions with cutoffs is bounded by
$\omega(k)$, by \eqref{expedition0012252} and orthonormality of the transverse polarization vectors.
Specifically,
\begin{equation}\label{expedition0012394}
\begin{aligned}
\sum_{\mathrm{j}_{\txtrad}\in\ringratint_2}
\abs{\faftr{\mathfrak{b}}_{\mathrm{i},x,A}^{\mathrm{j}_{\txtrad}}(k)}^2
&\leq
\fndef{k\in A}\omega(k),
\\ %%%%%%%%%%%%%%%%
\abs{\sum_{\mathrm{j}_{\txtrad}\in\ringratint_2}
\overline{\faftr{\mathfrak{b}}_{\mathrm{i},x,A}^{\mathrm{j}_{\txtrad}}(k)}
\faftr{\mathfrak{b}}_{\ell,y,B}^{\mathrm{j}_{\txtrad}}(k)}
&\leq
\fndef{k\in A\cap B}\omega(k)
\end{aligned}
\end{equation}
holds.
Using the finite-temperature kernel $K_{\sminvtemperature}(r,k)$ in \eqref{expedition0012238},
\eqref{expedition0012394} gives
\begin{equation}\label{expedition0012395}
\begin{aligned}
\abs{\fun{\opform{q}_{\txtrad,\txtbec,\sminvtemperature}^{\txteuclid}}
{j_t\mathfrak{b}_{\mathrm{i},x,A},
j_s\mathfrak{b}_{\ell,y,B}}}
\leq
C
\int_{A\cap B}
K_{\sminvtemperature}(t-s,k)
\omega(k)
\opdmsr{k}.
\end{aligned}
\end{equation}
We record the tail estimate needed after averaging over particle paths.
For pairs of magnetic-flux-density factors and for cross terms between the full-circle current source
and a magnetic-flux-density factor,
associate to a measurable set $A\subset \fldreal^3$ the quantity
\begin{equation}\label{expedition0012392}
\begin{aligned}
\mathfrak{P}(A)
&=
\rbk{\int_A
\rbk{\fndef{\omega(k)\leq1}
+\fndef{\omega(k)>1}
\frac{(1 + \omega(k))^2}
{\rbk{\omega(k)+\onehalf\omega(k)^2}^{3}}}
\opdmsr{k}}^{\onehalf}
\\ %%%%%%%%%%%%%%%%
&\quad+
\int_A
\rbk{\fndef{\omega(k)\leq1}
+\fndef{\omega(k)>1}
\frac{(1 + \omega(k))^2}
{\rbk{\omega(k)+\onehalf\omega(k)^2}^{3}}}
\opdmsr{k}.
\end{aligned}
\end{equation}
On the low-momentum side, the closed-loop compensation in Lemma \ref{expedition0012304}
gives a gain of $\abs{k}$ from the full-circle current factor.
On the high-momentum side, apply \eqref{expedition0012186}, \eqref{expedition0012190},
and \eqref{expedition0012192} of Lemma \ref{expedition0012142} for the Nelson model,
and the thermal-kernel estimate in Lemma \ref{expedition0012305} for the spinless Pauli--Fierz model,
componentwise to the transverse components.
Since the magnetic-flux-density factor contributes the $\omega(k)$ in \eqref{expedition0012394},
the current--magnetic-flux-density and magnetic-flux-density--magnetic-flux-density tail estimates
are controlled by $\mathfrak{P}(A)$.
Indeed, the following argument gives
\begin{equation}\label{expedition0012403}
\begin{aligned}
\mathfrak{P}(\set{k}{\abs{k}<\kappa})
&\leq
C\rbk{\kappa^{3/2}+\kappa^3}
\longrightarrow0,
\\ %%%%%%%%%%%%%%%%
\mathfrak{P}(\set{k}{\abs{k}>\Lambda})
&\leq
C\rbk{\Lambda^{-1/2}+\Lambda^{-1}}
\longrightarrow0.
\end{aligned}
\end{equation}
For arbitrary $\kappa>0$, split the low-momentum side into $\abs{k}\leq1$ and $\abs{k}>1$.
Using the fact that if $\kappa>1$ then $\kappa^3\geq1$, we obtain
$$\begin{aligned}
&\int_{\set{k}{\abs{k}<\kappa}}
\rbk{\fndef{\omega(k)\leq1}
+\fndef{\omega(k)>1}
\frac{(1 + \omega(k))^2}
{\rbk{\omega(k)+\onehalf\omega(k)^2}^{3}}}
\opdmsr{k}
\\ %%%%%%%%%%%%%%%%
&=
\int_{\set{k}{\abs{k}<\kappa\wedge1}}\opdmsr{k}
+
\fndef{\kappa>1}
\int_{\set{k}{1<\abs{k}<\kappa}}
\frac{(1 + \omega(k))^2}
{\rbk{\omega(k)+\onehalf\omega(k)^2}^{3}}
\opdmsr{k}
\\ %%%%%%%%%%%%%%%%
&\leq
4\pi\int_0^{\kappa\wedge1} r^2\opdmsr{r}
+
4\pi C\fndef{\kappa>1}
\int_1^\kappa r^{-2}\opdmsr{r}
\\ %%%%%%%%%%%%%%%%
&\leq
C\rbk{(\kappa\wedge1)^3+\fndef{\kappa>1}}
\leq
C\kappa^3.
\end{aligned}$$
Therefore, for arbitrary $\kappa>0$, the square-root term in \eqref{expedition0012392}
is bounded by $C\kappa^{3/2}$,
and the linear term is bounded by $C\kappa^3$.
Similarly, for arbitrary $\Lambda>0$, split the high-momentum side into $\abs{k}\leq1$ and
$\abs{k}>1$.
Using the fact that if $0<\Lambda<1$ then $\Lambda^{-1}\geq1$,
for $\omega(k)>1$,
$\frac{(1 + \omega(k))^2}
{\rbk{\omega(k)+\onehalf\omega(k)^2}^{3}}
\leq
C\omega(k)^{-4}$ holds, and in particular
$$\begin{aligned}
&\int_{\set{k}{\abs{k}>\Lambda}}
\rbk{\fndef{\omega(k)\leq1}
+\fndef{\omega(k)>1}
\frac{(1 + \omega(k))^2}
{\rbk{\omega(k)+\onehalf\omega(k)^2}^{3}}}
\opdmsr{k}
\\ %%%%%%%%%%%%%%%%
&\leq
\fndef{\Lambda<1}
4\pi\int_\Lambda^1 r^2\opdmsr{r}
+
4\pi C\int_{\Lambda\vee1}^{\infty} r^{-2}\opdmsr{r}
\\ %%%%%%%%%%%%%%%%
&\leq
C\rbk{\fndef{\Lambda<1}+(\Lambda\vee1)^{-1}}
\leq
C\Lambda^{-1}.
\end{aligned}$$
Therefore, for arbitrary $\Lambda>0$, the square-root term in \eqref{expedition0012392}
is bounded by $C\Lambda^{-1/2}$,
and the linear term is bounded by $C\Lambda^{-1}$.

For the cross term between the external field and the magnetic-flux-density tail, for $h
= \sum_{a=1}^{N_h}
j_{t_a}
f_a$, associate the external-field tail quantity
\begin{equation}\label{expedition0012404}
\begin{aligned}
\mathfrak{E}_{h}(A)
&=
\sum_{a=1}^{N_h}
\int_A
\frac{1+\napiernum^{-\sminvtemperature\omega(k)}}
{1-\napiernum^{-\sminvtemperature\omega(k)}}
\frac{\abs{\delta^\perp(k)\faftr{f_a}(k)}_{\fldcmp^3}}
{\sqrt{\omega(k)}}
\rbk{1+\frac{1}{\omega(k)+\onehalf\omega(k)^2}}
\opdmsr{k}.
\end{aligned}
\end{equation}
By \eqref{eq:PF-physical-test-space}, $\mathfrak{E}_{h}(\fldreal^3)<\infty$,
so absolute continuity gives
\begin{equation}\label{expedition0012405}
\begin{aligned}
\mathfrak{E}_{h}(\set{k}{\abs{k}<\kappa})
&\longrightarrow0,
&
\mathfrak{E}_{h}(\set{k}{\abs{k}>\Lambda})
&\longrightarrow0.
\end{aligned}
\end{equation}
Furthermore, by $K_{\sminvtemperature}$ in \eqref{expedition0012238},
\eqref{expedition0012388},
and \eqref{eq:PF-physical-cross-kernel}, for each $a$ and each $\mathrm{i}$,
\begin{equation}\label{expedition0012406}
\begin{aligned}
&\int_{S_{\sminvtemperature}}
\abs{\fun{\opform{q}_{\txtrad,\txtbec,\sminvtemperature}^{\txteuclid}}
{j_{t_a}f_a,j_\tau\mathfrak{b}_{\mathrm{i},\prbprocess_\tau,A}}}
\opdmsr{\tau}
\\ %%%%%%%%%%%%%%%%
&\leq
C
\int_A
\frac{1+\napiernum^{-\sminvtemperature\omega(k)}}
{1-\napiernum^{-\sminvtemperature\omega(k)}}
\frac{\abs{\delta^\perp(k)\faftr{f_a}(k)}_{\fldcmp^3}}
{\sqrt{\omega(k)}}
\rbk{1+\frac{1}{\omega(k)+\onehalf\omega(k)^2}}
\opdmsr{k}
\end{aligned}
\end{equation}
holds.

Summarizing the preceding argument,
after adjusting constants $C_{p,h},C_p>0$, for arbitrary $p\geq1$,
\begin{equation}\label{expedition0012407}
\begin{aligned}
\norm{\fun{\opform{q}_{\txtrad,\txtbec,\sminvtemperature}^{\txteuclid}}
{h,\mathsf{I}^{\txtpauli,\txtflip}_{\txtrad,j,A}}}_{\lp^p}
&\leq
C_{p,h}\mathfrak{E}_{h}(A),
\\ %%%%%%%%%%%%%%%%
\norm{\fun{\opform{q}_{\txtrad,\txtbec,\sminvtemperature}^{\txteuclid}}
{\mathsf{I}_{\txtrad,S_{\sminvtemperature},\kappa,\Lambda},
\mathsf{I}^{\txtpauli,\txtflip}_{\txtrad,j,A}}}_{\lp^p}
&\leq
C_p\mathfrak{P}(A),
\\ %%%%%%%%%%%%%%%%
\norm{\fun{\opform{q}_{\txtrad,\txtbec,\sminvtemperature}^{\txteuclid}}
{\mathsf{I}^{\txtpauli,\txtdiag}_{\txtrad,S_{\sminvtemperature},B},
\mathsf{I}^{\txtpauli,\txtflip}_{\txtrad,j,A}}}_{\lp^p}
&\leq
C_p\mathfrak{P}(A\cap B)
\leq
C_p\mathfrak{P}(A),
\\ %%%%%%%%%%%%%%%%
\norm{\fun{\opform{q}_{\txtrad,\txtbec,\sminvtemperature}^{\txteuclid}}
{\mathsf{I}^{\txtpauli,\txtflip}_{\txtrad,j,A},
\mathsf{I}^{\txtpauli,\txtflip}_{\txtrad,\ell,B}}}_{\lp^p}
&\leq
C_p\mathfrak{P}(A\cap B)
\leq
C_p\mathfrak{P}(A)
\end{aligned}
\end{equation}
holds.
The first estimate follows from \eqref{expedition0012406} and \eqref{expedition0012404},
and the second is the tail estimate of Lemma \ref{expedition0012305} with the closed-loop compensation
of Lemma \ref{expedition0012304} included.
The third and fourth estimates are obtained by applying to \eqref{expedition0012395}
the same thermal-kernel estimates as \eqref{expedition0012186}, \eqref{expedition0012190},
and \eqref{expedition0012192} of Lemma \ref{expedition0012142} for the Nelson model.

In particular, for arbitrary external field $h$ and arbitrary pair of Pauli sources,
\begin{equation}\label{expedition0012393}
\begin{aligned}
\sqfun{\prbexp_{\msrprb_{\txtparticle,\txtspin,\sminvtemperature}}}
{\abs{\fun{\opform{q}_{\txtrad,\txtbec,\sminvtemperature}^{\txteuclid}}
{h-\physcharge\mathsf{I}_{\txtrad,S_{\sminvtemperature},\kappa,\Lambda}
+
\mathsf{I}^{\txtpauli,\txtdiag}_{\txtrad,S_{\sminvtemperature},\kappa,\Lambda},
\mathsf{I}^{\txtpauli,\txtflip}_{\txtrad,j,A}}}}
&\leq
C_h\rbk{\mathfrak{E}_{h}(A)+\mathfrak{P}(A)},
\\ %%%%%%%%%%%%%%%%
\sqfun{\prbexp_{\msrprb_{\txtparticle,\txtspin,\sminvtemperature}}}
{\abs{\fun{\opform{q}_{\txtrad,\txtbec,\sminvtemperature}^{\txteuclid}}
{\mathsf{I}^{\txtpauli,\txtflip}_{\txtrad,j,A},
\mathsf{I}^{\txtpauli,\txtflip}_{\txtrad,\ell,A}}}}
&\leq
C\mathfrak{P}(A)
\end{aligned}
\end{equation}
there are constants $C_h,C>0$ satisfying these bounds.
The first estimate in \eqref{expedition0012393} summarizes the following three estimates:
\begin{equation}\label{expedition0012396}
\begin{aligned}
\sqfun{\prbexp_{\msrprb_{\txtparticle,\txtspin,\sminvtemperature}}}
{\abs{\fun{\opform{q}_{\txtrad,\txtbec,\sminvtemperature}^{\txteuclid}}
{h,\mathsf{I}^{\txtpauli,\txtflip}_{\txtrad,j,A}}}}
&\leq
C_h\mathfrak{E}_{h}(A),
\\ %%%%%%%%%%%%%%%%
\sqfun{\prbexp_{\msrprb_{\txtparticle,\txtspin,\sminvtemperature}}}
{\abs{\fun{\opform{q}_{\txtrad,\txtbec,\sminvtemperature}^{\txteuclid}}
{\mathsf{I}_{\txtrad,S_{\sminvtemperature},\kappa,\Lambda},
\mathsf{I}^{\txtpauli,\txtflip}_{\txtrad,j,A}}}}
&\leq
C\mathfrak{P}(A),
\\ %%%%%%%%%%%%%%%%
\sqfun{\prbexp_{\msrprb_{\txtparticle,\txtspin,\sminvtemperature}}}
{\abs{\fun{\opform{q}_{\txtrad,\txtbec,\sminvtemperature}^{\txteuclid}}
{\mathsf{I}^{\txtpauli,\txtdiag}_{\txtrad,S_{\sminvtemperature},\kappa,\Lambda},
\mathsf{I}^{\txtpauli,\txtflip}_{\txtrad,j,A}}}}
&\leq
C\mathfrak{P}(A).
\end{aligned}
\end{equation}
The new estimates needed in this proposition are the third and fourth estimates in \eqref{expedition0012407}.
The ultraviolet self term arising from the self term of the minimal-coupling current is separated in the
spinless kernel representations \eqref{expedition0012210} and \eqref{expedition0012211},
and is contained in the local semigroup limit of Proposition \ref{expedition0012165}
and Theorem \ref{expedition0012212}.
This proposition does not re-estimate that part;
it estimates only the tails containing the magnetic-flux-density factors in \eqref{expedition0012388},
through \eqref{expedition0012403}--\eqref{expedition0012407}.

We estimate terms arising from finitely many
$\mathsf{I}^{\txtpauli,\txtdiag}_{\txtrad,S_{\sminvtemperature},\kappa,\Lambda}$
and $\mathsf{I}^{\txtpauli,\txtflip}_{\txtrad,j,\kappa,\Lambda}$.
Factors corresponding to singleton sets are covariance forms between a current source or external field
and a magnetic-flux-density factor,
and factors corresponding to two-element sets are covariance forms between two magnetic-flux-density factors.
If these factors are denoted by $Q_{r,\kappa,\Lambda}$, then each $Q_{r,\kappa,\Lambda}$ is one of
\begin{equation}\label{expedition0012397}
\begin{aligned}
&\fun{\opform{q}_{\txtrad,\txtbec,\sminvtemperature}^{\txteuclid}}
{h,\mathsf{I}^{\txtpauli,\txtflip}_{\txtrad,j,A_{\kappa,\Lambda}}},
\quad
\fun{\opform{q}_{\txtrad,\txtbec,\sminvtemperature}^{\txteuclid}}
{\mathsf{I}_{\txtrad,S_{\sminvtemperature},\kappa,\Lambda},
\mathsf{I}^{\txtpauli,\txtflip}_{\txtrad,j,A_{\kappa,\Lambda}}},
\\ %%%%%%%%%%%%%%%%
&\fun{\opform{q}_{\txtrad,\txtbec,\sminvtemperature}^{\txteuclid}}
{\mathsf{I}^{\txtpauli,\txtdiag}_{\txtrad,S_{\sminvtemperature},A_{\kappa,\Lambda}},
\mathsf{I}^{\txtpauli,\txtflip}_{\txtrad,j,A_{\kappa,\Lambda}}},
\quad
\fun{\opform{q}_{\txtrad,\txtbec,\sminvtemperature}^{\txteuclid}}
{\mathsf{I}^{\txtpauli,\txtflip}_{\txtrad,j,A_{\kappa,\Lambda}},
\mathsf{I}^{\txtpauli,\txtflip}_{\txtrad,\ell,A_{\kappa,\Lambda}}}.
\end{aligned}
\end{equation}
Substituting \eqref{expedition0012390} into each factor for two cutoffs
$0<\kappa'<\kappa<\Lambda<\Lambda'<\infty$ gives
\begin{equation}\label{expedition0012408}
\begin{aligned}
&\norm{Q_{r,\kappa',\Lambda'}-Q_{r,\kappa,\Lambda}}_{\lp^p}
\\ %%%%%%%%%%%%%%%%
&\leq
C_{p,h}
\rbk{\mathfrak{E}_{h}(\set{k}{\kappa'\leq\abs{k}<\kappa})
+\mathfrak{E}_{h}(\set{k}{\Lambda<\abs{k}\leq\Lambda'})}
\\ %%%%%%%%%%%%%%%%
&\quad+
C_p
\rbk{\mathfrak{P}(\set{k}{\kappa'\leq\abs{k}<\kappa})
+\mathfrak{P}(\set{k}{\Lambda<\abs{k}\leq\Lambda'})}.
\end{aligned}
\end{equation}
By \eqref{expedition0012403} and \eqref{expedition0012405}, the right-hand side converges to $0$
as $\kappa,\kappa'\downarrow0$ and $\Lambda,\Lambda'\uparrow\infty$.
For factors without cutoff differences,
\begin{equation}\label{expedition0012409}
\begin{aligned}
\sup_{0<\kappa<\Lambda<\infty}
\norm{Q_{r,\kappa,\Lambda}}_{\lp^p}
+
\norm{Q_r}_{\lp^p}
\leq
C_{p,h}
\end{aligned}
\end{equation}
is obtained from \eqref{expedition0012407} with $A=\fldreal^3$.
The cutoff difference of products decomposes as
\begin{equation}\label{expedition0012398}
\begin{aligned}
\prod_{r=1}^{m}Q_{r,\kappa,\Lambda}
-\prod_{r=1}^{m}Q_{r}
=
\sum_{a=1}^{m}
\rbk{\prod_{r<a}Q_{r,\kappa,\Lambda}}
\rbk{Q_{a,\kappa,\Lambda}-Q_{a}}
\rbk{\prod_{r>a}Q_{r}}.
\end{aligned}
\end{equation}
Combining \eqref{expedition0012408} and \eqref{expedition0012409} with Hölder's inequality gives
\begin{equation}\label{expedition0012399}
\begin{aligned}
\norm{\prod_{r=1}^{m}Q_{r,\kappa,\Lambda}
-\prod_{r=1}^{m}Q_{r}}_{\lp^1}
\leq
C_m
\sum_{a=1}^{m}
\norm{Q_{a,\kappa,\Lambda}-Q_a}_{\lp^m}
\longrightarrow0.
\end{aligned}
\end{equation}
By \eqref{expedition0012409}, the product corresponding to a partition $\pi$ can be bounded by
the $m(\pi)$-th power of a cutoff-independent constant $C$,
and in the integration over spin jump times, the mass of the $n$-jump sector is bounded by
$\sminvtemperature^n/n!$.
The sum over partitions $\pi$ is bounded by the combinatorial estimate obtained from the
$n$-th absolute moment estimate for a centered Gaussian variable whose variance is bounded by $C$,
and
$$\sum_{n=0}^{\infty}
C^n
\sqrt{n!}
\frac{\sminvtemperature^n}{n!}
< \infty$$
holds.
This integrable dominating function is independent of the cutoffs.
This proves $\lp^{1}$ convergence and uniform integrability for the factors arising from the Pauli term.
\end{proof}

\begin{prop}[Cutoff Removal for the Interaction Factor with Spin]\label{expedition0012291}
For arbitrary finitely many times and arbitrary $f_j
\in \sphilb{D}_{\txtrad,\txtphys,\sminvtemperature}$, set $h=\sum_j j_{t_j}f_j$.
Then the interaction factor with spin in Definition \ref{expedition0012255} has the limit
$$\mathsf{D}
_{\txtpaulifierz,\txtspin,\sminvtemperature}
(h,\prbprocess,\prbprocess^{\txtspin})
=
\fun{\lp^{1}}{\msrprb_{\txtparticle,\txtspin,\sminvtemperature}}\text{-}
\lim_{\kappa\downarrow0,\ \Lambda\uparrow\infty}
\mathsf{D}_{\txtpaulifierz,\txtspin,\sminvtemperature,\kappa,\Lambda}
(h,\prbprocess,\prbprocess^{\txtspin}).$$
Furthermore, the cutoff family is uniformly integrable.
\end{prop}

\begin{proof}
Expand \eqref{expedition0012283} of Definition \ref{expedition0012255} on a fixed $n$-jump sector.
The term corresponding to the same partition $\pi$ as in \eqref{expedition0012391} has the form
$$\begin{aligned}
\mathsf{D}^{(n,\pi)}_{\kappa,\Lambda}(h)
=
c_{\pi}
\fnexp{\oneoverfour
\fun{\opform{q}_{\txtrad,\txtbec,\sminvtemperature}^{\txteuclid}}{h}}
\fnexp{-\oneoverfour
\fun{\opform{q}_{\txtrad,\txtbec,\sminvtemperature}^{\txteuclid}}
{\fun{G_{\kappa,\Lambda}}{h,\prbprocess,\prbprocess^{\txtspin}}}}
\prod_{r=1}^{m(\pi)}
Q_{r,\kappa,\Lambda},
\end{aligned}$$
where each $Q_{r,\kappa,\Lambda}$ is one of the four types in \eqref{expedition0012397}.
Comparing two cutoffs $0<\kappa'<\kappa<\Lambda<\Lambda'<\infty$,
Hölder's inequality,
\eqref{expedition0012398},
\eqref{expedition0012408},
\eqref{expedition0012409},
and the local Lipschitz estimate for the exponential function give
\begin{equation}\label{expedition0012411}
\begin{aligned}
&\norm{\mathsf{D}^{(n,\pi)}_{\kappa',\Lambda'}(h)
-\mathsf{D}^{(n,\pi)}_{\kappa,\Lambda}(h)}_{\lp^1}
\\ %%%%%%%%%%%%%%%%
&\leq
C_{n,\pi,h}
\norm{\mathsf{U}_{\txtpaulifierz,\sminvtemperature,\kappa',\Lambda'}
-\mathsf{U}_{\txtpaulifierz,\sminvtemperature,\kappa,\Lambda}}_{\lp^{m(\pi)+1}}
\\ %%%%%%%%%%%%%%%%
&\quad+
C_{n,\pi,h}
\sum_{a=1}^{N_h}
\norm{\mathsf{Y}_{\txtpaulifierz,\sminvtemperature,\kappa',\Lambda';f_a}
-\mathsf{Y}_{\txtpaulifierz,\sminvtemperature,\kappa,\Lambda;f_a}}
_{\lp^{m(\pi)+1}}
\\ %%%%%%%%%%%%%%%%
&\quad+
C_{n,\pi,h}
\rbk{\mathfrak{E}_{h}(\set{k}{\kappa'\leq\abs{k}<\kappa})
+\mathfrak{E}_{h}(\set{k}{\Lambda<\abs{k}\leq\Lambda'})}
\\ %%%%%%%%%%%%%%%%
&\quad+
C_{n,\pi,h}
\rbk{\mathfrak{P}(\set{k}{\kappa'\leq\abs{k}<\kappa})
+\mathfrak{P}(\set{k}{\Lambda<\abs{k}\leq\Lambda'})}.
\end{aligned}
\end{equation}
The first term on the right-hand side is the self term of the minimal-coupling current from
Proposition \ref{expedition0012165}(1) and Theorem \ref{expedition0012212}
for the spinless Pauli--Fierz model.
The second term on the right-hand side is the external-field--current cross term from
Proposition \ref{expedition0012165}(2) and Lemma \ref{expedition0012306}.
The third and fourth terms on the right-hand side are the Pauli-term tails from
\eqref{expedition0012403}, \eqref{expedition0012405}, and \eqref{expedition0012408}.
Therefore the right-hand side of \eqref{expedition0012411} converges to $0$ as
$\kappa,\kappa'\downarrow0$ and $\Lambda,\Lambda'\uparrow\infty$.
On a fixed $n$-jump sector the number of partitions is finite,
so $\mathsf{D}_{\txtpaulifierz,\txtspin,\sminvtemperature,\kappa,\Lambda}
(h,\prbprocess,\prbprocess^{\txtspin})$ is a Cauchy sequence in $\lp^{1}$.

We now discuss a dominating function for summing over all $n$.
Combining the partition-sum estimate in Proposition \ref{expedition0012290}
with the current-source moment estimate in Proposition \ref{expedition0012165},
the factors containing $h$ only increase a constant determined by finitely many $f_j$ and times $t_j$,
and this constant is independent of the cutoffs.
Thus, if the $n$-jump sector is denoted by $\mathsf{D}_{\kappa,\Lambda}^{(n)}(h)$,
$$\begin{aligned}
\norm{\mathsf{D}_{\kappa,\Lambda}^{(n)}(h)}_{\lp^1}
\leq
C_h^{n+1}\sqrt{n!},
\quad
\norm{\mathsf{D}_{\kappa,\Lambda}^{(n)}(h)
-\mathsf{D}^{(n)}(h)}_{\lp^1}
\longrightarrow 0
\end{aligned}$$
holds.
Since the mass of the $n$-jump sector of the closed spin path measure is at most
$\sminvtemperature^n/n!$,
$$\sum_{n=0}^{\infty}
C_h^{n+1}
\sqrt{n!}
\frac{\sminvtemperature^n}{n!}
< \infty$$
holds.
This series is independent of the cutoffs and gives $\lp^{1}$ convergence and uniform integrability of
$\mathsf{D}_{\txtpaulifierz,\txtspin,\sminvtemperature,\kappa,\Lambda}
(h,\prbprocess,\prbprocess^{\txtspin})$.
\end{proof}

\begin{prop}[KMS State of the Pauli--Fierz Model with Spin after Cutoff Removal]\label{expedition0012264}
Let $B$ be a bounded particle--spin function, and for finitely many $f_j
\in\sphilb{D}_{\txtrad,\txtphys,\sminvtemperature}$ set $h
= \sum_j j_{t_j}f_j$.
Then
$$\begin{aligned}
\fun{\oastate[\psi_{\txtpaulifierz,\txtspin,\sminvtemperature}]}
{B(\prbprocess,\prbprocess^{\txtspin})
\fnexp{\imunit\opfocksegalradiation(h)}}
=
\lim_{\kappa\downarrow0,\ \Lambda\uparrow\infty}
\fun{\oastate[\psi_{\txtpaulifierz,\txtspin,\sminvtemperature,\kappa,\Lambda}]}
{B(\prbprocess,\prbprocess^{\txtspin})
\fnexp{\imunit\opfocksegalradiation(h)}}
\end{aligned}$$
defines the finite-time correlation functions after cutoff removal.
In particular, for two-point exponential correlation functions,
\begin{equation}\label{expedition0012265}
\begin{aligned}
\fun{\oastate[\psi_{\txtpaulifierz,\txtspin,\sminvtemperature}]}
{\fnexp{\imunit\opfocksegalradiation(j_t f)}
\fnexp{\imunit\opfocksegalradiation(j_s g)}}
=
\fnexp{-\oneoverfour
\fun{\opform{q}_{\txtrad,\txtbec,\sminvtemperature}^{\txteuclid}}
{j_t f+j_s g}}
\mathsf{S}_{\txtpaulifierz,\txtspin,\sminvtemperature;t,s}(f,g)
\end{aligned}
\end{equation}
holds.
The two-point interaction factor after cutoff removal is defined by
$$\begin{aligned}
\smpartitionfunc_{\txtpaulifierz,\txtspin,\sminvtemperature}
&=
\sqfun{\prbexp_{\msrprb_{\txtparticle,\txtspin,\sminvtemperature}}}
{\mathsf{D}_{\txtpaulifierz,\txtspin,\sminvtemperature}
(0,\prbprocess,\prbprocess^{\txtspin})},
\\
\opdmsr{\widetilde{\msrprb}_{\txtparticle,\txtspin,\sminvtemperature}
(\prbprocess,\prbprocess^{\txtspin})}
&=
\frac{\mathsf{D}_{\txtpaulifierz,\txtspin,\sminvtemperature}
(0,\prbprocess,\prbprocess^{\txtspin})}
{\smpartitionfunc_{\txtpaulifierz,\txtspin,\sminvtemperature}}
\opdmsr{\msrprb_{\txtparticle,\txtspin,\sminvtemperature}
(\prbprocess,\prbprocess^{\txtspin})},
\\
\mathsf{S}_{\txtpaulifierz,\txtspin,\sminvtemperature;t,s}(f,g)
&=
\sqfun{\prbexp_{\widetilde{\msrprb}_{\txtparticle,\txtspin,\sminvtemperature}}}
{\frac{\mathsf{D}_{\txtpaulifierz,\txtspin,\sminvtemperature}
(j_t f+j_s g,\prbprocess,\prbprocess^{\txtspin})}
{\mathsf{D}_{\txtpaulifierz,\txtspin,\sminvtemperature}
(0,\prbprocess,\prbprocess^{\txtspin})}}.
\end{aligned}$$
\end{prop}

\begin{proof}
Take the limits of the numerator and denominator in \eqref{expedition0012259} separately.
In this proof, denote the numerator by
$$\begin{aligned}
\mathcal{N}_{\kappa,\Lambda}(B,h)
&=
\int_{\Omega_{\txtparticle,\txtspin,\sminvtemperature}}
B(\prbprocess,\prbprocess^{\txtspin})
\mathsf{D}_{\txtpaulifierz,\txtspin,\sminvtemperature,\kappa,\Lambda}
(h,\prbprocess,\prbprocess^{\txtspin})
\opdmsr{\msrprb_{\txtparticle,\txtspin,\sminvtemperature}}
(\prbprocess,\prbprocess^{\txtspin}),
\\ %%%%%%%%%%%%%%%%
\mathcal{N}(B,h)
&=
\int_{\Omega_{\txtparticle,\txtspin,\sminvtemperature}}
B(\prbprocess,\prbprocess^{\txtspin})
\mathsf{D}_{\txtpaulifierz,\txtspin,\sminvtemperature}
(h,\prbprocess,\prbprocess^{\txtspin})
\opdmsr{\msrprb_{\txtparticle,\txtspin,\sminvtemperature}}
(\prbprocess,\prbprocess^{\txtspin}).
\end{aligned}$$
By Proposition \ref{expedition0012291},
$\mathsf{D}_{\txtpaulifierz,\txtspin,\sminvtemperature,\kappa,\Lambda}(h)$ converges in $\lp^1$
to $\mathsf{D}_{\txtpaulifierz,\txtspin,\sminvtemperature}(h)$,
so for bounded $B$,
$$\begin{aligned}
\abs{\mathcal{N}_{\kappa,\Lambda}(B,h)-\mathcal{N}(B,h)}
&\leq
\norm{B}_{\infty}
\norm{\mathsf{D}_{\txtpaulifierz,\txtspin,\sminvtemperature,\kappa,\Lambda}(h)
-\mathsf{D}_{\txtpaulifierz,\txtspin,\sminvtemperature}(h)}_{\lp^1}
\longrightarrow
0
\end{aligned}$$
holds.

The denominator is obtained from the numerator by taking $B=1$ and $h=0$.
In the notation of Definition \ref{expedition0012257}, set
$$\begin{aligned}
\smpartitionfunc_{\txtpaulifierz,\txtspin,\sminvtemperature,\kappa,\Lambda}
=
\mathcal{N}_{\kappa,\Lambda}(1,0),
\quad
\smpartitionfunc_{\txtpaulifierz,\txtspin,\sminvtemperature}
=
\mathcal{N}(1,0).
\end{aligned}$$
Then, by Proposition \ref{expedition0012291} in the case $h=0$,
$$\begin{aligned}
\abs{\smpartitionfunc_{\txtpaulifierz,\txtspin,\sminvtemperature,\kappa,\Lambda}
-\smpartitionfunc_{\txtpaulifierz,\txtspin,\sminvtemperature}}
&\leq
\norm{\mathsf{D}_{\txtpaulifierz,\txtspin,\sminvtemperature,\kappa,\Lambda}(0)
-\mathsf{D}_{\txtpaulifierz,\txtspin,\sminvtemperature}(0)}_{\lp^1}
\longrightarrow
0
\end{aligned}$$
is obtained.
For each cutoff, $\smpartitionfunc_{\txtpaulifierz,\txtspin,\sminvtemperature,\kappa,\Lambda}
=
\operatorname{Tr}
\napiernum^{-\sminvtemperature
\physham_{\txtpaulifierz,\txtspin,\kappa,\Lambda}}
>0$ holds.
The limiting side is the diagonal integral of the local semigroup with spin after cutoff removal,
$$\begin{aligned}
\smpartitionfunc_{\txtpaulifierz,\txtspin,\sminvtemperature}
&=
\int_{\fldreal^3}
\sum_{\sigma\in\ringratint_2}
K_{\txtpaulifierz,\txtspin,\sminvtemperature}(x,\sigma;x,\sigma)
\opdmsr{x},
\end{aligned}$$
and by positivity of the local semigroup kernel and the semigroup property starting from the identity kernel,
the right-hand side is a finite positive number.
Therefore, for sufficiently wide cutoffs,
$\begin{aligned}
\smpartitionfunc_{\txtpaulifierz,\txtspin,\sminvtemperature,\kappa,\Lambda}
\geq
\onehalf
\smpartitionfunc_{\txtpaulifierz,\txtspin,\sminvtemperature}
>0
\end{aligned}$ holds.

\eqref{expedition0012259} becomes, in this notation,
$$\begin{aligned}
\fun{\oastate[\psi_{\txtpaulifierz,\txtspin,\sminvtemperature,\kappa,\Lambda}]}
{B(\prbprocess,\prbprocess^{\txtspin})
\fnexp{\imunit\opfocksegalradiation(h)}}
=
\fnexp{-\oneoverfour
\fun{\opform{q}_{\txtrad,\txtbec,\sminvtemperature}^{\txteuclid}}{h}}
\frac{\mathcal{N}_{\kappa,\Lambda}(B,h)}
{\smpartitionfunc_{\txtpaulifierz,\txtspin,\sminvtemperature,\kappa,\Lambda}}.
\end{aligned}$$
The free Gaussian factor is independent of the cutoffs.
Furthermore, the convergence of the numerator and denominator above and the lower bound on the denominator give
$$\begin{aligned}
&\abs{\frac{\mathcal{N}_{\kappa,\Lambda}(B,h)}
{\smpartitionfunc_{\txtpaulifierz,\txtspin,\sminvtemperature,\kappa,\Lambda}}
-\frac{\mathcal{N}(B,h)}
{\smpartitionfunc_{\txtpaulifierz,\txtspin,\sminvtemperature}}}
\\ %%%%%%%%%%%%%%%%
&\leq
\frac{2}{\smpartitionfunc_{\txtpaulifierz,\txtspin,\sminvtemperature}}
\abs{\mathcal{N}_{\kappa,\Lambda}(B,h)-\mathcal{N}(B,h)}
+\frac{2\abs{\mathcal{N}(B,h)}}
{\smpartitionfunc_{\txtpaulifierz,\txtspin,\sminvtemperature}^{2}}
\abs{\smpartitionfunc_{\txtpaulifierz,\txtspin,\sminvtemperature,\kappa,\Lambda}
-\smpartitionfunc_{\txtpaulifierz,\txtspin,\sminvtemperature}}
\longrightarrow
0.
\end{aligned}$$
Therefore the finite-time correlation functions after cutoff removal are defined by
$$\begin{aligned}
\fun{\oastate[\psi_{\txtpaulifierz,\txtspin,\sminvtemperature}]}
{B(\prbprocess,\prbprocess^{\txtspin})
\fnexp{\imunit\opfocksegalradiation(h)}}
=
\fnexp{-\oneoverfour
\fun{\opform{q}_{\txtrad,\txtbec,\sminvtemperature}^{\txteuclid}}{h}}
\frac{\mathcal{N}(B,h)}
{\smpartitionfunc_{\txtpaulifierz,\txtspin,\sminvtemperature}}.
\end{aligned}$$

For the two-point exponential correlation function, take $B
= 1$ and $h
= j_t f+j_s g$.
Then
$$\begin{aligned}
\mathsf{S}_{\txtpaulifierz,\txtspin,\sminvtemperature;t,s}(f,g)
=
\sqfun{\prbexp_{\widetilde{\msrprb}_{\txtparticle,\txtspin,\sminvtemperature}}}
{\frac{
\mathsf{D}_{\txtpaulifierz,\txtspin,\sminvtemperature}
(j_t f+j_s g,\prbprocess,\prbprocess^{\txtspin})
}{
\mathsf{D}_{\txtpaulifierz,\txtspin,\sminvtemperature}
(0,\prbprocess,\prbprocess^{\txtspin})}}
\end{aligned}$$
The free Gaussian factor $\fnexp{-\oneoverfour
\fun{\opform{q}_{\txtrad,\txtbec,\sminvtemperature}^{\txteuclid}}
{j_t f+j_s g}}$ is independent of the cutoffs,
so the desired formula \eqref{expedition0012265} follows.
\end{proof}

\subsection{Physical Radiation Field after Cutoff Removal}\label{physical-radiation-field-after-cutoff-removal}

This cutoff-removal construction for the model with spin uses the same Yukalov selection criterion for the field operator \cite{VIYukalov001}: after the one-point correction term is extracted from the interacting functional, the physical radiation field is the centered field whose expectation vanishes on the physical test-function space.

\begin{defn}[One-Point Correction Kernel for the Model with Spin]\label{expedition0012266}
For arbitrary $f
\in\sphilb{D}_{\txtrad,\txtphys,\sminvtemperature}$,
\begin{equation}\label{expedition0012267}
\mathsf{Y}_{\txtpaulifierz,\txtspin,\sminvtemperature,\kappa,\Lambda;f}
(\prbprocess,\prbprocess^{\txtspin})
=
\imunit
\frac{\fnrestr{\opod{s}
\mathsf{D}_{\txtpaulifierz,\txtspin,\sminvtemperature,\kappa,\Lambda}
(s j_0 f,\prbprocess,\prbprocess^{\txtspin})}
{s=0}}
{\mathsf{D}_{\txtpaulifierz,\txtspin,\sminvtemperature,\kappa,\Lambda}
(0,\prbprocess,\prbprocess^{\txtspin})}
\end{equation}
and define the correction kernel after cutoff removal
$\mathsf{Y}
_{\txtpaulifierz,\txtspin,\sminvtemperature;f}$ as the $\lp^{1}$ limit of the right-hand side
as $\kappa\downarrow0,\Lambda\uparrow\infty$.
\end{defn}

\begin{prop}[Physical Radiation Field of the Model with Spin after Cutoff Removal]\label{expedition0012268}
The one-point correction kernel in \eqref{expedition0012267} has an $\lp^{1}$ limit
$\mathsf{Y}
_{\txtpaulifierz,\txtspin,\sminvtemperature;f}$,
obtained by applying the dominating function of Proposition \ref{expedition0012291} to the finite sum
arising as the first derivative of \eqref{expedition0012283}.
Moreover, in $\lp^{1}$,
$$\begin{aligned}
\mathsf{Y}_{\txtpaulifierz,\txtspin,\sminvtemperature,\kappa,\Lambda;f}
\mathsf{D}_{\txtpaulifierz,\txtspin,\sminvtemperature,\kappa,\Lambda}(0)
\to
\mathsf{Y}_{\txtpaulifierz,\txtspin,\sminvtemperature;f}
\mathsf{D}_{\txtpaulifierz,\txtspin,\sminvtemperature}(0)
\end{aligned}$$
holds.
Then
$$\ell_{\txtpaulifierz,\txtspin,\sminvtemperature}(f)
=
-\sqfun{\prbexp_{\widetilde{\msrprb}_{\txtparticle,\txtspin,\sminvtemperature}}}
{\mathsf{Y}_{\txtpaulifierz,\txtspin,\sminvtemperature;f}
(\prbprocess,\prbprocess^{\txtspin})}$$
is defined and $\fun{\oastate[\psi_{\txtpaulifierz,\txtspin,\sminvtemperature}]}
{\opfocksegalradiation(j_0 f)}
=
\ell_{\txtpaulifierz,\txtspin,\sminvtemperature}(f)$ holds.
In particular, the expectation of
$$\opfocksegalradiation_{\txtphys,\txtspin,\sminvtemperature}(j_0 f)
=
\opfocksegalradiation(j_0 f)
-\ell_{\txtpaulifierz,\txtspin,\sminvtemperature}(f)$$
with respect to the KMS state vanishes.
\end{prop}

\begin{proof}
By \eqref{expedition0012267},
$$\fnrestr{\opod{s}
\mathsf{D}_{\txtpaulifierz,\txtspin,\sminvtemperature,\kappa,\Lambda}
(s j_0 f,\prbprocess,\prbprocess^{\txtspin})}
{s=0}
=
-\imunit
\mathsf{Y}_{\txtpaulifierz,\txtspin,\sminvtemperature,\kappa,\Lambda;f}
(\prbprocess,\prbprocess^{\txtspin})
\mathsf{D}_{\txtpaulifierz,\txtspin,\sminvtemperature,\kappa,\Lambda}
(0,\prbprocess,\prbprocess^{\txtspin})$$
is obtained.
Differentiating \eqref{expedition0012283} once in $s$ adds exactly one covariance form,
either between $j_0f$ and the current source,
between $j_0f$ and
$\mathsf{I}^{\txtpauli,\txtdiag}_{\txtrad,S_{\sminvtemperature},\kappa,\Lambda}$,
or between $j_0f$ and
$\mathsf{I}^{\txtpauli,\txtflip}_{\txtrad,j,\kappa,\Lambda}$,
to the partition sum handled in the proof of Proposition \ref{expedition0012291}.
The cross term between $j_0f$ and the current source is handled by Proposition \ref{expedition0012165}(2)
and Lemma \ref{expedition0012306},
and the cross term between $j_0f$ and a magnetic-flux-density factor is handled by the domination condition
of the physical test-function space and the one-factor estimate of Proposition \ref{expedition0012290}.
Therefore the absolute value on the $n$-jump sector is bounded by
$C_f^{n+1}\sqrt{(n+1)!}$ for a constant $C_f$.
Multiplying by the mass of the $n$-jump sector of the closed spin path measure gives
$\sum_{n=0}^{\infty}
C_f^{n+1}\sqrt{(n+1)!}\frac{\sminvtemperature^n}{n!}<\infty$.
This series gives a cutoff-independent $\lp^{1}$ dominating function for $\mathsf{Y}
_{\txtpaulifierz,\txtspin,\sminvtemperature,\kappa,\Lambda;f}$ and $\mathsf{Y}
_{\txtpaulifierz,\txtspin,\sminvtemperature,\kappa,\Lambda;f}
\mathsf{D}_{\txtpaulifierz,\txtspin,\sminvtemperature,\kappa,\Lambda}
(0,\prbprocess,\prbprocess^{\txtspin})$.
Since the cutoff difference of each covariance form converges to $0$ by
Proposition \ref{expedition0012165} and Proposition \ref{expedition0012290},
$\mathsf{Y}
_{\txtpaulifierz,\txtspin,\sminvtemperature,\kappa,\Lambda;f}$ converges in $\lp^{1}$ to
$\mathsf{Y}
_{\txtpaulifierz,\txtspin,\sminvtemperature;f}$,
and moreover
$$\mathsf{Y}_{\txtpaulifierz,\txtspin,\sminvtemperature,\kappa,\Lambda;f}
\mathsf{D}_{\txtpaulifierz,\txtspin,\sminvtemperature,\kappa,\Lambda}
(0,\prbprocess,\prbprocess^{\txtspin})
\longrightarrow
\mathsf{Y}_{\txtpaulifierz,\txtspin,\sminvtemperature;f}
\mathsf{D}_{\txtpaulifierz,\txtspin,\sminvtemperature}(0)$$
holds in $\lp^{1}$.

Set $h=sj_0f$ in \eqref{expedition0012259} and differentiate the right-hand side of
Definition \ref{expedition0012257} in $s$.
The derivative of the free Gaussian factor at $s
= 0$ vanishes,
and the denominator is independent of $s$.
The $s$-derivative computed above gives
$$\begin{aligned}
&\fnrestr{\opod{s}
\fun{\oastate[\psi_{\txtpaulifierz,\txtspin,\sminvtemperature,\kappa,\Lambda}]}
{\fnexp{\imunit s\opfocksegalradiation(j_0 f)}}}{s=0}
\\ %%%%%%%%%%%%%%%%
&=
\sqfun{\prbexp_{\widetilde{\msrprb}_{\txtparticle,\txtspin,\sminvtemperature,\kappa,\Lambda}}}
{\frac{
\fnrestr{\opod{s}
\mathsf{D}_{\txtpaulifierz,\txtspin,\sminvtemperature,\kappa,\Lambda}
(s j_0 f,\prbprocess,\prbprocess^{\txtspin})}
{s=0}
}{
\mathsf{D}_{\txtpaulifierz,\txtspin,\sminvtemperature,\kappa,\Lambda}
(0,\prbprocess,\prbprocess^{\txtspin})}}
\\ %%%%%%%%%%%%%%%%
&=
-\imunit
\sqfun{\prbexp_{\widetilde{\msrprb}_{\txtparticle,\txtspin,\sminvtemperature,\kappa,\Lambda}}}
{
\mathsf{Y}_{\txtpaulifierz,\txtspin,\sminvtemperature,\kappa,\Lambda;f}
(\prbprocess,\prbprocess^{\txtspin})}
\end{aligned}$$
and in particular
$$\fnrestr{\opod{s}
\fun{\oastate[\psi_{\txtpaulifierz,\txtspin,\sminvtemperature,\kappa,\Lambda}]}
{\fnexp{\imunit s\opfocksegalradiation(j_0 f)}}}{s=0}
=
\imunit
\fun{\oastate[\psi_{\txtpaulifierz,\txtspin,\sminvtemperature,\kappa,\Lambda}]}
{\opfocksegalradiation(j_0 f)}$$
is obtained.
Removing the cutoffs by the quotient limit in Proposition \ref{expedition0012264}
and the $\lp^{1}$ convergence above gives
$\ell
_{\txtpaulifierz,\txtspin,\sminvtemperature}(f)$.
Summarizing the preceding argument,
$$\begin{aligned}
\fun{\oastate[
\psi_{\txtpaulifierz,\txtspin,\sminvtemperature}]}
{\opfocksegalradiation_{\txtphys,\txtspin,\sminvtemperature}(j_0 f)}
=
\fun{\oastate[\psi_{\txtpaulifierz,\txtspin,\sminvtemperature}]}
{\opfocksegalradiation(j_0 f)}
-\ell_{\txtpaulifierz,\txtspin,\sminvtemperature}(f)
= 0
\end{aligned}$$
is obtained.
\end{proof}

\subsection{Off-Diagonal Long-Range Order, Order Parameters, and the No-Go Theorem for BEC}\label{off-diagonal-long-range-order-order-parameters-and-the-no-go-theorem-for-bec-2}

\begin{defn}[Order Parameter of the Pauli--Fierz Model with Spin]\label{expedition0012269}
Use the same $\mathsf{b}_{L,e}^{(\#)}$ as in Definition \ref{expedition0012168}. For the bounded-system Pauli--Fierz model with spin state $\psi_{\txtpaulifierz,\txtspin,\sminvtemperature,\smchemicalpotential,L}$, define
\begin{equation}\label{expedition0012439}
\begin{aligned}
\mathsf{o}_{\txtpaulifierz,\txtspin,\sminvtemperature,L,e}^{(\#)}
=
\imunit
\fun{\psi_{\txtpaulifierz,\txtspin,\sminvtemperature,\smchemicalpotential,L}}
{\fun{\oaresolvent}{1,\mathsf{b}_{L,e}^{(\#)}}}
\end{aligned}
\end{equation}
and call it the order parameter of the Pauli--Fierz model with spin.
\end{defn}

\begin{prop}[Off-Diagonal Long-Range Order of the Pauli--Fierz Model with Spin]\label{expedition0012418}
For the Pauli--Fierz-model-with-spin KMS state after cutoff removal constructed in Proposition \ref{expedition0012264} and $f,g \in \sphilb{D}_{\txtrad,\txtphys,\sminvtemperature}$, the two-point off-diagonal long-range order of the radiation field is given by
\begin{equation}\label{expedition0012419}
\lim_{\abs{x}\to\infty}
\fun{\oastate[\psi_{\txtpaulifierz,\txtspin,\sminvtemperature}]}
{\opfocksegalradiation(j_0 f)\opfocksegalradiation(j_0\tau_x g)}
=
\onehalf\fun{\opform{q}_{\txtrad,0,\sminvtemperature}}{f,g}.
\end{equation}
\end{prop}

\begin{proof}
Use the two-point exponential correlation-function representation of Proposition \ref{expedition0012264}. In the Gaussian factor of the free radiation field, as in the spinless model, the non-zero-mode cross term disappears by the Riemann--Lebesgue lemma, and only the zero-mode form remains independent of the spatial variable. The spin degrees of freedom and the Pauli term are contained in $\mathsf{D}_{\txtpaulifierz,\txtspin,\sminvtemperature}$, but the same oscillatory factor appears in the cross term with the translated external test function. Using the cutoff-removal estimate of Proposition \ref{expedition0012290} and the quotient limit of Proposition \ref{expedition0012264}, this cross term disappears in the long-distance limit. Differentiating the two-point characteristic functional gives \eqref{expedition0012419}.
\end{proof}

\begin{prop}[Order-Parameter Criterion for the Pauli--Fierz Model with Spin]\label{expedition0012420}
For the order parameter of Definition \ref{expedition0012269}, the following assertions hold. If the zero-mode density corresponding to the polarization direction $e$ satisfies $\smnumberdensity_{\txtrad,0}(\sminvtemperature)>0$, then
$$\lim_{L\to\infty}\mathsf{o}_{\txtpaulifierz,\txtspin,\sminvtemperature,L,e}^{(1)}
=
\int_0^\infty
\fnexp{-r-\frac12\smnumberdensity_{\txtrad,0}(\sminvtemperature)r^2}
\opdmsr{r}
<1,$$
and if $\smnumberdensity_{\txtrad,0}(\sminvtemperature)=0$, then $$\lim_{L\to\infty}\mathsf{o}_{\txtpaulifierz,\txtspin,\sminvtemperature,L,e}^{(1)}
=1.$$
\end{prop}

\begin{proof}
The finite-volume $\mathsf{b}_{L,e}^{(1)}$ detects only the zero-momentum mode of the specified transverse polarization. The minimal-coupling current source is closed on the full circle, and the magnetic-flux-density source in the Pauli term contains the momentum factor $k\times e^{\mathrm{j}_{\txtrad}}(k)$, so it does not cross with the zero-momentum mode. Therefore, in the Laplace-transform representation of the order parameter, only the zero-mode evaluation of the free radiation field remains, and the same limit formula as in Proposition \ref{expedition0012417} is obtained.
\end{proof}

\begin{thm}[No-Go Theorem for BEC in the Pauli--Fierz Model with Spin via Off-Diagonal Long-Range Order]\label{expedition0012270}
Assume that the physical test-function space $\sphilb{D}_{\txtrad,\txtphys,\sminvtemperature}$ can distinguish the zero mode in the sense of Definition \ref{expedition0012433}. Then the following assertions are equivalent for the cutoff-removed KMS state constructed in Proposition \ref{expedition0012264}.
\begin{enumerate}
\item
For all $f,g \in \sphilb{D}_{\txtrad,\txtphys,\sminvtemperature}$, the off-diagonal long-range order in \eqref{expedition0012419} is $0$.

\item
The equality $\opform{q}_{\txtrad,0,\sminvtemperature}=0$ holds.

\item
The condition $\smnumberdensity_{\txtrad,0}(\sminvtemperature)=0$ holds.

\item
The order parameter of Definition \ref{expedition0012269} satisfies $\lim_{L\to\infty}\mathsf{o}_{\txtpaulifierz,\txtspin,\sminvtemperature,L,e}^{(1)}=1$ for every polarization direction $e$.
\end{enumerate}
In particular, the absence of off-diagonal long-range order and the order-parameter criterion are equivalent to the disappearance of the BEC zero mode of the radiation field on the physical test-function space after cutoff removal.
\end{thm}

\begin{proof}
Proposition \ref{expedition0012418} gives the equivalence between assertions (1) and (2). Zero-mode distinguishability and the zero-momentum density representation of $\opform{q}_{\txtrad,0,\sminvtemperature}$ give the equivalence between assertions (2) and (3). Proposition \ref{expedition0012420} gives the equivalence between assertions (3) and (4).
\end{proof}

\section{Operator-Algebraic Formulation}\label{operator-algebraic-formulation}

Although the type of field and the finite-dimensional degrees of freedom attached to the particle differ, the treatment of infrared directions and BEC directions in the resolvent algebra reduces to the same abstract structure. We therefore discuss the resolvent-algebraic arguments for all models together in this section. The required inputs are the algebra of particle observables, the boson-field spaces \(\sphilb{X}_{\txtbsn,0}\supset \sphilb{X}_{\txtbsn,\txtphys}\) of the Nelson model, the radiation-field spaces \(\sphilb{X}_{\txtrad,0}\supset \sphilb{X}_{\txtrad,\txtphys}\) of the Pauli--Fierz models, and the corresponding zero-mode sesquilinear forms \(\opform{q}_{\txtbsn,0,\sminvtemperature}\) and \(\opform{q}_{\txtrad,0,\sminvtemperature}\).

The KMS states constructed by functional integrals in this paper start from a particle system with a confining potential \(V\), and are therefore generally not spatially translation invariant. This does not change even if we restrict the transformation to spatial translations of the Bose field alone. The Gaussian factor of the free field is invariant under spatial translations of the field, but in the interaction factors in \eqref{expedition0012351}, \eqref{expedition0012248}, and \eqref{expedition0012265}, cross terms remain between the translated test function and the source on the particle path. Since the particle path measure is fixed by the confining potential \(V\), this cross term is generally not invariant under spatial translations of the field alone. Hence the operator-algebraic criterion for spatial translations in the latter half of this section is not asserted to apply automatically to the constructed states with a confining potential. It is positioned as an abstract sufficient condition for the case where a spatially translation-invariant KMS state is given separately.

\subsection{Ideal Structure in the Resolvent Algebra}\label{ideal-structure-in-the-resolvent-algebra}

Introduce a field label \(\mathfrak{b}\), and set \(\mathfrak{b}=\txtbsn\) for the Nelson model and \(\mathfrak{b}=\txtrad\) for the Pauli--Fierz models. Define the field spaces by \[\begin{aligned}
\sphilb{X}_{\txtbsn,0}
&=
\sphilb{D}_{\txtbsn,0,\sminvtemperature},
&
\sphilb{X}_{\txtbsn,\txtphys}
&=
\sphilb{D}_{\txtbsn,\txtphys,\sminvtemperature},
\\
\sphilb{X}_{\txtrad,0}
&=
\sphilb{D}_{\txtrad,\sminvtemperature},
&
\sphilb{X}_{\txtrad,\txtphys}
&=
\sphilb{D}_{\txtrad,\txtphys,\sminvtemperature}.
\end{aligned}\] Here \(\sphilb{D}_{\txtrad,\txtphys,\sminvtemperature}\) is the physical test-function space defined in \eqref{eq:PF-physical-test-space-majorant}.

Let the model label \(\mathfrak{m}\) be one of \(\txtnelson\), \(\txtpaulifierz\), and \(\pairbk{\txtpaulifierz,\txtspin}\). Define the algebra of particle observables by \[\begin{aligned}
\oa{A}_{\mathfrak{m},\txtparticle}
=
\begin{cases}
\oa{A}_{\txtparticle},
& \mathfrak{m}=\txtnelson,\ \txtpaulifierz,
\\
\oa{A}_{\txtparticle,\txtspin},
& \mathfrak{m}=\pairbk{\txtpaulifierz,\txtspin}.
\end{cases}
\end{aligned}\] Set the corresponding field label by \(\mathfrak{b}(\txtnelson)=\txtbsn\) and \(\mathfrak{b}(\txtpaulifierz)=\mathfrak{b}(\pairbk{\txtpaulifierz,\txtspin})=\txtrad\). Then the full observable algebras are expressed in one formula as \[\begin{aligned}
\oa{A}_{\mathfrak{m},0,\sminvtemperature}
&=
\oa{A}_{\mathfrak{m},\txtparticle}
\otimes_{\max}
\oaresolventalgebra(\sphilb{X}_{\mathfrak{b}(\mathfrak{m}),0},\sigma),
\\
\oa{A}_{\mathfrak{m},\txtphys,\sminvtemperature}
&=
\oa{A}_{\mathfrak{m},\txtparticle}
\otimes_{\max}
\oaresolventalgebra(\sphilb{X}_{\mathfrak{b}(\mathfrak{m}),\txtphys},\sigma).
\end{aligned}\] With this notation, \(\oa{A}_{\txtnelson,0,\sminvtemperature}\), \(\oa{A}_{\txtnelson,\txtphys,\sminvtemperature}\), \(\oa{A}_{\txtpaulifierz,0,\sminvtemperature}\), \(\oa{A}_{\txtpaulifierz,\txtphys,\sminvtemperature}\), \(\oa{A}_{\txtpaulifierz,\txtspin,0,\sminvtemperature}\), and \(\oa{A}_{\txtpaulifierz,\txtspin,\txtphys,\sminvtemperature}\) are treated simultaneously.

\begin{defn}[Infrared Ideal for the Field]\label{expedition0012161}
For the field label $\mathfrak{b}$, define the infrared ideal for the field by $$\oaideal{J}_{\mathfrak{b},\txtirsingular}
=
\clos{\opideal
\set{\oaresolvent(\lambda,f)}
{\lambda\in\fldreal\setminus\setone{0},
f\in\sphilb{X}_{\mathfrak{b},0}\setminus\sphilb{X}_{\mathfrak{b},\txtphys}}},$$
and define the infrared ideal in the full observable algebra for the model $\mathfrak{m}$ by $$\oaideal{J}_{\mathfrak{m},\txtirsingular}
=
\oa{A}_{\mathfrak{m},\txtparticle}
\otimes_{\max}
\oaideal{J}_{\mathfrak{b}(\mathfrak{m}),\txtirsingular}.$$
\end{defn}

\begin{prop}[Infrared Quotient]\label{expedition0012162}
For each $\mathfrak{m}$, the universality of the resolvent algebra gives a quotient map $$\Theta_{\mathfrak{m},\txtirsingular}
\colon
\oa{A}_{\mathfrak{m},0,\sminvtemperature}
\to
\oa{A}_{\mathfrak{m},\txtphys,\sminvtemperature},$$
and its kernel is $\oaideal{J}_{\mathfrak{m},\txtirsingular}$. Hence $$\setquot{\oa{A}_{\mathfrak{m},0,\sminvtemperature}}{\oaideal{J}_{\mathfrak{m},\txtirsingular}}
\eqalgisom
\oa{A}_{\mathfrak{m},\txtphys,\sminvtemperature}$$ holds, and the algebra of particle observables $\oa{A}_{\mathfrak{m},\txtparticle}$ is not collapsed in the quotient.
\end{prop}

\begin{proof}
Since $\sphilb{X}_{\mathfrak{b}(\mathfrak{m}),\txtphys}\subset \sphilb{X}_{\mathfrak{b}(\mathfrak{m}),0}$ is a linear subspace, the universality of the resolvent algebra \cite{DetlevBuchholz001} determines a surjective $\ast$-homomorphism that restricts the generators on $\sphilb{X}_{\mathfrak{b}(\mathfrak{m}),0}$ to the generators on $\sphilb{X}_{\mathfrak{b}(\mathfrak{m}),\txtphys}$. Its kernel is the closed two-sided ideal generated by the generator resolvents corresponding to $\sphilb{X}_{\mathfrak{b}(\mathfrak{m}),0}\setminus \sphilb{X}_{\mathfrak{b}(\mathfrak{m}),\txtphys}$. By the property of the maximal tensor product, the kernel in the full observable algebra is the maximal tensor product of $\oa{A}_{\mathfrak{m},\txtparticle}$ and the infrared ideal for the field. In the quotient, $\oa{A}_{\mathfrak{m},\txtparticle}$ remains as the left factor, so the particle observables are not collapsed.
\end{proof}

\begin{defn}[BEC Directions and the BEC Ideal for the Field]\label{expedition0012163}
For the field label $\mathfrak{b}$, define the BEC direction set by $$\sphilb{X}_{\mathfrak{b},\txtbec,\sminvtemperature}
=
\set{f\in\sphilb{X}_{\mathfrak{b},\txtphys}}
{\fun{\opform{q}_{\mathfrak{b},0,\sminvtemperature}}{f}>0}.$$
Let $\oaideal{J}_{\mathfrak{b},\txtbec}$ be the closed two-sided ideal generated by the generator resolvents belonging to this set, and define the BEC ideal in the full observable algebra for the model $\mathfrak{m}$ by $$\oaideal{J}_{\mathfrak{m},\txtbec}
=
\oa{A}_{\mathfrak{m},\txtparticle}
\otimes_{\max}
\oaideal{J}_{\mathfrak{b}(\mathfrak{m}),\txtbec}.$$
\end{defn}

The infrared ideal removes by quotienting the directions in which the field operator after cutoff removal is not physically defined. In contrast, the BEC ideal is an auxiliary ideal recording directions in which the zero-mode covariance remains positive even after entering the physical test-function space. This feature is unchanged across the three models. The difference between the models is the type of field and the difference between the particle-observable algebras \(\oa{A}_{\txtparticle}\) and \(\oa{A}_{\txtparticle,\txtspin}\).

\begin{prop}[Vanishing of the BEC Ideal under Vanishing of the Zero-Mode Form]\label{expedition0012164}
Assume that the sesquilinear form $\opform{q}_{\mathfrak{b}(\mathfrak{m}),0,\sminvtemperature}$ is the zero form on $\sphilb{X}_{\mathfrak{b}(\mathfrak{m}),\txtphys}$. Then $$\sphilb{X}_{\mathfrak{b}(\mathfrak{m}),\txtbec,\sminvtemperature}
= \emptyset,
\quad
\oaideal{J}_{\mathfrak{b}(\mathfrak{m}),\txtbec}
= \setone{0},
\quad
\oaideal{J}_{\mathfrak{m},\txtbec}
=
\setone{0}$$ holds.
\end{prop}

\begin{proof}
By assumption, there is no positive BEC direction on $\sphilb{X}_{\mathfrak{b}(\mathfrak{m}),\txtphys}$. By Definition \ref{expedition0012163}, $\sphilb{X}_{\mathfrak{b}(\mathfrak{m}),\txtbec,\sminvtemperature}=\emptyset$, and the BEC ideal for the field is zero. Since the BEC ideal in the full observable algebra is $\oa{A}_{\mathfrak{m},\txtparticle}\otimes_{\max}\oaideal{J}_{\mathfrak{b}(\mathfrak{m}),\txtbec}$, it is also zero.
\end{proof}

\subsection{Operator-Algebraic Order-Parameter Criterion}\label{expedition0012421}

For the three models discussed here, we directly computed off-diagonal long-range order from exponential observables of the physical field. Here we organize the same zero-mode criterion from the viewpoint of order-parameter nets in the resolvent algebra and the center.

\begin{defn}[Resolvent Order-Parameter Net]\label{expedition0012422}
Assume that a directed family $\fml{b_{\nu}}{\nu}$ in the corresponding physical test-function space satisfies the following conditions.
\begin{enumerate}
\item
The family $\fml{b_{\nu}}{\nu}$ has the same zero-mode normalization as the $\mathsf{b}^{(1)}$-type order parameters in Definitions \ref{expedition0012159}, \ref{expedition0012168}, and \ref{expedition0012269}.

\item
For every physical test function $h$, $\sigma(b_{\nu},h)\to0$ holds.

\item
The norm of $b_{\nu}$ measured by the non-zero-mode form $\opform{q}_{\mathfrak{b},\txtnonzero,\sminvtemperature}$ converges to $0$.
\end{enumerate}
A family satisfying these conditions is called a resolvent order-parameter net. The collection of all order-parameter nets for each model is denoted by $\mathcal{N}_{\mathfrak{m}}^{\mathrm{res}}$.
\end{defn}

\begin{prop}[Asymptotic Centrality of Order Parameters]\label{expedition0012423}
For every $\fml{b_{\nu}}{\nu}\in\mathcal{N}_{\mathfrak{m}}^{\mathrm{res}}$ and every $A\in \oa{A}_{\mathfrak{m},\txtphys,\sminvtemperature}$,
\begin{equation}\label{expedition0012424}
\norm{\commutator{\idone\otimes \oaresolvent(1,b_{\nu})}{A}}\to0
\end{equation}
holds.
\end{prop}

\begin{proof}
It is enough to verify the assertion for resolvent generators $A_0\otimes\oaresolvent(\lambda,h)$. By the commutator formula in the resolvent relations, the right-hand side is bounded by a constant multiple of $\abs{\sigma(b_{\nu},h)}$, and this converges to $0$ by the second condition of Definition \ref{expedition0012422}. For finite products, use the Leibniz formula for commutators, and finally use the density of the resolvent generators and $\norm{\oaresolvent(1,b_{\nu})}\leq1$.
\end{proof}

\begin{defn}[$\lp^{1}$ Asymptotic Abelianness and Weak Clustering]\label{expedition0012425}
Consider a $\oacstar$-dynamical system $\pairbk{\oa{A},\tau}$ with spatial translations $\fml{\tau_x}{x\in\fldreal^3}$ and a dense $\ast$-subalgebra $\oa{A}_{\txtloc}\subset\oa{A}$.
\begin{enumerate}
\item
If $$\int_{\fldreal^3}
\norm{\commutator{A}{\tau_x(B)}}
\opdmsr{x}
< \infty$$ holds for all $A,B\in \oa{A}_{\txtloc}$, then $\pairbk{\oa{A},\tau}$ is said to satisfy $\lp^{1}$ asymptotic abelianness on $\oa{A}_{\txtloc}$.

\item
If a spatially translation-invariant state $\oastate[\psi]$ satisfies $$\lim_{L\to\infty}
\frac{1}{\abs{\Lambda_L}}
\int_{\Lambda_L}
\fun{\oastate[\psi]}{A\tau_x(B)}
\opdmsr{x}
=
\fun{\oastate[\psi]}{A}\fun{\oastate[\psi]}{B}$$ for all $A,B\in \oa{A}_{\txtloc}$, then $\oastate[\psi]$ is said to satisfy weak clustering on $\oa{A}_{\txtloc}$.
\end{enumerate}
Here $\Lambda_L$ is a Folner sequence of cubes in $\fldreal^3$.
\end{defn}

\begin{prop}[Scalarization of the Central Limit by Weak Clustering]\label{expedition0012426}
Consider a spatially translation-invariant state $\oastate[\psi]$ satisfying the weak clustering property of Definition \ref{expedition0012425}. Suppose that, in the GNS representation, the strong limit $\slim_{L\to\infty}\oarepn_{\psi}(B_L)=Z$ exists for the finite-volume averages $B_L=\frac{1}{\abs{\Lambda_L}}\int_{\Lambda_L}\tau_x(B)\opdmsr{x}$ and satisfies $Z\in \oacenter(\oa{M}_{\psi})$. Then $Z\oagnsvector_{\psi}=\fun{\oastate[\psi]}{B}\oagnsvector_{\psi}$ holds. If $\oastate[\psi]$ is a primary state, then $Z=\fun{\oastate[\psi]}{B}\idone_{\sphilb{H}_{\psi}}$.
\end{prop}

\begin{proof}
For every $A\in \oa{A}_{\txtloc}$, weak clustering gives $\bkt{\oarepn_{\psi}(\faadj{A})\oagnsvector_{\psi}}{Z \oagnsvector_{\psi}}=\fun{\oastate[\psi]}{A}\fun{\oastate[\psi]}{B}$. By cyclicity of the GNS vector, $Z\oagnsvector_{\psi}=\fun{\oastate[\psi]}{B}\oagnsvector_{\psi}$ holds. For a primary state, $\oa{M}_{\psi}$ is a factor, so the central operator $Z$ is equal to the displayed scalar operator.
\end{proof}

\begin{thm}[Araki--Haag--Kastler--Takesaki-Type No-Go Criterion for BEC in Spatially Translation-Invariant States]\label{expedition0012427}
Assume that a spatially translation-invariant KMS state $\oastate[\psi_{\mathfrak{m},\sminvtemperature}]$ corresponding to the time evolution of each model is given, and that the corresponding physical test-function space can distinguish the zero mode in the sense of Definition \ref{expedition0012433}. For the corresponding field $\mathfrak{b}=\mathfrak{b}(\mathfrak{m})$ and centered physical field $\Phi_{\mathfrak{b}}^{\circ}(f)$, define the zero-mode sesquilinear form $\opform{q}_{\mathfrak{b},0,\sminvtemperature}^{\psi}$ by the two-point long-distance limit
\begin{equation}\label{expedition0012434}
\lim_{\abs{x}\to\infty}
\fun{\oastate[\psi_{\mathfrak{m},\sminvtemperature}]}
{\Phi_{\mathfrak{b}}^{\circ}(f)\tau_x(\Phi_{\mathfrak{b}}^{\circ}(g))}
=
\onehalf\fun{\opform{q}_{\mathfrak{b},0,\sminvtemperature}^{\psi}}{f,g},
\end{equation}
and assume that this form has a density representation for the zero-mode evaluation in Definition \ref{expedition0012433}. Assume further that the following conditions hold on a suitable local subalgebra of the resolvent algebra.
\begin{enumerate}
\item
The $\lp^{1}$ asymptotic abelianness of Definition \ref{expedition0012425} holds.

\item
The state $\oastate[\psi_{\mathfrak{m},\sminvtemperature}]$ satisfies the weak clustering property of Definition \ref{expedition0012425}.

\item
The state $\oastate[\psi_{\mathfrak{m},\sminvtemperature}]$ is primary.

\item
For the two-point function of the centered physical field, the Cesaro average obtained from weak clustering is $0$.
\end{enumerate}
Then $\opform{q}_{\mathfrak{b},0,\sminvtemperature}^{\psi}=0$ holds. In particular, the corresponding BEC ideal defined by the same method as in Definition \ref{expedition0012163} from this zero-mode form is trivial by the same argument as Proposition \ref{expedition0012164}.
\end{thm}

\begin{proof}
By assumption (4), the Cesaro-averaged long-distance correlation of the centered physical field is $0$. On the other hand, by \eqref{expedition0012434}, the pointwise long-distance limit is equal to the zero-mode form $\opform{q}_{\mathfrak{b},0,\sminvtemperature}^{\psi}$. When the pointwise limit exists, the Cesaro limit has the same value, so the corresponding zero-mode form vanishes on the physical test-function space. By zero-mode distinguishability, the corresponding zero-mode density also vanishes. If the BEC directions and the BEC ideal are defined by the same method as in Definition \ref{expedition0012163}, then the BEC ideal is trivial by the same argument as Proposition \ref{expedition0012164}.

Assumption (1) is a sufficient condition ensuring the asymptotic centrality of the order parameters in Proposition \ref{expedition0012423}, and assumptions (2) and (3) scalarize the central limit in the primary representation by Proposition \ref{expedition0012426}. This is consistent with the zero-mode disappearance criterion by the absence of off-diagonal long-range order above.
\end{proof}

\bibliography{myref.bib}

\end{document}

%% file: mycommands.tex
\makeatletter
\providecommand*{\dashv}{\mathrel{\mathpalette\@Dashv\vDash}}
\newcommand*{\@dashv}[2]{\reflectbox{$\m@th#1#2$}}
\makeatother
\newcommand{\abscard}[1]{\abs{#1}} % 絶対値記号による濃度の記号
\newcommand{\absvol}[1]{\abs{#1}} % 体積
\newcommand{\abs}[1]{\left| #1 \right|} % 絶対値
\NewDocumentCommand{\weaknorm}{O{\dbk} m}{#1{#2}} % 弱(準)ノルム
\newcommand{\norm}[1]{\left\Vert #1 \right\Vert} % ノルム
\newcommand{\pnorm}[1]{\left\Vert #1 \right\Vert_p} % pノルム
\newcommand{\twonorm}[1]{\norm{#1}_2}
\newcommand{\bkt}[2]{\left\langle #1,\,#2 \right\rangle} % 内積
\newcommand{\dualbkt}[2]{\bkt{#1}{#2}_{\txtdual}} % 双対内積
\newcommand{\rbkt}[2]{\left( #1,\,#2 \right)} % 内積に利用
\newcommand{\slim}{\mathrm{s} \hyphen \lim} % 強極限
\newcommand{\isomto}{\mathrel{\rightarrowtail\kern-1.9ex\twoheadrightarrow}} % 全単射, 同型射, ホモロジー代数・圏論でよく使う
\newcommand{\cbk}[1]{\left\{ #1 \right\}} % 中括弧
\newcommand{\dbk}[1]{\left\langle #1 \right\rangle} % <f>形式の括弧, Dirac bracketの意味でdbk
\newcommand{\pairbk}[1]{\rbk{#1}} % 適当な対のための括弧
\newcommand{\rbkleft}[1]{\left( #1 \right.} % 片側括弧
\newcommand{\rbkright}[1]{\left. #1 \right)} % T片側括弧
\newcommand{\rbk}[1]{\left( #1 \right)} % 丸括弧, 共通
\newcommand{\sqbk}[1]{\left[ #1 \right]} % いわゆる大括弧
\newcommand{\vecbk}[1]{\rbk{#1}} % ベクトル用の括弧
\newcommand{\funrbk}[2]{\fun{\rbk{#1}}{#2}} % 関数の部分に丸括弧を入れてまとめる
\newcommand{\fun}[2]{#1 \rbk{#2}} % 関数テンプレート
\newcommand{\sqfuncond}[3]{\sqfun{#1}{#2 \middle| #3}} % 角括弧を使う条件つき期待値などに利用する関数
\newcommand{\sqfun}[2]{#1 \sqbk{#2}} % 角括弧を使う関数, 確率論の期待値や完全な熱力学関数に利用
\newcommand{\closedinterval}[2]{\sqbk{#1,\,#2}} % 閉区間
\newcommand{\commutator}[2]{\sqbk{#1,\,#2}} % 交換子
\DeclareMathOperator{\sgn}{sgn} % 符号, 共通
\NewDocumentCommand{\imunit}{O{\mathsf{i}}}{#1} % 虚数単位, 共通
\NewDocumentCommand{\placeholder}{O{\bullet}}{#1} % 関手やホモロジーなどでハイフンや点で表す要素
\NewDocumentCommand{\trace}{O{\operatorname{Tr}}}{#1} % トレース
\newcommand{\Ker}{\operatorname{Ker}} % 核の集合
\newcommand{\bigmiddleslash}[2]{\left. #1 \middle/ #2 \right.} % 主に商集合\setquotに利用
\newcommand{\clos}[1]{\overline{#1}} % 外測度など何らかの意味で拡張・閉包のような趣があるものの, 具体的に「---閉包」のような名前がない対象に利用
\newcommand{\cmpconj}[1]{\overline{#1}} % 複素共役
\newcommand{\diracdelta}{\delta} % ディラックのデルタ関数
\newcommand{\dom}{\operatorname{dom}} % 定義域
\newcommand{\ep}{\varepsilon} % エイリアス
\newcommand{\eqcsq}[1]{\sqbk{#1}} % 同値類の記号
\newcommand{\fml}[2]{\cbk{#1}_{#2}} % 族の基本系
\newcommand{\hyphen}{\hbox{-}} % 記号定義の補助
\newcommand{\idone}{1} % 1で表した恒等写像
\newcommand{\inv}[1]{#1^{-1}} % 逆像・逆写像, \invもよく使うため独立させる
\newcommand{\laplacian}{\Delta} % ラプラシアン
\newcommand{\napiernum}{\mathsf{e}} % ネイピア数, 自然対数の底, 素電荷など他にもeが出てきて紛らわしい場合に利用
\newcommand{\onehalf}{\frac{1}{2}} % 1/2
\newcommand{\oneoverfour}{\frac{1}{4}} % 1/4
\newcommand{\opideal}{\operatorname{Ideal}} % イデアル
\newcommand{\opimag}{\operatorname{Im}} % 複素数の虚部
\newcommand{\opod}[1]{\frac{d}{d #1}} % 一階の常微分作用素
\newcommand{\oppd}[1]{\frac{\partial}{\partial #1}} % 偏微分作用素
\newcommand{\opreal}{\operatorname{Re}} % 複素数の実部
\newcommand{\setSymbolDownLeft}[2]{{\vphantom{#2}}_{#1}{#2}} % 左下に添字をつけるvphantomのエイリアス
\newcommand{\setSymbolUpLeft}[2]{{\vphantom{#2}}^{#1}{#2}} % 左上に添字をつけるvphantomのエイリアス
\DeclareMathOperator{\eqalgisom}{\cong} % 代数的な同型
\DeclareMathOperator{\Rep}{Rep} % 表現の空間や圏
\NewDocumentCommand{\agvariety}{O{\mathcal}}{#1} % 代数多様体, 特にイデアルが生成する代数多様体
\NewDocumentCommand{\cmdrel}{O{\omega}}{#1} % 分散関係
\NewDocumentCommand{\dfsp}{O{A}}{#1} % 微分形式の空間
\NewDocumentCommand{\eqcpointed}{O{\eqcsq} m}{#1{#2}_{\ast}} % 基点つき同値類
\NewDocumentCommand{\fnheaviside}{O{H}}{#1} % ヘビサイド関数
\NewDocumentCommand{\fthol}{O{\mathcal{O}}}{#1} % 正則関数の空間用
\NewDocumentCommand{\ftmero}{O{\mathcal{M}}}{#1} % 有理型関数の空間用
\NewDocumentCommand{\grcentralizer}{O{Z}}{#1} % 中心化群, expedition0005395
\NewDocumentCommand{\grmetform}{O{2} m m}{\grmet[#1] \! \rbkt{#2}{#3}} % (一般)相対性理論, 計量の二次形式
\NewDocumentCommand{\grmet}{O{2}}{\setSymbolDownLeft{#1}{g}} % (一般)相対性理論, 計量の文字
\NewDocumentCommand{\grnormalizer}{O{N}}{#1} % 正規化群, expedition0005395
\NewDocumentCommand{\gropasym}{O{A}}{#1} % 反対称化作用素, expedition0001365
\NewDocumentCommand{\gropsym}{O{S}}{#1} % 対称化作用素, expedition0001365
\NewDocumentCommand{\grpermorderedpair}{O{\mathcal{P}}}{#1} % 偶数個の要素の異なる順序対への分解の全体, expedition0009453
\NewDocumentCommand{\grsym}{O{\mathfrak{S}} m}{#1_{#2}} % 対称群, expedition0001319
\NewDocumentCommand{\gtbase}{O{\mathcal}}{#1} % 位相の基底, 開集合の基底, expedition0000259
\NewDocumentCommand{\gtfilter}{O{\mathcal}}{#1} % フィルター, expedition0000349
\NewDocumentCommand{\gtfmlclosed}{O{\mathcal}}{#1} % 閉集合系だけではなく, ある集合上の一般の閉集合族を表す, expedition0000227
\NewDocumentCommand{\gtfmlopen}{O{\mathcal}}{#1} % 「位相」の意味での開集合系だけではなく, ある集合上の一般の開集合族を表す, expedition0000227
\NewDocumentCommand{\gtopenball}{O{U}}{#1} % 開球
\NewDocumentCommand{\gtopencover}{O{\mathcal}}{#1} % 開被覆
\NewDocumentCommand{\gtopennbh}{O{\mathcal}}{#1} % 開近傍系
\NewDocumentCommand{\gtpreopencover}{O{\mathcal}}{#1} % 内部を取ると開被覆になる集合の族, expedition0009107
\NewDocumentCommand{\gtsubbase}{O{\mathcal}}{#1} % 位相の準基, expedition0005835
\NewDocumentCommand{\gtvicinity}{O{\mathcal}}{#1} % 一様構造での近縁
\NewDocumentCommand{\lasp}{O{\mathcal}}{#1} % 線型空間
\NewDocumentCommand{\latprightrbk}{O{\top} m}{\rbk{#2}^{#1}} % 転置, expedition0009877
\NewDocumentCommand{\latpright}{O{\top} m}{#2^{#1}} % 転置, expedition0009877
\NewDocumentCommand{\latp}{O{t} m}{\setSymbolUpLeft{#1}{#2}} % 転置, expedition0009877
\NewDocumentCommand{\lpdistribution}{O{\mu} m}{#2_{\ast,#1}} % 分布関数, expedition0010310
\NewDocumentCommand{\lpmollifier}{O{\rho}}{#1} % フリードリクスの軟化子
\NewDocumentCommand{\lpofpositive}{O{\chi}}{#1} % 正型関数
\NewDocumentCommand{\manliederiv}{O{L}}{#1} % リー微分
\NewDocumentCommand{\mansmoothnbh}{O{\mathcal}}{#1} % 多様体上の滑らかな開近傍の族
\NewDocumentCommand{\mblfmldsysgenerated}{O{d} m}{\fun{#1}{#2}} % 生成された$d$-系
\NewDocumentCommand{\mblfmlgenerated}{O{\sigma} m}{\fun{#1}{#2}} % 生成された加法族
\NewDocumentCommand{\oacorrfn}{O{\Gamma}}{#1} % 汎関数積分表示とも関わる相関関数, expedition0011645
\NewDocumentCommand{\oagnsvector}{O{\Omega}}{#1} % GNS表現のベクトル状態
\NewDocumentCommand{\oaideal}{O{\mathcal}}{#1} % 作用素環のイデアル：環や代数のイデアルとは分けておく
\NewDocumentCommand{\oanumberoperator}{O{A}}{#1} % レゾルベント環での個数作用素
\NewDocumentCommand{\oaposcone}{O{\mathcal{P}}}{#1} % 冨田-竹崎理論での正錐
\NewDocumentCommand{\oapressure}{O{P}}{#1} % トレースによる通常の圧力は大文字でP: expedition0009842, トレース状態による修正圧力は小文字でp: expedition0009797
\NewDocumentCommand{\oarepn}{O{\pi}}{#1} % 作用素環の表現
\NewDocumentCommand{\oaspnormalstate}{O{N}}{#1} % 作用素環上の正規状態の空間
\NewDocumentCommand{\oasppurestate}{O{P}}{#1} % 作用素環上の純粋状態の空間
\NewDocumentCommand{\oaspstate}{O{E}}{#1} % 作用素環の状態がなす空間
\NewDocumentCommand{\oastatevector}{O{\Omega}}{#1} % GNS表現に付随する状態ベクトル
\NewDocumentCommand{\oastate}{O{\omega}}{#1} % 作用素環の状態
\NewDocumentCommand{\opdilation}{O{\delta}}{#1} % スケーリング(伸張)の作用素, expedition0010830
\NewDocumentCommand{\opdmat}{O{\rho}}{#1} % 密度行列(密度作用素)
\NewDocumentCommand{\opfockan}{O{a}}{#1} % 場の量子論での消滅作用素
\NewDocumentCommand{\opfockcran}{O{a}}{#1^{\#}} % 場の量子論での生成・消滅作用素をまとめて表す記法
\NewDocumentCommand{\opfockcrdagger}{O{a}}{#1^{\dagger}} % 場の量子論での生成作用素のバージョン
\NewDocumentCommand{\opfockcr}{O{a}}{#1^{\ast}} % 場の量子論での生成作用素
\NewDocumentCommand{\opfocknumber}{O{N}}{#1} % 個数作用素
\NewDocumentCommand{\opfocksegalconj}{O{\pi}}{#1} % 場の量子論でのシーガルの場の作用素, expedition0009962
\NewDocumentCommand{\opfocksegal}{O{\phi}}{#1} % 場の量子論でのシーガルの場の作用素
\NewDocumentCommand{\opspecmeas}{O{E}}{#1} % スペクトル測度の標準的な記号
\NewDocumentCommand{\opspec}{O{} m}{\fun{\sigma_{#1}}{#2}} % 作用素のスペクトル, expedition0001551
\NewDocumentCommand{\optransl}{O{\tau}}{#1} % 平行移動の作用素, expedition0010829
\NewDocumentCommand{\physaction}{O{\mathcal{A}}}{#1} % 物理の作用
\NewDocumentCommand{\physcharge}{O{e}}{\mathrm{#1}} % 電荷, 素電荷などの物理量を表す. e以外にもqに利用する
\NewDocumentCommand{\physcplconst}{O{\mathsf{g}}}{#1} % 結合定数, coupling constant
\NewDocumentCommand{\physelectrostaticcapasity}{O{\mathrm{Cap}}}{#1} % 静電容量, expedition0010917
\NewDocumentCommand{\physenergy}{O{E}}{#1} % エネルギー
\NewDocumentCommand{\physgse}{O{E}}{#1_{0}} % 基底エネルギー, ground state energy
\NewDocumentCommand{\physham}{O{H}}{#1} % ハミルトニアンの雛形, たいていは下つき添字で水素原子, 自由電磁場などを表す
\NewDocumentCommand{\physlagdensity}{O{\mathcal{L}}}{#1} % ラグランジアン密度の雛形
\NewDocumentCommand{\physlag}{O{L}}{#1} % ラグランジアンの雛形
\NewDocumentCommand{\physliouvilean}{O{L}}{#1} % リウビリアンの雛形
\NewDocumentCommand{\physmass}{O{m}}{#1} % 物質の質量
\NewDocumentCommand{\prbbrownmv}{O{B}}{#1} % ブラウン運動
\NewDocumentCommand{\prbcharfun}{O{\chi}}{#1} % 特性関数
\NewDocumentCommand{\prbdist}{O{\mathcal{P}}}{#1} % 確率分布
\NewDocumentCommand{\prbgaussianmeasure}{O{\msrcal{N}}}{#1} % ガウス測度
\NewDocumentCommand{\prbnormaldist}{O{N}}{#1} % 正規分布
\NewDocumentCommand{\prbpoissonprocess}{O{N}}{#1} % ポアソン過程
\NewDocumentCommand{\prbprocess}{O{X}}{#1} % 確率過程
\NewDocumentCommand{\prbqspace}{O{\mathcal{Q}}}{#1} % $Q$-表示の土台の集合を表す
\NewDocumentCommand{\prbspsample}{O{\Omega}}{#1} % 確率空間に対する標本空間, expedition0010114
\NewDocumentCommand{\psh}{O{\mathfrak}}{#1} % 前層
\NewDocumentCommand{\qtquantumchannel}{O{\mathcal{L}}}{#1} % 量子通信路, expedition0005561, expedition0006844
\NewDocumentCommand{\repn}{O{\pi}}{#1} % 表現
\NewDocumentCommand{\schattencls}{O{\mathbb{K}}}{#1} % コンパクト作用素またはシャッテンクラス
\NewDocumentCommand{\setfmlcylinder}{O{\mathcal{C}}}{#1} % 筒集合族, expedition0000313, expedition0004973
\NewDocumentCommand{\setfml}{O{\mathcal}}{#1} % 集合族
\NewDocumentCommand{\setindex}{O{\mathcal} m}{#1{#2}} % 添字集合
\NewDocumentCommand{\setlattice}{O{\Gamma}}{#1} % 格子, expedition0010278
\NewDocumentCommand{\setspecial}{O{\mathcal} m}{#1{#2}} % 特別な集合を表すための記号, 一点集合・二点集合などを太字で表すといった用途
\NewDocumentCommand{\shdiffform}{O{\sheaf{A}}}{#1} % 微分形式がなす層
\NewDocumentCommand{\sheaf}{O{\mathfrak}}{#1} % 層の記号
\NewDocumentCommand{\smchemicalpotential}{O{\mu}}{#1} % 化学ポテンシャル
\NewDocumentCommand{\smenergydensity}{O{\varrho}}{#1} % エネルギー密度, expedition0011356
\NewDocumentCommand{\smfluctuationwithdmat}{O{\beta} m}{\smuncertaintywithdmat[#1]{#2}^2} % 密度行列に付随する正規状態でのゆらぎ, expedition0011059
\NewDocumentCommand{\sminvtemperature}{O{\beta}}{#1} % 逆温度
\NewDocumentCommand{\smlocaldensityoperator}{O{\rho}}{#1} % 大正準状態に対する局所密度作用素, expedition0011294
\NewDocumentCommand{\smmicrocanonicalstate}{O{\beta} m}{\physmean{#2}_{#1}} % 密度行列に付随する正規状態, expedition0011059
\NewDocumentCommand{\smnumberdensity}{O{\rho}}{#1} % 粒子数密度, expedition0011351
\NewDocumentCommand{\smooth}{O{\mathcal{E}}}{#1} % 滑らかな関数の空間, 特に微分形式関連でも利用, expedition0010649
\NewDocumentCommand{\smparticlenumber}{O{N}}{#1} % 粒子粒子数, expedition0011351
\NewDocumentCommand{\smpressure}{O{p}}{#1} % 統計力学での圧力, expedition0011341
\NewDocumentCommand{\smspecificfreeenergy}{O{\bar{f}}}{#1} % (ヘルムホルツの)比自由エネルギー
\NewDocumentCommand{\smthermalvac}{O{\beta}}{\Omega_{#1}} % 熱的真空, expedition0011060
\NewDocumentCommand{\smuncertaintywithdmat}{O{\beta} m}{\rbk{\triangle #2}_{#1}} % 密度行列に付随する正規状態での不確かさ, expedition0011059
\NewDocumentCommand{\sphilb}{O{\mathcal}}{#1} % ヒルベルト空間
\NewDocumentCommand{\splowerhalf}{O{\mathbb{H}}}{#1_{\txtneg}} % 開下半空間
\NewDocumentCommand{\spupperhalf}{O{\mathbb{H}}}{#1_{\txtnonneg}} % 開上半空間
\NewDocumentCommand{\topmetric}{O{d}}{#1} % 位相空間上の距離関数
\NewDocumentCommand{\vaoutnormal}{O{\widehat}}{#1} % 外向き外法線
\newcommand{\category}[1]{\mathop{\mathsf{#1}}} % 圏を生成する
\newcommand{\catpresheaf}[1]{\category{PSh}} % 前層がなす圏
\newcommand{\conti}{C} % 連続または滑らかな関数の空間
\newcommand{\faadjpresharp}[1]{#1_{\#}} % 行列・作用素に対する前共役(前双対)
\newcommand{\faadj}[1]{#1^{\ast}} % 行列・作用素に対する共役・随伴. 線型代数でも内積の必要性からこれに統合. 双対空間にはdualをあてる
\newcommand{\faftr}[1]{\widehat{#1}} % フーリエ変換
\newcommand{\fldcmp}{\fld{C}} % 複素数体
\newcommand{\fldmultiplicativegroup}[1]{#1^{\times}} % 体の乗法群
\newcommand{\fldreal}{\fld{R}} % 実数体
\newcommand{\fld}[1]{\mathbb{#1}} % 実数体・複素数体・四元数体, またはこれらをまとめて表す体用のコマンド
\newcommand{\fndef}[1]{\boldsymbol{1}_{#1}} % 定義関数
\newcommand{\fnexp}[1]{\fun{\exp}{#1}} % 指数関数
\newcommand{\fnrestr}[2]{\left. #1 \right|_{#2}} % 写像の制限
\newcommand{\greuctr}[2]{\fldreal_{#1}^{#2}} % ユークリッド空間の空間並進群
\newcommand{\gtclos}[1]{\overline{#1}} % 一般位相の閉包
\newcommand{\lp}{L} % ルベーグ空間, L_{\txtloc}^p(X), L^p(X)
\newcommand{\mansphere}{\mathbb{S}} % 球面
\newcommand{\mblfmlborel}{\mblfml{B}} % ボレル集合族
\newcommand{\mblfml}[1]{\mathcal{#1}} % 可測集合族に関わる記号の基礎
\newcommand{\msrcal}[1]{\mathcal{#1}} % mathcalで表す測度
\newcommand{\msrlaw}{\mathrm{Law}} % 法則
\newcommand{\msrprb}{\mathrm{Pr}} % たまに役立つであろう, 立体のPrで表す確率測度
\newcommand{\msr}[1]{#1} % 測度の識別用
\newcommand{\nfoldvar}[2]{\underline{#1}_{#2}} % #2個の変数をまとめて下線つきで表す
\newcommand{\oacenter}{\mathcal{Z}} % 作用素環の中心
\newcommand{\oacstar}{C^{\ast}} % $C^{\ast}$-環
\newcommand{\oaresolventalgebra}{\oa{R}} % レゾルベント環
\newcommand{\oaresolvent}{R} % レゾルベント環の要素としてのレゾルベント
\newcommand{\oastaralgebra}{\operatorname{\ast \hyphen \mathrm{alg}}} % 指定した集合が生成する$\ast$-環
\newcommand{\oa}[1]{\mathcal{#1}} % 作用素環を表す記号用
\newcommand{\opdmsr}[1]{\mathop{d #1}} % 積分で現れる測度用のコマンド, dxのような単純な対象も含む
\newcommand{\opfocksegalradiation}{A} % 輻射場に対するシーガルの場の作用素, expedition0010034
\newcommand{\opfocksndqntdiff}{d \Gamma} % 第一種の第二量子化作用素, 微分版
\newcommand{\opfocksndqnt}{\Gamma} % 第二種の第二量子化作用素
\newcommand{\opfockweyl}{W} % フォック空間上のワイル作用素
\newcommand{\opformdomain}{\mathop{Q}} % 準双線型形式の定義域, expedition0009907
\newcommand{\opform}[1]{\mathsf{#1}} % 二次形式
\newcommand{\opspbddlin}[1]{\fun{\mathbb{B}}{#1}} % 有界線型作用素がなす空間
\newcommand{\opspecint}[1]{\mathcal{E}} % スペクトル積分, expedition0001621
\newcommand{\physmean}[1]{\dbk{#1}} % 物理, 特に統計力学によく出てくる平均の記号
\newcommand{\physmfdensity}{B} % 磁束密度, magnetic flux density
\newcommand{\physoptimeorder}{\operatorname{T}} % 時間順序作用素
\newcommand{\prbdbkquadvar}[1]{\dbk{#1}} % 山括弧の二次変分
\newcommand{\prbexp}{\mathbb{E}} % 期待値, 条件つき期待値にも応用
\newcommand{\qmpaulispin}{\sigma} % パウリ行列
\newcommand{\ringratint}{\mathbb{Z}} % 有理整数環
\newcommand{\seq}[2]{\if\relax\detokenize{#1}\relax \rbk{#1} \else \rbk{#1}_{#2} \fi} % 点列の基本系
\newcommand{\setisomorphism}[1]{\operatorname{Iso}} % 同型写像の集合
\newcommand{\setone}[1]{\cbk{#1}}
\newcommand{\setquot}[2]{\bigmiddleslash{#1}{#2}} % 商集合
\newcommand{\set}[2]{\left\{#1 \, \middle| \, #2\right\}}
\newcommand{\smpartitionfunc}{Z} % 分配関数
\newcommand{\spfock}{\mathcal{F}} % フォック空間
\newcommand{\splinspan}{\operatorname{span}} % 線型包
\newcommand{\txtbec}{\mathrm{BEC}} % BEC, ボース-アインシュタイン凝縮
\newcommand{\txtbrownbridge}{\mathrm{BB}} % ブラウン橋
\newcommand{\txtbsn}{\mathrm{b}} % ボソンのb
\newcommand{\txtdiag}{\mathrm{diag}} % 対角
\newcommand{\txtdual}{\mathrm{dual}} % 双対
\newcommand{\txteuclid}{\mathrm{E}} % ユークリッド, ユークリッド化の添字
\newcommand{\txtfin}{\mathrm{fin}} % 有限のfin
\newcommand{\txtflip}{\mathrm{flip}} % フリップ
\newcommand{\txtfock}{\mathrm{F}} % フォック
\newcommand{\txtfr}{\mathrm{fr}} % 自由場などを表すfr
\newcommand{\txtinteraction}{\mathrm{I}} % 相互作用
\newcommand{\txtirsingular}{\mathrm{irs}} % 赤外特異
\newcommand{\txtloc}{\mathrm{loc}} % 局所
\newcommand{\txtneg}{\mathrm{-}} % 負の部分
\newcommand{\txtnelson}{\mathrm{N}} % ネルソンの頭文字
\newcommand{\txtnonneg}{\mathrm{+}} % 正または非負の部分
\newcommand{\txtnonzero}{\mathrm{nz}} % 非零, 非零モードを表す
\newcommand{\txtparticle}{\mathrm{p}} % 粒子のp
\newcommand{\txtpaulifierz}{\mathrm{PF}} % パウリ-フィールツ
\newcommand{\txtpauli}{\mathrm{Pauli}} % パウリ
\newcommand{\txtphys}{\mathrm{phys}} % 物理
\newcommand{\txtrad}{\mathrm{rad}} % 動径方向
\newcommand{\txtreal}{\mathrm{real}} % 実
\newcommand{\txtregular}{\mathrm{reg}} % 正則な測度などの正則
\newcommand{\txtrenormalization}{\mathrm{ren}} % くり込み
\newcommand{\txtspin}{\mathrm{spin}} % スピン
\newcommand{\txtsym}{\mathrm{s}} % 対称, テンソル積に使う
\newcommand{\vagrad}{\operatorname{grad}} % 勾配作用素
\newcommand{\vainnprod}{\cdot} % ベクトル解析枠, 高校以来の内積の記号
\newcommand{\varot}{\operatorname{rot}} % 回転, expedition0002223